\documentclass[a4paper,10.5pt]{article}

\usepackage{amsmath,amsfonts,amssymb}
\usepackage{graphicx}
\usepackage[font={footnotesize}]{caption}
\usepackage{algorithmic}
\usepackage{algorithm}

\def\beq{\begin{equation}}
\def\eeq{\end{equation}}
\def\beqn{\begin{eqnarray}}
\def\eeqn{\end{eqnarray}}

\usepackage{fancyhdr}
\usepackage{bbold}

\usepackage{hyperref}
\usepackage{float}
\usepackage{authblk}
\usepackage{arydshln}
\usepackage{tikz}
\usetikzlibrary{shapes,arrows}
\usepackage{multirow}

\usepackage[utf8x]{inputenc}

\usepackage{subcaption}

\usepackage{sectsty}

\usepackage{appendix}

\usepackage{enumitem}   
   
\begin{document}

\title{Many-body effects in the excitations and dynamics of trapped Bose-Einstein condensates}

\author[1,2]{Ofir E. Alon \thanks{ofir@research.haifa.ac.il} }
\author[3]{Raphael Beinke \thanks{raphael.beinke@pci.uni-heidelberg.de} }
\author[3]{Lorenz S. Cederbaum \thanks{lorenz.cederbaum@pci.uni-heidelberg.de}}

\affil[1]{Department of Mathematics, University of Haifa, Haifa 3498838, Israel}
\affil[2]{Haifa Research Center for Theoretical Physics and Astrophysics, University of Haifa, Haifa 3498838, Israel}
\affil[3]{Theoretische Chemie, Physikalisch-Chemisches Institut, Universit\"at Heidelberg, Im Neuenheimer Feld 229, D-69120 Heidelberg, Germany}


\maketitle

\newpage
\begin{abstract}
This review explores the dynamics and the low-energy excitation spectra of Bose-Einstein condensates (BECs) of interacting bosons
in external potential traps putting particular emphasis on the emerging many-body effects beyond mean-field descriptions. To do so, methods have to be used that, in principle, can provide numerically exact results for both the dynamics and the excitation spectra in a systematic manner. Numerically
exact results for the dynamics are presented employing the well-established multicongurational time-dependent Hartree
for bosons (MCTDHB) method. The respective excitation spectra are calculated utilizing the more recently introduced
linear-response theory atop it (LR-MCTDHB). The latter theory  gives rise to an, in general, non-hermitian
eigenvalue problem. The theory and its newly developed implementation are described in detail and benchmarked
towards the exactly-solvable harmonic-interaction model. Several applications to BECs in one- and two-dimensional potential traps are discussed. With
respect to dynamics, it is shown that both the out-of-equilibrium tunneling dynamics and the dynamics of
trapped vortices are of many-body nature. Furthermore, many-body effects in the excitation spectra are
presented for BECs in different trap geometries. It is demonstrated that even for essentially-condensed systems,
the spectrum of the lowest-in-energy excitations computed at the many-body level can differ substantially
from the standard mean-field description. In general, it is shown that bosons carrying angular momentum are
more sensitive to many-body effects than bosons without. These effects are present in both the dynamics and the excitation spectrum.
\end{abstract}


\newpage
\tableofcontents
\appendixtitleon
\appendixtitletocon



\newpage
\section{Introduction}
In 1924, Albert Einstein discovered that the spectral energy distribution in certain systems of massive particles is different from the Maxwell-Boltzmann statistics. He was inspired by a work of Satayendra N. Bose who derived the thermal energy distribution of black-body radiation solely based on assumptions from quantum theory \cite{Intro_Bose}, which is slightly different to the previous semi-classical approach of Max Planck. Einstein realized the far-reaching consequences of this, namely that a composite system of massive particles at sufficiently low temperatures could behave as a matter wave due to all particles sharing the same single-particle state. Thus, he applied Bose's idea to a single-component ideal gas \cite{Intro_Einstein1,Intro_Einstein2}. This paved the way towards what is known as Bose-Einstein condensation. Consequently, all particles that obey the statistical distribution found by Bose and Einstein are termed bosons.

It took approximately 70 years until the first gaseous Bose-Einstein condensate (BEC) was realized in an experiment. The group of Eric A. Cornell and Carl E. Wieman produced a repulsive BEC of rubidium atoms \cite{Phantom_Anderson}, shortly followed by the group of Wolfgang Ketterle using repulsive sodium atoms in a magneto-optical trap \cite{Phantom_Davis}. The latter three scientists obtained the Nobel prize in physics for this achievement in 2001. The first attractive BEC was realized by the group of Randall G. Hulet using lithium atoms \cite{Phantom_Hulet}. In more recent experiments, condensation was observed in a large variety of systems, e.g., for exciton-polariton systems \cite{Intro_exc_pol1,Intro_exc_pol2}, magnons at room temperature \cite{Intro_magnons}, or, rather astonishingly because of the lack of a rest mass and a vanishing chemical potential, even for photons in a cavity \cite{Intro_photon_BEC1}.

Over the past two decades, the experimental control of BECs has grown remarkably. This includes both the interaction of the constituent particles by employing Feshbach resonances \cite{Intro_Feshbach1,Intro_Feshbach2,Intro_Feshbach3} as well as the trap geometry and dimensionality \cite{Intro_dim1}. This qualifies BECs for testing the fundamental principles of quantum mechanics, or very recently even to simulate certain processes in astrophysics \cite{Intro_Steinhauer1,Intro_Steinhauer2,Intro_Steinhauer3}.

Of particular interest is the understanding of the dynamics as well as the spectrum of excited states of trapped BECs made of interacting bosons. Theoretical descriptions for these were carried out many years before they were finally realized experimentally. The most prominent approach to obtain the ground state as well as the dynamical motion of condensates is the Gross-Pitaevskii (GP) equation \cite{Gross,Pitaevskii,Book_Pitaevskii,Book_Pethick}. In this theory it is assumed that actually all bosons in the system occupy the same single-particle state, making it a mean-field theory. The GP equation has been successfully applied in many cases \cite{Intro_GP_review2} and can be seen as the adequate starting point to solve many physical problems of BECs. Similarly, the linear-response (LR) theory atop the GP equation, resulting in the famous Bogoliubov-de Gennes (BdG) equations \cite{Bogoliubov,deGennes}, represents the commonly utilized approach to describe excitations of trapped ultracold bosons, e.g., for dark-bright soliton pairs \cite{Intro_Schmelcher}, in rotating BECs with dipole-dipole interaction \cite{Intro_BdG_appl_2}, or in a 2D hexagonal lattice with superimposed harmonic confinement \cite{Intro_BdG_appl_3}, to name only a few recent applications. Furthermore, the BdG equations were extensively used for the description of Cooper pairs in superconductors \cite{Intro_BdG_supercond} or for the understanding of the BCS-BEC crossover connecting superconductivity and superfluidity in correlated fermionic systems \cite{Intro_BdG_BCS_BEC}.  

A crucial limitation of the GP mean-field theory is the fact that it ignores correlations between the bosons, and thus does not account for condensate depletion, i.e., the fraction of particles outside the condensed mode due to correlations and interactions, even at zero temperature. In a highly-cited review article from 1999, the condensate depletion is found to be ``very small (less than $1\%$) in the presently available experimental conditions'' \cite{Intro_GP_review}, and thus the usage of mean-field methods like the BdG approach has been considered justified. However, the many-body methods developed in the following years were capable to demonstrate the remarkable impact of depletion and even fragmentation of a BEC onto its dynamics and its excitation spectrum. One of these many-body approaches is the multiconfigurational time-dependent Hartree method for bosons [MCTDHB, or MCTDHB($M$) with $M$ being the number of utilized single-particle states, called orbitals], which has been introduced \cite{MCTDHB_unified_view,MCTDHB_main}, benchmarked \cite{MCTDHB_benchmark}, and applied successfully to various systems in recent years. Being derived from a variational principle, its results can be improved systematically by increasing the number of orbitals $M$. Additionally, its LR theory, termed LR-MCTDHB($M$), has been introduced recently \cite{LR-MCTDHB1,LR-MCTDHB2}. The latter represents a full many-body description of excitations in trapped BECs.

This review has two main objectives. At first, the newly developed numerical implementation of LR-MCTDHB, capable of treating large systems also in $D>1$ spatial dimensions which has not been realized before, is introduced and benchmarked against an exactly-solvable model. Secondly, various applications of both MCTDHB and LR-MCTDHB to trapped BECs are discussed in which the depletion has a strong impact on the dynamics and on the spectrum of excited states. With regard to dynamics, it is shown that fragmentation can develop over time although a system is initially almost entirely condensed. With regard to excitation spectra, it is demonstrated that substantial many-body effects appear even if the ground-state depletion is of the order of the above mentioned $1\%$ or lower. Even for the case of marginal depletion where one might expect mean-field theory to accurately describe the physics, it is instructive to compare the results obtained from the many-body approaches described above to the ones obtained from the GP and BdG theories. It is important to note that for only a single orbital, $M=1$, MCTDHB reduces exactly to the GP equation. Correspondingly, LR-MCTDHB($1$) yields the BdG equations. Hence, the commonly used mean-field approaches are included in the utilized many-body theories as their simplest limiting cases.

To address these goals, the review is structured as follows. First, in Section \ref{CH_concepts}, the main theoretical concepts of many-boson physics are presented. This includes the many-body Schr{\"o}dinger equation, the framework of second quantization which is particularly useful for the many-body theory described in this work, the introduction of the Dirac-Frenkel variational principle from which the central equations of motion can be derived, as well as the definition of Bose-Einstein condensation. Second, in Section \ref{CH_MB_theory}, a detailed derivation of MCTDHB is given. The essential ingredients like the ansatz for the many-boson wave function are discussed. Then, its LR theory is derived, which represents the central theory utilized in this work to calculate low-energy excitations of trapped bosonic systems on the many-body level. Third, in Section \ref{Sec_Num}, the newly developed numerical implementation of LR-MCTDHB is described in detail. Therefore, general technical challenges and their solutions are explained. Furthermore, a detailed description of the numerical algorithms utilized is presented. Subsequently, in Section \ref{Ch_Benchmark}, the used implementations of MCTDHB and LR-MCTDHB are benchmarked against the exactly-solvable harmonic-interaction model, whose analytic solutions of the energies of the ground and excited states are presented beforehand. Applications of both implementations are then shown and their physics discussed in Section \ref{Ch_applications}. Concerning the dynamics, the focus is twofold. First, applications dealing with the phenomenon of tunneling of BECs between wells inside potential traps in one and two spatial dimensions are discussed, demonstrating the many-body nature of quantum tunneling of ultracold BECs. Second, several cases of dynamical fragmentation of \textit{initially coherent} bosonic systems are presented. For instance, the tunneling of a BEC to open space is analyzed. Furthermore, the connection between angular momentum, especially of quantized vortices, and the development of fragmentation is investigated. Concerning applications of LR-MCTDHB, the many-body effects in the low-energy spectra of BECs in one-dimensional multi-well traps and lattices are shown. Moreover, as a 2D application, the impact of angular momentum on the lowest-in-energy excitations of a rotating BEC is examined. Essential physical results and implications are provided. Finally, in Section \ref{Ch_summary}, a summary and an outlook of possible constituent future research as well as of further enrichment of the implementation of LR-MCTDHB is provided. 

Additionally, the review contains four appendices which broaden its scope and provide additional insight. In Appendix \ref{CH_MF_Theory}, the commonly-used mean-field theories, the GP theory and the constituent LR theory leading to the BdG equations, are described. A further mean-field approach for both the statics and dynamics of trapped BECs that has been developed prior to the many-body theories of Section \ref{CH_MB_theory} is presented, called the time-dependent multi-orbital mean field (TDMF) and the linear-response best-mean-field (LR-BMF). The latter can be seen as intermediate theories between the standard single-orbital mean-field approaches and the full many-body treatment and have their own merits. Appendix \ref{Appendix_BlockDiagonal} demonstrates that LR-MCTDHB can be written in block-diagonal form, which, in some cases, can be beneficial. Advantages and possible weaknesses are discussed. In Appendix \ref{Appendix_further_benchmarks}, additional results of the LR-MCTDHB benchmark are provided. Finally, in Appendix \ref{Appendix_variance}, the many-body variance, a quantity that turns out to be very sensitive to many-body effects, is introduced and derived for the example of the center-of-mass (c.m.) position of a BEC. Its relevance is discussed in several examples.


\newpage
\section{General concepts of many-boson physics}\label{CH_concepts}

In this section, the general concepts of describing trapped ultracold bosons are introduced. This includes the representation of the many-body Hamiltonian, the bosonic field as well as properties of the wave function for a system of identical bosons. Moreover, the definition of Bose-Einstein condensation and depletion which is used in the subsequent sections is presented.

\subsection{Many-body Hamiltonian and Schr{\"o}dinger equation}\label{Sec_MBSE}

The time-dependent Schr{\"o}dinger equation for a generic system of $N$ identical spinless bosons in $D\leq 3$ dimensions is given by 
\begin{equation}\label{MBSE}
	\hat{H}\Psi(\mathbf{r}_1,\mathbf{r}_2,...,\mathbf{r}_N;t)=i\frac{\partial}{\partial t}\Psi(\mathbf{r}_1,\mathbf{r}_2,...,\mathbf{r}_N;t)
\end{equation}
with the full many-body wave function $\Psi(\mathbf{r}_1,\mathbf{r}_2,...,\mathbf{r}_N;t)$ depending on the spatial coordinates $\{\mathbf{r}_i\}$ of all $N$ bosons and the time $t$. 

The Hamiltonian $\hat{H}$ reads
\begin{equation}\label{MB_HAM1}
	\hat{H}=\sum_{i=1}^N \hat{h}(\mathbf{r}_i)+\lambda_0\sum_{j>i}^N \hat{W}(\mathbf{r}_i,\mathbf{r}_j)
\end{equation} 
with the single-particle term 
\begin{align}\label{1body_HAM}
	\hat{h}(\mathbf{r}_i)&=\hat{T}(\mathbf{r}_i)+V(\mathbf{r}_i) \\
	\hat{T}(\mathbf{r}_i)&=-\frac{1}{2}\Delta_i=-\frac{1}{2}\frac{\partial^2}{\partial\mathbf{r}_i^{\,2}}
\end{align}
comprising the contributions of the kinetic energy $\hat{T}$ and the external potential $V$ which is assumed to be real. For simplicity, the terms are given in dimensionless units with $\hbar=m=1$. The two-body operator $\hat{W}(\mathbf{r_i},\mathbf{r_j})$ represents the interaction potential of strength $\lambda_0$. In this work, interaction potentials which solely depend on the distance $r_{ij}=|\mathbf{r_i}-\mathbf{r_j}|$ between two bosons are considered. In general, both $V$ and $\hat{W}$ can be time-dependent.

\subsection{Second quantization}
In this section, the formalism of second quantization is briefly explained, including the Fock state representation of an $N$-boson state, the symmetry properties of the wave function, and the Hamiltonian in second quantization.

\subsubsection{Fock states and commutation relations}\label{Sec_Fock}
As a consequence of the spin-statistics theorem \cite{Fierz,Pauli}, the wave function $\Psi(\mathbf{r}_1,...,\mathbf{r}_N)$ of $N$ identical bosons is a fully symmetrized $N$-particle state, i.e., it is symmetric under the permutation of two bosons with coordinates $\mathbf{r}_i$ and $\mathbf{r}_j$, i.e., 
\begin{equation}\label{wave_echchage_symm}
	\Psi(...,\mathbf{r}_i,...,\mathbf{r}_j,...)=+\Psi(...,\mathbf{r}_j,...,\mathbf{r}_i,...).
\end{equation}
In second quantization, this is achieved by using a Fock state (or number state) basis of the $N$-particle Hilbert space $\mathcal{H}^{(N)}$ for the wave function $\Psi$. This basis incorporates the previously mentioned symmetrization requirements as well as the commutation relations of identical bosons shown below.

For a complete set of orthonormalized single-particle states, also referred to as orbitals, $\{\phi_i:i=1,...,\infty\}$, a general Fock state reads
\begin{equation}\label{permanents}
	|n_1,n_2,...\rangle=\prod_i\left[\frac{1}{\sqrt{n_i!}}\left(\hat{b}_i^\dagger\right)^{n_i}\right]|\text{vac}\rangle
\end{equation}
where the integers $\{n_i\}$ denote the occupation numbers of the individual orbitals and $|\text{vac}\rangle$ the particle vacuum. In Eq. (\ref{permanents}), the creation and annihilation operators $\hat{b}_i^\dagger$ and $\hat{b}_i$, which create or annihilate a particle in the state $\phi_i$, were introduced. The action of these operators on a Fock state is given by
\begin{align*}
	\hat{b}_i\,|n_1,n_2,...,n_i,...\rangle=&\sqrt{n_i}\,|n_1,n_2,...,n_i-1,...\rangle \\
	\hat{b}_i^\dagger\,|n_1,n_2,...,n_i,...\rangle=&\sqrt{n_i+1}\,|n_1,n_2,...,n_i+1,...\rangle
\end{align*}
and thus the number operator for any single-particle state $\phi_i$ reads 
\begin{equation}\label{numberOp_phi}
	\hat{n}_i=\hat{b}_i^\dagger\hat{b}_i.
\end{equation} 
For a system with a fixed amount of bosons $N$, the occupation numbers of the orbitals sum up to the total number of particles $N$, i.e.,
\begin{equation}
	N=\sum_i\,n_i.
\end{equation} 
The creation and annihilation operators fulfill the conventional bosonic commutation relations  
\begin{equation}\label{comm_b}
	\left[ \hat{b}_i,\hat{b}_k\right]=0, \quad\quad \left[ \hat{b}^\dagger_i,\hat{b}^\dagger_k\right]=0, \quad\quad \left[ \hat{b}_i,\hat{b}_k^\dagger \right]=\delta_{ik}
\end{equation}
which ensure the exchange symmetry of the Fock state basis in Eq. (\ref{permanents}) and thus of the total wave function $\Psi$ as required in Eq. (\ref{wave_echchage_symm}).

\subsubsection{Bosonic field operator}
Utilizing the creation and annihilation operators of bosons for single-particle states from the previous section, one can define operators that create or annihilate a boson at position $\mathbf{r}$ at time $t$. These operators are the so-called field operators,  
\begin{align}
	\hat{\Psi}^\dagger(\mathbf{r})=\sum_i \hat{b}^\dagger_i(t) \phi_i^\ast(\mathbf{r},t)  \label{Bose_Field_a} \\
	\hat{\Psi}(\mathbf{r})=\sum_i \hat{b}_i(t) \phi_i(\mathbf{r},t), \label{Bose_Field_b}
\end{align}
where it is taken into account that the orbitals as well as the creation and annihilation operators can in general be time-dependent. Thus, $\hat{b}^\dagger(t)$ and $\hat{b}(t)$ can be expressed in terms of the field operators, 
\begin{equation}\label{creation_OP}
		\hat{b}_i^\dagger(t)=\int \phi_i(\mathbf{r},t) \hat{\Psi}^\dagger(\mathbf{r}) d\mathbf{r}, \quad\quad \hat{b}_i(t)=\int \phi_i^\ast(\mathbf{r},t) \hat{\Psi}(\mathbf{r}) d\mathbf{r}
\end{equation}
where it is assumed that the set $\{\phi_i(\mathbf{r},t)\}$ consists of orthonormalized functions. In the following, the time argument in the above quantities is suppressed. The field operators in Eqs. (\ref{Bose_Field_a}) and (\ref{Bose_Field_b}) fulfill the bosonic commutation relations
\begin{equation}\label{comm_bose_field}
	\left[\hat{\Psi}(\mathbf{r}),\hat{\Psi}(\mathbf{r}^{\,\prime})\right]=0, \quad\quad \left[\hat{\Psi}^\dagger(\mathbf{r}),\hat{\Psi}^\dagger(\mathbf{r}^{\,\prime})\right]=0, \quad\quad \left[\hat{\Psi}(\mathbf{r}),\hat{\Psi}^\dagger(\mathbf{r}^{\,\prime})\right]=\delta\left( \mathbf{r}-\mathbf{r}^{\,\prime} \right).
\end{equation}
The density operator is defined as
\begin{equation}\label{dens_op}
	\hat{\rho}(\mathbf{r})=\hat{\Psi}^\dagger(\mathbf{r})\hat{\Psi}(\mathbf{r})
\end{equation} 
such that the operator of the total number of bosons in the system is given by
\begin{equation}
	\hat{N}=\int\,d\mathbf{r}\, \hat{\rho}(\mathbf{r}).
\end{equation}

\subsubsection{One- and two-body operators}
In second quantization, a general one-body operator $\hat{A}^{(1)}$, i.e., an operator acting on a single particle, reduces to the form
\begin{equation}\label{one-body-op}
	\hat{A}^{(1)}=\sum_{i,j}a_{ij}\,\hat{b}_i^\dagger\hat{b}_j
\end{equation}
where the matrix elements $a_{ij}$ are given by
\begin{equation}\label{one-body-matrix}
	a_{ij}=\int\phi_i^\ast(\mathbf{r})\hat{A}^{(1)}\,\phi_j(\mathbf{r})\,d\mathbf{r}
\end{equation} 
or, equivalently in bra-ket notation, $a_{ij}=\langle i|\hat{A}^{(1)} |j\rangle$. Similarly, the general form of a two-body operator, i.e., an operator acting on two particles, is given by
\begin{equation}\label{two-body-op}
	\hat{A}^{(2)}=\frac{1}{2}\sum_{i,j,k,l}a_{ijkl}\,\hat{b}_i^\dagger\hat{b}^\dagger_j\hat{b}_k\hat{b}_l
\end{equation} 
with 
\begin{equation}\label{two-body-matrix}
	a_{ijkl}=\iint\phi_i^\ast(\mathbf{r})\phi_j^\ast(\mathbf{r}^{\,\prime})\hat{A}^{(2)}\,\phi_k(\mathbf{r})\phi_l(\mathbf{r}^{\,\prime})\,d\mathbf{r}d\mathbf{r}^{\,\prime},
\end{equation} 
or $a_{ijkl}=\langle i,j|\hat{A}^{(2)} |k,l\rangle$ in short. The factor $1/2$ in Eq. (\ref{two-body-op}) accounts for double counting of the individual terms.

\subsubsection{Hamiltonian in second quantization}
The Hamiltonian in Eq. (\ref{MB_HAM1}) can now be expressed within the framework of second quantization. Using the bosonic field operators in Eqs. (\ref{Bose_Field_a}) and (\ref{Bose_Field_b}) as well as the general expressions for one- and two-body operators, Eqs. (\ref{one-body-op}) and (\ref{two-body-op}), one obtains
\begin{align}
	\hat{H}&= \sum_{i=1}^N \hat{h}(\mathbf{r}_i)+\lambda_0\sum_{j>i}^N \hat{W}(\mathbf{r}_i,\mathbf{r}_j) \Rightarrow \nonumber \\
	\hat{H}&=\int d\mathbf{r}\left( \hat{\Psi}^\dagger(\mathbf{r})\hat{h}(\mathbf{r})\hat{\Psi}(\mathbf{r})+\frac{\lambda_0}{2}\int d\mathbf{r}^{\,\prime} \hat{\Psi}^\dagger(\mathbf{r})\hat{\Psi}^\dagger(\mathbf{r}^{\,\prime}) \hat{W}(\mathbf{r},\mathbf{r}^{\,\prime}) \hat{\Psi}(\mathbf{r})\hat{\Psi}(\mathbf{r}^{\,\prime}) \right)   \nonumber  \\
	&=\sum_{i,j}h_{ij}\,\hat{b}_i^\dagger\hat{b}_j+\frac{\lambda_0}{2}\sum_{i,j,k,l}W_{ijkl}\,\hat{b}_i^\dagger\hat{b}_j^\dagger\hat{b}_k\hat{b}_l   \label{Ham_2nd}
\end{align}
with the matrix elements
\begin{align}
	h_{ij}&=\int\phi_i^\ast(\mathbf{r})\,\hat{h}(\mathbf{r})\,\phi_j(\mathbf{r})\,d\mathbf{r} \label{hij} \\
	W_{ijkl}&=\iint\,\phi_i^\ast(\mathbf{r})\phi_j^\ast(\mathbf{r}^{\,\prime}) \hat{W}(\mathbf{r},\mathbf{r}^{\,\prime}) \, \phi_k(\mathbf{r})\phi_l(\mathbf{r}^{\,\prime})\,d\mathbf{r}\,d\mathbf{r}^{\,\prime} \label{Wijkl} .
\end{align}
To keep the notation simple, the time argument of the quantities in Eqs. (\ref{Ham_2nd})-(\ref{Wijkl}) is suppressed. The above representation of $\hat{H}$ from Eq. (\ref{Ham_2nd}) will be used in the many-body formalism presented in Section \ref{CH_MB_theory}.

\subsection{Dirac-Frenkel and least action variational principles}\label{Sec_DFVP}
The working equations of all utilized many-body methods in this work can be derived from the Dirac-Frenkel variational principle (DFVP) \cite{Kramer,Dirac,Frenkel}. Like for any other variational method, it yields a condition for the dynamical evolution for any parametrized ansatz of the wave function $\Psi$. Within the DFVP, the equation that determines the dynamics is given by 
\begin{equation}\label{DFVP}
	\left\langle \delta\Psi(t) \left|\hat{H}-i\hbar\frac{\partial}{\partial\,t} \right|\Psi(t)  \right\rangle=0
\end{equation}
which basically means that any allowed variation $|\delta\Psi(t)\rangle$ is orthogonal to the residual wave function $\left(\hat{H}-i\hbar\frac{\partial}{\partial\,t}\right) |\Psi(t)\rangle$. It is important to note that under certain conditions, the DFVP is fully equivalent to two other commonly used time-dependent variational principles, which are the variational principle due to MacLachlan \cite{MacLachlan} and the least action principle. See in this context Ref. \cite{Kucar}. The least action principle is based on the assumption that the wave function minimizes the action
\begin{equation}\label{functional_action}
	S=\int dt \, \left\langle\Psi\left|\hat{H}-i\hbar\frac{\partial}{\partial t} \right|\Psi\right\rangle,
\end{equation}
meaning that its variation with respect to $\Psi$ should vanish, i.e.,
\begin{equation}\label{least_action}
	\delta S=0.
\end{equation}
Eqs. (\ref{functional_action}) and (\ref{least_action}), in combination with normalization constraints, are used to derive the equations of motion for MCTDHB. The full derivation of the latter theory is carried out in detail in Section \ref{Sec_MCTDHB}. Furthermore, the least action principle can also be employed to derive the standard mean-field equation for the ground state and dynamics of a trapped BEC, i.e., the GP equation. Its derivation is presented in Appendix \ref{Sec_GP}.

\subsection{Definition of Bose-Einstein condensation in traps}\label{Sec_Def_BE_cond}
In 1956, Penrose and Onsager developed a theoretical criterion which is utilized in the definition of Bose-Einstein condensation and is based on standard quantities from statistical quantum mechanics \cite{Penrose}. It makes use of the one-body reduced density matrix (RDM) of an $N$-boson system which can be defined via the first order correlation function
\begin{align}\label{Eq_1BRDM}
	\mathbf{\rho}^{(1)}(\mathbf{r}|\mathbf{r}^{\,\prime})&=\left\langle \Psi\left|\hat{\Psi}^\dagger(\mathbf{r}^{\,\prime})\hat{\Psi}(\mathbf{r}) \right|\Psi \right\rangle \nonumber \\ 
	&=\sum_{i,j} \,\rho_{ij} \, \phi_i^\ast(\mathbf{r}^{\,\prime})\phi_j(\mathbf{r})
\end{align}
with
\begin{equation}
	\rho_{ij} = \langle \Psi|\hat{\rho}_{ij}|\Psi \rangle	,\quad \hat{\rho}_{ij}=\hat{b}_i^\dagger \hat{b}_j
\end{equation}
where the time argument is suppressed in all quantities. More details on RDMs are given in Appendix \ref{Appendix_variance}. For $\mathbf{r}=\mathbf{r}^{\,\prime}$, Eq. (\ref{Eq_1BRDM}) denotes the one-particle density. Diagonalizing the one-body RDM yields 
\begin{equation}\label{Eq_Definition_natural_occs}
	\mathbf{\rho}^{(1)}(\mathbf{r}|\mathbf{r}^{\,\prime})=\sum_i n_i^{(1)} \, \alpha_i^\ast(\mathbf{r}^{\,\prime})\alpha_i(\mathbf{r})
\end{equation}
where the eigenvalues $\left\{n_i^{(1)}|n_1^{(1)}\geq n_2^{(1)}\geq...\right\}$ and the eigenfunctions $\{\alpha_i(\mathbf{r})\}$ are the natural occupation numbers and the natural orbitals, respectively. Whenever possible, the superscript '$(1)$' of the natural occupation numbers will be omitted throughout this work. According to Ref. \cite{Penrose}, any $N$-boson system where only the largest eigenvalue is macroscopic, i.e., $n_1=\mathcal{O}(N)$, is said to be condensed. However, if more than one eigenvalue of the one-body RDM are macroscopic, the system is said to be depleted/fragmented \cite{Noziere,Legett}. In this work, the degree of depletion (or fragmentation) is quantified by  
\begin{equation}\label{Eq_depletion}
	f=\frac{1}{N}\sum_{i>1}n_i
\end{equation}
which is the aggregated occupation of all but the first natural orbital. Another classification that uses the above relation distinguishes the terms depletion and fragmentation more clearly and was used in Ref. \cite{Sakmann_PhD}. 

In Section \ref{Ch_applications}, the dynamics and excited states of BECs in different trap geometries in one and two spatial dimensions are studied. There, the degree of depletion/fragmentation will be an essential quantity to distinguish the many-body results obtained from the methods described in the subsequent Section \ref{CH_MB_theory} from the mean-field results obtained by using the theories described in the Appendices \ref{Sec_GP} and \ref{Sec_BdG}.



\section{Many-body theory}\label{CH_MB_theory}
In this section, the many-body approach utilized in this work to compute the ground state and the dynamics of trapped BECs as well as their low-energy excitation spectra is presented. First, an introduction to MCTDHB is made in Section \ref{Sec_MCTDHB}, followed by the application of LR theory atop it, termed LR-MCTDHB, in Section \ref{Sec_LRMCTDHB}.


\subsection{Multiconfigurational time-dependent Hartree method for bosons (MCTDHB)}\label{Sec_MCTDHB}
In this section, a detailed derivation of the MCTDHB [or MCTDHB($M$)] theory which has been originally introduced in Refs. \cite{MCTDHB_unified_view,MCTDHB_main}, is presented. It originates from the closely related MCTDH method \cite{Meyer,Manthe,mixtures_book} which is a widely used technique to compute the dynamics and excited states of nuclei in molecules. Further extensions of MCTDHB to the case of type conversion of particles, to mixtures of bosons and fermions, to Hubbard-type Hamiltonians of BECs on lattices, as well as to BECs with internal degrees of freedom can be found in Refs. \cite{MCTDHB_type_conversion,MCTDHB_mixtures,MCTDHB_Hubbard,MCTDHB_internal_degrees}. Other many-body theories for the ground state and excitations of BECs were utilized as well. One example is the density matrix renormalization group (DMRG) theory \cite{Intro_DMRG1}. It is particularly designed for applications in the field of one-dimensional condensed matter physics, but was extended beyond this to quantum chemistry and quantum information. An early review can be found in \cite{Intro_DMRG2}. Furthermore, DMRG has been extended and applied to two-dimensional systems as well \cite{Intro_DMRG3}. However, this review deals with MCTDHB and its LR theory.

The ansatz for the many-body wave function to solve the many-boson Schr{\"o}dinger equation, Eq. (\ref{MBSE}), is given by
\begin{equation}\label{Eq_MCTDHB_ansatz}
	|\Psi(t)\rangle=\sum_{\mathbf{n}}C_{\mathbf{n}}(t)\,|\mathbf{n};t\rangle,
\end{equation}
which is a superposition of permanents $\{|\mathbf{n};t\rangle\}$ comprised of $M$ single-particle orbitals $\{\phi_j(\mathbf{r},t):1 \leq j \leq M\}$ and expansion coefficients $\{C_{\mathbf{n}}(t)\}$ where $\mathbf{n}=(n_1,\ldots,n_M)^t$ is a vector carrying the individual occupation numbers of the orbitals for a given permanent. The summation in Eq. (\ref{Eq_MCTDHB_ansatz}) includes all $N_\text{conf}=\binom{N+M-1}{N}$ possibilities to distribute $N$ bosons onto $M$ single-particle orbitals. It is important to note that both the permanents and the coefficients are time-dependent, which is the main difference compared to the frequently used many-body approach of the full configuration interaction (FCI) method. In the latter theory, the amount of configurations for a given number $M$ of single-particle basis functions is the same as for MCTDHB. However, these basis states are fixed in shape, i.e., they remain unchanged and only the coefficients are time-dependent. For MCTDHB, the single-particle orbitals are time-adaptive and, as shown below, the corresponding set of equations of motion (EOMs) is coupled to the set of EOMs of the coefficients. In terms of numerical convergence, utilizing a time-adaptive instead of a shape-fixed basis dramatically improves the convergence towards exact results for both the ground state and the dynamics of trapped BECs, as discussed in Section \ref{Ch_Benchmark}.

The working equations of MCTDHB are determined by the least action principle of Eqs. (\ref{functional_action}) and (\ref{least_action}), which explicitly reads
\begin{equation}\label{MCTDHB_functional1}
	0=\delta S=\delta\left( \int dt \, L(t)  \right) 
\end{equation}
with 
\begin{equation}\label{MCTDHB_L}
    L(t)=  \left\langle\Psi(t)\left|\hat{H}-i\frac{\partial}{\partial t}\right|\Psi(t)\right\rangle-\sum_{i,j}\mu_{ij}(t)[\langle\phi_i|\phi_j\rangle-\delta_{ij}]
\end{equation}
and $\hbar=1$. The time-dependent Langrange multipliers $\{\mu_{ij}(t)\}$ account for the orthonormalization of the orbitals in time. Whenever possible, the time-argument of all quantities is suppressed in the following. Plugging in the ansatz for the wave function, Eq. (\ref{Eq_MCTDHB_ansatz}), and the expression for the Hamiltonian, Eq. (\ref{Ham_2nd}), into Eq. (\ref{MCTDHB_L}) results in
\begin{align}\label{MCTDHB_functional2}
	L(t)&=\sum_{k,q}^M\rho_{kq}\left[ h_{kq}-\left( i\frac{\partial}{\partial t}\right)_{kq} \right]+\frac{1}{2}\sum_{k,s,q,l}^M\rho_{ksql}W_{ksql} \nonumber \\ 
	&-i\sum_{\mathbf{n}}C_{\mathbf{n}}^\ast (t)\frac{\partial C_{\mathbf{n}}(t)}{\partial t}-\sum_{k,q}\mu_{kq}[\langle\phi_k|\phi_q\rangle-\delta_{kq}]
\end{align}
where the elements of the one- and two-body RDMs in this case read
\begin{align}
	\rho_{kq}&=\sum_{\mathbf{n},\mathbf{n}^{\,\prime}}C_{\mathbf{n}}^\ast(t) C_{\mathbf{n}^{\,\prime}}(t) \left\langle\mathbf{n};t \left|\hat{b}_k^\dagger \hat{b}_q \right|\mathbf{n}^{\,\prime};t\right\rangle \label{1b_rdm} \\
	\rho_{ksql}&=\sum_{\mathbf{n},\mathbf{n}^{\,\prime}}C_{\mathbf{n}}^\ast (t) C_{\mathbf{n}^{\,\prime}}(t) \left\langle\mathbf{n};t \left|\hat{b}_k^\dagger \hat{b}_s^\dagger \hat{b}_q \hat{b}_l \right|\mathbf{n}^{\,\prime};t \right\rangle \label{2b_rdm}.
\end{align}
The variation of $S$ with respect to $\phi_k^\ast(\mathbf{r},t)$ yields
\begin{equation}\label{MCTDHB_orb_variation}
	\frac{\delta S}{\delta\phi_k^\ast(\mathbf{r},t)}=\sum_{q=1}^M \left[ \rho_{kq}\left( \hat{h}-i\frac{\partial}{\partial t} \right)  -\mu_{kq} +\sum_{s,l=1}^M \rho_{ksql} \hat{W}_{sl} \right] |\phi_q\rangle
\end{equation}
with 
\begin{equation}
	\hat{W}_{sl}(\mathbf{r},t)=\int \phi_s^\ast(\mathbf{r}^{\,\prime},t)\,\hat{W}(\mathbf{r},\mathbf{r}^{\,\prime})\,\phi_l(\mathbf{r}^{\,\prime},t)\,d\mathbf{r}^{\,\prime}.
\end{equation}
According to the DFVP, this variation should vanish, which determines the EOMs of the orbitals given by
\begin{equation}\label{orb_EOM1}
	\sum_{q=1}^M \left[ \rho_{kq}\hat{h}   -\mu_{kq} +\sum_{s,l=1}^M \rho_{ksql} \hat{W}_{sl} \right] |\phi_q\rangle=i\sum_{q=1}^M \rho_{kq}\frac{\partial}{\partial t}|\phi_q\rangle.
\end{equation}
Taking the scalar product with $\langle\phi_i|$ in Eq. (\ref{orb_EOM1}) results in the expression for the Lagrange multipliers,
\begin{equation}\label{lagrange_multipliers}
	\mu_{ki}(t)=\sum_{q=1}^M \left( \rho_{kq} \left[ h_{iq}-\left(i\frac{\partial}{\partial t} \right)_{iq} \right]+\sum_{s,l=1}^M \rho_{ksql} W_{isql}  \right).
\end{equation}
By reinserting them into Eq. (\ref{orb_EOM1}), multiplying by the inverse of the one-body RDM $\{ \boldsymbol{\rho}^{-1}\}_{ik}$ from the left-hand side (LHS), and summing over $k$, one arrives at
\begin{equation}\label{orb_EOM2}
	i\hat{\mathbf{P}}\frac{\partial}{\partial t}|\phi_i\rangle=\hat{\mathbf{P}}\left[ \hat{h}|\phi_i\rangle+\sum_{k,s,l,q=1}^M \{ \boldsymbol{\rho}^{-1}\}_{ik} \, \rho_{ksql} \hat{W}_{sl}|\phi_q\rangle \right]
\end{equation}
with 
\begin{equation}\label{proj1}
	\hat{\mathbf{P}}=\mathbb{1}-\sum_{j^\prime=1}^M |\phi_{j^\prime}\rangle\langle\phi_{j^\prime}|
\end{equation}
being a projection operator on the tangential space of $\text{span}\left(\phi_1,...,\phi_M\right)$. Due to invariance properties of the ansatz for the wave function in Eq. (\ref{Eq_MCTDHB_ansatz}), the orbitals can be chosen without loss of generality to be orthogonal to their corresponding time-derivatives \cite{Meyer,Manthe}, i.e.,
\begin{equation}\label{inv_cond}
	\langle\phi_i|\dot{\phi_j}\rangle=0 \quad \forall i,j=1,...,M.
\end{equation}
The time evolution of the orbitals is therefore fully orthogonal and ensures the orthonormality of the orbitals in time. Hence, one can further simplify Eq. (\ref{orb_EOM2}) to read
\begin{equation}\label{orb_EOM_final}
	i\frac{\partial}{\partial t}|\phi_i\rangle=\hat{\mathbf{P}}\left[ \hat{h}|\phi_i\rangle+\sum_{k,s,l,q=1}^M \{ \boldsymbol{\rho}^{-1}\}_{ik} \, \rho_{ksql} \hat{W}_{sl}|\phi_q\rangle \right],
\end{equation}
and Eq. (\ref{orb_EOM_final}) is referred to as the orbitals' EOM in the following.

With respect to the coefficients, the least action principle
\begin{align}
	\frac{\delta S}{\delta C_{\mathbf{n}}^\ast(t)}=0
\end{align}
leads to 
\begin{equation}
	\sum_{\mathbf{n}^{\,\prime}}\left\langle\mathbf{n};t\left|\hat{H}-i\frac{\partial}{\partial t}\right|\mathbf{n}^{\,\prime};t\right\rangle C_{\mathbf{n}^{\,\prime}}(t)=i\frac{\partial}{\partial t}C_{\mathbf{n}}(t). 
\end{equation}
Thus, the EOM of the coefficients is given by
\begin{equation}\label{coeff_EOM}
	\mathbf{\mathcal{H}}(t)\mathbf{C}(t)=i\frac{\partial}{\partial t}\mathbf{C}(t)
\end{equation}
with the matrix elements
\begin{equation}\label{coeff_matrix_MCTDHB}
	\mathcal{H}_{\mathbf{n}\mathbf{n}^{\,\prime}}=\left\langle\mathbf{n};t\left|\hat{H}-i\frac{\partial}{\partial t}\right|\mathbf{n}^{\,\prime};t\right\rangle
\end{equation}
and $\mathbf{C}(t)$ being a vector whose entries are the $N_{\text{conf}}$ coefficients $\{C_{\mathbf{n}}(t)\}$. It is worth noting that $\mathbf{\mathcal{H}}(t)$ is hermitian. Eq. (\ref{coeff_EOM}) is referred to as the coefficients' EOM in the following. 

One observes that the sets of working equations for the orbitals and coefficients are coupled to each other. With regard to the orbitals' EOM, the coefficients appear in the expressions for the two-body RDM as well as in the inverse of the one-body RDM, see Eqs. (\ref{1b_rdm}), (\ref{2b_rdm}) and (\ref{orb_EOM_final}), respectively. With regard to the coefficients' EOM, the implicit dependence on the orbitals is incorporated in the matrix elements of Eq. (\ref{coeff_matrix_MCTDHB}) since the permanents appearing therein are assembled by the orbitals.

By employing imaginary time-propagation, i.e., by setting $it\rightarrow \tau$ in the working equations (\ref{orb_EOM_final}) and (\ref{coeff_EOM}), the MCTDHB theory reduces to the multiconfigurational Hartree for bosons (MCHB) method which has been introduced in Ref. \cite{MCHB1}. In that way, the ground state of a given system can be computed very efficiently, as will be shown in Section \ref{Sec_MCTDHB_benchmark}.

It can be inferred from Eq. (\ref{coeff_EOM}) that the evolution of $\mathbf{C}(t)$ is unitary and therefore its normalization is conserved in time. However, in comparison to the orbitals' EOMs in Eq. (\ref{orb_EOM_final}), the coefficient vector $\mathbf{C}(t)$ does not automatically propagate in an orthogonal manner. The latter can be achieved by introducing a time-dependent phase of the form
\begin{equation}\label{Coeffs_phase_evo}
	\mathbf{C}(t)\rightarrow \mathbf{C}(t)e^{-i\int^t dt^\prime \mathbf{C^\dagger}(t^\prime)\mathbf{\mathcal{H}}(t^\prime)\mathbf{C}(t^\prime)}
\end{equation}
which, by inserting it into Eq. (\ref{coeff_EOM}), yields 
\begin{equation}\label{Coeff_EOM_proj}
	i\frac{\partial}{\partial t}\mathbf{C}(t)=\mathbf{\mathcal{P}_c} \mathbf{\mathcal{H}}(t)\mathbf{C}(t)
\end{equation}
with the projector that maps onto the orthogonal space of the coefficients is given by
\begin{equation}\label{coeff_proj}
	\mathbf{\mathcal{P}_c}=\mathbb{1}-\mathbf{C}(t)\mathbf{C^\dagger}(t).
\end{equation}
Including the projector $\mathbf{\mathcal{P}_c}$ is however redundant in the sense that the evolution of the coefficient vector is unitary already without it, see Eq. (\ref{coeff_EOM}). On the contrary, this is not the case for the orbitals, where the projector $\hat{\mathbf{P}}$ that stems from the Lagrange multipliers in Eq. (\ref{MCTDHB_functional1}) ensures that the orbitals stay normalized in time. Henceforth, only the latter projector is considered in the subsequent derivation of LR-MCTDHB (Section \ref{Sec_LRMCTDHB}), whereas the coefficients' projector $\mathbf{\mathcal{P}_c}$ is set to unity in the following without any loss of generality. 

It is important to note that for a single orbital, i.e., $M=1$, the MCTDHB theory reduces to the GP mean-field equation \cite{Gross,Pitaevskii,Book_Pitaevskii,Book_Pethick}, meaning that GP$\equiv$MCTDHB(1). The derivation of the latter theory using the variational principle is described in Appendix \ref{Sec_GP}.

\subsection{Many-body linear response}\label{Sec_LRMCTDHB_Kapitel}

\subsubsection{The connection between linear response and excitation spectra}\label{Sec_LR_general}
Before carrying out the derivation of the many-body LR theory atop MCTDHB, a rather general perspective of applying LR to the time-dependent Schr{\"o}dinger equation, Eq. (\ref{MBSE}), in order to calculate the excitation spectrum of a trapped BEC is discussed. The goal is to demonstrate that in general, a LR analysis atop an exact eigenstate of the unperturbed Hamiltonian $\hat{H}^0$, typically the ground state, yields the exact excitation spectrum of the many-particle system. The following derivation is closely related to the one in Ref. \cite{LR-MCTDHB2}.

One considers the time-dependent Hamiltonian 
\begin{align}
	\hat{H}(t)&=\hat{H}^0+\hat{H}_{\text{ext}}(t),  \label{Ham_linresp} \\
	\hat{H}_{\text{ext}}(t)&=\hat{f}^+(\mathbf{r})e^{-i\omega t}+\hat{f}^-(\mathbf{r})e^{i\omega t} \label{ext_pert}
\end{align}
where a weak time-dependent external field with amplitudes $\hat{f}^+$ and $\hat{f}^-$ and oscillation frequency $\omega$ is added to the stationary hermitian Hamiltonian $\hat{H}^0$. The latter has eigenenergies $E_k,\,k\in \mathbb{N}$, and the ground-state energy is denoted by $\varepsilon^0$. The projected time-dependent Schr{\"o}dinger equation with $\hbar=1$,
\begin{equation}\label{projected_SE}
	\hat{\mathbf{P}}_\Psi\hat{H}(t)\Psi(t)=i\dot{\Psi}(t), \quad \hat{\mathbf{P}}_\Psi=\mathbb{1}-|\Psi(t)\rangle\langle\Psi(t)|,
\end{equation}
which can be obtained by introducing the phase 
\begin{equation}\label{Phase_Psi}
	\Psi(t)\rightarrow\Psi(t)e^{-i\int^t dt^\prime \langle\Psi(t^\prime)|\hat{H}(t^\prime)\rangle\Psi(t^\prime)}
\end{equation}
to the system's wave function $\Psi(t)$ and inserting it into Eq. (\ref{MBSE}), ensures that the time evolution is orthogonal in the sense that
\begin{equation}\label{projected_orthonormality}
	\langle\Psi(t)|\dot{\Psi}(t)\rangle=0
\end{equation}
at any time $t$. To solve Eq. (\ref{projected_SE}), the ansatz 
\begin{equation}\label{proj_SE_ansatz}
	\Psi(\mathbf{r},t)=e^{-\varepsilon^0 t}\left(\Psi_0(\mathbf{r})+u(\mathbf{r})\,e^{-i \omega t}+v^\ast(\mathbf{r})\,e^{i \omega t}\right)
\end{equation} 
with is employed. The ground state $\Psi_0$ obtained by $\hat{H}^0\Psi_0=\varepsilon^0 \Psi_0$ as well as the correction amplitudes $u$ and $v$ are assumed to be stationary. The attached time-dependent oscillations have the same frequency $\omega$ as the external perturbing field. Utilizing the orthogonality condition, Eq. (\ref{projected_orthonormality}), one obtains
\begin{align}
	\langle\Psi_0|u\rangle=0 \quad\text{and}\quad \langle\Psi_0|v^\ast\rangle=0,
\end{align}
meaning that the correction amplitudes are orthogonal to the ground-state wave function and thus
\begin{align}
	\hat{\mathbf{P}}_{\Psi_0}|u\rangle=|u\rangle \quad\text{and}\quad \hat{\mathbf{P}}_{\Psi_0}|v^\ast\rangle=|v^\ast\rangle
\end{align}
for the operator projecting onto the tangential space of $\Psi_0$, i.e., $\hat{\mathbf{P}}_{\Psi_0}=\mathbb{1}-|\Psi_0\rangle\langle\Psi_0|$. Inserting the ansatz (\ref{proj_SE_ansatz}) into Eq. (\ref{projected_SE}), one arrives at
\begin{align}
	&\hat{\mathbf{P}}_{\Psi_0}\hat{H}^0(u\,e^{-i \omega t}+v^\ast\,e^{i \omega t})+\hat{\mathbf{P}}_{\Psi_0}(\hat{f}^+e^{-i\omega t}+\hat{f}^-e^{i\omega t})\Psi_0 \nonumber \\
	&=(\omega+\varepsilon^0)(u\,e^{-i \omega t}-v^\ast\,e^{i \omega t})
\end{align}
where only terms linear to $u$ and $v$ are kept. This can be written in matrix form by ordering terms proportional to $e^{-i\omega t}$ and $e^{i\omega t}$, respectively. The result reads
\begin{equation}\label{Gen_LR_inhomo_eq}
	\left[\begin{pmatrix} \hat{\mathbf{P}}_{\Psi_0}(\hat{H}^0-\varepsilon^0) & 0 \\ 0 & -\hat{\mathbf{P}}^\ast_{\Psi_0}(\hat{H}^{0,\ast}-\varepsilon^0) \end{pmatrix}-\omega\right] \begin{pmatrix} u \\ v \end{pmatrix} = \begin{pmatrix} -\hat{\mathbf{P}}_{\Psi_0}\hat{f}^+ \, \Psi_0 \\ \hat{\mathbf{P}}^\ast_{\Psi_0} \hat{f}^{-,\ast}\, \Psi_0  \end{pmatrix}
\end{equation}
where complex conjugation to the lower equation has been applied. Defining the LR matrix $\mathcal{L}$ as
\begin{equation}
	\mathcal{L}=\begin{pmatrix} \hat{\mathbf{P}}_{\Psi_0}(\hat{H}^0-\varepsilon^0)\hat{\mathbf{P}}_{\Psi_0} & 0 \\ 0 & -\hat{\mathbf{P}}^\ast_{\Psi_0}(\hat{H}^{0,\ast}-\varepsilon^0)\hat{\mathbf{P}}^\ast_{\Psi_0} \end{pmatrix}
\end{equation}
where redundantly the projectors $\hat{\mathbf{P}}_{\Psi_0}$ and $\hat{\mathbf{P}}^\ast_{\Psi_0}$ have been inserted from the right-hand side (RHS) to the upper and lower blocks, the homogeneous version of Eq. (\ref{Gen_LR_inhomo_eq}) is given by
\begin{equation}\label{eigenvalue_eq_general}
	\mathcal{L}\begin{pmatrix} u \\ v \end{pmatrix}=\omega\begin{pmatrix} u \\ v \end{pmatrix}.
\end{equation} 
For the upper block, one can immediately deduce that the spectrum of corresponding eigenvalues is given by $\omega_k=E_k-\varepsilon^0$, meaning that one obtains the eigenvalues of the unperturbed Hamiltonian $\hat{H}^0$. Moreover, one obtains $\omega_k=-(E_k-\varepsilon^0)$ for the lower block.

This is an important result because it means that the LR due to an external perturbation, as described in Eq. (\ref{ext_pert}), yields, if applied to an exact eigenstate of $\hat{H}^0$ (not necessarily the ground state), the full spectrum of exact excitation energies. It is worth stressing the generality of the above derivation because no physical properties of the system were specified, i.e., whether $\hat{H}^0$ describes a single- or multi-particle system, bosons or fermions (or even mixtures), or identical or distinguishable particles. Thus, LR represents a generic method for obtaining the spectrum of excited states of a quantum system. Section \ref{Sec_LRMCTDHB}, explicitly deals with the many-body linear response of systems consisting of identical bosons. 
 
This section is closed by referring to systems where no exact eigenstate of $\hat{H}^0$, e.g., the ground state, is known. This is naturally the most common case since the number of systems where analytic solutions are available is very small. Hence, in order to obtain accurate results for the excitation energies from a LR analysis, one is in need of a very accurate approximation of the system's ground state. In other words, increasing the quality of the underlying  ground-state approximation will increase the accuracy of the  spectrum of excited states. From a numerical point of view, Section \ref{Ch_Benchmark} shows that MCTDHB turns out to be a very powerful approach to calculate the ground state of a trapped BEC to very high accuracy, and that it is clearly superior to other widely used methods like the GP mean-field equation or the FCI method. As a result, LR-MCTDHB leads to highly accurate excitation spectra, for which numerical evidence is presented in Section \ref{Sec_LRMCTDHB_benchmark}.

\subsubsection{Linear-response MCTDHB (LR-MCTDHB)}\label{Sec_LRMCTDHB}
In this section, LR-MCTDHB [or LR-MCTDHB($M$)] is derived. To this end, a trapped $N$-boson system is considered and its ground-state orbitals and coefficients, $\{\phi_k^0:1\leq k \leq M\}$ and $\mathbf{C}^0$, are calculated by using imaginary time-propagation of the MCTDHB working equations \cite{MCHB1}, Eqs. (\ref{orb_EOM_final}) and (\ref{coeff_EOM}). Afterwards, the same time-dependent external perturbation as given in Eq. (\ref{ext_pert}) is added to the Hamiltonian, such that the latter is finally given by Eqs. (\ref{Ham_linresp}). The EOMs are linearized according to this. The subsequent derivation of the LR equations follows Refs. \cite{LR-MCTDHB1,LR-MCTDHB2}. 

The ans{\"a}tze for the response orbitals and coefficients due to the external perturbation are
\begin{align}
	\phi_k(\mathbf{r},t)&\approx\phi_k^0(\mathbf{r})+\delta\phi_k(\mathbf{r},t) \label{del_phi1} \\
	\delta\phi_k(\mathbf{r},t)&=u_k(\mathbf{r})e^{-i\omega t}+v_k^\ast(\mathbf{r})e^{i\omega t} \label{del_phi2} \\
	\mathbf{C}(t)&\approx e^{-i\epsilon^0 t}[\mathbf{C}^0+\delta\mathbf{C}(t)] \label{del_C1} \\
	\delta\mathbf{C}(t)&=\mathbf{C}_u e^{-i\omega t}+\mathbf{C}_v^\ast e^{i\omega t} \label{del_C2}
\end{align}
with $1\leq k \leq M$. The $k$-th response orbital is thus given by the zeroth-order orbital, typically the time-independent ground-state orbital $\phi_k^0(\mathbf{r})$, plus a weak, time-dependent perturbation $\delta\phi_k(\mathbf{r},t)$ with oscillation frequency $\omega$ and stationary response amplitudes $u_k(\mathbf{r})$ and $v_k(\mathbf{r})$. The response coefficients consist of similar parts and are additionally multiplied with a time-dependent phase $e^{-i\varepsilon^0 t}$ to which it is referred below. The time-dependence is therefore essentially incorporated in the perturbation part. In the following, the linearization of the orbitals' and coefficients' EOMs are discussed separately.

\vspace*{4ex}
For linearizing the orbitals' EOMs, it is instructive to utilize the version in Eq. (\ref{orb_EOM1}) given by
\begin{equation}\label{LR_MCTDHB_StartPoint}
	i\sum_{q=1}^M\rho_{kq}|\dot{\phi}_q(t)\rangle=\sum_{q=1}^M[\hat{Z}_{kq}-\mu_{kq}(t)]|\phi_q(t)\rangle,
\end{equation} 
with the replacement
\begin{equation}\label{Z_kq}
	\hat{Z}_{kq}=\rho_{kq}[\hat{h}+\hat{H}_\text{ext}(t)]+\sum_{s,l=1}^M\rho_{ksql}\hat{W}_{sl}. 
\end{equation}
Inserting Eq. (\ref{del_phi1}) into Eq. (\ref{LR_MCTDHB_StartPoint}), one can group the resulting expression in terms of perturbation orders. In zeroth order, one obtains
\begin{equation}\label{LR-MCTDHB_zero}
	0=\sum_q [\hat{Z}_{kq}^0-\mu^0_{kq}]|\phi_q^0\rangle 
\end{equation}
with 
\begin{equation}\label{Z_kq_zero}
	\hat{Z}^0_{kq}=\rho^0_{kq}\hat{h}+\sum_{s,l=1}^M\rho^0_{ksql}\hat{W}^0_{sl}.
\end{equation}
Here and in the following, quantities with the superscript '0' denote that they only contain zeroth-order orbitals and coefficients. Eq. (\ref{LR-MCTDHB_zero}) represents the orbital part of the MCHB working equations, see Ref. \cite{MCHB1} in this context.
For the first order, the LHS of Eq. (\ref{LR_MCTDHB_StartPoint}) yields
\begin{equation}\label{LHS_first_order}
	i\sum_q (\delta\rho_{kq}\underbrace{|\dot{\phi}_q^0\rangle}_{=0}+\rho_{kq}^0|\dot{\delta\phi_q}(t)\rangle)=i\sum_q \rho_{kq}^0|\dot{\delta\phi_q}(t)\rangle.
\end{equation}
The RHS is given by
\begin{equation}\label{RHS_first_order1}
	\sum_q [\hat{Z}_{kq}^0-\mu_{kq}^0]\,|\delta\phi_q(t)\rangle+\sum_q [\delta\hat{Z}_{kq}-\delta\mu_{kq}(t)]\,|\phi_q^0\rangle
\end{equation}
with
\begin{align}\label{delta_Z_kq1}
	\delta\hat{Z}_{kq}&=\delta\left( \rho_{kq}\hat{h}+\sum_{s,l}\rho_{ksql}\hat{W}_{sl} \right)	+ \rho_{kq}^0\hat{H}_\text{ext}(t) \nonumber \\
	&=\delta\rho_{kq}\hat{h}+\sum_{s,l}\rho_{ksql}^0\delta\hat{W}_{sl}+\sum_{s,l}\delta\rho_{ksql}\hat{W}_{sl}^0+ \rho_{kq}^0\hat{H}_\text{ext}(t).
\end{align}
To compute the RHS further, the variation of the chemical potential $\delta\mu_{kq}(t)$, given by
\begin{align}\label{delta_mu_kq}
	\delta\mu_{kq}(t)&=\left\langle\delta\phi_q \left|\sum_j\left[\hat{Z}_{kj}^0-i\rho_{kj}\frac{\partial}{\partial t}\right]\,\right|\phi_j^0\right\rangle+\left\langle\phi_q^0 \left|\sum_j \delta\left[\hat{Z}_{kj}^0-i\rho_{kj}\frac{\partial}{\partial t}\,\right|\phi_j\right\rangle\right] \nonumber \\
	&\overset{\text{Eq.} (\ref{LR-MCTDHB_zero})}{=} \sum_j\mu_{kj}^0\left\langle\delta\phi_q|\phi_j^0\right\rangle + \left\langle\phi_q^0\left|\sum_j\rho_{kj}^0\hat{H}_\text{ext}(t)\right|\phi_j^0\right\rangle \nonumber \\
	&+ \left\langle\phi_q^0\left|\left( \sum_j \delta\rho_{kj}\hat{h}+\sum_{j,s,l}(\rho_{ksjl}^0\delta\hat{W}_{sl}+\delta\rho_{ksjl}\hat{W}_{sl}^0) \right)\right|\phi_j^0\right\rangle \nonumber \\
	&+\left\langle\phi_q^0\left|\sum_j \left( \hat{Z}_{kj}^0-i\rho_{kj}^0\frac{\partial}{\partial t} \right) \right|\delta\phi_j\right\rangle\,,
\end{align}
is employed. Thus, Eq. (\ref{RHS_first_order1}) can be written as
\begin{align}\label{RHS_first_order2}
	&\sum_q [ \hat{Z}_{kq}^0-\mu_{kq}^0]\,|\delta\phi_q\rangle+\sum_q\left\lbrace \delta\rho_{kq}\hat{h}+\rho_{kq}^0\hat{H}_\text{ext}(t)+\sum_{s,l}(\rho_{ksql}\,\delta\hat{W}_{sl}+\delta\rho_{ksql}\,\hat{W}_{sl}^0) \right\rbrace|\phi_q^0\rangle \nonumber \\
	&-\sum_q \left\lbrace \sum_j(-\mu_{kj}^0)\left\langle\phi_q^0|\delta\phi_j\right\rangle + \left\langle\phi_q^0\left|\sum_j\rho_{kj}^0\hat{H}_\text{ext}(t)\right|\phi_j^0\right\rangle \right.  \nonumber \\
	&+ \left. \left\langle\phi_q^0\left|\left( \sum_j \delta\rho_{kj}\hat{h}+\sum_{j,s,l}(\rho_{ksjl}^0\delta\hat{W}_{sl}+\delta\rho_{ksjl}\hat{W}_{sl}^0) \right)\right|\phi_j^0\right\rangle \right. \nonumber \\
	&+\left.  \left\langle\phi_q^0\left|\sum_j \left( \hat{Z}_{kj}^0-i\rho_{kj}^0\frac{\partial}{\partial t} \right) \right|\delta\phi_j\right\rangle\   \right\rbrace |\phi_q^0\rangle
\end{align}
which, combined with the LHS, Eq. (\ref{LHS_first_order}), leads to
\begin{align}\label{LR_EOM1}
	&\sum_q \hat{\mathbf{P}} \left\lbrace  \left( \hat{Z}_{kq}^0-\mu_{kq}^0 \right)\,|\delta\phi_q\rangle + \left( \delta\rho_{kq}\hat{h}+\rho_{kq}^0\hat{H}_\text{ext}(t) \right.\right. \nonumber \\
	&+\sum_{s,l}\delta\rho_{ksql}\,\hat{W}_{sl}^0+\sum_{s,l} \left.\left. \rho_{ksql}^0\,\delta\hat{W}_{sl} \right) |\phi_q^0\rangle  \right\rbrace \nonumber \\
	&= i \hat{\mathbf{P}} \sum_q \rho_{kq}^0 |\delta\dot{\phi}_q\rangle
\end{align}
where the projection operator $\hat{\mathbf{P}}=\mathbb{1}-\sum_{j^\prime} |\phi_{j^\prime}^0\rangle \langle \phi_{j^\prime}^0|$ projects on the tangential space of the stationary orbitals. From the invariance condition, Eq. (\ref{inv_cond}), one can deduce
\begin{align}\label{cond_LR}
	0&=\delta\langle\phi_k|\dot{\phi}_q\rangle=\langle\delta\phi_k\underbrace{|\dot{\phi}_q^0\rangle}_{=0}+\langle\phi_k^0|\delta\dot{\phi}_q\rangle \nonumber \\
	&=-i\omega\langle\phi_k^0|u_q\rangle e^{-i \omega t}+i\omega\langle\phi_k^0|v^\ast_q\rangle e^{i \omega t} \nonumber \\
	&\Rightarrow \langle\phi_k^0|u_q\rangle=\langle\phi_k^0|v^\ast_q\rangle=0 \quad\quad \forall\, 1\leq k,q \leq M
\end{align}
such that
\begin{equation}
	\hat{\mathbf{P}}|\delta\dot{\phi}_q\rangle=\left(\mathbb{1}-\sum_{j^\prime=1}^M |\phi^0_{j^\prime}\rangle\langle\phi^0_{j^\prime}|\right)|\delta\dot{\phi}_q\rangle=|\delta\dot{\phi}_q\rangle
\end{equation}
which makes the projection operator on the RHS of Eq. (\ref{LR_EOM1}) redundant. One further utilizes
\begin{align}
\delta\hat{W}_{sl}\left[u_l,u_s^\ast,v_l^\ast,v_s \right] &=\int d\mathbf{r}^{\,\prime}\left\lbrace \delta\phi_s^\ast(\mathbf{r}^{\,\prime},t)\,\hat{W}(\mathbf{r},\mathbf{r}^{\,\prime})\,\phi_l^0(\mathbf{r}^{\,\prime}) \right. \nonumber \\
&+ \left. \phi^{0,\ast}_s(\mathbf{r}^{\,\prime})\,\hat{W}(\mathbf{r},\mathbf{r}^{\,\prime})\,\delta\phi_l(\mathbf{r}^{\,\prime},t) \right\rbrace \label{delta_Wsl} \\
\delta\rho_{kq}\left[\mathbf{C}_u,\mathbf{C}_v,\mathbf{C}_u^\ast,\mathbf{C}_v^\ast\right]&=\langle\delta\mathbf{C}(t)|\hat{b}^\dagger_k\hat{b}_q|\mathbf{C}^0\rangle + \langle\mathbf{C}^0|\hat{b}^\dagger_k\hat{b}_q|\delta\mathbf{C}(t)\rangle \label{delta_rho_kq} \\
\delta\rho_{ksql}\left[\mathbf{C}_u,\mathbf{C}_v,\mathbf{C}_u^\ast,\mathbf{C}_v^\ast\right]&=\langle\delta\mathbf{C}(t)|\hat{b}^\dagger_k \hat{b}^\dagger_s\hat{b}_q \hat{b}_l|\mathbf{C}^0\rangle + \langle\mathbf{C}^0|\hat{b}^\dagger_k \hat{b}^\dagger_s\hat{b}_q \hat{b}_l|\delta\mathbf{C}(t)\rangle \label{delta_rho_ksql} 
\end{align}
where the action of, e.g., $\hat{b}^\dagger_k\hat{b}_q$ on $|\mathbf{C}\rangle$ results in a new vector of coefficients for the same Fock states, i.e.,
\begin{equation}
	\hat{b}^\dagger_k\hat{b}_q|\mathbf{C}\rangle=\sum_\mathbf{n}C_{\mathbf{n}}(t)\hat{b}^\dagger_k\hat{b}_q|\mathbf{n};t\rangle=\sum_{\mathbf{n}^\prime}C_{\mathbf{n}^\prime}(t)|\mathbf{n}^\prime;t\rangle\equiv|\mathbf{C}^\prime\rangle,
\end{equation}
and inserts the above equations, together with Eqs. (\ref{ext_pert}), (\ref{del_phi2}), and (\ref{del_C2}), into Eq. (\ref{LR_EOM1}). Collecting all terms proportional to $e^{-i\omega t}$ results in
\begin{align}\label{LR_orb_-iomegat}
	&\sum_q \hat{\mathbf{P}} \left\lbrace  \left( \hat{Z}_{kq}^0-\mu_{kq}^0 \right)\,|u_q\rangle + \left( \delta\rho_{kq}\left[\mathbf{C}_u,\mathbf{C}_v\right]\hat{h}+\rho_{kq}^0\,\hat{f}^+ \right.\right. \nonumber \\
	&+\sum_{s,l}\delta\rho_{ksql}\left[\mathbf{C}_u,\mathbf{C}_v\right]\,\hat{W}_{sl}^0+\sum_{s,l} \left.\left. \rho_{ksql}^0\,\delta\hat{W}_{sl}[u_l,v_s] \right) |\phi_q^0\rangle  \right\rbrace \nonumber \\
	&= \omega \sum_q \rho_{kq}^0 |u_q\rangle.
\end{align}
Similarly, collecting all terms proportional to $e^{i\omega t}$ yields
\begin{align}\label{LR_orb_+iomegat}
	&\sum_q \hat{\mathbf{P}} \left\lbrace  \left( \hat{Z}_{kq}^0-\mu_{kq}^0 \right)\,|v^\ast_q\rangle + \left( \delta\rho_{kq}\left[\mathbf{C}_u^\ast,\mathbf{C}_v^\ast\right]\hat{h}+\rho_{kq}^0\,\hat{f}^- \right.\right. \nonumber \\
	&+\sum_{s,l}\delta\rho_{ksql}\left[\mathbf{C}_u^\ast,\mathbf{C}_v^\ast\right]\,\hat{W}_{sl}^0+\sum_{s,l} \left.\left. \rho_{ksql}^0\,\delta\hat{W}_{sl}[u_s^\ast,v_l^\ast] \right) |\phi_q^0\rangle  \right\rbrace \nonumber \\
	&= -\omega \sum_q \rho_{kq}^0 |v_q^\ast\rangle \nonumber \\
	\overset{\text{C.C.}}{\iff} &\sum_q \hat{\mathbf{P}}^\ast \left\lbrace  \left( \hat{Z}_{kq}^{0,\ast}-\mu_{qk}^0 \right)\,|v_q\rangle + \left( \delta\rho_{qk}\left[\mathbf{C}_u,\mathbf{C}_v\right]\hat{h}^\ast+\rho_{qk}^0\,\hat{f}^{-,\ast} \right.\right. \nonumber \\
	&+\sum_{s,l}\delta\rho_{qlks}\left[\mathbf{C}_u,\mathbf{C}_v\right]\,\hat{W}_{ls}^0+\sum_{s,l} \left.\left. \rho_{qlks}^0\,\delta\hat{W}_{ls}[u_s,v_l] \right) |\phi_q^{0,\ast}\rangle  \right\rbrace \nonumber \\
	&= -\omega \sum_q \rho_{qk}^0 |v_q\rangle
\end{align}
where 'C.C.' denotes complex conjugation of the entire equation. Here, it is used that the matrix $\{\mu_{ij}\}$ of the Lagrange multipliers is hermitian. Furthermore, the two-body interaction potential is assumed to be real throughout this work.

For linearizing the coefficients' EOMs, one obtains in zeroth order the stationary equation
\begin{equation}\label{Coeff_zeroth_order}
	\boldsymbol{\mathcal{H}^0}\mathbf{C}^0=\varepsilon^0\,\mathbf{C}^0
\end{equation}
with the matrix elements $\{\mathcal{H}^0_{\mathbf{n}\mathbf{n}^{\,\prime}}\}=\langle\mathbf{n}|\hat{H}^0|\mathbf{n}^{\,\prime}\rangle$. Eq. (\ref{Coeff_zeroth_order}) is the working equation for the coefficients within the MCHB theory. In first order, one arrives at
\begin{equation}\label{LR_MCTDHB_orb_StartPoint}
	i\frac{\partial}{\partial t}\delta\mathbf{C}(t)=\delta\boldsymbol{\mathcal{H}}\mathbf{C}^0+\left(\boldsymbol{\mathcal{H}^0}-\varepsilon^0\right)\delta\mathbf{C}(t)
\end{equation}
with
\begin{equation}\label{delta_H}
	\delta\boldsymbol{\mathcal{H}}=\sum_{k,q}\delta h_{kq}\,\hat{b}^\dagger_k\hat{b}_q+\frac{1}{2}\sum_{k,s,q,l}\delta W_{ksql}\,\hat{b}^\dagger_k \hat{b}^\dagger_s\hat{b}_q \hat{b}_l
\end{equation}
and
\begin{align}\label{delta_hkq_Wksql}
	\delta h_{kq}&=\int\,d\mathbf{r}\,\delta\phi^\ast_k(\mathbf{r},t)\hat{h}\phi_q^0(\mathbf{r})+\int\,d\mathbf{r}\,\phi^{0,\ast}_k(\mathbf{r})\hat{h}\delta\phi_q(\mathbf{r},t) \nonumber \\
	&+\underbrace{\int\,d\mathbf{r}\,\phi^{0,\ast}_k(\mathbf{r})\hat{H}_\text{ext}(\mathbf{r},t)\phi_q^0(\mathbf{r})}_{\equiv(H_\text{ext})_{kq}} \\
	\delta W_{ksql}&=\iint\,d\mathbf{r}d\mathbf{r}^{\,\prime}\,\delta\phi^\ast_k(\mathbf{r},t)\phi_s^{0,\ast}(\mathbf{r}^{\,\prime})\hat{W}(\mathbf{r},\mathbf{r}^{\,\prime})\phi_q^0(\mathbf{r})\phi_l^0(\mathbf{r}^{\,\prime}) \nonumber \\
	&+\iint\,d\mathbf{r}d\mathbf{r}^{\,\prime}\,\phi^{0,\ast}_k(\mathbf{r})\delta\phi_s^\ast(\mathbf{r}^{\,\prime},t)\hat{W}(\mathbf{r},\mathbf{r}^{\,\prime})\phi_q^0(\mathbf{r})\phi_l^0(\mathbf{r}^{\,\prime}) \nonumber \\
	&+\iint\,d\mathbf{r}d\mathbf{r}^{\,\prime}\,\phi^{0,\ast}_k(\mathbf{r})\phi_s^{0,\ast}(\mathbf{r}^{\,\prime})\hat{W}(\mathbf{r},\mathbf{r}^{\,\prime})\delta\phi_q(\mathbf{r},t)\phi_l^0(\mathbf{r}^{\,\prime}) \nonumber \\
	&+\iint\,d\mathbf{r}d\mathbf{r}^{\,\prime}\,\phi^{0,\ast}_k(\mathbf{r})\phi_s^{0,\ast}(\mathbf{r}^{\,\prime})\hat{W}(\mathbf{r},\mathbf{r}^{\,\prime})\phi_q(\mathbf{r})\delta\phi_l(\mathbf{r}^{\,\prime},t).
\end{align}
Collecting all terms proportional to $e^{-i\omega t}$ yields
\begin{align}\label{LR_coeff_-iomegat}
	\omega\mathbf{C}_u &=(\boldsymbol{\mathcal{H}^0}-\varepsilon^0)\mathbf{C}_u+\left[\sum_{k,q}\delta h_{k,q}[v_k,u_q,\hat{f}^+]\hat{b}^\dagger_k\hat{b}_q \right. \nonumber \\
	&+ \left. \frac{1}{2}\sum_{k,s,q,l}\delta W_{ksql}[v_k,v_s,u_q,u_l]\hat{b}^\dagger_k\hat{b}^\dagger_s\hat{b}_q\hat{b}_l\right]\mathbf{C}^0,
\end{align}
whereas collecting all terms proportional to $e^{+i\omega t}$ results in
\begin{align}\label{LR_coeff_+iomegat}
	-\omega\mathbf{C}_v^\ast &=(\boldsymbol{\mathcal{H}^0}-\varepsilon^0)\mathbf{C}_v^\ast+\left[\sum_{k,q}\delta h_{k,q}[u^\ast_k,v^\ast_q,\hat{f}^-]\hat{b}^\dagger_k\hat{b}_q \right. \nonumber \\
	&+ \left. \frac{1}{2}\sum_{k,s,q,l}\delta W_{ksql}[u^\ast_k,u^\ast_s,v^\ast_q,v^\ast_l]\hat{b}^\dagger_k\hat{b}^\dagger_s\hat{b}_q\hat{b}_l\right]\mathbf{C}^0 \nonumber \\
	\overset{\text{C.C.}}{\iff} -\omega\mathbf{C}_v&=(\boldsymbol{\mathcal{H}^{0,\ast}}-\varepsilon^0)\mathbf{C}_v+\left[\sum_{k,q}\delta h_{q,k}[u_k,v_q,\hat{f}^{-,\ast}]\left(\hat{b}^\dagger_k\hat{b}_q\right)^\ast \right. \nonumber \\
	&+ \left. \frac{1}{2}\sum_{q,l,k,s}\delta W_{qlks}[u_k,u_s,v_q,v_l]\left(\hat{b}^\dagger_k\hat{b}^\dagger_s\hat{b}_q\hat{b}_l\right)^\ast\right]\mathbf{C}^{0,\ast}.
\end{align}

\vspace*{4ex}
After the linearization of both the orbitals' and coefficients' EOMs, one can combine the sets of LR equations by casting Eqs. (\ref{LR_orb_-iomegat}), (\ref{LR_orb_+iomegat}), (\ref{LR_coeff_-iomegat}) and (\ref{LR_coeff_+iomegat}) into matrix form, leading to
\begin{equation}\label{LR_eigenvalue_eq1}
	\left(\mathbf{\mathcal{P}}\mathcal{L}-\mathcal{M}\omega\right)
	  \begin{pmatrix}\mathbf{u}\\\mathbf{v}\\\mathbf{C}_u\\ \mathbf{C}_v \end{pmatrix}
	 =\mathbf{\mathcal{M}}\,\mathbf{\mathcal{P}}\,\mathbf{\mathcal{R}}[\hat{f}^+,\hat{f}^-]
\end{equation}
with $\mathbf{u}=(u_1,...,u_M)^t$, $\mathbf{v}=(v_1,...,v_M)^t$ and the LR matrix
\begin{equation}\label{LR_matrix}
	\mathcal{L}=\begin{pmatrix} \mathbf{\mathcal{L}}_{oo} & \mathbf{\mathcal{L}}_{oc}	\\ \mathbf{\mathcal{L}}_{co} & \mathbf{\mathcal{L}}_{cc}	\end{pmatrix}
\end{equation}
where the submatrices refer to the couplings between the orbitals and the coefficients. In Eq. (\ref{LR_eigenvalue_eq1}), $\mathbf{\mathcal{P}}$ is a projection matrix, $\mathbf{\mathcal{M}}$ is a metric which contains the one-body RDM, and the RHS $\mathbf{\mathcal{R}}$ contains the amplitudes of the external perturbation, $\hat{f}^+$ and $\hat{f}^-$. In the following, the submatrices in Eq. (\ref{LR_matrix}) are explicitly derived. 

The $(2M)$-dimensional square orbital matrix $\mathcal{L}_{oo}$ consists of four blocks and reads
\begin{equation}\label{Loo_mat}
	\mathcal{L}_{oo}=\begin{pmatrix} \mathcal{L}^u_{oo} & \mathcal{L}^v_{oo}	\\ -\left(\mathcal{L}^v_{oo}\right)^\ast & -\left(\mathcal{L}^u_{oo}\right)^\ast	\end{pmatrix}.
\end{equation}
The upper part can be obtained from Eq. (\ref{LR_orb_-iomegat}) by collecting all terms proportional to $\mathbf{u}$, yielding
\begin{equation}\label{Loo_u}
	\mathcal{L}^u_{oo}=\boldsymbol{\rho^0}\hat{h}-\boldsymbol{\mu^0}+\mathbf{\Omega^0}+\boldsymbol{\kappa}^1
\end{equation}
with $\boldsymbol{\rho^0}=\{\rho^0_{kq}\}$, $\boldsymbol{\mu^0}=\{\mu^0_{kq}\}$, $\mathbf{\Omega^0}=\{\Omega^0_{kq}\}=\sum_{s,l}\rho^0_{ksql}\hat{W}^0_{sl}$ and $\boldsymbol{\kappa}^1=\{\kappa^1_{kq}\}=\sum_{s,l}\rho^0_{ksql}\hat{K}^0_{sl}$. Therein, the exchange interaction operator is defined as 
\begin{equation}\label{K_sl}
	\hat{K}_{sl}=\int d\mathbf{r}^{\,\prime}\phi_s^\ast(\mathbf{r}^{\,\prime})\hat{W}(\mathbf{r},\mathbf{r}^{\,\prime})\hat{\mathcal{P}}_{\mathbf{r}\mathbf{r}^{\,\prime}}\phi_l(\mathbf{r}^{\,\prime})
\end{equation}
where $\hat{\mathcal{P}}_{\mathbf{r}\mathbf{r}^{\,\prime}}$ permutes the coordinates $\mathbf{r}$ and $\mathbf{r}^{\,\prime}$, i.e., 
\begin{equation}
	\hat{\mathcal{P}}_{\mathbf{r}\mathbf{r}^{\,\prime}}f(\mathbf{r})=f(\mathbf{r}^{\,\prime})
\end{equation}
for any function $f(\mathbf{r})$. The action of $\hat{K}_{sl}$ on an arbitrary function $f(\mathbf{r})$ therefore reads
\begin{equation}\label{K_sl_action}
	\hat{K}_{sl}\,f(\mathbf{r})=\hat{W}_{sf}\,\phi_l(\mathbf{r}).
\end{equation}
Furthermore, collecting all terms proportional to $\mathbf{v}$ yields
\begin{equation}\label{Loo_v}
	\mathcal{L}^v_{oo}=\boldsymbol{\kappa}^2,\quad\boldsymbol{\kappa}^2=\{\kappa^2_{kq}\}=\sum_{s,l}\rho_{kqsl}^0\hat{K}_{l^\ast s}
\end{equation}
for the upper right block of $\mathcal{L}_{oo}$. The lower two submatrices can be obtained from Eq. (\ref{LR_orb_+iomegat}) and turn out to be the negative complex conjugate of the upper submatrices. 

The upper right block of $\mathcal{L}$, i.e., the $2(M\times N_{\text{conf}})$-matrix $\mathcal{L}_{oc}$, has a similar structure than $\mathcal{L}_{oo}$, meaning it consists of four separate blocks given by
\begin{equation}\label{Loc_mat}
	\mathcal{L}_{oc}=\begin{pmatrix} \mathcal{L}^u_{oc} & \mathcal{L}^v_{oc}	\\ -\left(\mathcal{L}^v_{oc}\right)^\ast & -\left(\mathcal{L}^u_{oc}\right)^\ast	\end{pmatrix}.
\end{equation}
The upper two submatrices can again be extracted from Eq. (\ref{LR_orb_-iomegat}) and yield, collecting all terms proportional to $\mathbf{C_u}$,
\begin{equation}\label{Loc_u}
	\mathcal{L}^u_{oc}=\sum_q\left\lbrace \hat{h}\left( \hat{\rho}_{qk}\mathbf{C^0} \right)^\dagger +\sum_{s,l}\hat{W}_{sl}^0\left(\hat{\rho}_{qlks} \mathbf{C^0} \right)^\dagger \right\rbrace \phi_q^0\,,	
\end{equation}
whereas collecting all terms proportional to $\mathbf{C_v}$ results in
\begin{equation}\label{Loc_v}
	\mathcal{L}^v_{oc}=\sum_q\left\lbrace \hat{h}\left( \hat{\rho}_{kq}\mathbf{C^0} \right)^t +\sum_{s,l}\hat{W}_{sl}^0\left(\hat{\rho}_{ksql} \mathbf{C^0} \right)^t \right\rbrace \phi_q^0\,.
\end{equation}
The lower two submatrices can be extracted from Eq. (\ref{LR_orb_+iomegat}), see Ref. \cite{LR-MCTDHB2}.

With respect to the $2(N_{\text{conf}}\times M)$-matrix $\mathcal{L}_{co}$, one finds 
\begin{equation}\label{Lco_mat}
	\mathcal{L}_{co}=\begin{pmatrix} \mathcal{L}^u_{co} & \mathcal{L}^v_{co}	\\ -\left(\mathcal{L}^v_{co}\right)^\ast & -\left(\mathcal{L}^u_{co}\right)^\ast	\end{pmatrix}
\end{equation}
with
\begin{equation}\label{Lco_u}
	\mathcal{L}_{co}^u=\sum_k \phi_k^{0,\ast}\left\lbrace \left( \hat{\rho}_{kq}\mathbf{C^0} \right)\hat{h}+\sum_{sl}  \left( \hat{\rho}_{klqs}\mathbf{C^0} \right) \hat{W}_{ls}^0 \right\rbrace
\end{equation}
from collecting all terms proportional to $u_q$ in Eq. (\ref{LR_coeff_-iomegat}) and
\begin{equation}\label{Lco_v}
	\mathcal{L}_{co}^v=\sum_k \phi_k^0\left\lbrace \left( \hat{\rho}_{qk}\mathbf{C^0} \right)\hat{h}^\ast+\sum_{sl}  \left( \hat{\rho}_{qskl}\mathbf{C^0} \right) \hat{W}_{sl}^0 \right\rbrace
\end{equation}
from collecting all terms proportional to $v_q$ in Eq. (\ref{LR_coeff_+iomegat}). It can thus be seen that
\begin{equation}\label{Loc_Lco_relation}
	\left( \mathcal{L}_{oc}^u \right)^\dagger=\mathcal{L}_{co}^u, \quad  \left( \mathcal{L}_{oc}^v \right)^t=\mathcal{L}_{co}^v.
\end{equation}

Furthermore, one can extract the $(2N_{\text{conf}})$-dimensional square matrix $\mathcal{L}_{cc}$ from Eqs. (\ref{LR_coeff_-iomegat}) and (\ref{LR_coeff_+iomegat}), which gives
\begin{equation}\label{Lcc_mat}
	\mathcal{L}_{cc}=\begin{pmatrix} \boldsymbol{\mathcal{H}^0}-\varepsilon^0 & \mathbf{0_{cc}} \\ \mathbf{0_{cc}}	& -\left( \boldsymbol{\mathcal{H}^{0,\ast}}-\varepsilon^0\right)  \end{pmatrix}
\end{equation}
where $\mathbf{0_{cc}}$ is the ($N_{\text{conf}}$)-dimensional zero matrix.

The projection operator $\mathbf{\mathcal{P}}$ is given by 
\begin{equation}\label{P_mat}
	\mathbf{\mathcal{P}}=\begin{pmatrix} \boldsymbol{\mathcal{P}}_{oo} & \mathbf{0_{oc}} \\ \mathbf{0_{co}} & \boldsymbol{\mathbb{1}_{cc}}   \end{pmatrix}
\end{equation}
with
\begin{equation}\label{P_oo}
	\boldsymbol{\mathcal{P}}_{oo}=\begin{pmatrix} \hat{\mathbf{P}} & 0 \\ 0 & \mathbf{\hat{P}^\ast}   \end{pmatrix}
\end{equation}
and $\mathbf{\mathbb{1}_{cc}}$ being the $(2N_{\text{conf}})$-dimensional unit matrix. The matrices $\mathbf{0_{oc}}$ and $\mathbf{0_{co}}$ denote the $2(M\times N_{\text{conf}})$- and $2(N_{\text{conf}}\times M)$-dimensional zero matrices, respectively.

The metric $\mathbf{\mathcal{M}}$ reads
\begin{equation}\label{metric_mat}
	\mathbf{\mathcal{M}}=\begin{pmatrix}	\boldsymbol{\rho}_{oo} & \mathbf{0_{oc}} \\ \mathbf{0_{co}} & \mathbf{\mathbb{1}_{cc}}	\end{pmatrix}
\end{equation}
where $\mathbf{\rho}_{oo}$ is given by
\begin{equation}\label{metric_oo}
	\boldsymbol{\rho}_{oo}=\begin{pmatrix}	\boldsymbol{\rho^0} & \mathbf{0_o} \\ \mathbf{0_o} & \boldsymbol{\rho^{0,\ast}}	\end{pmatrix}
\end{equation}
and $\mathbf{0_o}$ being the ($M \times M$)-dimensional zero matrix.

In the following, the response amplitudes and coefficients are written in bra-ket notation for reasons that become clear below. In order to render Eq. (\ref{LR_eigenvalue_eq1}) an eigenvalue equation, it is helpful to notice that only the term proportional to the external frequency $\omega$ has no projector, which implies 
\begin{equation}\label{proj_invar_LRmat}
	\mathbf{\mathcal{P}}\mathbf{\mathcal{M}}\begin{pmatrix}|\mathbf{u}\rangle\\ |\mathbf{v}\rangle \\ |\mathbf{C}_u\rangle \\ |\mathbf{C}_v\rangle \end{pmatrix}=\mathbf{\mathcal{M}}\begin{pmatrix}|\mathbf{u}\rangle\\ |\mathbf{v}\rangle \\ |\mathbf{C}_u\rangle \\ |\mathbf{C}_v\rangle \end{pmatrix}
\end{equation}
and thus one may add a redundant projector $\mathbf{\mathcal{P}}\mathcal{L}\Rightarrow\mathbf{\mathcal{P}}\mathcal{L}\mathbf{\mathcal{P}}$ to the left term of the LHS of Eq. (\ref{LR_eigenvalue_eq1}). Multiplying now with $\mathbf{\mathcal{M}^{-1/2}}$ from the left yields
\begin{equation}
	\left(\mathbf{\mathcal{M}^{-1/2}} \mathbf{\mathcal{P}}\mathcal{L}\mathbf{\mathcal{P}} \mathbf{\mathcal{M}^{-1/2}}  -\omega \right) \mathbf{\mathcal{M}^{+1/2}} \begin{pmatrix}|\mathbf{u}\rangle\\ |\mathbf{v}\rangle \\ |\mathbf{C}_u\rangle \\ |\mathbf{C}_v\rangle \end{pmatrix} = \mathbf{\mathcal{M}^{+1/2}} \mathbf{\mathcal{P}} \mathbf{\mathcal{R}}[\hat{f}^+,\hat{f}^-] 
\end{equation}
or, in a more compact notation,
\begin{equation}\label{LR_eigenvalue_eq2}
	\left(\bar{\mathcal{L}}-\omega \right) \begin{pmatrix}|\mathbf{\overline{u}}\rangle \\ |\mathbf{\overline{v}}\rangle \\  |\overline{\mathbf{C}_u}\rangle \\ |\overline{\mathbf{C}_v}\rangle \end{pmatrix} = \mathbf{\overline{\mathcal{R}}}[\hat{f}^+,\hat{f}^-]
\end{equation}
where
\begin{align}\label{Eq_final_LR_matrix}
	\bar{\mathcal{L}}&=\mathbf{\mathcal{M}^{-1/2}} \mathbf{\mathcal{P}}\mathcal{L}\mathbf{\mathcal{P}} \mathbf{\mathcal{M}^{-1/2}} \nonumber \\
	&=\begin{pmatrix} \boldsymbol{\rho_{oo}}^{-1/2}\, \boldsymbol{\mathcal{P}}_{oo}\, \mathcal{L}_{oo} \, \boldsymbol{\mathcal{P}}_{oo} \, \boldsymbol{\rho_{oo}}^{-1/2} & \quad  \boldsymbol{\rho_{oo}}^{-1/2}\, \boldsymbol{\mathcal{P}}_{oo}\,\mathcal{L}_{oc} \\ \mathcal{L}_{co} \,\boldsymbol{\mathcal{P}}_{oo} \,\boldsymbol{\rho_{oo}}^{-1/2} & \mathcal{L}_{cc} \end{pmatrix},
\end{align}
as well as $\left(\mathbf{\overline{u}},\mathbf{\overline{v}}, \overline{\mathbf{C}_u},\overline{\mathbf{C}_v}\right)^t=\mathbf{\mathcal{M}^{+1/2}}\left(\mathbf{u},\mathbf{v}, \mathbf{C}_u,\mathbf{C}_v\right)^t$ and $\mathbf{\overline{\mathcal{R}}}=\mathbf{\mathcal{M}^{+1/2}} \mathbf{\mathcal{P}} \mathbf{\mathcal{R}}$ were introduced. To keep the notation simple, the bar over the individual quantities in Eq. (\ref{LR_eigenvalue_eq2}) will be omitted in the following. For the homogeneous case, i.e., with $\mathbf{\mathcal{R}}=0$, one arrives at the central equation of this work, given by
\begin{equation}\label{LR_eigenvalue_final}
	(\mathcal{L}-\omega) \begin{pmatrix}|\mathbf{u}\rangle\\ |\mathbf{v}\rangle \\ |\mathbf{C}_u\rangle \\ |\mathbf{C}_v\rangle \end{pmatrix}=0
\end{equation}
which is a standard eigenvalue problem of the LR matrix $\mathcal{L}$. Its solutions, i.e., the set of eigenvalues $\{\omega_k=E_k-\varepsilon^0\}$ and eigenvectors $\{(|\mathbf{u}^k\rangle,|\mathbf{v}^k\rangle,|{\mathbf{C}_u}^k\rangle,|{\mathbf{C}_v}^k\rangle)^t\}$ where $k\in\mathbb{N}$ labels the excitations, can be used to solve the inhomogeneous problem of Eq. (\ref{LR_eigenvalue_eq2}) by employing the ans{\"a}tze
\begin{equation}\label{inhomo_ansatz1}
	\begin{pmatrix}|\mathbf{u}\rangle\\ |\mathbf{v}\rangle \\ |\mathbf{C}_u\rangle \\ |\mathbf{C}_v\rangle \end{pmatrix}=\sum_k c_k \begin{pmatrix}|\mathbf{u}^k\rangle\\ |\mathbf{v}^k\rangle \\ |{\mathbf{C}_u}^k\rangle \\ |{\mathbf{C}_v}^k\rangle \end{pmatrix}
\end{equation}
and 
\begin{equation}\label{inhomo_ansatz2}
	\mathbf{\mathcal{R}}=-\sum_k \gamma_k \begin{pmatrix}|\mathbf{u}^k\rangle\\ |\mathbf{v}^k\rangle \\ |{\mathbf{C}_u}^k\rangle \\ |{\mathbf{C}_v}^k\rangle \end{pmatrix}
\end{equation}
where the minus sign is chosen arbitrarily. One finally obtains
\begin{equation}\label{inhomo_eq}
	\sum_k c_k(\omega_k-\omega)\begin{pmatrix}|\mathbf{u}^k\rangle\\ |\mathbf{v}^k\rangle \\ |{\mathbf{C}_u}^k\rangle \\ |{\mathbf{C}_v}^k\rangle \end{pmatrix}=-\sum_k \gamma_k \begin{pmatrix}|\mathbf{u}^k\rangle\\ |\mathbf{v}^k\rangle \\ |{\mathbf{C}_u}^k\rangle \\ |{\mathbf{C}_v}^k\rangle \end{pmatrix}
\end{equation}
which means that the expansion coefficients $c_k$ in Eq. (\ref{inhomo_ansatz1}) are given by
\begin{equation}\label{c_k}
	c_k=\frac{\gamma_k}{\omega-\omega_k}\,.
\end{equation}
It is stressed that the response weights $\{\gamma_k\}$ are the only quantities that depend on the perturbing fields $\hat{f}^+$ and $\hat{f}^-$ and thus on the actual shape of the perturbation. All other quantities like the LR matrix $\mathcal{L}$, the eigenenergies $\{\omega_k\}$ and the eigenvectors $\{(|\mathbf{u}^k\rangle,|\mathbf{v}^k\rangle,|{\mathbf{C}_u}^k\rangle,|{\mathbf{C}_v}^k\rangle)^t\}$ are obtained from the homogeneous eigenvalue problem of Eq. (\ref{LR_eigenvalue_final}) where the amplitudes $\hat{f}^+$ and $\hat{f}^-$ of the perturbing field do not appear. 

Multiplying Eq. (\ref{inhomo_ansatz2}) by $\left(\langle\mathbf{u}^k|,-\langle\mathbf{v}^k|,\langle {\mathbf{C}_u}^k|,-\langle {\mathbf{C}_v}^k|\right)$ from the left, taking into account the orthonormality conditions for the perturbed amplitudes and coefficients given by
\begin{align}\label{orthonormality_u_v}
	\langle \mathbf{u}^k|\mathbf{u}^{k^\prime}\rangle-\langle \mathbf{v}^k|\mathbf{v}^{k^\prime}\rangle+\langle {\mathbf{C}_u}^k|{\mathbf{C}_u}^{k^\prime}\rangle-\langle {\mathbf{C}_v}^k|{\mathbf{C}_v}^{k^\prime}\rangle&=\delta_{kk^\prime} \\
	\langle \mathbf{v}^k|\mathbf{u}^{k^\prime,\ast}\rangle-\langle \mathbf{u}^k|\mathbf{v}^{k^\prime,\ast}\rangle+\langle {\mathbf{C}_v}^k|{\mathbf{C}_u}^{k^\prime,\ast}\rangle-\langle {\mathbf{C}_u}^k|{\mathbf{C}_v}^{k^\prime,\ast}\rangle&=0,
\end{align} 
leads to the response weights
\begin{align}\label{gamma_k}
	\gamma_k &= -\left(\langle\mathbf{u}^k|,-\langle\mathbf{v}^k|,\langle {\mathbf{C}_u}^k|,-\langle {\mathbf{C}_v}^k|\right)\,\mathbf{\mathcal{R}} \nonumber \\
	&=-\left(\langle\mathbf{u}^k|,-\langle\mathbf{v}^k|,\langle {\mathbf{C}_u}^k|,-\langle {\mathbf{C}_v}^k|\right) \begin{pmatrix} -(\boldsymbol{\rho^0})^{1/2}\,\hat{\mathbf{P}} \hat{f}^+ |\boldsymbol{\phi^0}\rangle \\ (\boldsymbol{\rho^{0,\ast}})^{1/2}\, \hat{\mathbf{P^\ast}} \hat{f}^{-,\ast} |\boldsymbol{\phi^{0,\ast}}\rangle \\ -\sum_{i,j} \langle\phi_i|\hat{f}^+|\phi_j\rangle\,\hat{b}_i^\dagger\hat{b}_j \,|\mathbf{C^0}\rangle \\ \sum_{i,j} \langle\phi_j|\hat{f}^{-,\ast}|\phi_i\rangle\,\left(\hat{b}_i^\dagger\hat{b}_j\right)^\ast \, |\mathbf{C^{0,\ast}}\rangle \end{pmatrix} \nonumber \\
	&=\langle\mathbf{u}^k|\hat{f}^+ (\boldsymbol{\rho^0})^{1/2}|\boldsymbol{\phi^0}\rangle + \langle\mathbf{v}^k|\hat{f}^{-,\ast} (\boldsymbol{\rho^{0,\ast}})^{1/2}|\boldsymbol{\phi^{0,\ast}}\rangle \nonumber \\
	&+\sum_{i,j}\langle\phi_i|\hat{f}^+|\phi_j\rangle\langle {\mathbf{C}_u}^k|\hat{b}_i^\dagger\hat{b}_j \,|\mathbf{C^0}\rangle 
	+ \sum_{i,j}\langle\phi_j|\hat{f}^{-,\ast}|\phi_i\rangle\left\langle {\mathbf{C}_v}^k\left|\left(\hat{b}_i^\dagger\hat{b}_j\right)^\ast\right|\mathbf{C^{0,\ast}}\right\rangle
\end{align}
where the vector $\boldsymbol{|\phi^{0}\rangle}=\left(|\phi_1^0\rangle,...,|\phi_M^0\rangle\right)^t$ collects the stationary ground-state orbitals. Each response weight $\gamma_k$ quantifies how strong the $k$-th excitation contributes to the overall response of the system to an external perturbation with amplitudes $\hat{f}^+$ and $\hat{f}^-$. For instance, it may happen that for a chosen shape of the perturbing external fields, some excited states do not respond at all and thus appear to be transparent.

The response density of the $k$-th excited state due to the perturbation can in general be expressed as 
\begin{equation}\label{dens_resp_gen}
	\Delta\rho^k(\mathbf{r})=\Delta\rho_o^k(\mathbf{r})+\Delta\rho_c^k(\mathbf{r})
\end{equation}
where $\Delta\rho_o^k(\mathbf{r})$ and $\Delta\rho_c^k(\mathbf{r})$ denote the orbitals' and coefficients' contribution, respectively. Those are given by
\begin{align}
		\Delta\rho_o^k&=\langle\boldsymbol{\phi^{0}}|(\boldsymbol{\rho^0})^{1/2}\left( \mathbf{|u^k\rangle}+\mathbf{|v^{k,\ast}\rangle}\right)  \nonumber \\
		&+\left( \mathbf{|u^{k}\rangle}+\mathbf{|v^{k,\ast}\rangle}\right)^\dagger(\boldsymbol{\rho^{0,\ast}})^{-1/2}\boldsymbol{\rho^0}\, \boldsymbol{|\phi^0\rangle} 
\end{align}
and
\begin{equation}
		\Delta\rho_c^k=\sum_{i,j=1}^M \, \phi_i^{0,\ast}\phi_j^{0} \left( \langle \mathbf{C}^0 |\hat{b}_i^\dagger \hat{b}_j| {\mathbf{C}_u}^k \rangle+\langle {\mathbf{C}_v}^{k,\ast} |\hat{b}_i^\dagger \hat{b}_j| \mathbf{C}^0 \rangle \right).
\end{equation}

To summarize, the many-body LR theory based on the MCTDHB working equations, termed LR-MCTDHB, was introduced and derived. For a given $N$-bosons system in a potential trap $V$ and a two-body interaction $\hat{W}$, the central equation in order to obtain the excitation spectrum is Eq. (\ref{LR_eigenvalue_final}). Although this equation appears to be rather straightforward to solve, it in general represents a demanding task because firstly, $\mathcal{L}$ is not hermitian and secondly, it can become extremely large in size when the number of bosons $N$ and/or the number of orbitals $M$ grows. In Appendix \ref{Appendix_BlockDiagonal}, it is shown that the LR matrix can be brought to block-diagonal form by using a complex transformation, which halves the dimensionality of the eigenvalue problem. However, it involves matrix-matrix products of the individual blocks of $\mathcal{L}$, which can be numerically very expensive. It hence remains a system-dependent question whether it is beneficial to perform this transformation or not.

In the next section, Section \ref{Sec_Num}, details on the numerical implementation of LR-MCTDHB are presented. First, general rather technical challenges of the eigenvalue problem in Eq. (\ref{LR_eigenvalue_final}) are discussed before explaining the code structure and the numerical methods utilized. Finally, it is once again stressed that for $M=1$, LR-MCTDHB reduces to the (particle-conserving) BdG equations, i.e., BdG$\equiv$LR-GP$\equiv$LR-MCTDHB(1). The derivation of the BdG theory is described in Appendix \ref{Sec_BdG}. Since the BdG is commonly used to calculate excitations of trapped BECs, its results for certain applications (see Section \ref{Ch_applications}) are compared to the many-body results for $M>1$. 




\section{Numerical implementation of many-body linear response theory}\label{Sec_Num}
In this section, information about the LR-MCTDHB implementation are presented. Before dealing with the technical and numerical details in Sections \ref{Sec_NUM_Constr} and \ref{Sec_NUM_Diag}, a brief overview about the general structure of the code is given in Section \ref{Sec_Num_General}, together with a description of technical challenges that were to be solved.

\subsection{Code structure and challenges}\label{Sec_Num_General}
As a starting point, the structure of the numerical implementation, consisting of several thousand lines of code, is explained. The entire task of calculating the energies of excited states and the corresponding correction amplitudes of a trapped interacting BEC, i.e., solving the eigenvalue problem of Eq. (\ref{LR_eigenvalue_final}), is separated into two major parts. Those are (i) the construction of the LR matrix $\bar{\mathcal{L}}$ (simply denoted by $\mathcal{L}$ in the following) as described in Eq. (\ref{Eq_final_LR_matrix}) and (ii) finding a desired amount of eigenpairs, i.e., eigenvalues and the corresponding eigenvectors, of the low-energy spectrum. To this end, two running modes were implemented. Mode 1 is the construction mode. As an input, it requires a stationary state of a system of $N$ bosons in a trap $V$ with interaction $\hat{W}$, where the bosons can occupy $M$ single-particle orbitals. Typically, this stationary state is the ground state of the system, and it can therefore be obtained from a previous MCTDHB calculation using imaginary time propagation. The output of mode 1 is the LR matrix $\mathcal{L}$. Afterwards, the code can be restarted in mode 2, the diagonalization mode, which takes  $\mathcal{L}$ as an input and calculates a desired amount of eigenpairs in the low-energy part of the excitation spectrum. Furthermore, the response weights and response densities, described in Section \ref{Sec_LRMCTDHB}, are calculated.

The main challenge of both the construction and the eigenvalue calculation is the memory consumption of $\mathcal{L}$, and, especially in mode 1, the size of intermediate matrices that appear during the construction, like the projector $\mathbf{\mathcal{P}}$ or the individual blocks $\mathcal{L}_{oo},\,\mathcal{L}_{oc},\,\mathcal{L}_{co}$ and $\mathcal{L}_{cc}$. Although the memory consumption might be a minor issue for systems in 1D with small values of $N$ and $M$, it can become a massive problem for larger systems. To give an example, one can consider a system with $N=100$ particles and $M=4$ orbitals. In such a case, already the coefficient matrix $\mathcal{L}_{cc}$ has a dimensionality of $353702$, which means it would need approximately $4$ terabytes of memory to store all matrix elements, assuming those are in general complex numbers of 32 bytes. However, the available working memory (RAM) of compute nodes on modern high performance computation clusters is typically much smaller, being of the order of $128$ or $256$ gigabytes. One possible solution of this problem is to store all (large) matrices that appear during the construction of $\mathcal{L}$, as well as $\mathcal{L}$ itself, in a sparse matrix storage format. There are multiple possibilities to do that, e.g., the compressed-sparse-row (CSR) or compressed-sparse-column (CSC) format (see, e.g., Ref. \cite{Saad2} for additional formats).

To obtain eigenpairs of $\mathcal{L}$ by running the code in mode 2, one is in need of a numerical method that is capable of dealing with sparse and large non-hermitian matrices. The method of choice is the \textit{implicitly restarted Arnoldi method} (IRAM), implemented in the ARPACK numerical library \cite{Arpack_HP}. In the current implementation, the parallel version, termed PARPACK, is utilized \cite{Parpack}. The IRAM is presented in detail in Section \ref{Sec_NUM_Diag}. The PARPACK routines have a significant advantage over eigenvalue routines from other libraries like ScaLAPACK \cite{ScaLAPACK} because they do not require the matrix that one wishes to diagonalize as an input, not even in any sparse format. It is only required to hand over the result of matrix-vector multiplications (matvecs) as input, and the actual calculation of these matvecs takes place outside of any PARPACK routine, meaning that the implementation of the matvecs is completely independent. It can be carried out separately, utilizing routines from other numerical libraries, e.g., the Intel Math Kernel Library (MKL) \cite{MKL}, that calculate matvecs of large and sparse matrices in a very efficient manner. Moreover, calculating matvecs it is a process that can be easily parallelized which further improves the code performance.

Due to the large amount of operations in both running modes, especially for large $N$ and $M$ in two- or three-dimensional space, sequential runs of the code would require very long runtimes. Therefore, a fully MPI-parallelized implementation of the code was carried out for both running modes. Whereas this was a comparatively straightforward task for mode 2 -- since only the matvecs needed to be parallelized -- it turned out to be a demanding challenge with respect to mode 1. A parallelization scheme that is a compromise between taking care of the least possible memory consumption on the one hand, and distributing the workload among all participating MPI processes (PEs) as equal as possible on the other hand, is employed. In Section \ref{Sec_NUM_Constr}, the overall construction process is explained in more detail.

\subsection{Construction of the LR matrix}\label{Sec_NUM_Constr}
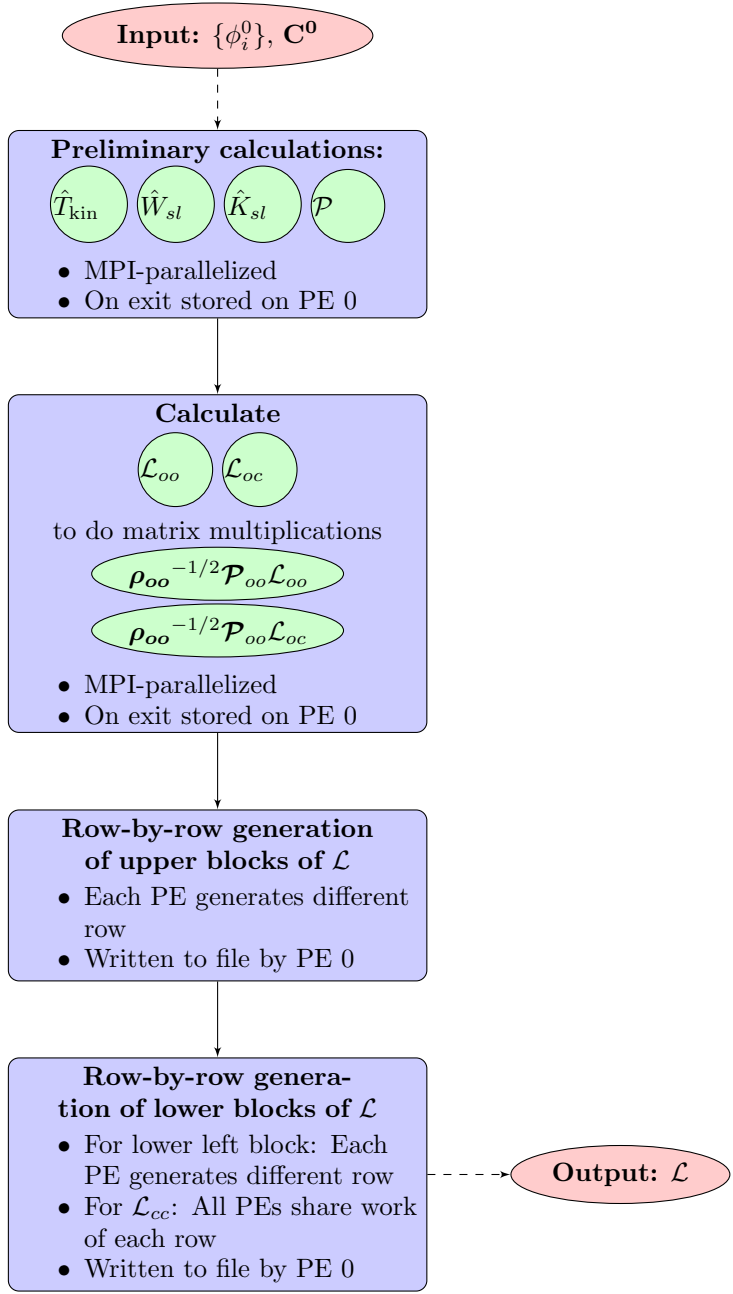
\begin{figure}[pt!]
 \begin{minipage}[c]{0.5\textwidth}
\tikzstyle{decision} = [diamond, draw, fill=blue!20, 
    text width=4.5em, text badly centered, node distance=3cm, inner sep=0pt]
\tikzstyle{block} = [rectangle, draw, fill=blue!20, 
    text width=5em, text centered, rounded corners, minimum height=4em]
\tikzstyle{largeblock} = [rectangle, draw, fill=blue!20, 
    text width=15em, text centered, rounded corners, minimum height=4em]
\tikzstyle{largeblock2} = [rectangle, draw, fill=blue!20, 
    text width=30em, text centered, rounded corners, minimum height=4em]
\tikzstyle{line} = [draw, -latex']
\tikzstyle{cloud} = [draw, ellipse,fill=red!20, node distance=3cm,
    minimum height=2em]
\tikzstyle{bubble} = [draw, circle,fill=green!20, node distance=1cm, text width = {width("blabla")},
    minimum height=2em, inner sep=0pt]
\tikzstyle{ellipse_2} = [draw, ellipse,fill=green!20, node distance=3cm, text width = {width("blablablablabla")},
    minimum height=2em, inner sep=0pt] 
    
\begin{tikzpicture}[auto]
    \node [cloud] (input) {\textbf{Input:} $\{\phi_i^0\},\,\mathbf{C^0}$};
    \node [largeblock, below of=input,node distance = 2.5cm] (pre) {\textbf{Preliminary calculations:}

    \tikz\node[bubble] (Tkin){$\hat{T}_{\text{kin}}$};
    \tikz\node[bubble, right of=Tkin] (Wsl){$\hat{W}_{sl}					$}; 
    \tikz\node[bubble, right of=Wsl] (Ksl){$\hat{K}_{sl}					$};
    \tikz\node[bubble, right of=Ksl] (projector){$\mathbf{\mathcal{P}}$};
    
    	\vspace*{-0.3cm}
    \begin{itemize}
   		\item{MPI-parallelized} \vspace*{-0.3cm}
   		\item{On exit stored on PE 0}
   	\end{itemize}
   
    };

    \node [largeblock, below of=pre,node distance = 4.5cm] (Loo) {\textbf{Calculate} \\
    \vspace*{0.1cm}
    \tikz\node[bubble, right of=projector] (Loo1){$\mathcal{L}_{oo}$}; 
    \tikz\node[bubble, right of=Loo1] (Loc1){$\mathcal{L}_{oc}$};
    
    to do matrix multiplications
    
 	\tikz\node[ellipse_2] (matmulLoo){$\boldsymbol{\rho_{oo}}^{-1/2} \boldsymbol{\mathcal{P}}_{oo}\mathcal{L}_{oo}$};  	
 	
 	\tikz\node[ellipse_2] (matmulLoc){$\boldsymbol{\rho_{oo}}^{-1/2} \boldsymbol{\mathcal{P}}_{oo}\mathcal{L}_{oc}$};
   
   	\vspace*{-0.3cm}
   	\begin{itemize}
   		\item{MPI-parallelized} \vspace*{-0.3cm}
   		\item{On exit stored on PE 0}
   	\end{itemize}
 
    };
   
     \node [largeblock, below of=Loo,node distance = 4.4cm] (upper) {\textbf{Row-by-row generation of upper blocks of $\mathcal{L}$} \\
     
     	\vspace*{-0.3cm}
   	\begin{itemize}
   		\item{Each PE generates different row} \vspace*{-0.3cm}
   		\item{Written to file by PE 0}
   	\end{itemize}
   
    };
 
 	\node [largeblock, below of=upper,node distance = 3.7cm] (lower) {\textbf{Row-by-row generation of lower blocks of $\mathcal{L}$} \\
     
     	\vspace*{-0.3cm}
   	\begin{itemize}
  	 	\item{For lower left block: Each PE generates different row} \vspace*{-0.3cm}
   		\item{For $\mathcal{L}_{cc}$: All PEs share work of each row} \vspace*{-0.3cm}
   		\item{Written to file by PE 0}
   	\end{itemize}

    };

    \node [cloud, right of=lower, node distance=5.3cm] (output) {\textbf{Output:} $\mathcal{L}$};
    \path [line] (pre) -- (Loo);
    \path [line] (Loo) -- (upper);
    \path [line] (upper) -- (lower);

 
    \path [line,dashed] (input) -- (pre);
    \path [line,dashed] (lower) -- (output);

\end{tikzpicture}
\end{minipage} \hfill
\begin{minipage}[c]{0.4\textwidth}
\caption[Schematic diagram of the MPI-parallelized implementation for the construction of the LR matrix]{Schematic diagram of the MPI-parallelized implementation for the construction of $\mathcal{L}$. The parallelization scheme shares the amount of calculations equally between all PEs. Before $\mathcal{L}$ is generated, preliminary calculations of important intermediate matrices are performed. To minimize the memory consumption, those are only stored on PE 0 in a sparse storage format. For the row-by-row construction of the upper and lower blocks of $\mathcal{L}$, PE 0 distributes all necessary information to the other PEs, collects the results afterwards and writes them to a file. See text for details. }
\label{fig:Fig_Constr}
\end{minipage}
\end{figure}

In this section, details on the implementation of the LR matrix construction are presented. Figure \ref{fig:Fig_Constr} shows the structure of the code and its main steps. As input, the code requires the stationary state to which the LR analysis should be applied, usually the ground state of a trapped BEC. In particular, one needs to hand over the set of orbitals $\{\phi_i^0:1 \leq i \leq M\}$ as well as the corresponding coefficients $\mathbf{C^0}$. Afterwards, intermediate matrices are calculated which are essential to built up the individual blocks of $\mathcal{L}$ as described in Eq. (\ref{LR_matrix}). This includes the kinetic energy operator $\hat{T}_{\text{kin}}$, the direct and exchange interaction operators $\hat{W}_{sl}$ and $\hat{K}_{sl}$, as well as the projection operator $\mathbf{\mathcal{P}}$. All of these calculations are done in parallel, and the final result is stored in a sparse storage format on the master process PE 0.

Then, the matrices $\mathcal{L}_{oo}$ and $\mathcal{L}_{oc}$ are calculated. Whereas $\mathcal{L}_{oo}$ is a $2(M\cdot N_{\text{grid}})$ square matrix with $N_{\text{grid}}$ denoting the number of grid points, $\mathcal{L}_{oc}$ is a $2(M\cdot N_{\text{grid}} \times N_{\text{conf}})$ matrix. It is stressed that constructing these two matrices is usually the most time-consuming part of the entire calculation of $\mathcal{L}$ due to the large amount of operations involved. On the one hand, for problems where the grid is large (e.g., in 2D and 3D) and $M\geq 7$, the construction of $\mathcal{L}_{oo}$ represents the essential part of the calculations. On the other hand, if the number of particles is large such that $N_{\text{conf}}=\mathcal{O}(10^5)$ or higher, the construction of $\mathcal{L}_{oc}$ becomes the most demanding task. Additionally, both matrices need to be multiplied by $\boldsymbol{\rho_{oo}}^{-1/2} \boldsymbol{\mathcal{P}}_{oo}$ from the left, which further increases the amount of operations significantly. Because of this, both the construction of $\mathcal{L}_{oo}$ and $\mathcal{L}_{oc}$ are MPI-parallelized, as well as the subsequent multiplication with the metric and projector matrices. As for the preliminary calculation described above, the final results of $(\boldsymbol{\rho_{oo}}^{-1/2} \boldsymbol{\mathcal{P}}_{oo}\mathcal{L}_{oo})$ and $(\boldsymbol{\rho_{oo}}^{-1/2} \boldsymbol{\mathcal{P}}_{oo}\mathcal{L}_{oc})$ are stored on PE 0 in a sparse storage format.

Next, the generation of $\mathcal{L}$ starts. Therefore, the problem is separated into the construction of the upper and lower part. With respect to the upper blocks, this means that only the upper left block needs to be multiplied by $\boldsymbol{\mathcal{P}}_{oo}\,\boldsymbol{\rho_{oo}}^{-1/2}$ from the right, whereas the upper right block is already fully calculated and can be written to the output file directly. To efficiently calculate the upper-left block, a row-by-row parallelization scheme is employed where each PE calculates a different row at a time. Therefore, PE 0 distributes all necessary matrix elements of $(\boldsymbol{\rho_{oo}}^{-1/2} \boldsymbol{\mathcal{P}}_{oo}\mathcal{L}_{oo})$ and $\mathbf{\mathcal{P}}$ to the other PEs. Finally, all PEs send back their calculated rows (i.e., the non-zero elements and their coordinates) to PE 0, which writes the results to the output file. 

Concerning the construction of the lower part, again a row-by-row parallelization scheme is utilized. For the lower-left block, each PE generates a different row at a time. However, in the case of $\mathcal{L}_{cc}$, all PEs work on the same row at a time. Upon finalization, the results are collected by PE 0 and written to the output file, again in a sparse storage format. 

This section is ended by briefly discussing the issue of small but non-zero matrix elements. As already stated, the matrices produced are typically sparse, i.e., most elements are zero. Moreover, many other elements can be very small but nonzero. Their impact on the subsequent eigenvalue calculation, which is described in the next sections, is very often negligible. However, for a problem where the dimensionality of $\mathcal{L}$ is very large, those elements can become challenging in terms of memory consumption. Thus, it is important to define a tolerance value $\tau$ that is used to minimize the amount of non-zero elements which need to be stored. In the current implementation, a matrix element is treated as zero and is therefore not stored if its magnitude is smaller than $\tau$. For very large matrices, the value of $\tau$ has to be defined such that on the one hand not too many matrix elements are set to zero in order to obtain numerically accurate results for the eigenvalues and eigenvectors, but, on the other hand, it has to be large enough because too many very small elements can significantly decrease the performance of the numerical method for calculating eigenpairs of $\mathcal{L}$. The latter numerical procedure will be explained in the next section.

\subsection{Calculation of eigenvalues}\label{Sec_NUM_Diag}
In order to calculate the lowest eigenvalues of a given LR matrix, the IRAM is employed. In the following, the main ideas of this approach are presented by first discussing the basic ingredients (the Arnoldi method, the QR-iteration, and polynomial filtering) before dealing with the IRAM explicitly. In short, it generates a Krylov subspace of a given matrix whose basis vectors are used to build a much smaller matrix. The eigenvalues of this matrix are also approximate eigenvalues of the original matrix. After each iteration of the IRAM, the convergence of a set of wanted eigenvalues is checked, and if it is still not converged, the entire process is restarted with an initial vector where the directions of unwanted eigenvalues are projected out. Thus, the convergence with respect to the wanted eigenvalues is gradually enhanced. Below, the individual steps of the IRAM are explained in more detail.

\subsubsection{The Arnoldi algorithm}\label{Sec_Arnoldi}
The first step is to calculate an orthogonal projection of a sparse, complex, in general non-hermitian matrix $A\in \mathbb{C}^{n\times n}$. Therefore, the Arnoldi algorithm is used which generates an orthonormal basis (ONB) of the $k$-dimensional subspace given by
\begin{equation}
	\mathcal{K}^k(A,v)=\text{span}(v,Av,...,A^{k-1}v)
\end{equation}
where $v\in \mathbb{C}^n$ is an initial vector and $k \leq n$, typically even $k \ll n$. $\mathcal{K}^k$ is called the $k$-dimensional Krylov subspace of $A$. The entire Arnoldi algorithm is given by Algorithm \ref{Arnoldi_alg}. The ONB is obtained by performing a Gram-Schmidt-orthonormalization at each Arnoldi step, i.e., making the vector $v_j$ in the $j$-th step orthogonal to all previous basis vectors $v_i,\,i\leq j,$ and normalizing it afterwards. The elements of the ONB are called the Arnoldi vectors. If the resulting Arnoldi vector $v_{j+1}$ of the $j$-th step is non-zero, its norm is stored in the matrix element $h_{j+1,j}$ of the matrix $\tilde{H}_{k+1}\in \mathbb{C}^{(k+1)\times k}$. If $j\leq k$, the next Arnoldi step is initialized. After the step for $j=k$, the iteration stops and returns the $k$-dimensional ONB of  $\mathcal{K}^k$, plus an additional orthonormal vector $v_{k+1}\in \mathbb{C}^n$, together with the full matrix $\tilde{H}_{k+1}$. It is possible that for a certain step $j<k$, the norm of the Arnoldi vector $v_{j+1}$ is zero, meaning that one has found an ONB of an invariant Krylov subspace of $A$ with dimensionality $j$. In this case, the Arnoldi algorithm finishes immediately.

\begin{algorithm}[t]
\begin{algorithmic}[1]
\STATE $v_1 \gets v,\, ||v||_2=1$ 
\FOR{$j = 1, \, j\leq k$}
	\STATE $z \gets A v_j$
	\FOR{$i=1,\,i \leq j$}
		\STATE $h_{ij} \gets \langle v_i,z\rangle$
		\STATE $z \gets z-h_{ij}v_i$
	\ENDFOR
	\STATE $h_{j+1,j} \gets ||z||_2$
	\IF{$h_{j+1,j}=0$}
		\STATE quit
	\ELSE
		\STATE $v_{j+1} \gets z/h_{j+1,j}$
	\ENDIF
\ENDFOR
\end{algorithmic}
\caption[The Arnoldi algorithm]{The Arnoldi algorithm}
\label{Arnoldi_alg}
\end{algorithm}

As an essential result, one obtains the Arnoldi decomposition of $A$ given by 
\begin{equation}\label{Arnoldi_decomp}
	A V_k=V_k H_k+h_{k+1,k}\,v_{k+1}\,e_k^t
\end{equation}
where $V_k \in \mathbb{C}^{n\times k}$ contains the Arnoldi basis as columns and $H_k\in \mathbb{C}^{k\times k}$ is an upper Hessenberg matrix, i.e., $\{(H_k)_{ij}\}=0$ if $i>j+1$, built by the first $k$ rows of $\tilde{H}_{k+1}$. The second term in Eq. (\ref{Arnoldi_decomp}), which contains the matrix element $h_{k+1,k}$, the additional Arnoldi vector $v_{k+1}$ and the transpose of the $k$-th unit vector, i.e., $e_k=(0,0,...,0,1)^t\in\mathbb{C}^k$, takes care of the fact that the orthonormal projection of $A$, 
\begin{equation}\label{Arnoldi_orth_proj}
	A\approx V_k H_k V_k^\dagger,
\end{equation}
is in general not exact. However, for the case of having found an invariant Krylov subspace of $A$ of lower dimensionality $j \leq k$, it becomes exact. It is straightforward to realize that the eigenvalues of $H_k$ are approximate, or, in the case of an invariant subspace, exact eigenvalues of A. This is due to the cyclic permutation symmetry of the determinant that defines the characteristic polynomial of $A$, i.e., 
\begin{align}
	0&=\det(A-\lambda I) \overset{\text{Eq.\,}(\ref{Arnoldi_orth_proj})}{\approx} \det(V_k H_k V_k^\dagger-\lambda I) \nonumber \\
	&=\det(V_k [H_k-\lambda I] V_k^\dagger) \nonumber \\
	&=\det( \underbrace{V_k^\dagger V_k}_{=I} [H_k-\lambda I]) \nonumber \\
	&=\underbrace{\det(I)}_{=1}\,\det(H_k-\lambda I) \nonumber \\
	&=\det(H_k-\lambda I)
\end{align}
with $\lambda\in\mathbb{C}$ and $I$ denoting the identity matrix. 

If $u_i\in\mathbb{C}^k$ is an eigenvector of $H_k$ and $\theta_i\in\mathbb{C}$ the corresponding eigenvalue, the pair $(\theta_i,V_k\,u_i)$ is called a Ritz pair of $A$ with Ritz value $\theta_i$ and the associated Ritz vector $V_k\,u_i\in\mathbb{C}^n$. In order to estimate how accurate this Ritz pair approximates an exact eigenpair of $A$, one can calculate the residual norm $r$ defined as
\begin{align}\label{res_norm}
	r&\equiv ||A(V_k u_i)-\theta_i (V_k u_i)||_2  \overset{\text{Eq.\,}(\ref{Arnoldi_decomp})}{=}  ||(V_k H_k+h_{k+1,k}\,v_{k+1}\,e_k^t) u_i-\theta_i (V_k u_i)||_2  \nonumber \\
	&=||\theta_i (V_k u_i)+h_{k+1,k}\,v_{k+1}\,e_k^t\,u_i-\theta_i (V_k u_i)||_2 \nonumber \\
	&=|h_{k+1,k}|\cdot\underbrace{|v_{k+1}|}_{=1}\cdot|e_k^t u_i| \nonumber \\
	&=|h_{k+1,k}|\cdot|e_k^t u_i|,
\end{align}
i.e., $r$ is the product of the magnitudes of the matrix element $h_{k+1,k}$ of $\tilde{H}_{k+1}$ and the $k$-th element of $u_i$. The smaller it is, the better does the Ritz pair $(\theta_i,V_k u_i)$ approximate the exact $i$-th eigenpair of $A$. 

To summarize, the eigenvalue problem of a typically large matrix $A$ can be mapped onto an eigenvalue problem of a much smaller matrix $H_k$ whose eigenvalues are believed to be good approximations of the eigenvalues of $A$. Due to the Hessenberg form of $H_k$, its eigenvalues can efficiently be calculated by utilizing the QR-iteration, which is explained in the next section. Additional details on the Arnoldi algorithm can be found in Ref. \cite{Mehl_num}.

\subsubsection{The QR-iteration}\label{Sec_QR}
The QR-iteration is a technique to calculate the eigenvalues and eigenvectors of a given matrix $A\in\mathbb{C}^{n\times n}$. It is based on the so-called QR-decomposition which reads
\begin{equation}\label{QR_decomp}
	A=QR, \quad Q,R\in\mathbb{C}^{n\times n}
\end{equation}
where $Q$ is a unitary matrix, i.e., $QQ^\dagger=I$, and $R$ is upper triangular. This decomposition exists for any matrix \cite{Mehl_num}. Plugging the ansatz of Eq. (\ref{QR_decomp}) into the characteristic polynomial gives
\begin{align}
	0&=\det(A-\lambda I) \nonumber \\
	&=\det(QR-\lambda I) \nonumber \\
	&=\det(QQ^\dagger(QR)QQ^\dagger-Q\lambda IQ^\dagger) \nonumber \\
	&=\det(Q(RQ-\lambda I)Q^\dagger) \nonumber \\
	&=\det(RQ-\lambda I)
\end{align}
which means that the eigenvalues of $A$ are the same as the eigenvalues of the matrix $RQ$. Furthermore, 
\begin{equation}
	A=QR \iff Q^\dagger A Q=RQ,
\end{equation}
i.e., $RQ$ is just a unitary transformation of $A$. 

The QR-iteration is described in Algorithm \ref{QR_iter}. At each step $k$, the QR-decomposition of the current matrix $A_k$ is used to define the matrix of the next step as $A_{k+1}=R_kQ_k=Q_k^\dagger A_k Q_k $. It can be shown that this sequence converges to a triangular matrix, i.e., $\lim_{k\rightarrow\infty} A_k=\tilde{A}$ with $\{\tilde{A}_{ij}\}=0$ if $i>j$, and the eigenvalues of a triangular matrix are simply given by its diagonal elements. One can show that the eigenvalues of $A_k$ are the same as the ones of $A$ by considering again the characteristic polynomial 
\begin{align}
	0&=\det(A_k-\lambda I) \nonumber \\
	&=\det(Q_{k-1}^\dagger A_{k-1} Q_{k-1}-\lambda I) \nonumber \\
	&=\det(Q_{k-1}^\dagger Q_{k-2}^\dagger A_{k-2} Q_{k-2}Q_{k-1}-\lambda I) \nonumber \\
	&=...=\det(Q_{k-1}^\dagger Q_{k-2}^\dagger...Q_{0}^\dagger \underbrace{A_0}_{=A} Q_0... Q_{k-2}Q_{k-1}-\lambda I) \nonumber \\
	&=\det(A-\lambda I)
\end{align}
where the cyclic permutation symmetry of the determinant is used in the last step. 

Although the QR-iteration possibly requires a large amount of operations per step for a general matrix $A$, it turns out to be of very low computational cost with respect to upper Hessenberg matrices like the matrix $H_k$ obtained from a $k$-step Arnoldi process described in Algorithm \ref{Arnoldi_alg}. In such a case, one can apply $(k-1)$ Givens rotations $G_i$, $i\in\{1,...,k-1\}$, which are all unitary operations such that the overall QR-decomposition of $H_k$ at each QR-step $j$ is given by 
\begin{equation}
	(H_k)_j= Q_j R_j=(G_{k-1})_j...(G_1)_j R_j.
\end{equation}
Further details on Gives rotations can be found in Ref. \cite{Saad}. With respect to the IRAM, the QR-iteration turns out to be important not only because it is used to calculate the eigenvalues of $H_k$, but also in terms of restarting the Arnoldi iteration with an initial vector that incorporates shifts that filter out the direction of unwanted eigenvectors of $H_k$. Those shifts are obtained by utilizing QR-steps, which are explained in detail in Section \ref{Sec_IRAM}. In Ref. \cite{Mehl_num}, further details on the QR-iteration are presented.  

\begin{algorithm}[t]
\begin{algorithmic}[1]
\STATE $A_0 \gets A$
\STATE $\text{Calculate the}\,\text{QR-decomposition}\,A_0 = Q_0 R_0$ 
\FOR{$k = 1\,\text{until convergence}$}
	\STATE $A_k \gets R_{k-1} Q_{k-1}$
	\IF{\text{$A_k$ "triangular enough"}}
		\STATE quit
	\ELSE
		\STATE $\text{Calculate the}\,\text{QR-decomposition}\,A_k = Q_k R_k$ 
	\ENDIF
\ENDFOR
\end{algorithmic}
\caption[The QR-iteration]{The QR-iteration}
\label{QR_iter}
\end{algorithm}

\subsubsection{Polynomial filtering}
Before dealing with the IRAM in particular, a brief discussion on polynomial filtering techniques of the initial vector $v$ of a Krylov subspace iteration, like the Arnoldi algorithm of Section \ref{Sec_Arnoldi}, is realized. In general, the target of applying a filter to a vector $v$ is to make the contribution of unwanted directions very small, or even zero. 

To understand how polynomial filtering works in general, a square matrix $A\in\mathbb{C}^{n\times n}$ whose eigenpairs are given by the set $\{(\lambda_i,v_i)|1\leq i \leq n\}$ with $\lambda_i\in\mathbb{C}$ and $v_i\in\mathbb{C}^n$ is considered. Suppose the eigenvectors form a basis. Each arbitrary vector $v\in\mathbb{C}^n$ can then be written as $v=\sum_{j=1}^n c_j v_j$ with complex expansion coefficients $\{c_j\}$. Assuming that one wants to filter out the contribution of the $k$-th eigenvector $v_k$, one can define 
\begin{equation}\label{pol_filter}
	p(t)\equiv t-\lambda_k
\end{equation}
and apply this very simple polynomial onto $v$:
\begin{align}
	p(A)v&=\sum_{j=1}^n c_j p(A) v_j=\sum_{j=1}^n c_j (A-\lambda_k) v_j \nonumber \\
	&=\sum_{j=1}^n c_j (\lambda_j-\lambda_k) v_j \nonumber \\
	&=\sum_{j\neq k}^n c_j (\lambda_j-\lambda_k) v_j.
\end{align}
If $A$ is hermitian, the overlap with $v_k$ vanishes, i.e., 
\begin{equation}
	\langle v_k|p(A)v\rangle =\sum_{j\neq k}^n c_j (\lambda_j-\lambda_k) \underbrace{\langle v_k|v_j\rangle}_{=\delta_{kj}} = 0,
\end{equation}
because in that case the eigenvectors of $A$ would be orthogonal. Thus, the vector $p(A)v$ is orthogonal to $v_k$, and any additional power iteration of this vector with respect to $A$ preserves this property, meaning that
\begin{equation}
	\langle v_k|A^n p(A)v\rangle =\sum_{j\neq k}^n c_j \lambda_j^n(\lambda_j-\lambda_k) \underbrace{\langle v_k|v_j\rangle}_{=\delta_{kj}} = 0.
\end{equation}
Therefore, the Arnoldi iteration, applied to an initial vector $p(A)v$, yields a Krylov basis which is orthogonal to $v_k$, and also the orthogonal projection $H_k$ would not contain $\lambda_k$ in its spectrum of eigenvalues. If $A$ is non-hermitian, the overlap of $p(A)v$ and $v_k$ is not strictly zero since the eigenvectors $\{v_i\}$ can be linearly dependent. However, within the IRAM, the overall contribution of $v_k$ gradually decreases upon increasing the number of restarts with properly shifted initial vectors. Nevertheless, this shows that the LR-matrix $\mathcal{L}$, which is in general non-hermitian, leads to additional numerical effort which would be absent in the case of a hermitian matrix.

It is also possible to do multiple shifts at once by using a filter given by
\begin{equation}\label{pol_filter2}
	p(t)= \prod_{i=1}^N  \left(t-\lambda_{\pi(i)}\right)
\end{equation}
where $N\leq n$ is the number of shifts and $\pi(i)\in\{1,...,n\}\,\forall \,i$. In this case, the directions of $N$ different eigenvectors of $A$ are filtered out. The above choices of $p(t)$ do only reflect the simplest possibilities of filter polynomials, but also more sophisticated ones like the Chebyshev polynomials can be chosen \cite{Saad}.

\subsubsection{The implicitly restarted Arnoldi method (IRAM)}\label{Sec_IRAM}

\begin{algorithm}[t!]
\begin{algorithmic}[1]
\STATE Take an initial vector $v_1 \gets v$, $||v||_2=1$
\STATE Perform a $k$-step Arnoldi iteration to get \\ $\quad\quad A V_k=V_k H_k+f_k\,e_k^t$ with $f_k:=h_{k+1,k}\,v_{k+1}$
\STATE Calculate the eigenvalues $\theta_1,...,\theta_k$ of $H_k$ by using the QR-iteration
\WHILE{$m$ wanted eigenvalues of $H_k$ not converged}
	\STATE Select $p=k-m$ unwanted eigenvalues $\theta_1,...,\theta_p$ 
 	\STATE Perform $p$ QR-steps with unwanted eigenvalues as shifts: \vspace{0.1cm} \\ $\quad\quad [H^{(p)}_k,Q]:=QR[H_k,\theta_1,...,\theta_p], \quad V^{(p)}_k:=V_kQ $ \vspace{0.1cm}
 	\STATE Set $f_m:=h_{k+1,k}\,Q_{kp}\,v_{k+1}+h^{(p)}_{m+1,m}\,v^{(p)}_{m+1} $ \vspace{0.1cm}
 	\STATE Define $H_m:=H^{(p)}_k(1:m,1:m)$, $V_m:=V^{(p)}_k(1:n,1:m)$
 	\STATE Perform $p$ additional Arnoldi steps onto $V_m,\, H_m$ and $f_m$ to obtain \\  $\quad\quad$a $k$-step Arnoldi decomposition
 	\STATE Calculate the eigenvalues of the new $H_k$ by using the QR-iteration
\ENDWHILE
\end{algorithmic}
\caption[The implicitly restarted Arnoldi method (IRAM)]{The implicitly restarted Arnoldi method (IRAM)}
\label{Alg_IRAM}
\end{algorithm}

After the introduction of the three essential ingredients of the IRAM, namely the Arnoldi algorithm, the QR-iteration, and polynomial filtering, one can finally discuss the IRAM itself in detail. The full algorithm is given in Algorithm \ref{Alg_IRAM}. At first, a $k$-step Arnoldi iteration is performed, yielding an Arnoldi decomposition of $A$ as described in Eq. (\ref{Arnoldi_decomp}). Then, the eigenvalues $(\theta_1,...,\theta_k)$ are calculated utilizing the QR-iteration. From this set, $p$ unwanted eigenvalues $(\theta_1,...,\theta_p)$ are selected. If one for example seeks the eigenvalues of $A$ that have the smallest magnitude, an appropriate selection would be the $p$ eigenvalues of $H_k$ that have the largest magnitude. Afterwards, a sequence of $p$ QR-shifts with the unwanted eigenvalues is performed. For the first shift, the Arnoldi decomposition reads
\begin{align}
	(A-\theta_1 I)V_k&=V_k(H_k-\theta_1 I)+h_{k+1,k}\,v_{k+1}\,e_k^t  \\
	\iff (A-\theta_1 I)V_k&=V_kQ_1R_1+h_{k+1,k}\,v_{k+1}\,e_k^t \label{IRAM_eq_1} \\
	\iff AV_kQ_1&=V_kQ_1(R_1Q_1+\theta_1 I)+h_{k+1,k}\,v_{k+1}\,e_k^t \,Q_1
\end{align}
where the QR-decomposition $H_k-\theta_1 I=Q_1R_1$ is used in Eq. (\ref{IRAM_eq_1}). Setting
\begin{align}\label{IRAM_def_step1}
	H_k^{(1)}:=R_1Q_1+\theta_1 I, \quad V_k^{(1)}:=V_kQ_1, \quad \left(e_k^{(1)}\right)^t:=e_k^tQ_1
\end{align}
the next QR-shift can be applied:
\begin{align}
	(A-\theta_2 I)V_k^{(1)}&=V_k^{(1)}(H^{(1)}_k-\theta_2 I)+h_{k+1,k}\,v_{k+1}\,\left(e_k^{(1)}\right)^t  \\
	\overset{H^{(1)}_k-\theta_2 I=Q_2R_2}{\iff} AV_k^{(1)}Q_2&=V_k^{(1)} Q_2 (R_2Q_2+\theta_2 I)+h_{k+1,k}\,v_{k+1}\,\left(e_k^{(1)}\right)^t  \,Q_2  	
\end{align}
where one can define, similar to Eq. (\ref{IRAM_def_step1}),
\begin{align}\label{IRAM_def_step2}
	H_k^{(2)}:=R_2Q_2+\theta_2 I, \quad V_k^{(2)}:=V_k^{(1)}Q_2, \quad \left(e_k^{(2)}\right)^t:=\left(e_k^{(1)}\right)^tQ_2.
\end{align}
This procedure is continued until all $p$ QR-shifts are applied.

Multiplying Eq. (\ref{IRAM_eq_1}) by $e_1$ yields
\begin{align}
	(A-\theta_1 I)V_ke_1&=V_kQ_1R_1e_1+h_{k+1,k}\,v_{k+1}\,e_k^t e_1 \nonumber \\
	\iff (A-\theta_1 I)v_1&=V_kQ_1r_{11}e_1 \nonumber \\ 	\iff (A-\theta_1 I)v_1&=v^{(1)}_1r_{11}
\end{align}
where $r_{11}$ is the upper left element of $R_1$ and $v_1^{(1)}$ is the first column of $V_kQ_1$. This in turn means that the first vector of the new basis, i.e., $v^{(1)}_1$, is proportional to $(A-\theta_1 I)v_1$. As discussed in the previous section, this polynomial filter reduces the contribution of the unwanted eigenvector corresponding to $\theta_1$ in $v^{(1)}_1$, or, in the optimal case, makes its contribution even vanish completely. This holds for all successive vectors in $V_k^{(1)}$. After having applied the second QR-shift, one can show that $v_1^{(2)}\sim (A-\theta_2 I)(A-\theta_1 I)v_1$. Continuing in the same manner until the $p$-th QR-shift is applied yields $v_1^{(p)}\sim \prod_{i=1}^p(A-\theta_i I)v_1$. Then, by taking the first $m=k-p$ columns of $V_k^{(p)}$ as the new Arnoldi basis $V_m$, as well as taking the leading principal $(m\times m)$-matrix of $H_k^{(p)}$ as the new orthogonal projection $H_m$ of $A$, one obtains an $m$-step Arnoldi decomposition $A V_m=V_m H_m+f_m\,e_m^t$ with an error term given by
\begin{equation}
	f_m=h_{k+1,k}\,Q_{kp}\,v_{k+1}+h^{(p)}_{m+1,m}\,v^{(p)}_{m+1}
\end{equation}
where it is used that the unitary matrix $Q$ is of Hessenberg form. Based on this, one performs $p$ additional Arnoldi steps yielding again a $k$-step Arnoldi decomposition. The essential difference to the previous $k$-step Arnoldi decomposition is that now all basis vectors in $V_k$ are approximately or even fully orthogonal to the eigenvectors of the associated unwanted eigenvalues. The accuracy of the eigenvalues and eigenvectors of the newly computed matrix $H_k$ can be estimated with the residual norm $r$ introduced in Section \ref{Sec_Arnoldi}, Eq. (\ref{res_norm}). If it is below a predefined tolerance for a desired amount of wanted eigenpairs, the algorithm finishes. Otherwise, the set of eigenvalues of $H_k$ is again separated into wanted and unwanted eigenvalues, and the procedure of QR-shifting and refining the given Arnoldi basis restarts.

By employing the above described shift strategy without restarting the Arnoldi iteration from the very beginning, one obtains the same result as if an explicit restart of the entire $k$-step Arnoldi process with the initial vector $v_1^{(p)}$ had been performed. Because of this similarity, the overall algorithm is said to be \textit{implicitly} restarted. The computational effort in terms of matvecs with $A$ is given by $(k+p)$ operations in both cases, i.e., for explicit and implicit restarting. However, implicit restarting turned out to be numerically more stable, and has additional advantages which are beyond the scope of this work. 

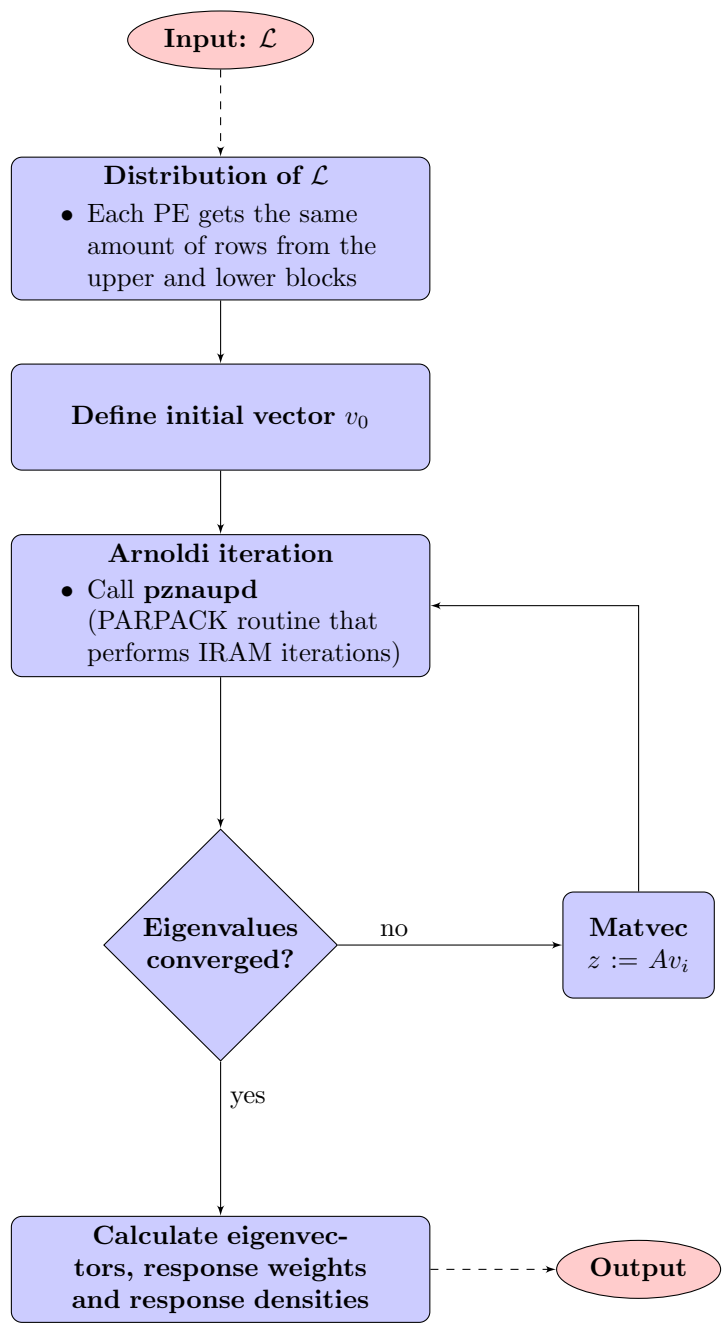
\begin{figure}[ht!]
 \begin{minipage}[c]{0.5\textwidth}
\tikzstyle{decision} = [diamond, draw, fill=blue!20, 
    text width=4.5em, text badly centered, node distance=3cm, inner sep=0pt, text width = {width("blablablablabla")}]
\tikzstyle{block} = [rectangle, draw, fill=blue!20, 
    text width=5em, text centered, rounded corners, minimum height=4em]
\tikzstyle{largeblock} = [rectangle, draw, fill=blue!20, 
    text width=15em, text centered, rounded corners, minimum height=4em]
\tikzstyle{largeblock2} = [rectangle, draw, fill=blue!20, 
    text width=30em, text centered, rounded corners, minimum height=4em]
\tikzstyle{line} = [draw, -latex']
\tikzstyle{cloud} = [draw, ellipse,fill=red!20, node distance=3cm,
    minimum height=2em]
\tikzstyle{bubble} = [draw, circle,fill=green!20, node distance=1cm, text width = {width("blabla")},
    minimum height=2em, inner sep=0pt]
\tikzstyle{ellipse_2} = [draw, ellipse,fill=green!20, node distance=3cm, text width = {width("blablablabla")},
    minimum height=2em, inner sep=0pt] 
    
\begin{tikzpicture}[auto]
    \node [cloud] (input) {\textbf{Input:} $\mathcal{L}$};
    \node [largeblock, below of=input,node distance = 2.5cm] (dist) {\textbf{Distribution of $\mathcal{L}$} 
    \vspace*{-0.2cm}
  	\begin{itemize}
   		\item{Each PE gets the same amount of rows from the upper and lower blocks} 
   	\end{itemize}
   
    };

    \node [largeblock, below of=dist,node distance = 2.5cm] (init) {\textbf{Define initial vector $v_0$} };
   
    \node [largeblock, below of=init,node distance = 2.5cm] (Arnoldi) {\textbf{Arnoldi iteration} \\
     
     	\vspace*{-0.3cm}
   	\begin{itemize}
   		\item{Call \textbf{pznaupd} \\ (PARPACK routine that performs IRAM iterations)} 
   	\end{itemize}
   
    };
 
 	\node [decision, below of=Arnoldi,node distance = 4.5cm] (dec) {\textbf{Eigenvalues converged?}};
   
   	\node [block, right of=dec,node distance = 5.5cm] (matvec) {\textbf{Matvec $z:=Av_i$}};
   	
   	\node [largeblock, below of=dec,node distance = 4.3cm] (EVcalc) {\textbf{Calculate eigenvectors, response weights and response densities}};

    \node [cloud, right of=EVcalc, node distance=5.5cm] (output) {\textbf{Output}};
    \path [line] (dist) -- (init);
    \path [line] (init) -- (Arnoldi);
  	\path [line] (Arnoldi) -- (dec);
	\path [line] (dec) -- node [near start] {no} (matvec);
 	\path [line] (matvec) |- (Arnoldi);
    \path [line,dashed] (input) -- (dist);
    \path [line] (dec) -- node [near start] {yes} (EVcalc);
    \path [line,dashed] (EVcalc) -- (output);

\end{tikzpicture}
\end{minipage} \hfill
\begin{minipage}[c]{0.4\textwidth}
\vspace*{-10cm}\caption[Schematic diagram of the MPI-parallelized implementation for the diagonalization of the LR matrix]{Schematic diagram of the MPI-parallelized implementation for the diagonalization of $\mathcal{L}$. Different chunks of non-zero elements of $\mathcal{L}$ are distributed among the PEs. The Arnoldi iteration is carried out by the routine $\textbf{pznaupd}$ of Parallel ARPACK (PARPACK) \cite{Parpack}, which either finishes upon convergence of a desired amount of eigenvalues or expects the user to perform a matvec and return the result to it. Finally, the eigenvalues, eigenvectors, as well as response weights and response densities are calculated and written to output files. See text for details.}
\label{fig:Fig_Diag_code}
\end{minipage}
\end{figure}

Figure \ref{fig:Fig_Diag_code} shows the general structure of the code for the matrix diagonalization utilizing the implementation of IRAM in the PARPACK library \cite{Parpack}. Essential for the performance is the efficient parallel implementation of the matvecs which are done outside of the PARPACK routine \textbf{pznaupd}. The latter carries out the IRAM iterations in parallel \cite{Parpack}.  It is important to note that the IRAM, although it is particularly efficient for sparse matrices, would also work for dense matrices. Further details are presented in Refs. \cite{Saad,Arpack_HP,Mehl_num}.

To summarize, the newly developed implementation of LR-MCTDHB was presented in this section and the code was explained in detail. Furthermore, the IRAM, a very powerful method for computing eigenpairs of very large and sparse matrices, was described. In the next section, the code is benchmarked against an exactly-solvable model.

\clearpage



\newpage
\section{Comparison with an exactly-solvable model}\label{Ch_Benchmark}
After the presentation of the theoretical and numerical frameworks utilized, along with the structure of the implemented code, the newly developed implementation of LR-MCTDHB is benchmarked in the following. Furthermore, since it builds the basis for the utilized LR theory, results from a benchmark of the MCTDHB implementation used in this work \cite{package}, which makes use of a sophisticated mapping of bosonic operators in Fock space \cite{Streltsov_combi}, are presented as well. This implementation is now also available with a recently developed graphical user interface and works on all common operating systems \cite{MCTDHB-lab}. Other codes became available during the last years, e.g., the MCTDH-X software package \cite{MCTDH-X} which contains a separate graphical user interface. Furthermore, the multi-layer (ML-)MCTDHB approach \cite{ML_first,ML-MCTDHB-impl,ML-MCTDHB-extension}, which is particularly suited for the description of Bose-Bose or Bose-Fermi mixtures, is now implemented. 

As a reference system, the exactly-solvable harmonic-interaction model (HIM) is used. In the latter model, both the single-particle and interaction potentials are of harmonic type. This model system has been used recently to benchmark the fermionic counterpart of MCTDHB, termed MCTDHF \cite{MCTDHF_Lode_implementation}, as well as the time-dependent restricted-active-space  self-consistent-field theory for bosons (TD-RASSCF-B) \cite{Summary_Leveque} and its variant for bosonic mixtures \cite{Leveque_mixtures}. In the first part of this section, an introduction to the HIM and the analytic results of its energy spectrum are given (Section \ref{Sec_HIM_Intro}). Afterwards, an overview of the MCTDHB benchmark against the HIM, which has been carried out in great detail in Ref. \cite{MCTDHB_benchmark}, is presented (Section \ref{Sec_MCTDHB_benchmark}). This includes both the ground state and the out-of-equilibrium quantum dynamics of a bosonic many-particle system. Finally, a comparison of the numerical results of the low-energy spectrum obtained from LR-MCTDHB and the analytic values, both for the 1D and 2D cases, is made (Section \ref{Sec_LRMCTDHB_benchmark}).

\subsection{The harmonic-interaction model (HIM)}\label{Sec_HIM_Intro}
The HIM Hamiltonian contains the harmonic trap potential 
\begin{equation}
V(\mathbf{r})=\frac{1}{2}\left(\Omega_x^2 x^2+\Omega_y^2 y^2+\Omega_z^2 z^2 \right), \quad \mathbf{r}=(x,y,z)^t
\end{equation}
where the boson mass $m$ is set to unity. In $D>1$ dimensions, one distinguishes between the isotropic and anisotropic HIM, where the former refers to the case where all frequencies $\{\Omega_i\}$ are equal, whereas the latter refers to the case where at least one frequency differs from the others. 

The interaction between the $i$-th and $j$-th boson is described as 
\begin{equation}
\hat{W}(\mathbf{r}_i,\mathbf{r}_j)=\lambda_0|\mathbf{r}_i-\mathbf{r}_j|^2
\end{equation}
where a positive (negative) interaction strength $\lambda_0$ denotes attraction (repulsion) between the bosons. 

As described in Ref. \cite{Cohen}, the following coordinate transformation yields the separation of the relative and center-of-mass (c.m.) coordinates,
\begin{equation}
	\mathbf{Q}_k=\frac{1}{\sqrt{k(k+1)}}\sum_{i=1}^k(\mathbf{r}_{k+1}-\mathbf{r}_{i}), \quad 1\leq k \leq N-1
\end{equation}
and
\begin{equation}
	\mathbf{Q}_N=\frac{1}{\sqrt{N}}\sum_{i=1}^k \mathbf{r}_i,
\end{equation}
where $N$ is the number of bosons in the system. This leads to the Hamiltonian in the c.m. frame given by
\begin{align}
\hat{H}&=\hat{H}_{rel}+\hat{H}_{c.m.} \\
\hat{H}_{rel}&=\frac{1}{2}\sum_{j=x,y,z}\sum_{k=1}^{N-1} (p_{j,k}^2+\delta_j^2 Q_{j,k}^2) \label{HIM_Hrel} \\
\hat{H}_{c.m.}&=\frac{1}{2} \sum_{j=x,y,z}(p_{j,N}^2+\Omega_j^2 Q_{j,N}^2) \label{HIM_Hcom}
\end{align} 
with momenta $p_{j,N}=(1/i)\partial_{Q_{j,N}}$ of the c.m. coordinates, $p_{j,k}=(1/i)\partial_{Q_{j,k}}$ of the relative coordinates, and the parameters $\delta_{j}^2=\Omega_{j}^2\pm2N|\lambda_0|$ where again the plus (minus) sign denotes attraction (repulsion) between the bosons. For simplicity $\hbar=1$ is assumed. It is instructive to interpret the Hamiltonian as a composite model system of $(N-1)$ identical non-interacting particles in a harmonic confinement with trap frequencies $\mathbf{\delta}=(\delta_x,\delta_y,\delta_z)^t$ and the c.m. particle which moves in the harmonic confinement with the original trap frequencies $\{\Omega_i\}$.

The ground-state wave function in the c.m. frame takes the form 
\begin{equation}\label{HIM_GS_WF}
	\Psi_0(\mathbf{Q}_1,...,\mathbf{Q}_N)= \prod_{j=x,y,z}\left( \frac{\delta_j}{\pi} \right)^{\frac{N-1}{4}} \left( \frac{\Omega_j}{\pi} \right)^{\frac{1}{4}} \exp\left(-\frac{1}{2\delta_j} \sum_{k=1}^{N-1}Q_{j,k}^2-\frac{\Omega_j}{2}  Q_{j,N}^2\right),
\end{equation}
i.e., it is basically a product of $N$ harmonic oscillator (HO) ground-state wave functions in each spatial direction \cite{Cohen}. However, in the laboratory frame, the representation of the ground state is much more involved. Already for $N=2$ bosons in 1D, it is an infinite sum of Hartree products of HO eigenfunctions instead of a single product of two HO ground states in the c.m. frame \cite{MCTDHB_benchmark}. Numerical convergence towards this state is therefore a very demanding challenge in the laboratory frame.

The exact solution for the energy levels reads
\begin{align}
	E\left(\{n_j\},\{m_j\}\right)&=\sum_{j=x,y,z}\left\{\left(n_j+\frac{N-1}{2}\right)\delta_j+\left(m_j+\frac{1}{2}\right)\Omega_j\right\}	
\end{align}
where the $\{n_j\}$ denote the quantum numbers of excitations of the relative coordinates and the $\{m_j\}$ denote the quantum numbers of c.m. excitations in the $x$-, $y$- and $z$-directions. In particular, this means that the ground-state energy is given by
\begin{equation}\label{HIM_GS_Energy}
	E_0\equiv\varepsilon^0=\sum_{j=x,y,z}\left\{\frac{(N-1)}{2}\delta_j+\frac{1}{2}\Omega_j\right\}
\end{equation}
whereas the energy distance between any excited state and the ground state reads
\begin{equation}\label{HIM_Energy_dist}
	\omega({\{n_j\},\{m_j\}})=\sum_{j=x,y,z}\left\{ n_j\delta_j+\Omega_jm_j \right\}.
\end{equation}
Eqs. (\ref{HIM_GS_Energy}) and (\ref{HIM_Energy_dist}) will be used for the comparison to the numerical results obtained from (LR-)MCTDHB. The degeneracies of excited states is discussed in \cite{Yan}. It is worth noting that for any number of bosons $N$, there is no solution for an excitation corresponding to $\sum_{j=x,y,z}n_j=1$, i.e., a single-particle excitation with respect to the relative coordinates. This is due to the even permutation symmetry for identical bosons, Eq. (\ref{wave_echchage_symm}), and it can be shown by mathematical induction as follows. For the sake of simplicity, the proof is restricted to the 1D case, i.e., $\mathbf{Q}_i=Q_i$, which does not limit the generality since in higher dimensions, the Hamiltonian decouples into the $x$-, $y$- and $z$-components, see Eqs. (\ref{HIM_Hrel}) and (\ref{HIM_Hcom}). The wave function corresponding to a first order excitation of the relative coordinates, denoted in the following by $\Psi^\prime$, obeys
\begin{equation}\label{HIM_Proof1}
	\Psi^\prime(Q_1,...,Q_N)\sim \frac{\partial}{\partial Q_k}\Psi_0(Q_1,...,Q_N),\quad 1\leq k \leq N-1.
\end{equation}
For $N=2$, this yields
\begin{align}
	\Psi^\prime(Q_1,Q_2)\sim Q_1\Psi_0\sim (x_2-x_1)\Psi_0,
\end{align}
see Eq. (\ref{HIM_GS_WF}). Because the wave function needs to be fully symmetric under the exchange $x_1 \leftrightarrow x_2$, one finds
\begin{align}
	\Psi^\prime\sim \left\{(x_2-x_1)+(x_1-x_2)\right\}\Psi_0=0,
\end{align}
meaning that there is no first order relative excited state for two bosons. Assuming that this is also true for $N$ bosons, i.e., 
\begin{equation}\label{HIM_Proof2}
	\Psi^\prime(Q_1,...,Q_N)\sim \frac{\partial}{\partial Q_k}\Psi_0(Q_1,...,Q_N)=0 \quad \forall \,1\leq k \leq N-1,
\end{equation}
one obtains for $(N+1)$ bosons and $K\leq N$
\begin{align}\label{HIM_deriv_proof} 
	\Psi^\prime(Q_1,...,Q_{N+1}) &\sim \frac{\partial}{\partial Q_K} \Psi_0(Q_1,...,Q_{N+1}) \nonumber \\ 
	&\sim\sum_{i=1}^K(x_{K+1}-x_i) \Psi_0 \nonumber \\
	&=\left(K x_{K+1}-\sum_{i=1}^K x_i\right)\Psi_0  \nonumber \\ 
	&=\left(K x_{K+1}-x_K-\sum_{i=1}^{K-1} x_i\right)\Psi_0 \nonumber \\
	&=\left(K x_{K+1}-x_K-(K-1)x_K+\underbrace{(K-1)x_K-\sum_{i=1}^{K-1} x_i}_{\sim Q_{K-1}}\right)\Psi_0 \nonumber \\
	&\sim K\left( x_{K+1}-x_K\right)\Psi_0+Q_{K-1}\Psi_0
\end{align}
where the argument of $\Psi_0$ is suppressed from the second line onward. For the second term in Eq. (\ref{HIM_deriv_proof}), one can make use of the assumption from Eq. (\ref{HIM_Proof2}) that the contribution of $Q_{K-1}\Psi_0$ to $\Psi^\prime$ vanishes after symmetrization of the wave function. Moreover, the first term is proportional to the relative coordinate of only two bosons, for which it was already shown that its contribution to $\Psi^\prime$ vanishes as well after symmetrization. Thus, one infers that also for $(N+1)$ bosons 
\begin{equation}\label{HIM_Proof3}
	\Psi^\prime(Q_1,...,Q_{N+1})=0 
\end{equation}
which ends the proof.

\subsection{Ground state and dynamics}\label{Sec_MCTDHB_benchmark}
At first, numerical results obtained from MCTDHB for the ground state of the 1D HIM for different particle numbers and the exact values from Eq. (\ref{HIM_GS_Energy}) are compared. The index '$x$' is suppressed in the following for simplicity. The trap frequency is set to $\Omega=1$ and the two-particle interaction strength $\lambda_0$ is adjusted such that the mean-field interaction parameter $\Lambda=\lambda_0(N-1)$ is kept fixed at $\Lambda=0.5$. Results are shown in Table \ref{table:MCTDHB_GS_bench}, which originally appeared in Ref. \cite{MCTDHB_benchmark} where it has been demonstrated that the MCTDHB method is much more efficient and accurate for the description of the ground state and dynamics of a many-boson system than the widely used FCI method. One observes that MCTDHB gives highly accurate results for the ground-state energy for all considered particle numbers. Furthermore, one needs less self-consistent orbitals $M$ in order to converge to the exact ground-state energy when $N$ increases. This is anticipated, since in the Hartree limit where $N\rightarrow\infty$ for fixed $\Lambda$, the energy per particle converges to the GP value. What is however surprising is the degree of accuracy. As an example, for $N=1000$ the relative error $(\varepsilon^0_{M=3}-\varepsilon^0_{\text{exact}})/\varepsilon^0_{\text{exact}}$ is lower than $10^{-10}\,\%$. The fact that MCTDHB can very accurately assemble the ground state of a large many-boson system with only very few orbitals has far-reaching consequences for the accuracy of excitation energies obtained by LR-MCTDHB. As discussed in Section \ref{Sec_LR_general}, the latter will be closer to the exact values if the ground state to which LR is applied is very accurately described. Further examples for the ground-state energies in the 1D and 2D HIM will be given in Section \ref{Sec_LRMCTDHB_benchmark} and in Appendix \ref{Appendix_further_benchmarks}, where in particular the convergence of excitation energies is discussed.

With regard to the dynamics, comparisons of numerical and exact results were carried out in Ref. \cite{MCTDHB_benchmark} for (i) the out-of-equilibrium quench dynamics and (ii) the dynamics due to time-dependent driving of the trapping frequency and the interaction strength. Concerning (i), Fig. \ref{fig_MCTDHB_benchmark_fig2} compares the exact, FCI and MCTDHB results for the oscillations of the density $\rho$ of $N=2$ bosons after a sudden quench of the interaction parameter from $\Lambda=0$ to $\Lambda=0.5$. The initial state $\Psi_0$ is therefore the ground state of the non-interacting system. In principle, the time-evolution is given by
\begin{align}
	|\Psi(t)\rangle &=\sum_{n=0}^\infty c_j\, e^{-i E_j t} |\alpha_j\rangle, \label{time_evo_HIM} \\ 
	c_j &=\langle\alpha_j|\Psi(0)\rangle \label{time_evo_HIM_overlaps}
\end{align}
where the $\{|\alpha_j\rangle\}$ are an arbitrary basis of the two-particle Hilbert space and the $\{E_j\}$ the corresponding eigenenergies. One possible choice of this basis are the solutions of the time-independent problem. It remains to compute the overlap integrals in Eq. ({\ref{time_evo_HIM_overlaps}). In practice, the infinite sum in Eq. (\ref{time_evo_HIM}) has to be truncated, and for the case described in Fig. \ref{fig_MCTDHB_benchmark_fig2}, numerical convergence at each time step of the expansion in Eq. (\ref{time_evo_HIM}) is ensured by including the $60$ lowest energy eigenstates of the time-independent problem. However, utilizing only $M=4$ time-adaptive orbitals, the MCTDHB calculations lead to a density oscillation that can no longer be distinguished from the exact result. Already for $M=3$ the results are highly accurate. If a fixed set of orbitals is used (FCI), even 8 orbitals do not lead to the same accuracy than MCTDHB(4). This remarkably shows the numerical benefit of using a time-adaptive, self-consistent basis set.

\begin{table}[t!]

\begin{tabular*}{\textwidth}{p{2.5cm} p{4cm} p{4cm} p{3cm}}
\hline\hline \rule{0pt}{3ex}  
		$M$	& $N=10$	& $N=100$ & $N=1000$	 \\ \hline \rule{-3pt}{3ex}  
	1	& 	7.0\underline{71067811865483}	&	70.\underline{71067811865483} 	&	707.\underline{1067811865483} 	\\
	2	&	7.038\underline{769026303168}	&	70.6801\underline{6951747168}	&	707.0764\underline{334257315}	\\
	3	&	7.0383\underline{50652406389}	&	70.680125\underline{41218675}	&	707.07642898\underline{71865}	\\
	4	&	7.0383484\underline{24909910}	&	70.6801253917\underline{4549}	&									\\
	5	&	7.0383484153\underline{49058}	&	70.680125391737\underline{62}	&									\\
	6	&	7.038348415311\underline{494}	&									&									\\
	7	&	7.03834841531101\underline{8}	&									&									\\
	$\varepsilon^0_{exact}$	&	7.038348415311011		&	70.68012539173752				&	707.0764289869851				\\
		
\hline\hline

\end{tabular*}\normalsize
\caption[Ground-state energies of $N=10,100,$ and $1000$ bosons in the 1D HIM]{Ground-state energies of $N=10,100,$ and $1000$ bosons in the 1D HIM with interaction parameter $\Lambda=0.5$ and trap frequency $\Omega=1.0$. Numerical results are shown for different numbers of orbitals $M$. Exact results are obtained from Eq. (\ref{HIM_GS_Energy}). The table is adapted from Ref. \cite{MCTDHB_benchmark}.}
\label{table:MCTDHB_GS_bench}
\end{table}

\begin{figure}[b!]
	\centering
  \includegraphics[width=0.85\textwidth]{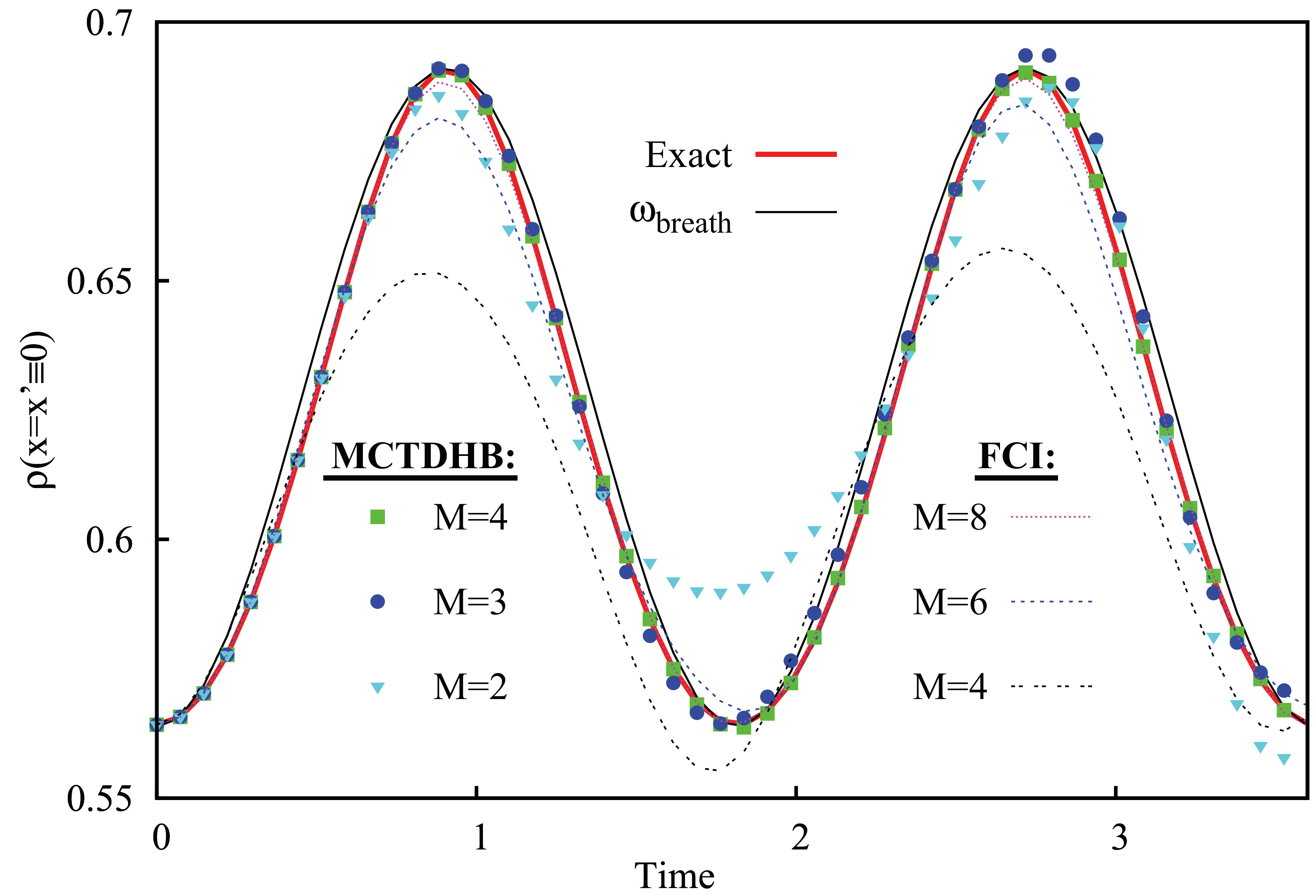}
  \caption[Out-of-equilibrium dynamics of the density $\rho$ after an interaction quench in the 1D HIM]{Out-of-equilibrium dynamics of the one-particle density $\rho(x=0)$ for $N=2$ bosons after an interaction quench from $\Lambda=0$ to $\Lambda=0.5$. Results are shown for different numbers of orbitals $M$. The trap frequency is $\Omega=1.0$. For the time scale shown, the MCTDHB(4) dynamics cannot be distinguished from the exact dynamics. The FCI method is less accurate even for $8$ orbitals. All quantities are dimensionless. See text for details. The figure is taken from Ref. \cite{MCTDHB_benchmark}.}
  \label{fig_MCTDHB_benchmark_fig2}
\end{figure}

\begin{figure}[ht!]
\centering	
  \includegraphics[width=0.8\textwidth]{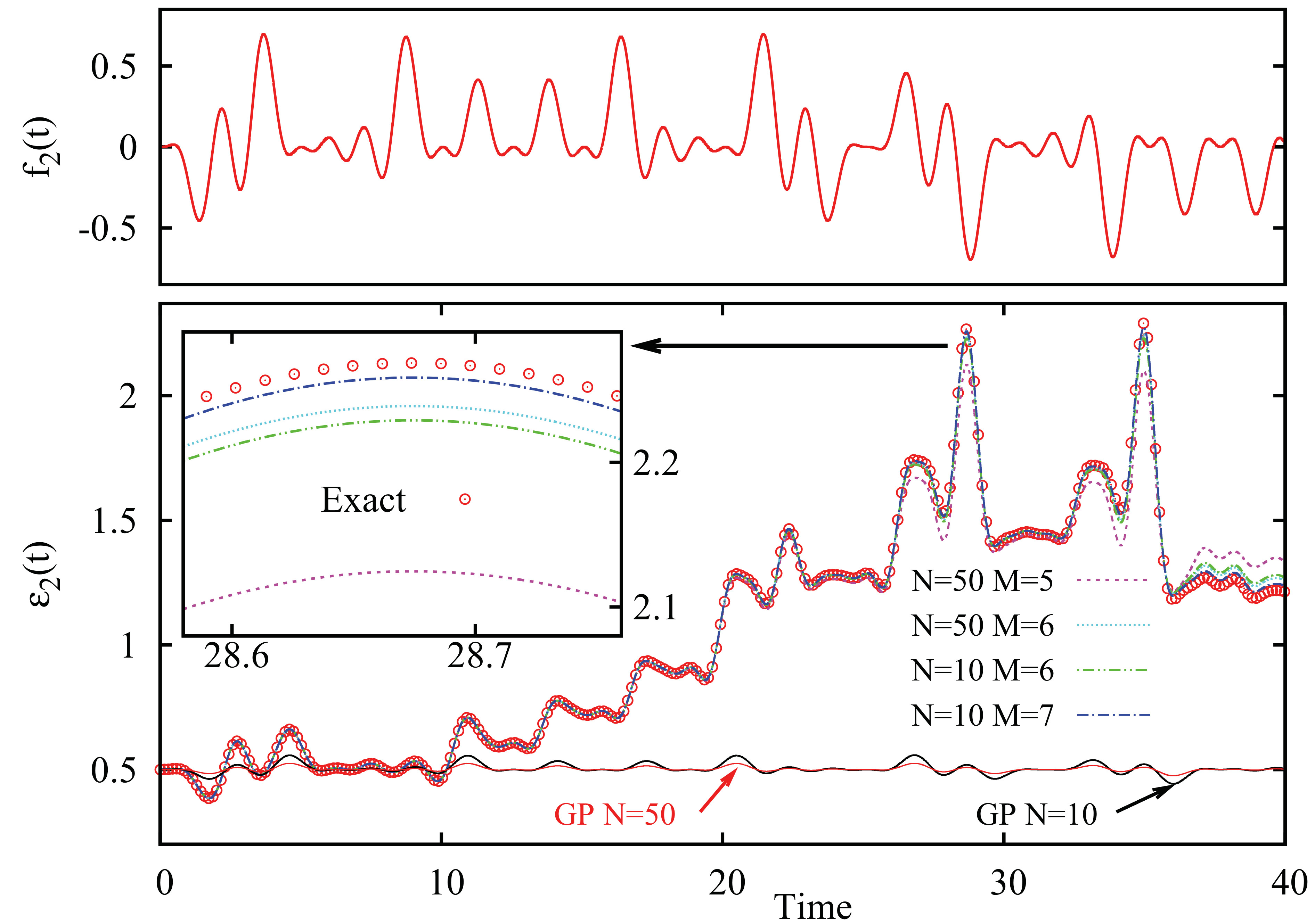}
  \caption[Out-of-equilibrium dynamics of the center-of-mass energy $\varepsilon_2 (t)$ due to a time-dependent driving of the trapping frequency in the 1D HIM]{Dynamics in the HIM due to a time-dependent driving $f_2(t)$ of the trapping frequency $\Omega(t)=\Omega_0\sqrt{1+f_2(t)}$ and the interaction strength $\lambda(t)=\lambda_0\left[1-\frac{\Omega_0^2}{2N\lambda_0}f_2(t)\right]$. The upper panel shows the complicated time-dependent profile of the driving frequency $f_2(t)$. The lower panel shows the evolution of the c.m. energy $\varepsilon_2(t)$ for different levels of MCTDHB($M$). For $N=10$ bosons, the curve of MCTDHB(7) can hardly be distinguished from the exact result for the shown time interval. The same holds true for $M=6$ and $N=50$ bosons. Even for less orbitals, MCTDHB yields an accurate description of the evolution of $\varepsilon_2(t)$ up to $t\approx 35$ for $N=10,\,M=6$ and $t\approx 25$ for $N=50,\,M=5$. On the contrary, the GP results do only follow the exact curves for very short times and clearly fail to describe the long-time behavior, even qualitatively. All quantities are dimensionless. See text for details. The figure is taken from Ref. \cite{MCTDHB_benchmark}.}
  \label{fig_MCTDHB_benchmark_fig3}
\end{figure}

Fig. \ref{fig_MCTDHB_benchmark_fig3} shows the time-evolution of the c.m. energy $\epsilon_2(t)$ when both the trapping frequency and the two-body interaction strength are time-dependent and given by $\Omega(t)=\Omega_0\sqrt{1+f_2(t)}$ with a constant $\Omega_0$ and $\lambda(t)=\lambda_0\left[1-\frac{\Omega_0^2}{2N\lambda_0}f_2(t)\right]$. The function $f_2(t)$ denotes a time-dependent driving frequency. In that way, the trapping frequency of the $(N-1)$ relative coordinates remains time-independent, i.e., 
\begin{equation}
\delta=\sqrt{\Omega_0^2[1-f_2(t)]+2N\lambda_0 \left[1+\frac{\Omega_0^2}{2N\lambda_0}f_2(t) \right]}=\sqrt{\Omega_0^2+2N\lambda_0}.
\end{equation}
Thus, although in the laboratory frame the bosons move in a time-dependent trap and interact via a time-dependent potential, the same system in the c.m. frame appears to be much less complicated because only the trapping potential of the c.m. particle is time-dependent due to $f_2(t)$. Because the c.m. and relative motions separate, the time-dependent energy of the system is given by $E=\frac{N-1}{2}\delta+\varepsilon_2(t)$ where $\varepsilon_2(t)$ is determined by the energy expectation value
\begin{equation}
	\varepsilon_2(t)=\langle\Psi_N(t)|\hat{H}_{c.m.}(t)|\Psi_N(t)\rangle	
\end{equation}
of the c.m. particle. The state $|\Psi_N(t)\rangle$ is obtained from the effective single-particle Schr{\"o}dinger equation $i\partial_t|\Psi_N(t)\rangle=\hat{H}_{c.m.}(t)|\Psi_N(t)\rangle$. From Fig. \ref{fig_MCTDHB_benchmark_fig3}, one can deduce that MCTDHB is capable to account for the dynamical evolution of $\varepsilon_2(t)$ very accurately, even for longer times. Furthermore, one observes that for a higher number of particles $N$ a smaller number of time-adaptive orbitals $M$ is required to obtain the same level of numerical accuracy. In contrast to that, the GP evolution of $\varepsilon_2(t)$ only follows the exact curves of $N=10$ and $50$ bosons for a very short time up to $t \approx 1$, and becomes remarkably inaccurate afterwards. For the long-time behavior, it cannot even reproduce qualitatively the profile of $\varepsilon_2(t)$.    

To sum up, the previous examples clearly indicate that the MCTDHB theory qualifies for describing the ground state, the dynamics due to a sudden quench of system parameters as well as the dynamics due to time-dependent trapping and interaction potentials, and that it is clearly superior to the FCI many-body and GP mean-field methods. The full benchmark, together with additional valuable details, can be found in Ref. \cite{MCTDHB_benchmark}.

\addtocontents{toc}{\protect\enlargethispage{\baselineskip}}
\subsection{Excited states}\label{Sec_LRMCTDHB_benchmark}
In this section, the applicability of LR-MCTDHB to obtain highly accurate results for the energies of excited states of trapped multi-boson systems is discussed. As for the benchmark concerning the ground state and dynamics, the analytic results of the HIM from Eq. (\ref{HIM_Energy_dist}) are used to benchmark the numerical results. In the following, LR-MCTDHB is first benchmarked against the 1D HIM for repulsive bosons, i.e., with $\lambda_0<0$. Then, for the 2D case, the isotropic HIM with $\Omega_x=\Omega_y$ is considered. In Appendix \ref{Appendix_further_benchmarks}, further benchmarks against the 1D HIM with attractive bosons, against the anisotropic case in 2D, as well as against the isotropic HIM in the rotating frame of reference are presented. 


\begin{table}[t!]
\centering
\begin{tabular*}{\textwidth}{p{1.8cm} p{2.4cm} p{2.4cm} p{2.4cm} p{2.2cm} c}

 \hline\hline \rule{0pt}{3ex} 
 	&	$M=1$	&	$M=3	$	&	$M=4$	& $(m_x,n_x)$	&	Exact  	 \\ 
 \hline  \rule{-4pt}{3ex} 
$\varepsilon^0$	&	4.52\underline{7693}		&	 4.524922	&	4.524922		&	$(0,0)$	&	4.524922		 \\ 	
$\omega_1$	&	1.000000		&	1.000000	&	1.000000		&	$(1,0)$	&	1.000000		 \\ 
$\omega_2$	&	1.\underline{811077}	&	1.78885\underline{5}	&	1.788854		&	$(0,2)$	&	1.788854		 \\ 
$\omega_3$	&	n/a	&	2.00000\underline{5}	&	2.000000	    &	$(2,0)$	&	2.000000		\\ 
$\omega_4$	&	2.\underline{716616}	&	2.683282	&	2.683282 	&	$(0,3)$	&	2.683282	\\ 
$\omega_5$	&	n/a	&	2.78885\underline{8}	&	2.788854	 &	$(1,2)$	&	2.788854 \\ 
$\omega_6$	&	n/a	&	3.00000\underline{9}	& 3.000000 &	$(3,0)$	 &	3.000000	\\ 
$\omega_7$	&	3.\underline{622154}	&	3.577\underline{961}	&	3.57\underline{3164}	&	$(0,4)$	&	3.577709 \\ 
$\omega_8$	&	n/a	&	3.5\underline{80111}	& 3.5777\underline{12} &	 $(0,4)$	&	3.577709	\\ 
$\omega_9$	&	n/a	&	3.68\underline{5038}	& 3.68\underline{1347} &	 $(1,3)$	&	3.683282	\\ 
$\omega_{10}$	&	n/a	&	3.7\underline{90078}	& 3.788\underline{483} &	 $(2,2)$	&	3.788854	\\ 
$\omega_{11}$	&	n/a	&	4.00\underline{2097}	& \underline{3.999997} &	 $(4,0)$	&	4.000000	\\ \hline\hline
\end{tabular*}
\caption[Low-energy spectrum of $N=10$ repulsive bosons in the 1D HIM]{Benchmark of the LR-MCTDHB implementation to the repulsive 1D HIM with $N=10$ bosons. Shown are the ground-state energy $\varepsilon^0$ and the energies $\omega_i$ of the first few excitations for different values of $M$. The trapping frequency is $\Omega=1.0$ and the interaction strength is $\lambda_0=-0.01$, yielding a marginal ground-state depletion of $f\approx 0.03\%$ for $M=4$. Whereas the BdG ($M=1$) spectrum misses many states, the many-body spectra for $M=3$ and $4$ contain all low-lying excitations. Even the two-fold degenerate state $(0,4)$, as proposed in Ref. \cite{Yan}, is obtained (see $\omega_7$ and $\omega_8$). The assignment of the corresponding single GP state to $\omega_7$ is arbitrary, and one would need to analyse for instance the response density to identify whether it correlates with the 7th or 8th many-body excitation. The overall accuracy of excitation energies increases with the number of orbitals. Underlined digits denote deviations from the exact values from Eqs. (\ref{HIM_GS_Energy}) and (\ref{HIM_Energy_dist}). All quantities are dimensionless. See text for more details.}
\label{table_1d_HIM_rep}

\end{table}

For the benchmark against the 1D repulsive HIM, the system parameters are set to $\Omega=1$ for the trap frequency and $\lambda_0=-0.01$ for the interaction strength, which yields a slight depletion of $f\approx 0.03\%$ for $N=10$ bosons and $M=4$ orbitals. Calculations were carried out on a grid from $[-9,9)$ with 128 grid points. Results for the excitation energies relative to the ground-state energy, $\omega_i=E_i-\varepsilon^0$, are shown in Table \ref{table_1d_HIM_rep}. There are several comments to be made. At first, it can be observed that the mean-field BdG calculation ($M=1$) misses several excited states in the low-energy spectrum. This is due to the fact that the BdG theory by construction does not account for excitations were several particles at a time are excited from the ground-state manifold, so-called multi-particle excitations. For further details see Appendix \ref{Sec_GP}. The second observation is that the numerical accuracy is enhanced if more self-consistent orbitals are included into the ground-state description. Also the two-fold degenerate excitation $(0,4)$, as proposed in Ref. \cite{Yan}, is obtained. The accuracy of the BdG spectrum, apart from the c.m. excitation $(1,0)$, is clearly lower than for the many-body results of $M=3$ and $4$ orbitals. In addition, the convergence towards the exact ground-state energy is observed.

Table \ref{table_2d_HIM_iso} shows excitation energies obtained for the isotropic 2D HIM with trap frequencies $\Omega_x=\Omega_y=1$ and $N=100$ repulsive bosons with interaction parameter $\lambda_0=-0.001$. All computations were carried out on a $[-9,9)\times[-9,9)$ grid with $64\times 64$ grid points. At first, one sees that the BdG approach yields the energy of the lowest c.m. excited state, which is two-fold degenerate, to perfect accuracy. All other states obtained are pure relative excitations. Higher c.m. excited states, as well as combinations of those with relative excitations, are missing in the BdG spectrum.


\begin{table}[t!]
\centering
\begin{tabular*}{\textwidth}{p{2.0cm} p{2.0cm} p{2.0cm} p{2.0cm} c c}
\hline\hline \rule{0pt}{3ex} 
	$\omega_i(\gamma)$		& $M=1$	& $M=2$ & $M=3$	&	\small{$(m_x,m_y,n_x$+$n_y)$}	&  Exact 	\\ \hline \rule{-4pt}{3ex}
$\varepsilon^0(1)$	& 	89.5\underline{54453}	&	89.5\underline{51373}	& 89.54829\underline{3}	&		$(0,0,0)$	&	89.548292			\\ 	 \rule{-4pt}{3ex}
$\omega_1(2)$	&	1.000000	& 1.000000 & 	1.000000	&		$(1,0,0)$		&		1.000000		\\ 
	&	1.000000	& 0.\underline{999792} & 	1.000000	&		$(0,1,0)$		&		1.000000			\\ 	 \rule{-4pt}{3ex}
$\omega_2(3)$	&	1.7\underline{91089}	& 1.78885\underline{9} & 	1.78885\underline{9}	&		$(0,0,2)$			&	1.788854			\\ 
	&	1.7\underline{91089}	& 1.7\underline{91089} & 	1.7888\underline{61}	&		$(0,0,2)$			&		1.788854		\\
	&	1.7\underline{91089} 	& 1.7\underline{91143} &	1.78885\underline{9}	&		$(0,0,2)$			&		1.788854		\\ 	 \rule{-4pt}{3ex}
$\omega_3(3)$	&	n/a	& 2.000\underline{438}	&	2.000\underline{439}	&			$(2,0,0)$		&	2.000000			\\
	&	n/a & n/a	&	2.000\underline{657}	&				$(1,1,0)$	&	2.000000			\\ 
	&	n/a & n/a	&	2.000\underline{439}	&		$(0,2,0)$			&	2.000000			\\ 	 \rule{-4pt}{3ex}	
$\omega_4(4)$	&	2.68\underline{6634}	& 2.6832\underline{92} & 	2.6832\underline{26}	&		$(0,0,3)$			&	2.683282			\\ 
	&	2.68\underline{6634}	& 2.68\underline{4452} &	2.6832\underline{26}	&			$(0,0,3)$		&	2.683282			\\ 
	&	2.68\underline{6634}	& 2.68\underline{5519} &	2.683\underline{312}	&			$(0,0,3)$		&	2.683282			\\
	&	2.68\underline{6634}	& 2.68\underline{6634} &	2.683\underline{312}	&			$(0,0,3)$		&	2.683282			\\ 	 \rule{-4pt}{3ex}
$\omega_5(6)$	&	n/a	& 2.78\underline{9191}	&	2.78\underline{7088}	&			$(1,0,2)$		&	2.788854			\\ 
	&	n/a & 2.7\underline{91314}	&	2.78\underline{7088}	&			$(1,0,2)$		&	2.788854		\\ 
	&	n/a & 2.7\underline{93711}	&	2.78\underline{9270}	&		$(1,0,2)$		&		2.788854	\\ 
    &	n/a	& n/a	&	2.78\underline{9270}	&				$(0,1,2)$	&	2.788854			\\ 
	&	n/a & n/a	&	2.7\underline{91536}	&			$(0,1,2)$	&	2.788854			\\ 
	&	n/a & n/a	&	2.7\underline{91535}	&			$(0,1,2)$	&	2.788854		\\  \hline\hline

\end{tabular*}
\caption[Low-energy spectrum of $N=100$ repulsive bosons in the 2D isotropic HIM]{Benchmark of the LR-MCTDHB implementation to the isotropic HIM in 2D. Results are presented for $N=100$ bosons and different numbers of orbitals $M$. The trapping frequencies are $\Omega_x=\Omega_y=1.0$, and the interaction strength is $\lambda_0=-0.001$. The depletion is marginal with $f\approx 0.003\%$ for $M=2$ and $f\approx 0.006\%$ for $M=3$. Shown are the energies of the ground state $\varepsilon^0$ and the first $5$ excited states, $\omega_i=E_i-\varepsilon^0$, together with their multiplicities $\gamma$. In the BdG case, $M=1$, only the first pure c.m. excitation is found, together with all pure excitations of relative coordinates. For LR-MCTDHB(2), one obtains more c.m. excited states in the $x$-direction, together with combinations of excitations in the relative coordinates. For LR-MCTDHB(3), one obtains the missing c.m. excited states in the $y$-direction and their combinations with excitations of the relative coordinates. The overall accuracy increases with $M$, with few exceptions in the spectrum for $M=2$ which are discussed in the text. Underlined digits denote deviations from the exact values from Eqs. (\ref{HIM_GS_Energy}) and (\ref{HIM_Energy_dist}). All quantities are dimensionless. See text for more details.}
\label{table_2d_HIM_iso}
\end{table}

In contrast to the BdG approach, LR-MCTDHB(2) gives access to more excited states. Before discussing those, it is worth examining the two underlying ground-state orbitals. The first one is a Gaussian in the trap center, whereas the second one resembles a $p_x$-orbital. The degree of depletion is marginal with $f\approx 0.003\%$. In addition to the states obtained for $M=1$, the spectrum contains the second-order c.m. excitation in the $x$-direction, as well as three out of six states corresponding to $\omega_5$, which are excited states combining a first-order c.m. excitation in the $x$-direction and relative excitations. The accuracy of levels that were already accessible within the BdG approach has increased, e.g., for the first state of $\omega_2$ or the states of $\omega_4$. Nevertheless, some states have lost accuracy. The most remarkable example is the second first-order c.m. excitation of $\omega_1$, which is a c.m. excitation in the $y$-direction. Naturally the question arises why this is the case. The reason for this is that the second ground-state orbital, as discussed above, resembles a $p_x$-orbital. This means that the ground-state description for $M=2$ orbitals shows a preferred direction, which in this case is the direction along the $x$-axis. Apparently, the consequence of this is that excitations in the $y$-direction either lose in accuracy or remain inaccessible. A possible reason for the loss in accuracy is that, compared to the BdG case, a small amount of probability weight of the first orbital has been transferred to the second orbital that effectively only covers the $x$-direction. It is stressed that rotating the second orbital by an arbitrary angle, e.g., $\pi/4$ or $\pi/2$, leads to exactly the same spectrum, which means that it does not matter for the numerical results which direction is preferred. 

Including $M=3$ orbitals significantly enhances the obtained spectrum. One observes that both problems of the spectrum for $M=2$, i.e., the loss in accuracy for certain states as well as the absence of excitations in the low-energy spectrum, are solved. The third ground-state orbital resembles a $p_y$-orbital, i.e., it is perpendicular to the preferred direction of the $M=2$ ground state. One can deduce from this observations a fundamental difference between the computation of the ground state of a given $N$-boson system and its excited states. Whereas increasing the number of orbitals always leads to a lower ground state energy, it can happen that the spectrum for ($M+1$) orbitals contains states that are less accurate than for only $M$ orbitals. This is in particular likely to happen in 2D and 3D because preferred directions may occur. As a result, in order to obtain accurate excitation spectra, one has to pay much more attention to the symmetry of the problem in higher spatial dimensions.

To summarize, it was demonstrated for the HIM that both the MCTDHB and LR-MCTDHB implementations utilized in this work are well-suited to obtain accurate results for the spectrum and dynamics of trapped interacting bosons. Furthermore, the many-body approaches clearly improve the GP and BdG mean-field methods. As mentioned above, further benchmarks are given in Appendix \ref{Appendix_further_benchmarks}.





\section{Applications to many-body dynamics and excitations of BECs}\label{Ch_applications}

Since the advent of MCTDHB in 2007, the amount of scientific publications where it is utilized grew substantially. As one of the first applications, the splitting of 1D repulsive, fully-coherent BECs by a potential barrier has been investigated \cite{Splitting1,Splitting2}, showing that it can lead to fragmented condensates. Similar observations were made by splitting a radially-symmetric BEC in 2D with a ring-shaped barrier \cite{Klaiman}. Also the reverse is possible, meaning that two initially independent and thus fragmented BECs with no overlap in space can collide, interfere and build up coherence \cite{Coherence_buildup}. Also for attractive BECs, the splitting due to a Gaussian-shaped barrier results in the formation of a superposition of two distinct bosonic clouds, a so-called caton, which is not describable at the GP level \cite{Splitting_attract}. Moreover, it has been shown for the first time that coherent attractive BECs in 1D can fragment even without a potential barrier, namely when the energy exceeds a threshold value \cite{Splitting3}. Fragmentation persists and may even intensify when more orbitals are included in the calculation \cite{Brand_convergence}. In the attractive case full convergence is difficult to achieve as it requires more orbitals, see also \cite{Ofir_MBVar_5}. When an attractive BEC propagates towards a potential barrier and scatters from it, the system also evolves to be highly fragmented although it was initially condensed \cite{Castin_Weiss,Splitting3b}. The degree of fragmentation depends on the degree of transmission through the barrier. Even bright-soliton trains, i.e., stable multi-hump matter waves of attractive bosons, whose dynamics were believed to be fully describable at the mean-field level, are shown to lose their initial coherence \cite{Alexej_brightsol}.

The splitting of BECs has been further investigated with respect to the generation of number- or phase-squeezed states applying optimal control theory \cite{Schmiedmayer1,Schmiedmayer2,Schmiedmayer2b,Schmiedmayer3,Schmiedmayer6}. Optimal control has also been utilized in 1D bosonic Josephson junctions (BJJ), on the one hand to steer the system parameters such that an enhancement of the Shapiro effect was observed \cite{Schmiedmayer4}, and on the other hand to drive a BEC from an initial state to a target state at the quantum speed limit \cite{Brouzos}. Moreover, it has been demonstrated that Mach-Zehnder interferometry with ultracold bosons is relatively robust against the nonlinear interaction between the particles \cite{Schmiedmayer5}.

A remarkable correspondence between the onset of wave chaos at the GP level and the onset of fragmentation at the many-body level has been found for BECs that scatter from either shallow periodic or disordered potential landscapes \cite{Wave_chaos1, Wave_chaos2}, but the result holds also for other potentials. Thus, the development of wave chaos, i.e., the exponential separation of initially close states, can be seen as an indication that a many-body treatment is necessary. The distance between states is measured utilizing the $L^2$ norm between two Hilbert space vectors. 

Of potentially high relevance is the demonstration how the positions of individual particles can be constructed from the many-body wave function by simulating single shots  \cite{Kaspar_single_shot}. The latter are connected to experiments on trapped BECs where typically an image of the density represents a histogram of a single shot.

Over the last years, MCTDHB has been applied to a growing number of systems, e.g., to study the breathing dynamics in 1D harmonic traps due to a quantum quench \cite{ML_apps_1} or the many-body effects in solitonic excitations of ultracold BECs \cite{ML_apps_2,ML_apps_3}. Other examples are the correlated dynamics of a single atom coupled to a trapped BEC by collisions \cite{ML_apps_4} or the examination of the structure of mesoscopic molecular ions \cite{ML_apps_5}, which contributes to the currently very active research in this field.

Recently, MCTDHB has also been applied to dipolar BECs in a double well \cite{App_Yi} and in 1D lattices \cite{Axel_dipolar1,Axel_dipolar2}. In addition, the coupling of one- and two-component BECs in a cavity to a radiation field has been of interest \cite{Axel_cavity1,Axel_cavity2,Axel_cavity3}. One of the main results of the latter research was the finding of a fragmented superradiant phase when the pump power of the laser exceeds a critical threshold. This fragmented superradiance cannot be explained by the Dicke model which assumes the BEC to be a simple two-level system. 

In the remaining part of this section, the focus is laid on several applications where the tunneling dynamics in traps (Section \ref{Sec_Appl_Tunneling_in_traps}) and the dynamical fragmentation of initially coherent BECs (Section \ref{Sec_Appl_dyn_frag}) is investigated, both in 1D and 2D systems. Afterwards, all applications of LR-MCTDHB that exist up to now are presented in detail (Section \ref{Sec_Applications_LR-MCTDHB}). Special emphasis for the 2D systems under consideration is laid on the impact of angular momentum on the appearance of many-body effects in the dynamics and excitation spectra.


\subsection{Tunneling dynamics in traps}\label{Sec_Appl_Tunneling_in_traps}
Subsequently, the tunneling dynamics of BECs held in double-well traps in both one and two spatial dimensions are discussed in this section. The main purpose is to apply MCTDHB to solve the many-boson Schr{\"o}dinger equation and to discover novel phenomena that are not captured by the standard methods.

\subsubsection{Exact tunneling dynamics in a bosonic Josephson junction}\label{App_Dyn1}
The quantum dynamics of a tunneling BEC in a 1D BJJ have constituted a large research field in the past decades. Before the first experimental realizations \cite{Albiez_1DBJJ,Levy_1DBJJ}, theoretical predictions concerning Josephson oscillations and self-trapping, i.e., the suppression of tunneling between the wells, were obtained by a (multi-mode) GP description \cite{Milburn_1DBJJ,Smerzi_1DBJJ,Raghavan_1DBJJ,Ostrovskaya_1DBJJ,Ananikian_1DBJJ,Jia_1DBJJ}, a path-integral formulation \cite{Zhou_1DBJJ}, a two-mode Bose-Hubbard (BH) approach \cite{Ferrini_1DBJJ} or a Fock-space WKB method \cite{Shchesnovich_1DBJJ}. A review of various theoretical and experimental results is given in \cite{Gati_1DBJJ}. A more recent study dealt with the tunneling dynamics in a controlled BJJ, where the spin state of an ionic impurity controls the tunneling between the wells \cite{Schmelcher_BJJ_ion2}. The latter extends the findings on the dynamics of trapped ultracold BECs due to the presence of an ion \cite{Schmelcher_BJJ_ion1b,Schmelcher_BJJ_ion1}. Furthermore, the orbital Josephson effect due to a time-dependent driving potential was investigated numerically by employing MCTDHB, not only in a double-well system \cite{Carr_OJE}.

The main objective of this section is however to study the out-of-equilibrium tunneling dynamics of a single-component BEC in a BJJ from a many-body perspective, and the method of choice is MCTDHB. As described below, the latter approach allows for the observation of fundamentally new physics dealing with, e.g., reduced self-trapping and dynamical fragmentation of initially fully-coherent bosonic clouds. A comparison to other popular theoretical tools like the GP approach and the BH model is made. The findings presented below were recently published in Refs. \cite{Kaspar_1D_Bjj_1,Kaspar_1D_Bjj_2,Thesis_Kaspar,Kaspar_1D_Bjj_3}, where further details can be found. Recently, results of a comparable study where both the BH model and MCTDHB were used to analyze the tunneling of few-boson systems in asymmetric double-wells, investigating the interaction blockade that isolates the motion of a single particle in the vicinity of others, became available \cite{Brand_asymmetric}. 

The trapping confinement considered here reflects two harmonic wells of the form
\begin{equation}\label{Pot_1DBJJ}
	V_{\pm}(x)=\frac{1}{2}(x\pm 2)^2
\end{equation}
which are connected by a cubic spline in the interval $|x|\leq 0.5$ that generates a potential barrier between the wells. The plus sign refers to the left well, whereas the minus sign refers to the right well. The repulsion between the bosons is modeled by the contact interaction potential, $\lambda_0\hat{W}(x,x^\prime)=\lambda_0\delta(x-x^\prime)$. The repulsion strength is given by the mean-field parameter $\Lambda=\lambda_0(N-1)$. An important quantity with respect to the tunneling dynamics is the occupation probability of the left and right wells, where the former is given by
\begin{equation}\label{Eq_P_L}
	P_L(t)=\frac{1}{N}\int_{-\infty}^0\rho(x,t)dx
\end{equation}
where $\rho(x,t)$ describes the time-dependent one-body density, see Eq. (\ref{Eq_1BRDM}), and $N$ refers to the number of particles in the trap.

Since the MCTDHB results are compared to the BH model below, it is instructive to define the left- and right-well orbitals, $\phi_L$ and $\phi_R$, with the ground and the first excited state of the trap, denoted by $\phi_g$ and $\phi_u$ due to their gerade and ungerade symmetry. These two states are energetically well below the barrier. The orbitals $\phi_L$ and $\phi_R$ are defined via 
\begin{equation}\label{Eq_phi_L_R}
	\phi_{L,R}(x)=\frac{\phi_g(x)\pm\phi_u(x)}{\sqrt{2}}.
\end{equation}
The BH parameters for on-site interaction $U$ and hopping $J$ are derived from $\phi_L$ and $\phi_R$ by 
\begin{equation}\label{Eq_BH_parameters}
	U=\lambda_0\int |\phi_L(x)|^4 \,dx,\quad J=-\int\phi_L^\ast(x)\hat{h}(x)\phi_R(x)\,dx
\end{equation}
and utilized to define the BH interaction parameter $\Lambda_{\text{BH}}=UN/(2J)$. The two-site BH Hamiltonian then reads
\begin{equation}\label{BH_Hamiltonian}
	\hat{H}_{\text{BH}}=-J(\hat{b}_L^\dagger\hat{b}_R+\hat{b}_R^\dagger\hat{b}_L)+\frac{U}{2}(\hat{b}_L^\dagger\hat{b}_L^\dagger\hat{b}_L\hat{b}_L+\hat{b}_R^\dagger\hat{b}_R^\dagger\hat{b}_R\hat{b}_R)
\end{equation}
where the operators $\hat{b}_{L,R}^{(\dagger)}$ annihilate (create) a boson in the left and right well, respectively. Within the two-mode BH approach, the eigenvalues and eigenfunctions of the one-body RDM
\begin{equation}\label{Eq_1b_RDM_BH}
	\rho^{(1)}=\begin{pmatrix}
	\rho_{LL} & \rho_{LR} \\ \rho_{RL} & \rho_{RR}
	\end{pmatrix}
\end{equation}
with $\rho_{LL}=\langle N,0|e^{+i\hat{H}_{\text{BH}}t}\,\hat{b}^\dagger_L\hat{b}_L \,e^{-i\hat{H}_{\text{BH}}t} |N,0\rangle$ (with $\hbar=1$) and the other elements defined analogously, represent the natural occupations and natural orbitals of the BH method. Therefore, the population of the second natural orbital can be used as a measure for fragmentation.

In the following, the tunneling dynamics when the BEC is initially trapped in the left well, i.e., $P_L(0)=1$,  will be explored for different repulsion strengths. Most importantly, for a two-mode GP description of the system, self-trapping of the bosons is predicted for $\Lambda_{\text{BH}}>\Lambda_c=2$ when the magnitude of the initial imbalance between the occupations of the two wells, denoted by $z=\frac{N_L-N_R}{N}$, is unity \cite{Milburn_1DBJJ,Smerzi_1DBJJ,Raghavan_1DBJJ}.

\begin{figure}[ht!]
  \centering
  \includegraphics[width=0.5\textwidth]{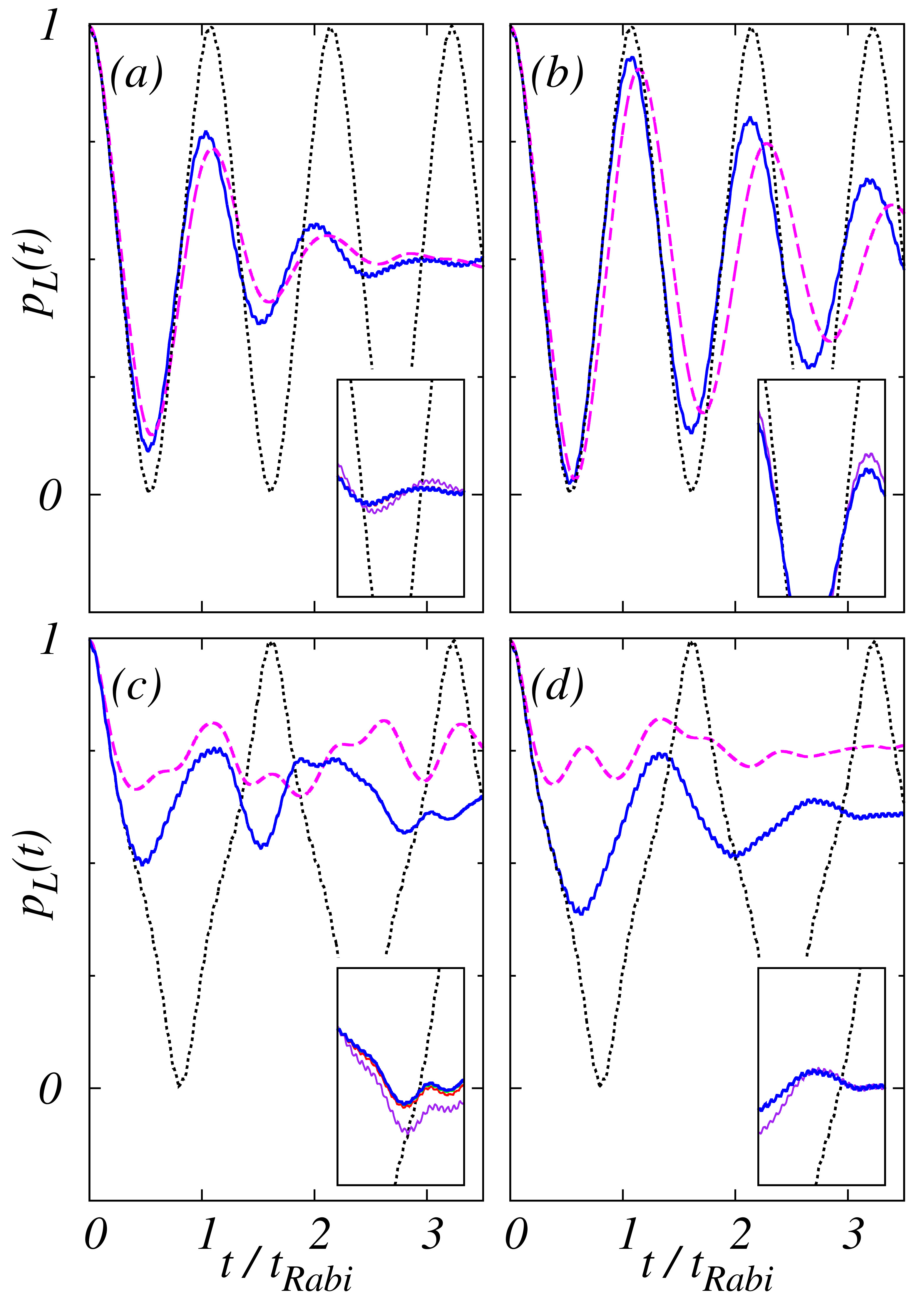}
  \caption[Short-time dynamics for weak repulsion in a 1D BJJ]{Short-time dynamics of the occupation probability of the left well, $P_L(t)$, for (a) $N=20$, $\Lambda=0.152$, (b) $N=100$, $\Lambda=0.152$ ($\Lambda_{\text{BH}}<\Lambda_c$) (c) $N=20$, $\Lambda=0.245$, and (d) $N=100$, $\Lambda=0.245$ ($\Lambda_{\text{BH}}>\Lambda_c$). The GP [dotted black], BH [dashed magenta] and numerically exact results from MCTDHB($M$) [solid blue] are compared. For the weak interaction in panels (a) and (b), the GP theory predicts Rabi oscillations of the full density. At the many-body level, a collapse of these oscillations occurs after approximately three cycles. For stronger interaction in panels (c) and (d), GP still predicts that all bosons tunnel between the wells, whereas at the many-body level tunneling is clearly suppressed. However, MCTDHB and the BH model yield different dynamics in all cases. The deviations become more pronounced for stronger repulsion. The inset shows the numerical convergence with respect to $M$ for 2, 4, 6 and 8 orbitals [(a) and (c)] and 2 and 4 orbitals [(b) and (d)]. All quantities are dimensionless. See text for details. The figure is taken from Ref. \cite{Kaspar_1D_Bjj_1}.}
  \label{fig:1D_BJJ_0}
\end{figure}

\begin{figure}[ht!]
  \centering
  \includegraphics[width=0.7\textwidth]{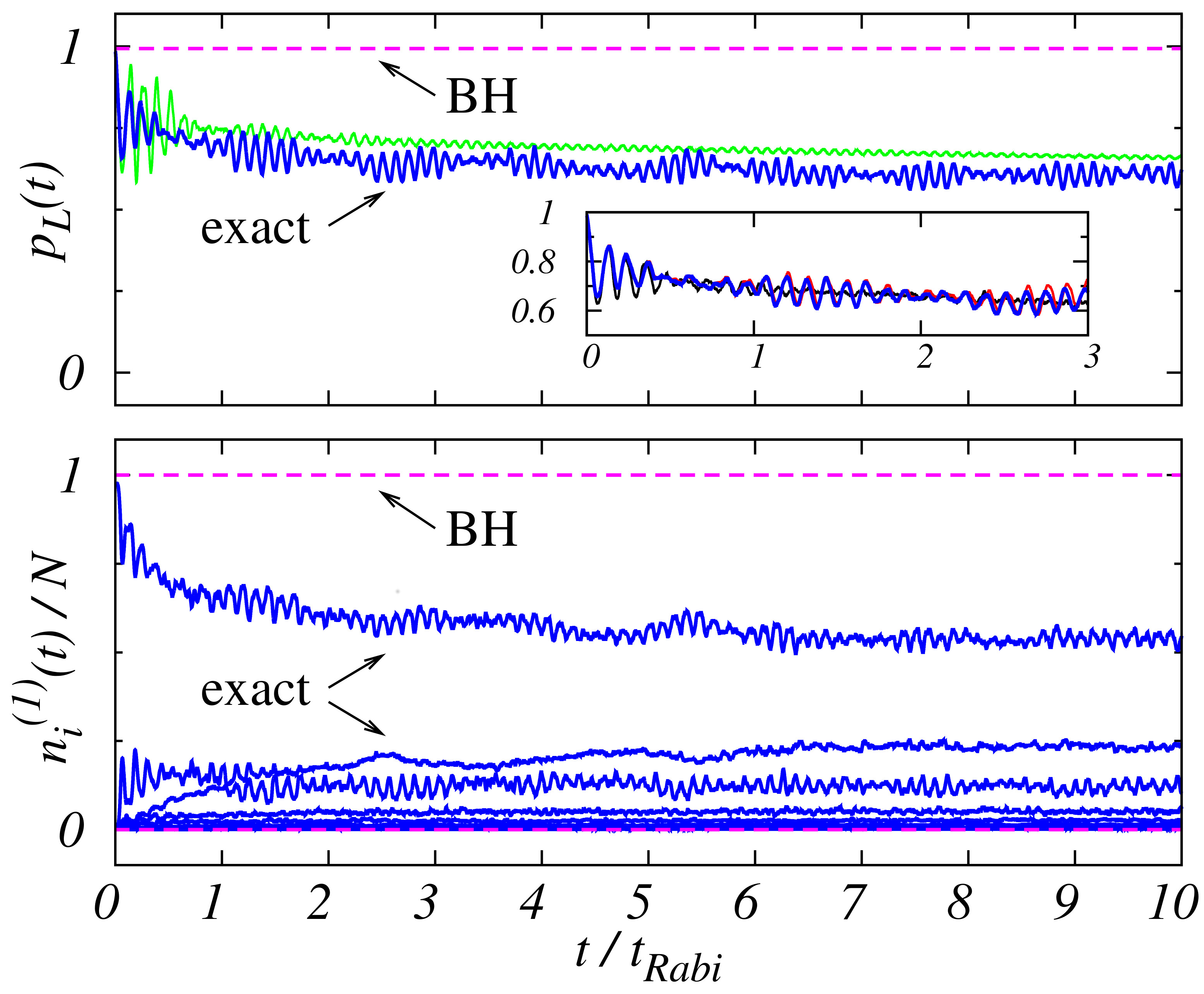}
  \caption[Short-time dynamics for very strong repulsion in a 1D BJJ]{Upper panel: Dynamics of $P_L(t)$ for very strong repulsion. The interaction parameter $\Lambda_{\text{BH}}=47.8$ ($43.4$) for $N=10$ [solid blue] ($N=100$, solid green) is clearly above the critical value of $\Lambda_c=2$. While the BH predicts full self-trapping of the bosons [dashed magenta], the exact results show tunneling between the wells. Lower panel: Evolution of the natural occupations $n_i^{(1)}(t)$. At the BH level, the system remains condensed throughout the propagation time, while it fragments at the exact many-body level where essentially four orbitals become macroscopically occupied. All quantities are dimensionless. See text for details. The figure is adapted from Ref. \cite{Kaspar_1D_Bjj_1}.}
  \label{fig:1D_BJJ_1}
\end{figure}

\begin{figure}[ht!]
  \centering
  \includegraphics[width=\textwidth]{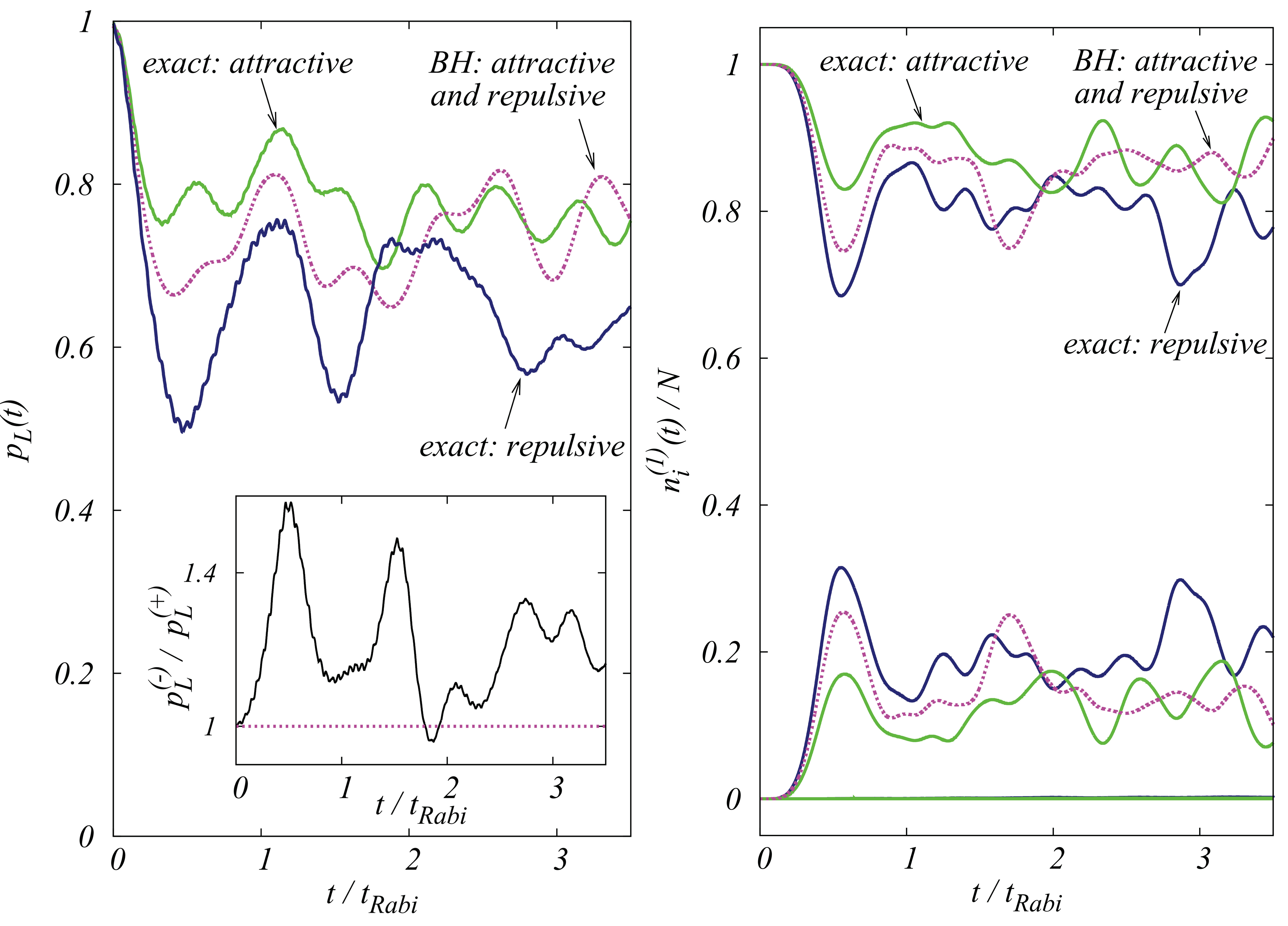}
  \caption[Bose-Hubbard versus exact dynamics for both attractive and repulsive BECs in a 1D BJJ]{BH versus exact dynamics for both attractive and repulsive BECs with system parameters $N=20$, $\frac{|U|}{J}=0.226$ and $|\lambda_0|=0.0129$. Left panel: Occupation probability $P_L(t)$ of the left well. While the BH predicts the same evolution for attraction and repulsion, differences can be observed at the exact many-body level. Inset: Time-evolution of the ratio of $P_L(t)$ for attraction and repulsion. Right panel: The same for the first two natural occupation numbers $n_i^{(1)}(t)$. Again, the BH model predicts equivalent dynamics for attraction and repulsion, whereas the exact results show clear differences. All quantities are dimensionless. See text for details. The figures are taken from Ref. \cite{Kaspar_1D_Bjj_2}.}
  \label{fig:1D_BJJ_2}
\end{figure}

Fig. \ref{fig:1D_BJJ_0} shows a comparison between the short-time dynamics of $P_L(t)$ for BECs with $N=20$ ($100$) bosons obtained from MCTDHB (solid blue), BH (dashed magenta) and GP (dotted black) for two different interaction strengths where one is below and one is above the critical value $\Lambda_c$. One observes that for the case of $20$ bosons with weak interaction parameter $\Lambda_{\text{BH}}<\Lambda_c$ [panel (a)], the GP prediction is totally different than the BH and numerically exact MCTDHB predictions. According to GP, Rabi oscillations of the entire cloud between the wells occur, whereas the BH model and MCTDHB predict a density collapse after approximately three Rabi cycles, resulting in roughly $50\%$ of the bosons in each well. The time for one Rabi cycle is close to the theoretical prediction of $t_{\text{Rabi}}=\pi/J$. Although the BH curve is clearly superior to the GP curve, it also deviates from the numerically exact result after half a Rabi cycle. In the case of $100$ bosons [panel (b)], similar observations can be made. However, the density collapse occurs after a longer time. 

As mentioned above, for stronger repulsion with $\Lambda_{\text{BH}}>\Lambda_c$, tunneling should be entirely suppressed due to self-trapping according to a two-mode GP description. It can be deduced from the exact dynamics of $20$ bosons [panel (c)] that indeed less bosons tunnel between the wells, meaning that the oscillation amplitude is clearly smaller than for the weaker repulsion. As predicted by MCTDHB, at most $50$ bosons tunnel into the right well after half a Rabi cycle, and even less in the BH dynamics. Nevertheless, tunneling is not completely suppressed. This means that the effect of self-trapping, as predicted by the two-mode GP theory, is obviously reduced at the accurate many-body level. The deviations between the MCTDHB and BH results are more pronounced for stronger than for weaker repulsion. The dynamics of $N=100$ bosons yields similar observations [panel (d)]. In contrast to that, the GP theory predicts tunneling of the whole cloud. 

The reason why GP fails to describe the dynamics correctly is the development of fragmentation, already on the short time scales shown. For the case of weak repulsion in the upper panels of Fig. \ref{fig:1D_BJJ_0}, the initially condensed BEC with $f=10^{-4}$ ($10^{-5}$) for $N=20$ ($100$) evolves to be fragmented by $33\%$ ($26\%$) after three Rabi cycles. Interestingly, for stronger repulsion, the degree of fragmentation of the again initially fully-coherent BEC is less after three Rabi cycles than for the case of weaker repulsion, yielding approximately $28\%$ and $18\%$ for $20$ and $100$ bosons, respectively. The predicted GP dynamics differ both quantitatively and qualitatively from the exact results. Most important, for all system parameters used in Fig. \ref{fig:1D_BJJ_0}, GP predicts that essentially all bosons tunnel back and forth between the wells. This is clearly not the case at the many-body level, neither for BH nor for MCTDHB. It is stressed that although the chosen system parameters are within the expected regime of validity for both the GP and BH approaches, both theories fail to describe the short-time dynamics of the BECs.

To further illustrate the effect of reduced self-trapping as well as the degree of fragmentation, Fig. \ref{fig:1D_BJJ_1} shows the evolution of $P_L(t)$ and of the occupation of the first few natural orbitals $n_i^{(1)}(t)$, both for MCTDHB and the BH model. The utilized repulsion parameter exceeds the critical value significantly with $\Lambda_{\text{BH}}=47.8\,(43.4)$ for $N=10\,(100)$. From the top panel, it can be inferred that the BH model predicts complete self-trapping for both values of $N$, meaning that all bosons remain in the left well throughout the time-evolution. On the contrary, the predictions of MCTDHB show that indeed bosons tunnel from left to right, and one can anticipate that $P_L(t)$ approaches the long-time average of $0.5$. With respect to the natural occupations, one observes that the BEC remains fully-condensed in time at the BH level, while the MCTDHB results show that fragmentation sets in already during the first Rabi cycle. Moreover, more than ten orbitals are necessary to accurately describe the dynamics. The different results are associated with a quick loss of coherence between the bosons that is only described by MCTDHB and not at the BH level. In this context, see Ref. \cite{Kaspar_1D_Bjj_1} for more details.

Another interesting aspect is the symmetry of the two-mode BH Hamiltonian in Eq. (\ref{BH_Hamiltonian}) under the unitary transformation $\hat{R}=\{\hat{b}_L\rightarrow\hat{b}_L,\hat{b}_R\rightarrow-\hat{b}_R\}$, leading to $\hat{R}\hat{H}_{\text{BH}}(U)\hat{R}=-\hat{H}_{\text{BH}}(-U)$. A consequence of this is the equivalence of the occupation of the wells for attractive and repulsive interactions, i.e.,
\begin{equation}
	P_{L,R}(t;U)=P_{L,R}(t;-U).
\end{equation}
Moreover, it can be shown that the eigenvalues of the BH one-body RDM in Eq. (\ref{Eq_1b_RDM_BH}) do not depend on the sign of the interaction. Thus, $P_L(t)$ and the occupations $n_i^{(1)}(t)$ will be the same for attraction and repulsion of the same magnitude. Naturally, the full many-body Hamiltonian of the system does not possess such a symmetry, and one can therefore expect different dynamics for attraction and repulsion between the bosons. 

The left panel of Fig. \ref{fig:1D_BJJ_2} shows the short-time dynamics of $p_L(t)$ for a BEC with $N=20$ bosons with interaction parameter $|U|/J=0.226$, corresponding to $|\lambda_0|=0.0129$. As anticipated, the evolution at the BH level is identical for attraction and repulsion. However, the results obtained by MCTDHB show differences between the cases of positive and negative $\lambda_0$. On the time scale shown, the BH model mainly underestimates tunneling for the repulsive case, whereas it overestimates tunneling for the attractive case. The right panel of Fig. \ref{fig:1D_BJJ_2} shows the evolution of the first few natural occupation numbers for the same system as in the left panel, both at the BH and full many-body level. As expected, the BH model predicts the same evolution for $n_1^{(1)}(t)$ and $n_2^{(1)}(t)$, irrespective of the sign of the interaction parameter. In contrast to that, the results obtained from MCTDHB show that the evolution of the different occupations are distinct for the attractive and repulsive cases. Compared to the exact results, the BH model mainly overestimates the fragmentation of the attractive BEC, whereas it mainly underestimates it for the repulsive BEC.

\begin{figure}[ht!]
  \centering
  \includegraphics[width=0.5\textwidth]{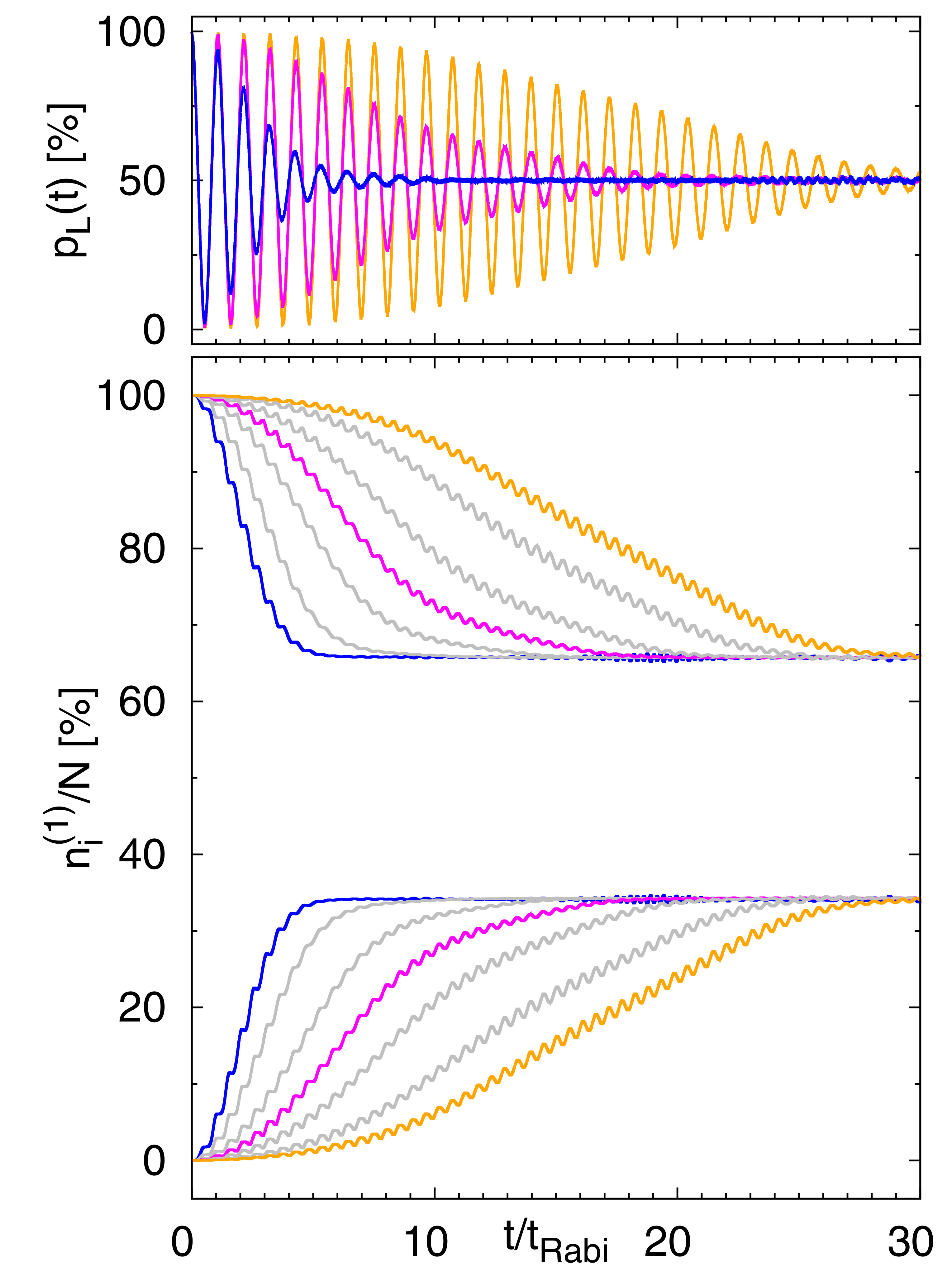}
  \caption[Universality of fragmentation in a 1D BJJ]{Universality of the degree of fragmentation. Upper panel: Evolution of $P_L(t)$ for different numbers of bosons ($N=100$ in blue, $N=1000$ in red and $N=10000$ in orange) where the interaction parameter $\Lambda=\lambda_0(N-1)=0.152$ is kept fixed. Intermediate particle numbers ($N=200,\,500,\,2000$ and $5000$) are shown in gray. The less particles are contained in the BEC, the faster the density collapses. Lower panel: Evolution of the first two natural occupations. Irrespective of the number of bosons, the degree of fragmentation upon the density collapse is the same. All quantities are dimensionless. See text for details. The figure is taken from \cite{Kaspar_1D_Bjj_3}.}
  \label{fig:1D_BJJ_3}
\end{figure}

The analysis of the dynamics in a 1D BJJ is closed by discussing the compelling feature of universal fragmentation. The latter means that all BECs with identical interaction parameter $\Lambda$ fragment to the same value after the density has collapsed. Fig. \ref{fig:1D_BJJ_3} shows exactly this behavior for $\Lambda=0.152$ and BECs with $N=100$, $1000$, and $10000$ particles where initially all bosons are kept in the left well, i.e., $P_L(0)=1$. From the upper panel, one sees that the density oscillations between the wells collapse after a few Rabi cycles, leading to the equal distribution of $50\%$ of the bosons in each well. In general, the less bosons are contained in the BEC, the faster the density collapses. Coming along with the damping of the oscillations, the BEC starts to fragment, reaching its maximal value after the collapse. Again, the less bosons are in the system, the faster the system fragments. Most importantly, irrespective of $N$, the final degree of fragmentation is the same. The fact that the degree of fragmentation appears to be universal is unexpected, since usually fragmentation strongly depends on the number of particles in the BEC. Moreover, the degree of fragmentation upon the density collapse, denoted by $f_{\text{col}}$, depends highly on the initial imbalance $z$ of bosons in the left and right wells. The larger $z$, the stronger does the system fragment in time. This holds true even in the limit of very weak interactions, $\Lambda\ll 1$, meaning that there is no weak-interaction regime where GP yields the correct dynamics. It is important to note that the universality is not an artifact of any approximation, since the result is obtained by solving the full many-boson Schr{\"o}dinger equation numerically exact. Interestingly, the two-mode BH approach can be used to derive an analytic prediction of $f_{\text{col}}$ yielding results close to the exact ones obtained by MCTDHB. Additional details on the analytic expression can be found in Ref. \cite{Kaspar_1D_Bjj_3}. 

To summarize, it was demonstrated that the out-of-equilibrium tunneling dynamics of BECs in a 1D BJJ show significant many-body features that are not accounted for at the GP mean-field level. This includes the appearance of a density collapse in the oscillations between the wells as well as dynamical fragmentation which the GP theory cannot capture by construction. The degree of fragmentation is universal for fixed $\Lambda=\lambda_0(N-1)$ for a wide range of initial states. Applying the commonly used BH model, it was shown that it is superior in certain aspects in comparison to the GP mean-field approach. However, it does also not predict the numerically exact tunneling dynamics as obtained by MCTDHB, e.g., on the short-time dynamics where the interaction parameter is in the vicinity of its critical value or for very strong repulsion where it predicts complete self-trapping of the bosonic cloud. However, the BH model can be utilized to derive an analytic expression for the universal degree of fragmentation that is in good agreement with the exact results.

\subsubsection{Many-body tunneling dynamics in a two-dimensional radial double well}\label{App_Dyn2}
After having shown the many-body nature of the tunneling dynamics of BECs in a 1D BJJ, the question of how a repulsive BEC behaves in a radially symmetric 2D system with a double-well structure is addressed. In particular, as for the previous section, the out-of-equilibrium tunneling dynamics are of interest, and it is studied whether many-body effects can be observed. The results presented in this section were published in Ref. \cite{BeinkeMBTun}, and additional details can be found therein.

\begin{figure}[ht!]
	\begin{minipage}[c]{0.54\textwidth}
 	\begin{subfigure}{\textwidth}
 	  \centering
 	  \includegraphics[width=.95\textwidth]{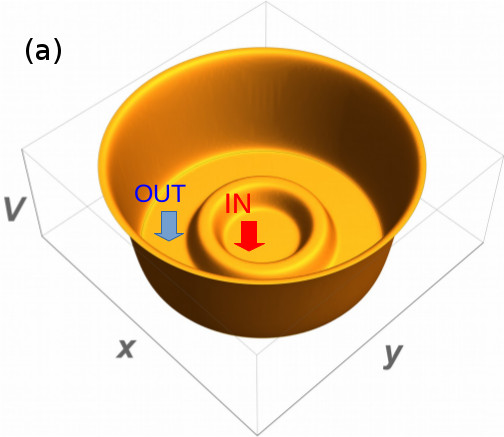}
 	\end{subfigure}
 	\begin{subfigure}{\textwidth}
 	  \centering
 	  \includegraphics[angle=-90,width=\textwidth]{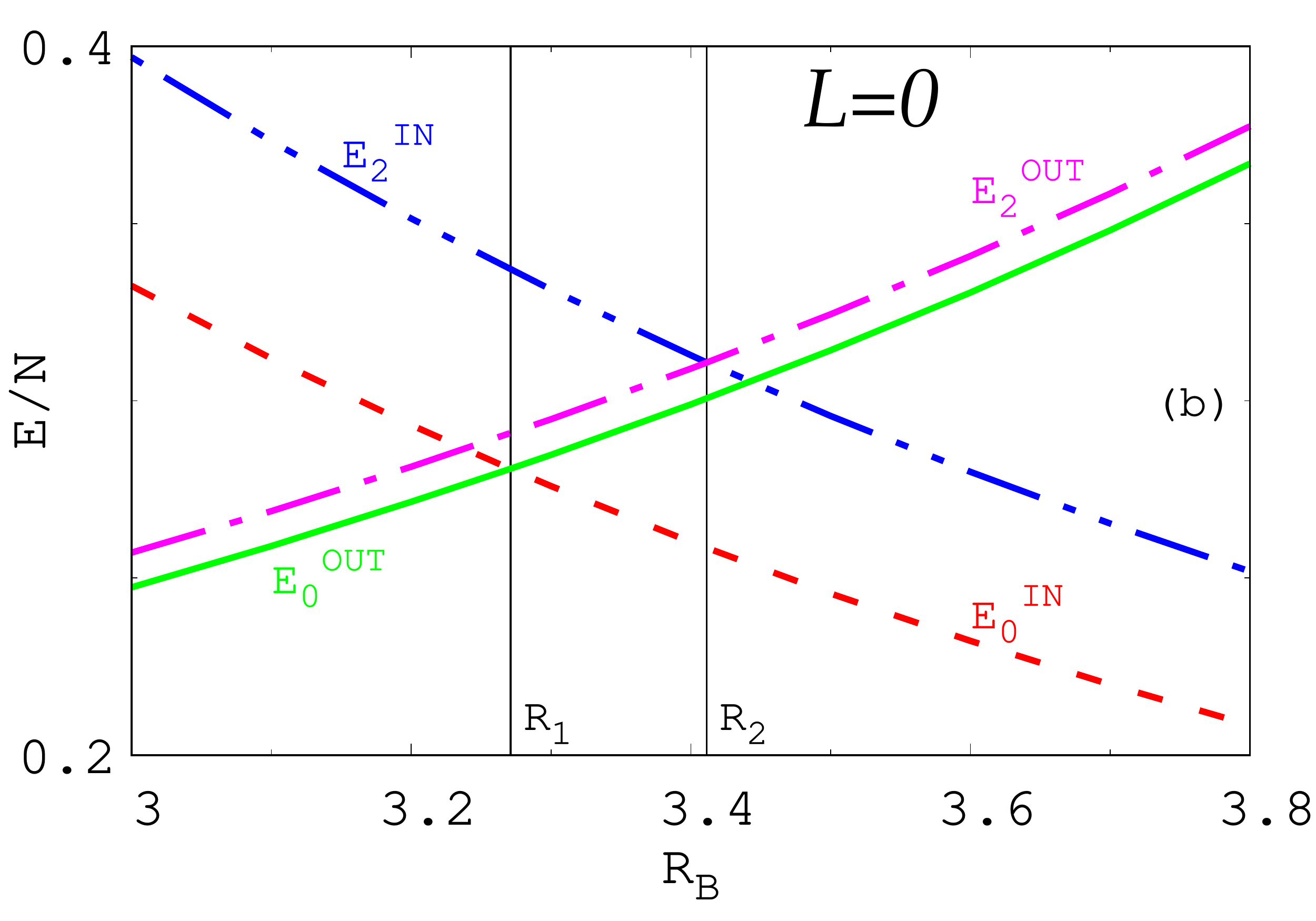}
	\end{subfigure}
	\begin{subfigure}{\textwidth}
	  \centering
 	  \includegraphics[angle=-90,width=\textwidth]{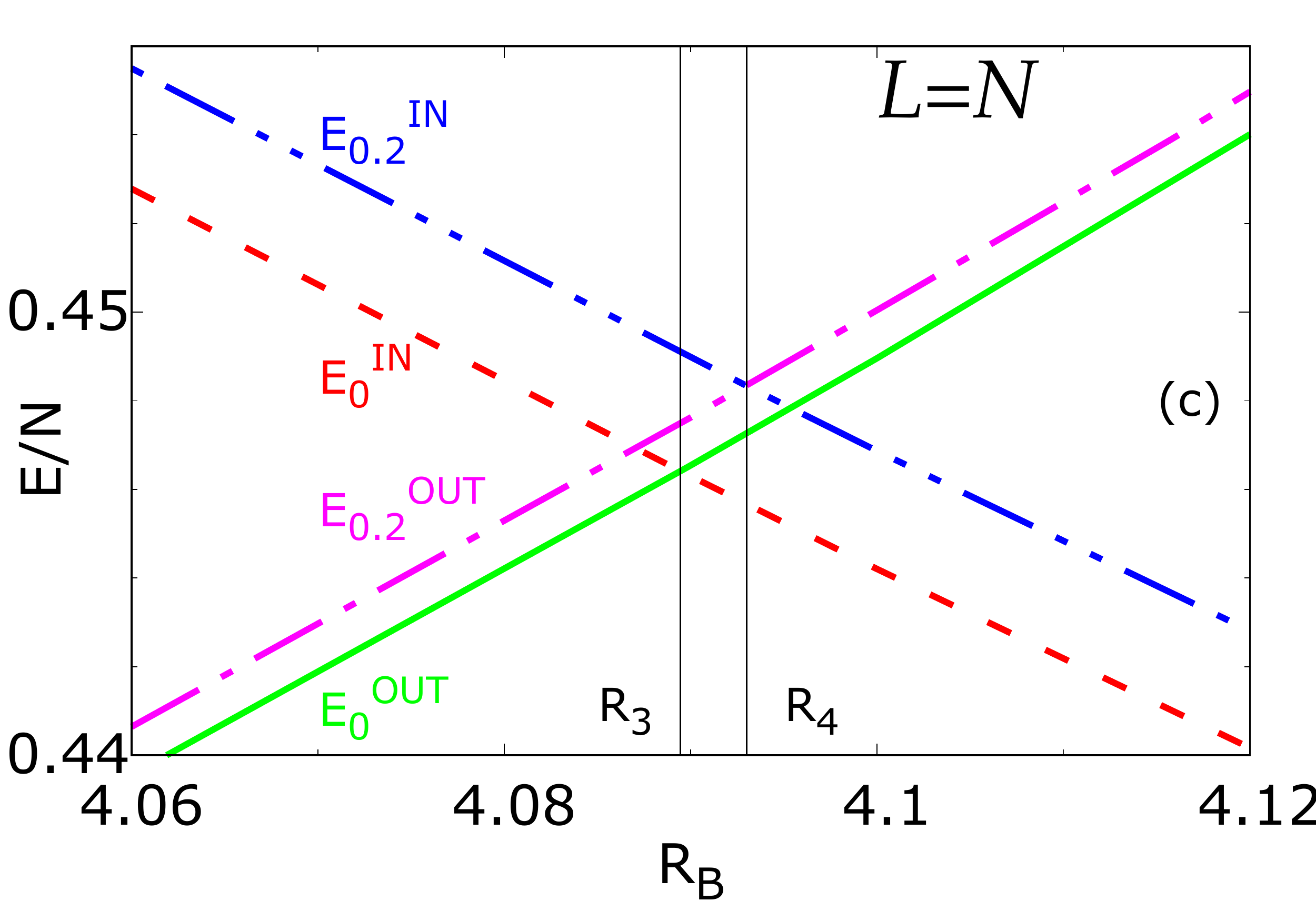} 
 	\end{subfigure}
 	\end{minipage}
\hfill
 	\begin{minipage}[t!]{0.44\textwidth}
 	\vspace*{-3cm}\caption[2D radial double well and ground-state energies for different positions of the central barrier]{(a) Schematic plot of the radial double well. The IN region denotes the trap center, the OUT region denotes the outer annulus. Both are separated by the central ring-shaped barrier. (b) Ground-state energy per particle of a BEC with $N=100$ bosons and total angular momentum $L=0$ for different positions $R_B$ of the radial barrier. The results are calculated at the mean-field GP level ($M=1$). The quantities $E_0^\text{IN}$ (short-dashed red) and $E_0^\text{OUT}$ (solid green) denote the energies for the IN and OUT regions for interaction strength $\Lambda=0$, whereas $E_2^\text{IN}$ (dash-double-dotted blue) and $E_2^\text{OUT}$ (dash-dotted magenta) denote the energies for $\Lambda=2$. The crossing points, i.e., the radii at which the energies for IN and OUT are alike, are located at $R_1=3.271$ ($\Lambda=0$) and $R_2=3.412$ ($\Lambda=2$), see the vertical black lines. Larger repulsion between the bosons shifts the crossing point to larger radii. (c) Same analysis as in (b) but for total angular momentum $L=N$, meaning that each particle carries on average one unit of angular momentum. The location of the crossing points are $R_3=4.089$ ($\Lambda=0$) and $R_4=4.093$ ($\Lambda=0.2$). The angular momentum has a significant effect on the position of the crossing points. Increasing $L$ shifts them to larger radii, which is qualitatively similar as for increasing $\Lambda$. However, one would need to highly increase the interaction strength in order to shift the crossing points as much as adding one unit of angular momentum per particle does. All quantities are dimensionless. See text for details. Figs. (b) and (c) are adapted from Ref. \cite{BeinkeMBTun}.}
 	 	\label{fig:Trap+Crossings}
 	\end{minipage}
\end{figure}

To study tunneling phenomena in 2D geometries has been of high interest in recent years, especially for trapped vortices, i.e., macroscopic quantum states with definite angular momentum, characterized by a density node at its core and phase discontinuities around it. Research concerning tunneling of vortices has been carried out for example in 2D superfluids \cite{arovas2008quantum}. Others dealt with vortices tunneling through a Gaussian-shaped barrier \cite{martin2007transmission}, in pinning potentials \cite{fialko2012quantum,Thouless,Auerbach}, or between two Gaussian wells \cite{salgueiro2009vortex}, to mention only a few. However, there has been no investigation of the tunneling dynamics of vortices in a radially-symmetric double well, in particular not at the many-body level. The latter geometry is very interesting to study many-body dynamics since the trap symmetry preserves the angular momentum of the bosonic cloud. It has been utilized recently to study many-body vortices which are separated in space due to the central barrier \cite{Ofir_MB_vortices}. The following results, recently published in Ref. \cite{BeinkeMBTun}, especially deal with the impact of angular momentum on the tunneling dynamics.

Accurate many-body dynamics are obtained by solving the full many-body Schr{\"o}dinger equation utilizing MCTDHB and compared to the corresponding mean-field dynamics from the GP theory.

For the dynamics, the radial double well in which a BEC with $N=100$ spinless bosons is confined reads
\begin{equation}\label{eq_crater_potential}
V(r)=
\begin{cases}
	B\,e^{-2(r-R_B)^4}+C\,e^{-0.5\,(r-R_C)^4}     \,\,  &\mbox{if } r=(x^2+y^2)^{1/2}\leq R_C \\
 	C \,\, &\mbox{if } r>R_C
\end{cases}
\end{equation}  
where $B$ is the height of the central ring-shaped barrier located at the radius $R_B$, and $C$ is the height of the crater wall at radius $R_C$. A schematic illustration of $V$ is shown in Fig. \ref{fig:Trap+Crossings}(a). The region enclosed by the barrier is denoted by IN, the external rim, i.e., the region between the barrier and the crater wall, is denoted by OUT. For the numerical calculations, $C=200.0$ and $B=1.0$ are set which ensures on the one hand that the energy per particle is sufficiently small such that the BEC stays trapped in the crater, and on the other hand allows for tunneling between IN and OUT, as described below. The radius of the crater is set to $R_C=9.0$.

The short-range repulsion between the particles is modeled by a Gaussian that has been employed recently \cite{Klaiman,Christensson},
\begin{equation}\label{eq_Gauss_rep}
 \lambda_0\hat{W}(\mathbf{r}_i,\mathbf{r}_j)=\frac{\lambda_0}{2\pi\sigma^2}\,e^{-|\mathbf{r}_i-\mathbf{r}_j|^2/2\sigma^2},
\end{equation}
which circumvents the regularization problems of the contact interaction \cite{Doganov}. The repulsion strength will again be expressed in terms of the mean-field parameter $\Lambda=\lambda_0(N-1)$ in the following. The effect of the range of the interaction on the degree of fragmentation has been investigated in detail for repulsive BECs in 1D and 2D single wells \cite{Fischer_Fragmentation}, essentially showing that it becomes increasingly dependent on the density the more long-ranged it is. 

The analysis starts with the ground-state energy of two distinct cases, namely for (i) having all particles in the IN region with closed external rim and (ii) having all particles in the OUT region with closed trap center. The notation $E^\text{IN}_\Lambda$ and $E^\text{OUT}_\Lambda$, where the superscript denotes whether the BEC is held in the IN or OUT region and the subscript denotes the repulsion strength $\Lambda$, is utilized. The ground-state energies for different radial positions of the barrier $R_B$ are calculated using the GP equation. The corresponding results are shown in Fig. \ref{fig:Trap+Crossings}(b) for total angular momentum $L=0$ and Fig. \ref{fig:Trap+Crossings}(c) for $L=N$, i.e., where each boson on average carries one unit of angular momentum. One observes that for all shown parameters ($\Lambda=0,\,2.0$ for $L=0$ and $\Lambda=0,\,0.2$ for $L=N$) the curves for $E^\text{IN}_\Lambda$ and $E^\text{OUT}_\Lambda$ intersect at certain radii which are referred to as the crossing points or simply the crossings in the following. These are defined as the positions where the ground-state energies of the BEC in the IN and OUT regions are identical. Putting the barrier onto the crossing point is the 2D analogue of the symmetric double well in 1D. However, the IN and OUT regions are only energetically equivalent, whereas their geometries are totally different. It is found that both the interaction strength $\Lambda$ and the angular momentum $L$   push the crossing point to larger radii once they are increased. Though, the effect of $L$ is significantly stronger since already for $\Lambda=0$, the transition from $L=0$ to $L/N=1$ leads to a shift of the crossing point from $R_B=3.271$ to $R_B=4.089$. In order to obtain a similar shift just by increasing $\Lambda$, the repulsion needs to be largely increased, and thus one would certainly leave the weakly-interacting regime. The depletion of the ground states with $M=4$ orbitals are $f\approx 10^{-3}$ for $L=0$ and $\Lambda=2.0$ and $f\approx 10^{-9}$ for $L=N$ and $\Lambda=0.2$, respectively. This means that in both cases, the systems' ground states are essentially condensed.

\begin{figure}[ht!]
 	\begin{subfigure}{0.5\textwidth}
 	  \centering	
 	  \includegraphics[angle=-90,width=\textwidth]{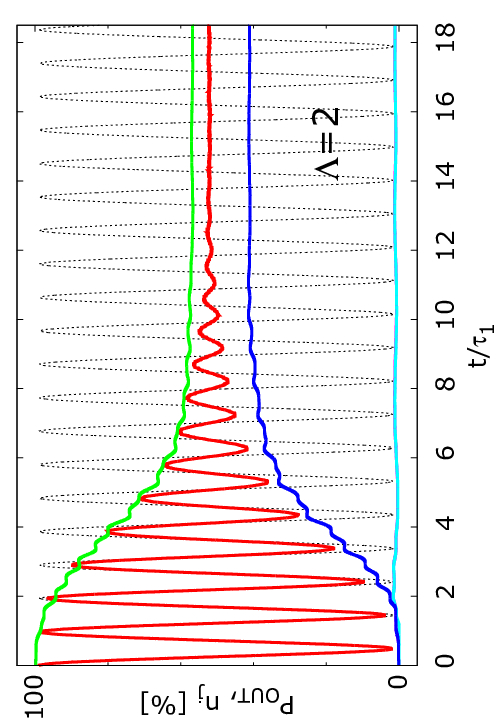} 
 	  \caption{\label{fig_2DMBTun_L0}}
 	\end{subfigure}
 	\begin{subfigure}{0.5\textwidth} 
 	  \centering
 	  \includegraphics[angle=-90,width=\textwidth]{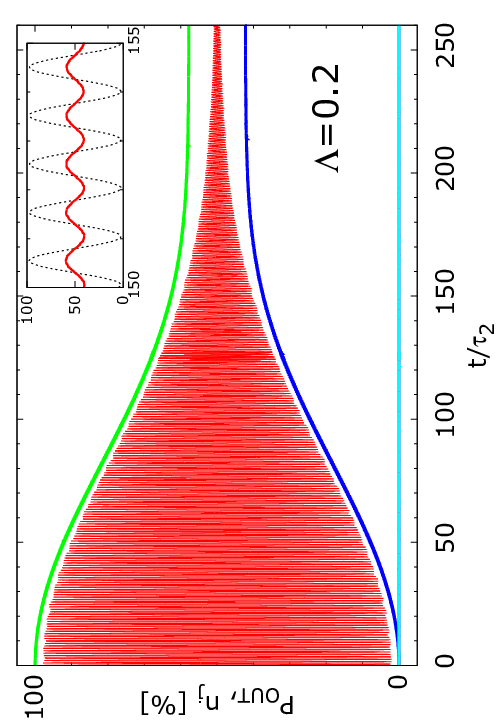}
 	  \caption{\label{fig_2DMBTun_L1}}
	\end{subfigure}
	\vskip\baselineskip\vspace*{-0.2cm}
	\begin{subfigure}{0.5\textwidth}
	  \centering
 	  \includegraphics[width=\textwidth]{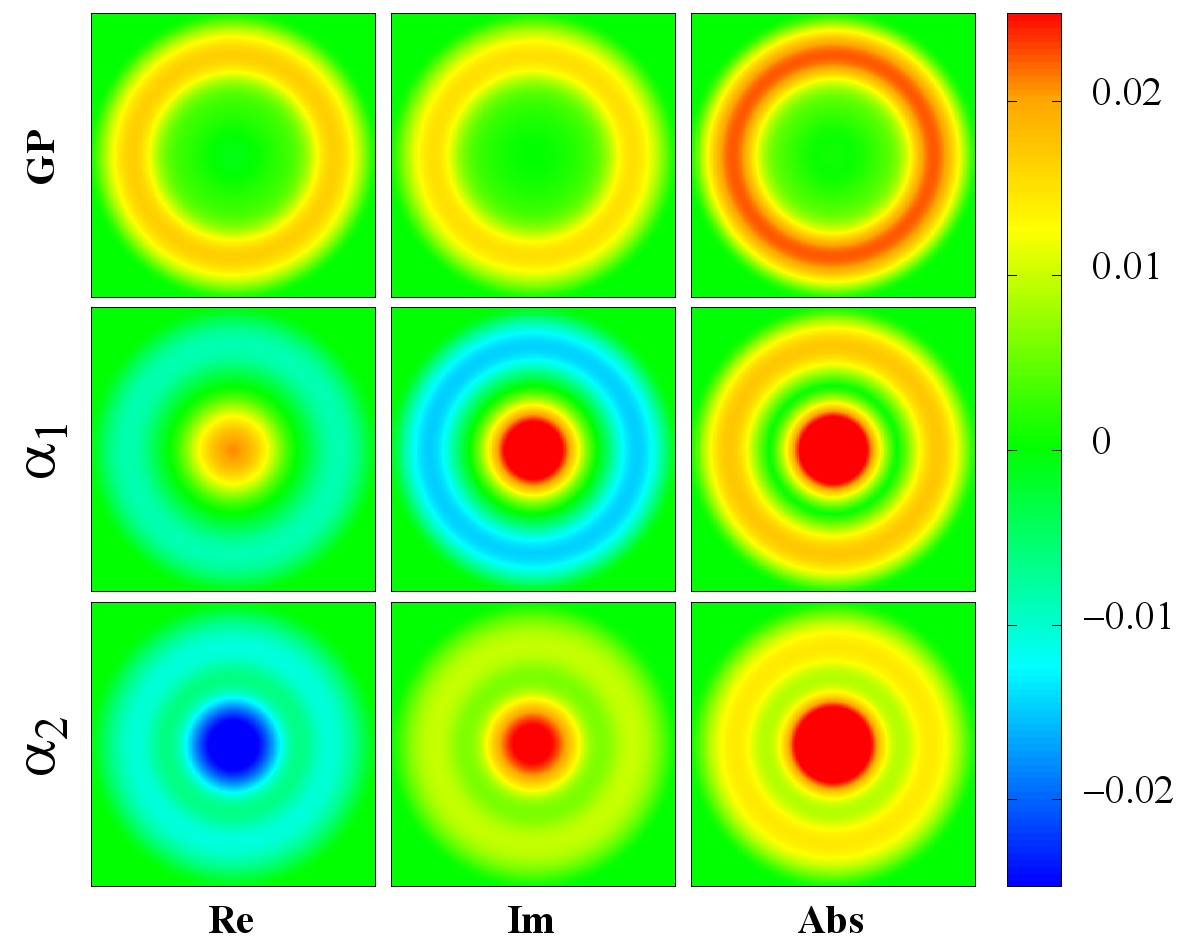} 
 	  \caption{\label{fig_2DMBTun_dens_L0}}
 	\end{subfigure}
 	\begin{subfigure}{0.5\textwidth} 
 	  \centering
 	  \includegraphics[width=\textwidth]{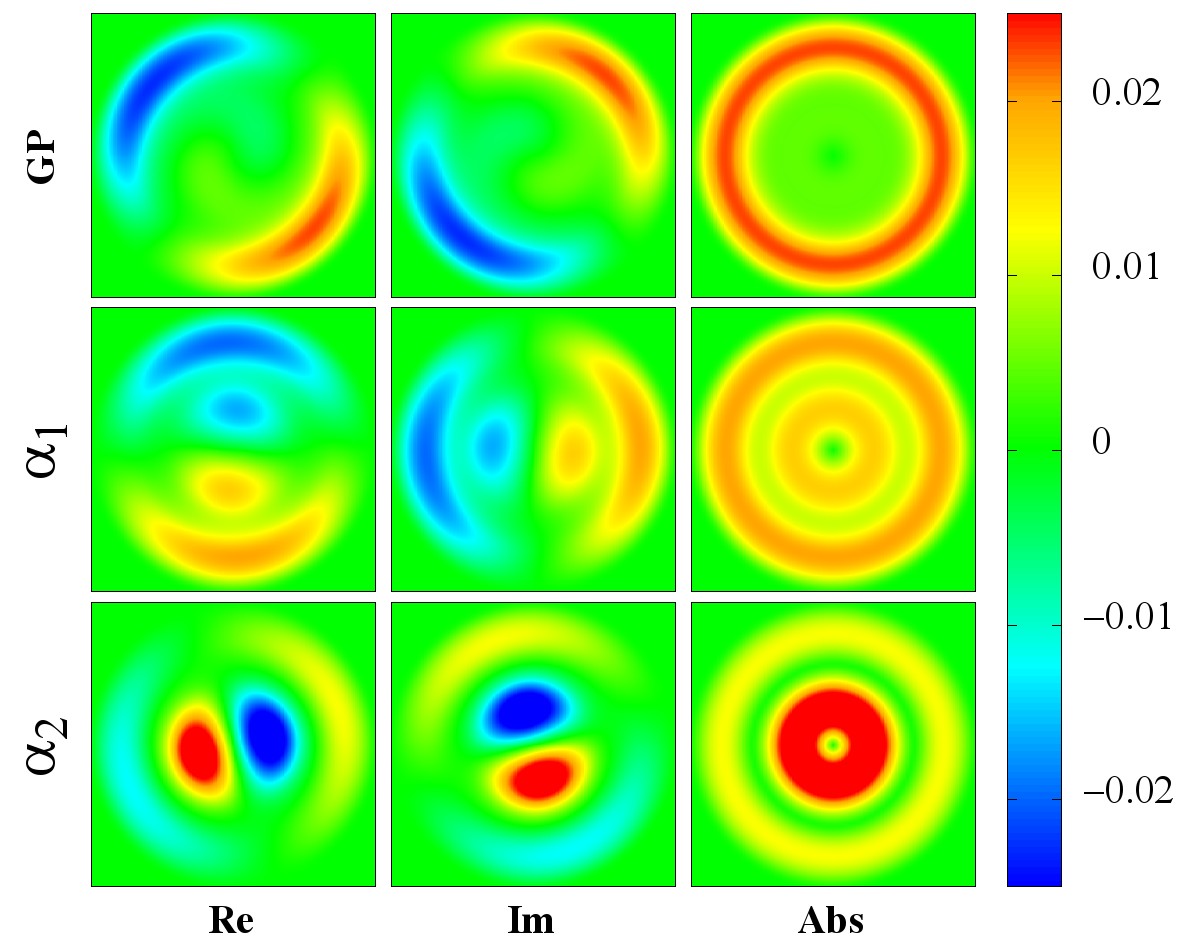}
 	  \caption{\label{fig_2DMBTun_dens_L1}}
	\end{subfigure}
	\caption[Out-of-equilibrium tunneling dynamics in the 2D radial double well]{(a) Time evolution of the occupation probability of the external rim, $P_\text{OUT}(t)$, for $N=100$ bosons with $L=0$ and $\Lambda=2.0$. For $M=4$, initially almost all bosons tunnel. The oscillation amplitude is damped over time and effectively collapses after 12 Rabi cycles with approximately half of the bosons in each part (solid red). The system evolves from being almost fully condensed to two-fold fragmented (see solid green and dark blue curves for $n_1$ and $n_2$). The occupations of the third and fourth orbital are identical and negligible (solid light blue). On the contrary, the GP theory predicts undamped oscillations (dotted gray). (b) Same as in (a) but for $L=N$ and $\Lambda=0.2$. Again, the system evolves from being fully coherent to two-fold fragmented, and the particles finally occupy both the IN and OUT parts by roughly 50$\%$. Mind the different time scales ($\tau_1>\tau_2$). Inset: Detailed view between $150 \tau_2 \leq t \leq 155 \tau_2$. The GP dynamics again do not show the density collapse.  (c) Real, imaginary, and absolute value of the first two natural orbitals $\alpha_1$ and $\alpha_2$ at $t= 18.4\,\tau_1$ where the density is already collapsed. The GP case is shown in the top panels. For $\alpha_1$ and $\alpha_2$, the density occupies both the IN and OUT regions. At the same time, the GP orbital only covers the external part. (d) Same as in (c) but for the dynamics described in (b). At the many-body level, the nodal structure of the vortex state is clearly visible in the trap center. All quantities are dimensionless. See text for details. All figures are adapted from Ref. \cite{BeinkeMBTun}.}
	\label{fig_2DMBTun_2}
\end{figure}

To study the out-of-equilibrium dynamics for different values of $\Lambda$ and $L$, the focus is laid on the situation where the barrier is located on the corresponding crossing points. First, the case of zero angular momentum, i.e., $L=0$, is analyzed. As an initial state, the ground state of trapping the BEC in the OUT region with closed central part is chosen. To trigger the dynamics, the system is quenched by suddenly opening the trap center, such that the entire potential is given by Eq. (\ref{eq_crater_potential}). It was found numerically that there is no difference in starting from either the IN or OUT region when the barrier is located at the crossing point. Fig. \ref{fig_2DMBTun_2}(a) shows the dynamical evolution of the occupation probability of the external rim, 
\begin{equation}\label{P_OUT} 
 P_\text{OUT}(t)=1-P_\text{IN}(t)=\frac{1}{N}\int_{R_B<r\leq R_C} \rho(\mathbf{r};t)d\mathbf{r},
\end{equation}
where $\rho(\mathbf{r},t)$ denotes the single-particle density. For the first two Rabi cycles of $\tau_1=110.23$, nearly the whole cloud tunnels between IN and OUT. Then, the occupation of the second natural orbital starts to grow, accompanied by a damping of the tunneling oscillations. After approximately 12 cycles, the density oscillations have collapsed and the particles are almost equally distributed between IN and OUT. The BEC is now highly (two-fold) fragmented, with occupations of 56.7\% and 41.1\% of the first two natural orbitals $\alpha_1$ and $\alpha_2$, respectively. In contrast to that, the GP dynamics do not show any damping of the oscillations, not even for the long-time behavior (not shown). Fig. \ref{fig_2DMBTun_2}(c) shows the real and imaginary parts as well as the density of the GP orbital (top panels) and of the first two natural orbitals in the many-body case (middle and lower panels) at an instant in time at which the density oscillations have already collapsed. Whereas there are no bosons in the trap center for the GP case, there is significant population of the IN region at the many-body level.

Similar observations can be made for $L=N$ and $\Lambda=0.2$, presented in Fig. \ref{fig_2DMBTun_2}(b). The many-body dynamics show that the density collapses and that approximately 50 bosons are occupying the IN and OUT regions finally. The BEC is two-fold fragmented, the occupation of higher natural orbitals is negligible. The GP dynamics (see inset) do not account for the damped oscillations. With regard to the density of the GP and natural orbitals in Fig. \ref{fig_2DMBTun_2}(d), it can be seen that $\alpha_1$ and $\alpha_2$ clearly show the characteristic nodal vortex structure in the center, whereas at the GP level there are no bosons in the IN region at the chosen instant of time $t=183.3\,\tau_2$ with $\tau_2=66.79$. 

It was further found that many-body effects like the density collapse or the onset of fragmentation are more likely to occur when the particles carry angular momentum. For $L=N$ and $\Lambda=2.0$, the system quickly becomes four-fold fragmented, i.e., at least four orbitals are necessary to account for the many-body dynamics. For $L=0$, the dynamics are essentially governed by only two orbitals, as can be seen from Fig. \ref{fig_2DMBTun_2}(c). Thus, the weakly-interacting regime is apparently smaller for the case of $L>0$, meaning that vortices are more sensitive to many-body effects. In Section \ref{App_RotBEC}, this will be discussed again when excitations in rotating BECs are investigated.

To summarize, the many-body nature of the tunneling dynamics of BECs in a 2D radial double well was demonstrated . The GP mean-field theory does not account for the collapse of the density oscillations between IN and OUT, and therefore fails to give an accurate description of the dynamics. Furthermore, the initially condensed BECs fragment. It was found that bosons carrying angular momentum are more sensitive to many-body effects than non-rotating ones because fragmentation sets in already for weaker interaction strengths. In agreement with these findings, a recent work demonstrates analytically that a many-body approach is necessary for describing the conservation of angular momentum in a 2D radially-symmetric system \cite{Sakmann_analytic}.

\subsection{Dynamical fragmentation}\label{Sec_Appl_dyn_frag}

\subsubsection{Tunneling to open space}\label{App_Dyn3}
The obtained findings on the many-body tunneling process of initially trapped bosons to open space in 1D are summarized in this section. The tunneling process of a single particle was understood already at the beginning of the 20th century \cite{OpenTun_Razavy,OpenTun_Kramers,OpenTun_Condon1,OpenTun_Condon2} and is still of current interest, e.g., with respect to the exit time and momentum of an electron in the ionization process \cite{OpenTun_Teeny1,OpenTun_Teeny2}. Tunneling of a many-boson system, which has also been studied \cite{OpenTun_Carr1,OpenTun_Lenz,OpenTun_Brand,OpenTun_Buchleitner}, is naturally more complicated due to the interactions and correlations between the constituent particles. The following results, published in Refs. \cite{Lode_Tun_Open_1st,Lode_Tun_Open_1st_Corrigendum,Lode_PNAS,Lode_Controlling,Lode_thesis}, explain the tunneling mechanism to open space from a many-body point-of-view utilizing MCTDHB. The main points that are addressed deal with the questions whether the particles tunnel one by one or if several particles tunnel simultaneously, and whether the entire process is indeed of many-body nature.

To study the dynamics of the tunneling to open space, a system of $N$ bosons confined in a harmonic trap of the form $V(x,t=0)=\frac{1}{2}x^2$ is considered. For $t>0$, the trap is quenched such that it becomes open to one side. The region where the bosons are trapped initially and the open space are separated by a potential barrier. The full potential reads
\begin{equation}
V(x,t>0)=\theta(x_{c1}-x)\frac{1}{2}x^2+\theta(x-x_{c1})\theta(x_{c2}-x)P(x)+\theta(x-x_{c2})T
\end{equation}
where $\theta(x)$ denotes the Heaviside step function, $P(x)$ denotes a third order polynomial that ensures a smooth connection between the harmonic trap, the potential barrier and the threshold potential $T$ whose role will be discussed below. The coordinates $x_{c1}$ and $x_{c2}$ refer to connection points, where the former connects the harmonic trap and the barrier and the latter connects the barrier and the threshold. The bosons repel each other via the contact interaction potential, and its strength is measured via the mean-field parameter $\Lambda=\lambda_0(N-1)$.

Fig. \ref{fig:open_space_1} shows a schematic plot of the trapping potential for $t>0$ (solid black). The maximum height of the barrier separates the IN and OUT regions, denoted by the vertical red line. Moreover, the colored horizontal lines mark the total energies of states with different numbers of bosons $N$ in the trap. Depending on the threshold value $T$, certain states can become bound, meaning that no additional particles can escape from the trap via tunneling through the barrier. How the threshold can be utilized to control the tunneling-to-open-space dynamics will be discussed below.

\begin{figure}[h!]
  \centering
  \includegraphics[width=0.8\textwidth]{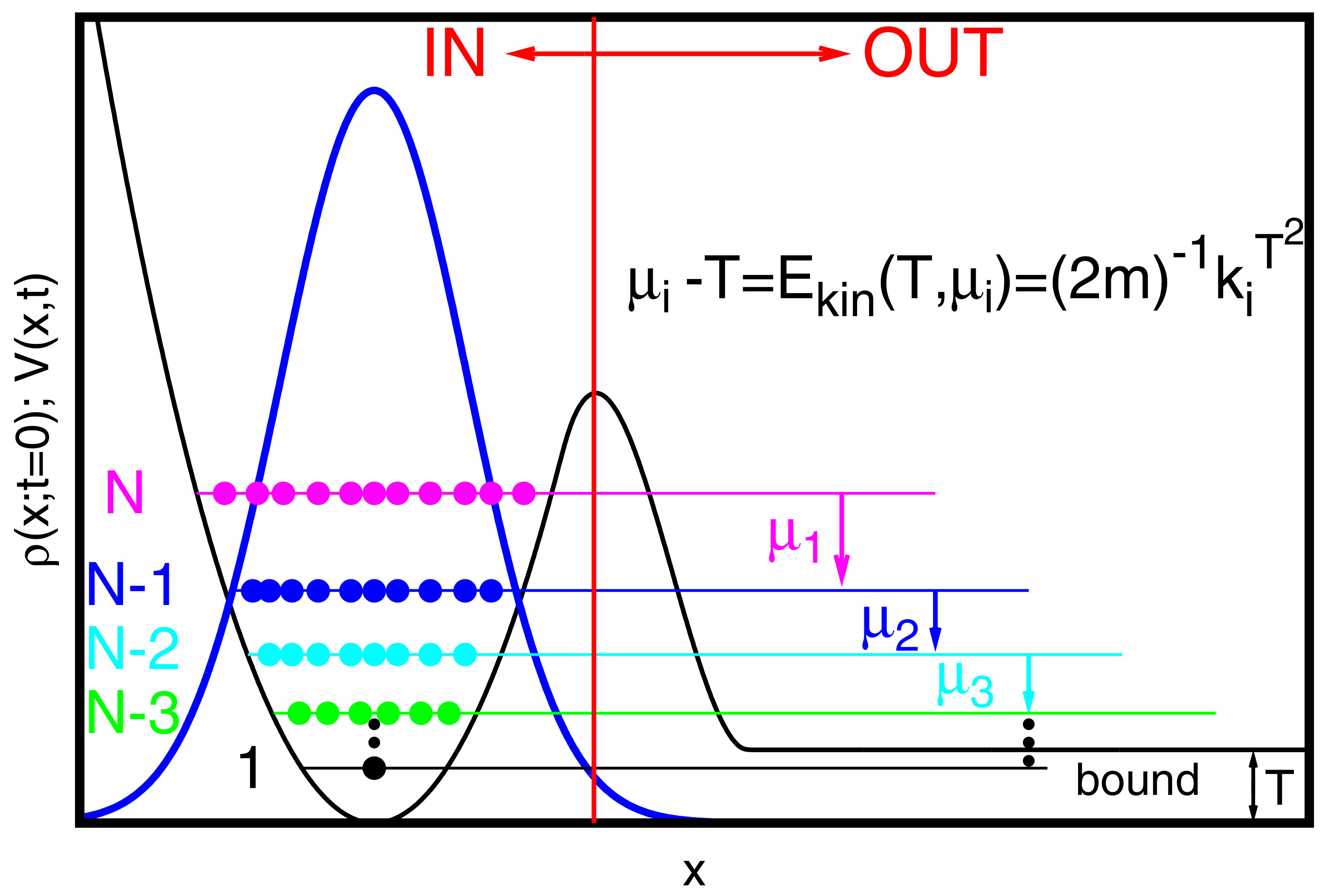}
  \caption[Schematic plot of the potential for tunneling to open space]{Schematic plot of the potential $V$ for $t>0$ (solid black). The initial Gaussian-shaped density $\rho(x,t=0)$ is shown in solid blue. The IN and OUT regions are separated at the position of the maximum of the potential barrier (see solid red lines and labels). Colored horizontal lines denote the energies of states with different numbers of bosons $N$ in the trap. The energy differences between those states are denoted by the chemical potentials $\mu_i$. The value $T$ denotes the height of the threshold potential in the OUT region. All quantities are dimensionless. See text for details. The figure is taken from Ref. \cite{Lode_Controlling}.}
  \label{fig:open_space_1}
\end{figure}

Important quantities for the analysis are the density $\rho(x,t)$ [see Eq. (\ref{Eq_1BRDM})], as well as the density in momentum space $\rho(k,t)$, or simply the momentum distribution, which is obtained by performing a Fourier transformation (FT) of the one-particle reduced density in real space. Furthermore, the degree of fragmentation $f$, defined in Eq. (\ref{Eq_depletion}), and especially its evolution in time, is of major interest. The above quantities are computed at the many-body level employing MCTDHB.

The analysis starts by looking onto the shape of the momentum distribution in general. Fig. \ref{fig:open_space_2} shows $\rho(k,t)$ for a system of $N=101$ bosons with interaction parameter $\Lambda=0.3$ at four different instants in time $t_1<t_2<t_3<t_4$. It consists of two parts, a Gaussian background and a peak structure. The former refers to the particles that did not escape from the trap yet, since the initial real-space density is also a Gaussian. On the contrary, the peaks refer to the momenta of the already escaped particles. The heights of the peaks are time-dependent, which can be seen from the growth of the dominant peak. The position of the peaks can be understood from a static model introduced below.

\begin{figure}[h!]
  \centering
  \includegraphics[width=0.7\textwidth]{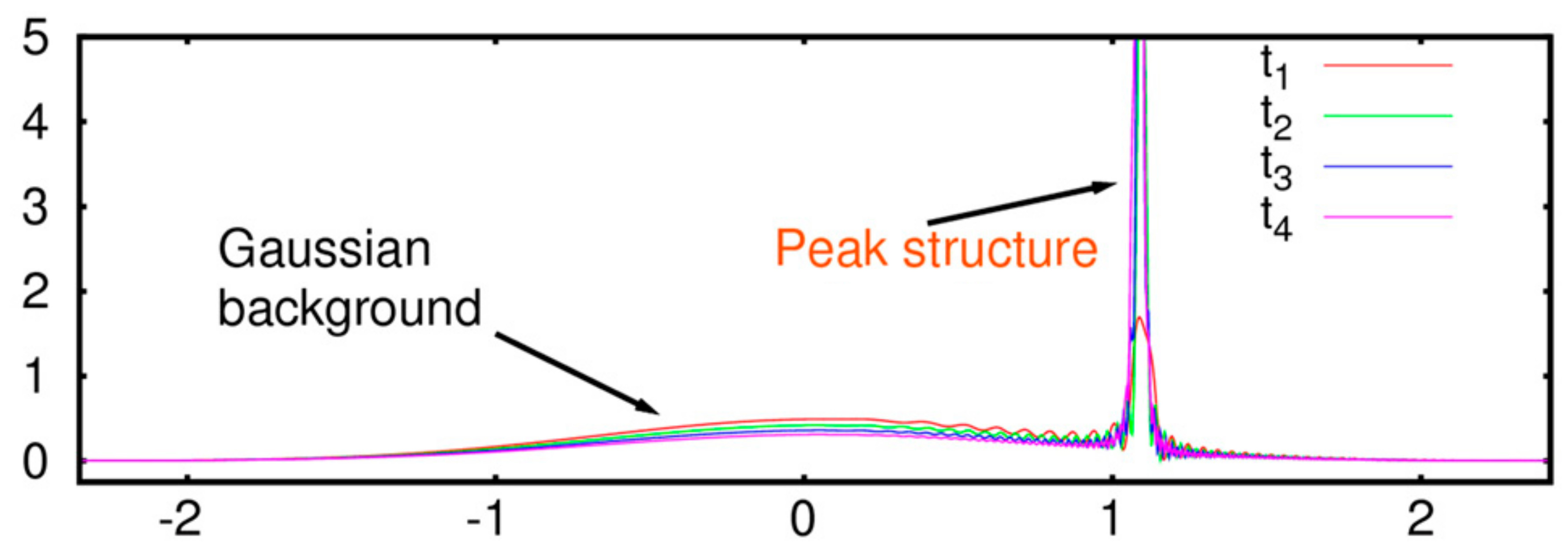}
  \caption[Momentum distribution for $N=101$ bosons at different times in the tunneling-to-open-space dynamics]{The momentum distribution $\rho(k,t)$ for $N=101$ bosons with repulsion strength $\Lambda=0.3$ at four instants in time with $t_1<t_2<t_3<t_4$. The Gaussian background refers to the bosons remaining in the trap, whereas the peaks refer to the momenta of the emitted bosons. The height of the dominant peak is clearly growing in time. All quantities are dimensionless. See text for details. The figure is adapted from Ref. \cite{Lode_PNAS}.}
  \label{fig:open_space_2}
\end{figure}

The minimal total energy $E_\text{TOT}$ of the system is comprised of the energy of the bosons that remain in the IN region, denoted by $E_{\text{HO}}(\lambda_0,N_{\text{IN}})$ indicating its dependence on the two-body repulsion strength $\lambda_0$ and the number of particles in the trap $N_{\text{IN}}$, as well as on the minimal energy of the $N_\text{OUT}$ bosons that tunneled through the barrier into open space. Their minimal energy is the height of the threshold $T$.  One thus obtains 
\begin{equation}
	E_\text{TOT}=E_{\text{HO}}(\lambda_0,N_{\text{IN}})+N_\text{OUT}\,T.
\end{equation}
Assuming that the interaction among the emitted bosons is negligible, the kinetic energy of the $i$-th escaped boson is given by
\begin{equation}
	E_\text{kin}(T,\mu_i)=\mu_i-T
\end{equation}
where the $\{\mu_i\}$ are chemical potentials that refer to the energy to bring an additional boson into the IN region with already $N-i$ bosons. The corresponding momenta read
\begin{equation}\label{Eq_OpenTun_model_k}
	k_i^T=\sqrt{2(\mu_i-T)}
\end{equation}
where $m=\hbar=1$ is assumed. It is stressed that this model describes the tunneling of a single boson.

\begin{figure}[h!]
  \centering
  \includegraphics[width=\textwidth]{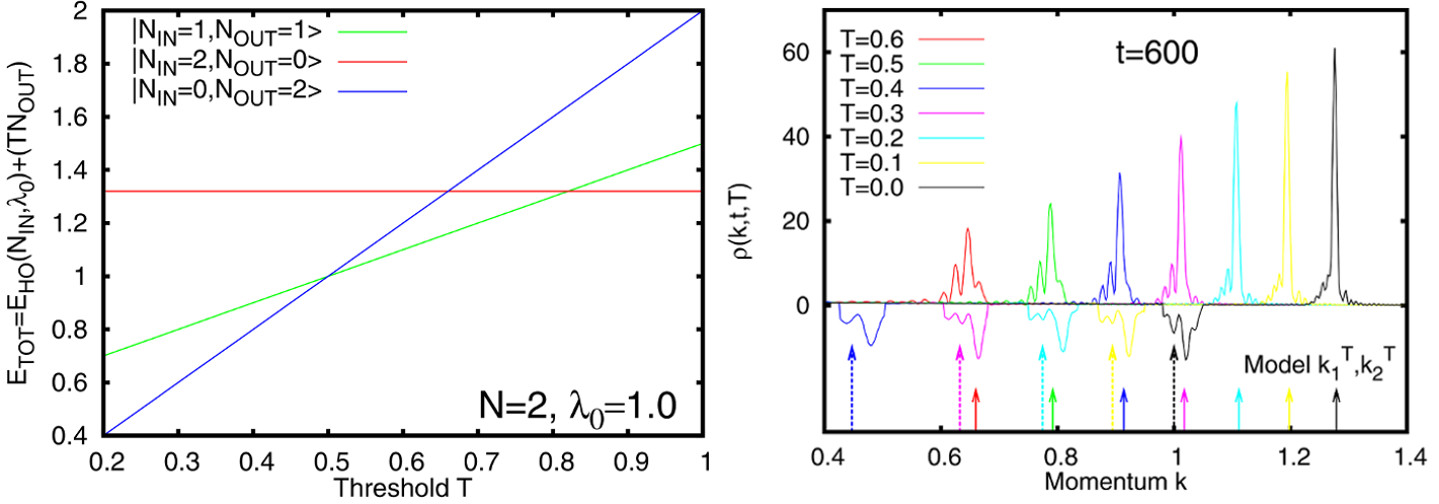}
  \caption[Static considerations and momentum distribution for $N=2$ bosons in the tunneling-to-open-space dynamics]{The tunneling to open space of $N=2$ bosons. Left panel: Static consideration of the energy dependence of the possible final states $|N_{\text{IN}},N_{\text{OUT}}\rangle$ on the threshold $T$. For $T\lesssim 0.5$, both particles are expected to tunnel, whereas for $0.5\lesssim T \lesssim 0.8$, only one boson is expected to tunnel. For even larger $T$, the bosons should remain trapped. Right panel: The obtained peaks in the momentum distribution $\rho(k,t=600,T)$ for different $T$. For $T\leq 0.4$, two peaks appear at $k_{1,2}^T$, for $T=0.5$ and $0.6$, only one peak appears. This is in agreement with the expected number of ejected bosons from the trap. For larger $T$, the peaks are shifted to smaller momenta since the bosons have to overcome a larger threshold. All quantities are dimensionless. See text for details. The figures are adapted from Ref. \cite{Lode_Controlling}.}
  \label{fig:open_space_3}
\end{figure}

These model assumptions are tested against a system with $N=2$ bosons and two-body repulsion strength $\lambda_0=1$. The left panel of Fig. \ref{fig:open_space_3} shows the dependence of $E_\text{TOT}$ on the threshold value $T$ for the three possible final states $|N_\text{IN},N_\text{OUT}\rangle$, given by $|2,0\rangle$, $|1,1\rangle$ and $|0,2\rangle$. The utilized notation $|N_\text{IN},N_\text{OUT}\rangle$ should not be confused with a Fock state and is just used to count the particles in the two regions IN and OUT. For $T\lesssim 0.5$, the lowest-in-energy state is $|0,2\rangle$, i.e., the final state where both bosons have tunneled to open space. For $0.5\lesssim T \lesssim 0.8$, the state $|1,1\rangle$ is the lowest, meaning that only one boson is expected to tunnel in the long-time dynamics. For even higher $T$, tunneling is suppressed completely because the state $|2,0\rangle$ corresponds to the lowest energy. This is in full agreement with the results shown in the right panel of Fig. \ref{fig:open_space_3}. There, the peaks in the momentum distribution $\rho(k,t=600)$, corresponding to the momenta $k_1^T$ and $k_2^T$ of the emitted particles, are presented for different values of the threshold $T$. The $k_2^T$-peaks are plotted upside-down for better visibility. The first observation is that increasing $T$ shifts the peaks to smaller momenta, which is due to the higher energy that is necessary to overcome the threshold. Secondly, two peaks appear up to $T=0.4$, whereas for $T=0.5$ and $0.6$ only one peak appears. This can be understood in terms of the static considerations from the left panel, where it was observed that for $T \lesssim 0.5$ both particles are expected to tunnel through the barrier, whereas only one boson is expected to tunnel for $0.5\lesssim T \lesssim 0.8$. Thus, the threshold can be utilized to control the number of tunneling bosons. The arrows in the bottom of the right panel mark the calculated positions of the peaks with respect to the model of Eq. (\ref{Eq_OpenTun_model_k}). The good agreement with the positions of the peaks of the momentum distribution suggests that the tunneling of single particles can be well described by the above model.

\begin{figure}[h!]
  \centering
  \begin{subfigure}{0.5\textwidth}
  \includegraphics[width=\textwidth]{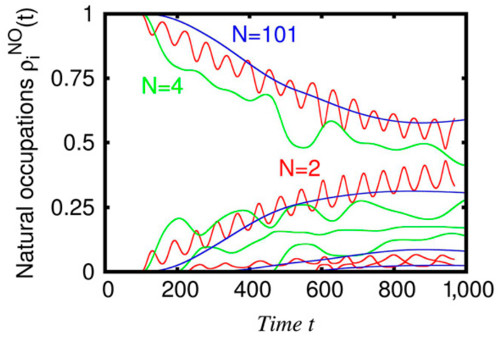}
  \end{subfigure}
  \begin{subfigure}{0.48\textwidth}
  \includegraphics[width=\textwidth]{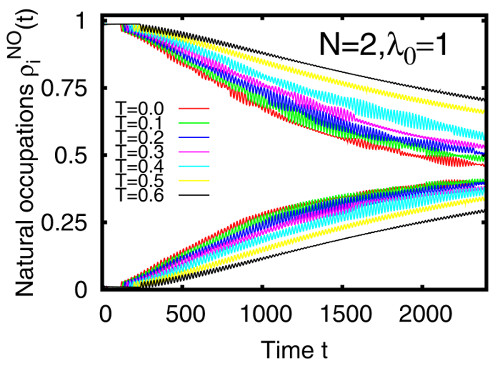}
  \end{subfigure}
  \caption[Evolution of the natural occupation numbers for $N=2$ bosons and different values of the threshold]{Evolution of the natural occupations of the first few natural orbitals. Left panel: The case of zero threshold, $T=0$, utilizing $M=4$ orbitals. The interaction strength is $\lambda_0=0.3$. For all particle numbers shown, the system evolves from being condensed to being multi-fold fragmented. Right panel: The occupation of the first two natural orbitals for $N=2$ particles and $\lambda_0=1$, but for different heights of the threshold $T$. In general, the higher the threshold the later the development of fragmentation sets in. Moreover, the system with larger $T$ fragment less than the ones with smaller $T$. All quantities are dimensionless. See text for details. The figures are taken from Refs. \cite{Lode_PNAS,Lode_Controlling}.}
  \label{fig:open_space_4}
\end{figure}

It remains the question whether the bosons do only tunnel one by one, or if it is also possible that several bosons tunnel at once. To answer this question, one has to consider that the peaks for $T\lesssim 0.5$ in Fig. \ref{fig:open_space_3} (right panel) appear sequentially, i.e., first the dominant peak develops, and then the second, smaller peak appears. The fact that the height of the peaks are time-dependent suggests that the two processes, namely the separate ejection of the first and second boson, are not independent. If however a multi-boson ejection happened, peaks at higher momenta than the ones observed would appear. For instance, in the limit of large $N$, the chemical potential corresponding to a two-boson ejection would be $\mu^{2b}\approx 2\mu_1$, which, for the case of $T=0$, yields $k^{2b,0}\approx\sqrt{2}k_1^0$. This is simply not observed in the Fourier spectrum of $\rho(k,t)$. Thus, it can be deduced that the bosons tunnel to open space one at a time. Nevertheless, as mentioned above, the individual processes of the ejection of single bosons are not independent, meaning that they quantum-mechanically interfere. 

Finally, it is discussed whether the overall process of tunneling to open space is indeed a many-body process or whether it can also be described at the mean-field level. To this end, the time-evolution of the occupations of the first few natural orbitals for systems with different numbers of bosons and different heights of the threshold is studied in Fig \ref{fig:open_space_4}. In the weakly-interacting regime considered here, all initial states of the bosonic clouds are essentially condensed. For the case of zero threshold (left panel), one observes that for all numbers of bosons considered the system becomes multi-fold fragmented. Especially for $N=4$ bosons, at least four orbitals become macroscopically occupied. With respect to the case of $N=2$ particles and non-zero thresholds, the behavior is similar, meaning that the initially coherent BEC becomes highly fragmented. In general, the onset of fragmentation happens at later times when $T$ is larger. Furthermore, the degree of fragmentation is largest in the absence of any threshold.

To summarize, it has been found that the tunneling-to-open-space dynamics of repulsive bosons from a harmonic trap are of many-body nature. With  time, the system evolves from being coherent to become highly fragmented, and thus a mean-field description is inapplicable even in the weakly-interacting regime. The bosons tunnel one by one, meaning that no multi-boson tunneling occurs where more than a single boson at the same time is emitted jointly from the trap. However, the individual tunneling processes are not independent, they quantum-mechanically interfere. The threshold in the open space can be utilized to control the number of emitted particles. In particular, it can be used to suppress tunneling completely. \enlargethispage{\baselineskip} A detailed analysis on the coherence of the system, showing that the emitted bosons lose coherence with the bosons in the trap and among themselves, can be found in Refs. \cite{Lode_Tun_Open_1st,Lode_Tun_Open_1st_Corrigendum,Lode_PNAS,Lode_Controlling,Lode_thesis}, as well as additional details.

\subsubsection{Phantom vortices}\label{App_Dyn4}
A recently found vortex type in a rotating and repulsive 2D single-component BEC, termed phantom vortex, is presented in this section. Different to the common understanding of a quantum vortex that, among other features, has a density node (or core) and a phase discontinuity surrounding it, a phantom vortex cannot be detected in the condensate density. Although they were not found in single-component condensates so far, coreless vortices have been predicted earlier for multi-component condensates, e.g., spinor condensates where the core of one species, i.e., bosons with a certain spin, is filled by another species \cite{Phantom_Kasamatsu,Phantom_Lovegrove}. Since the experimental realization of gaseous BECs \cite{Phantom_Anderson,Phantom_Hulet,Phantom_Davis,Phantom_Cornell,Phantom_Ketterle}, the nucleation of vortices in rotating BECs was of interest and it was found that there is a critical rotation frequency below which no quantized vortices appear \cite{Rokhsar,Phantom_AboShaeer,Madison1,Madison2,Madison3,Phantom_Dagnino}. In contrast to that, phantom vortices appear also below this critical rotation frequency. Instead of a node in the density, they rather manifest themselves in nodes of the densities of underlying natural orbitals. Thus, phantom vortices are a many-body object that cannot be described at the mean-field level. Moreover, their occurrence is accompanied by the onset of fragmentation. The following results were published in Ref. \cite{Phantom_Lode}. Earlier results on phantom vortices in a stirred BEC where, e.g., the fragmentation entropy is analyzed, can be found in Ref. \cite{Lode_stirred_BEC}.

The system under consideration is a BEC of $N=100$ spinless bosons. They interact via the same Gaussian repulsion as in Eq. (\ref{eq_Gauss_rep}) with interaction parameter $\Lambda=\lambda_0(N-1)=17.1$. They are confined in a time-dependent trap that is varied stepwise. It reads
\begin{equation}\label{Eq_trap_phantom}
	V(\mathbf{r},t)=\frac{1}{2} \left(x(t)^2+y(t)^2 \right)+\frac{1}{2}\eta(t)\left( x(t)^2-y(t)^2\right)
\end{equation}
where $\eta$ is the parameter of the anisotropy given by the second term in Eq. (\ref{Eq_trap_phantom}). The $x-$ and $y-$coordinates are varied in time according to
\begin{equation}\label{Eq_phantom_coord}
	\begin{pmatrix} x(t) \\ y(t)	\end{pmatrix}= \begin{pmatrix}
	\cos(\omega t) & \sin(\omega t) \\ -\sin(\omega t) & \cos(\omega t)
	\end{pmatrix}
\end{equation}
where the frequency $\omega$ is set to $0.78$. Initially, the trapping potential is isotropic and harmonic, meaning that $\eta(t=0)=0$, and the corresponding ground state is computed both at the mean-field level using the GP equation as well as at the many-body level using imaginary time-propagation of the MCTDHB equations. Afterwards, a slight anisotropy is added by increasing $\eta$ from zero to its maximal value of $0.1$ until $t=80$. The resulting elliptic trap is kept between $80<t\leq 300$, before $\eta$ is ramped down to zero again between $ 300< t \leq 380$. For the last interval until $t=500$, the isotropic harmonic trap is kept constant. The entire procedure is shown in the upper panel of Fig. \ref{fig:phantom_vortices_1}, together with representative densities of the BEC for the individual intervals. These densities already suggest that at no point in time there is a vortex contained in it, as discussed in more detail below.

\begin{figure}[h!]
  \centering
  \includegraphics[width=0.8\textwidth]{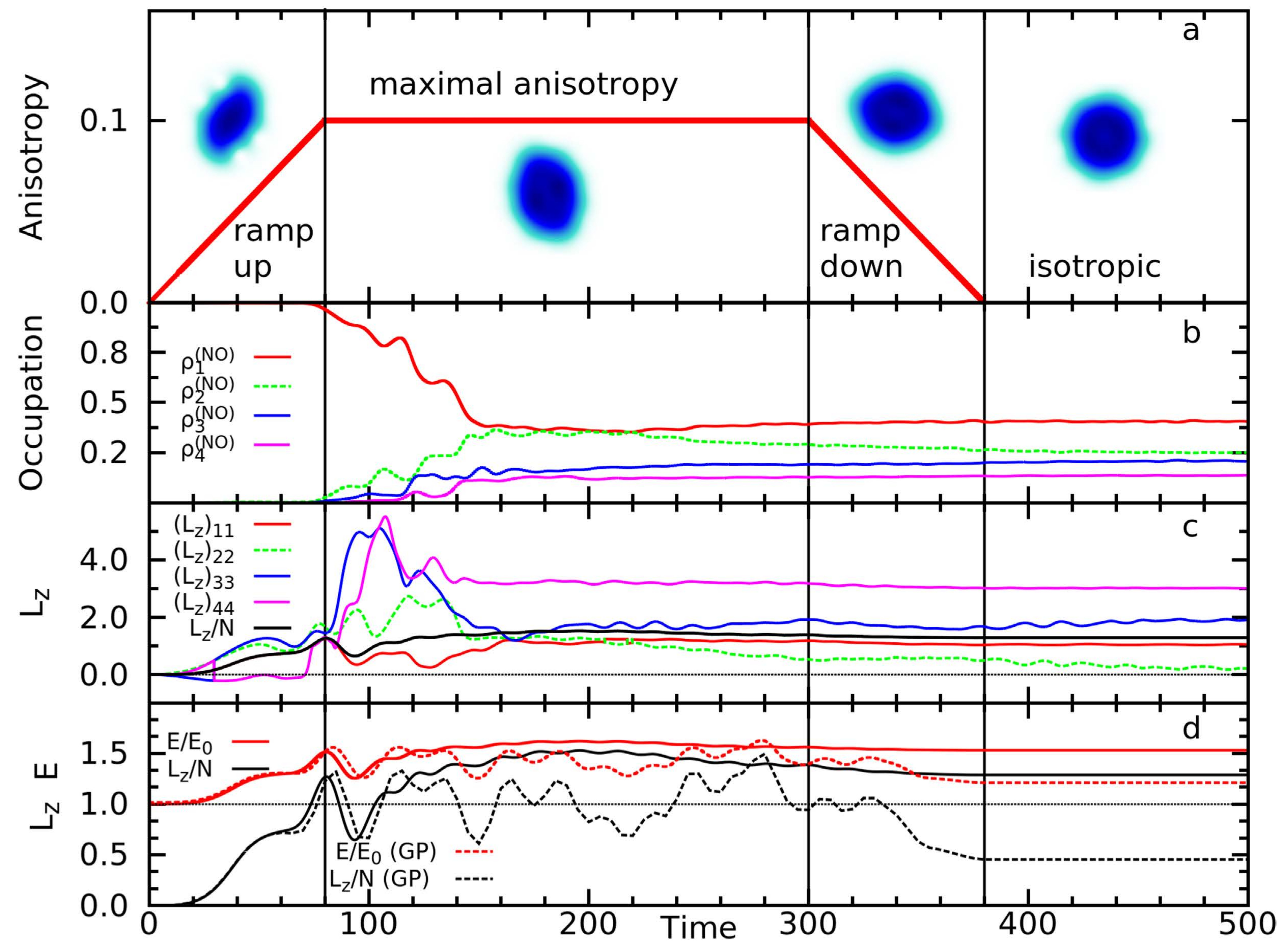}
  \caption[Dynamical protocol of the anisotropy and evolution of important quantities during the nucleation of phantom vortices]{Top panel: Time-evolution of the anisotropy parameter $\eta(t)$ (red line) for the 4 distinct intervals discussed in the text, together with representative densities (blue). The dynamic protocol is divided into a period of 'ramp up', 'maximal anisotropy', 'ramp down' and 'isotropic'. Second panel: Time-evolution of the first 4 natural occupation numbers $\rho_i^{(\text{NO})}(t)$. Fragmentation sets in at the end of the ramp up of the anisotropy. Third panel: Time-evolution of the orbital angular momenta $(L_z)_{ii}(t)$ and the total angular momentum per particle, $L_z(t)/N$. Fluctuations are maximal when fragmentation sets in. Bottom panel: Comparison between the time-evolution of the total energy $E$ relative to the initial energy $E_0$ and $L_z(t)/N$. The GP theory underestimates both quantities, especially the angular momentum is approximately three times less than at the many-body level. All quantities are dimensionless. See text for details. The figure is taken from Ref. \cite{Phantom_Lode}.}
  \label{fig:phantom_vortices_1}
\end{figure}

Of particular interest for the analysis will be, apart from the density $\rho(\mathbf{r})$, the natural occupation numbers, denoted here by $\rho_i^{(\text{NO})}$ but with the same meaning as the $n_i$ in Eq. (\ref{Eq_Definition_natural_occs}), and the densities $|\phi_i|^2$ of the natural orbitals (mind the different notation as compared to Section \ref{CH_MB_theory} where the $\phi_i$ denote the working orbitals of MCTDHB). Furthermore, the time-evolution of the expectation values of the orbital angular momenta, given by $(L_z)_{ii}=\langle\phi_i|\hat{L}_z|\phi_i\rangle$, and of the total angular momentum per particle, $L_z/N=\frac{1}{N}\sum_{i,j=1}^M \rho_{ij}\,(L_z)_{ij}$, are analyzed. The time-argument is suppressed in the above mentioned quantities. Computations are carried out for orbital numbers $M=4$ (many-body) and $M=1$ (GP). Additionally, signatures of phantom vortices can also be seen in the phase of the first-order correlation function $g^{(1)}(\mathbf{r}|\mathbf{r}^\prime;t)$ [see Eq. (\ref{Eq_pth_corr_func}) in Appendix \ref{Appendix_variance}] defined by
\begin{equation}\label{Eq_phase_first_order_corr}
	S_g(\mathbf{r}|\mathbf{r}^\prime;t)=\text{arg}[g^{(1)}(\mathbf{r}|\mathbf{r}^\prime;t)]
\end{equation}
as well as in the phases of the natural orbitals given by
\begin{equation}\label{Eq_phase_nat_orb}
	S_i(\mathbf{r},t)=\text{arg}[\phi_i(\mathbf{r};t)].
\end{equation}
The discussion of the results is split into two parts. At first, the appearance of phantom vortices is analyzed at an instant in time when the trap is already isotropic again. Afterwards, the degree of fragmentation in the system is discussed.

\begin{figure}[h!]
  \centering
  \includegraphics[width=\textwidth]{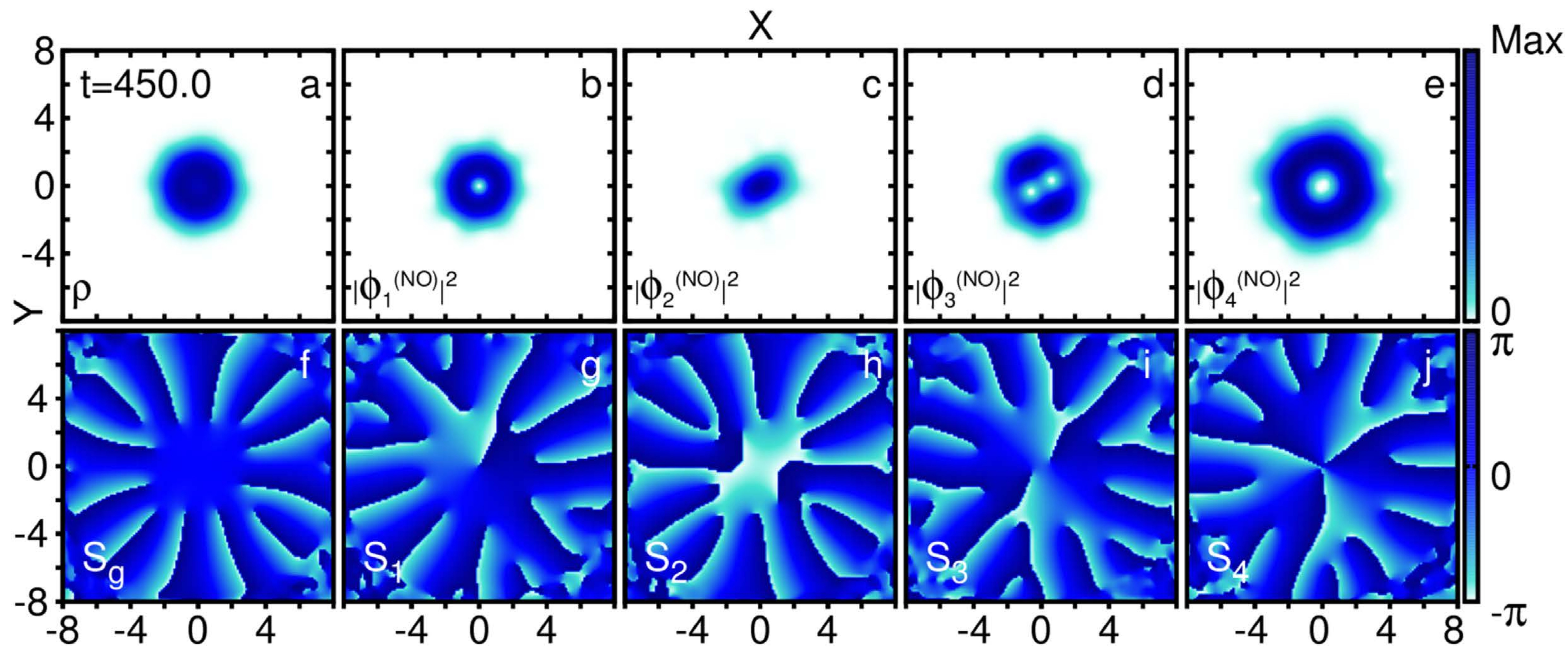}
  \caption[Total and orbital densities and phases during the nucleation of phantom vortices]{Top panels: Total density $\rho(\mathbf{r})$ (a) and densities of the natural orbitals (b)-(e) at $t=450$, i.e., when the trap is already isotropic again. Although no vortex is observed in the total density, phantom vortices in the orbitals $\phi_1$, $\phi_3$ and $\phi_4$ appear. Bottom panels: Phases of the first-order correlation function (f) and of the natural orbitals (g)-(j). Phase discontinuities are observed for all phantom vortices. All quantities are dimensionless. See text for details. The figure is taken from Ref. \cite{Phantom_Lode}.}
  \label{fig:phantom_vortices_2}
\end{figure}

With respect to the first part, the total density and the densities of the natural orbitals at $t=450$, i.e., when the anisotropy is ramped down to zero again, are examined in Fig. \ref{fig:phantom_vortices_2}. One observes that the total density [panel (a)] does not exhibit any vortex. On the contrary, several phantom vortices appear in the densities of the natural orbitals. The densities of the first and fourth natural orbitals contain a charge-1 and a charge-3 phantom vortex, respectively [panels (b) and (e)]. The density of the third orbital contains a pair of charge-1 phantom vortices [panel (d)]. The mechanism responsible for the nucleation of the phantom vortices in $\phi_1$ and $\phi_4$ is different to the one responsible for the pair in $\phi_3$. The former mechanism is called node mutation, and is explained for the charge-1 phantom vortex in $\phi_1$. This phantom vortex originates from the node of the initial second orbital $\phi_2$, which resembles a 2$p_x$ orbital that has an angular node. When angular momentum is added to the system due to the anisotropy, the lobes of the $2p_x$ orbital spread out and tend to close this angular node, leaving an elliptic density node in the center. After the trap is isotropic again, a phantom vortex is left at the center, which now appears in the density of $\phi_1$ since the first and second orbital have switched labels at $t\approx 220$ because of their changing occupation numbers [see second panel from the top in Fig. \ref{fig:phantom_vortices_1}]. Thus, the remaining phantom vortex was nucleated from an initial node. A similar process happens for the charge-3 phantom vortex in the density of $\phi_4$. It originates from $\phi_3(t=0)$ which resembles a $2p_y$ orbital. There, the angular node mutated first to three single phantom vortices that finally merged into a larger charge-3 phantom vortex. The fact that it finally appears in the fourth natural orbital is due to the label switching of $\phi_3$ and $\phi_4$ during the ramp-up of the anisotropy.

\begin{figure}[h!]
  \centering
  \includegraphics[width=0.6\textwidth]{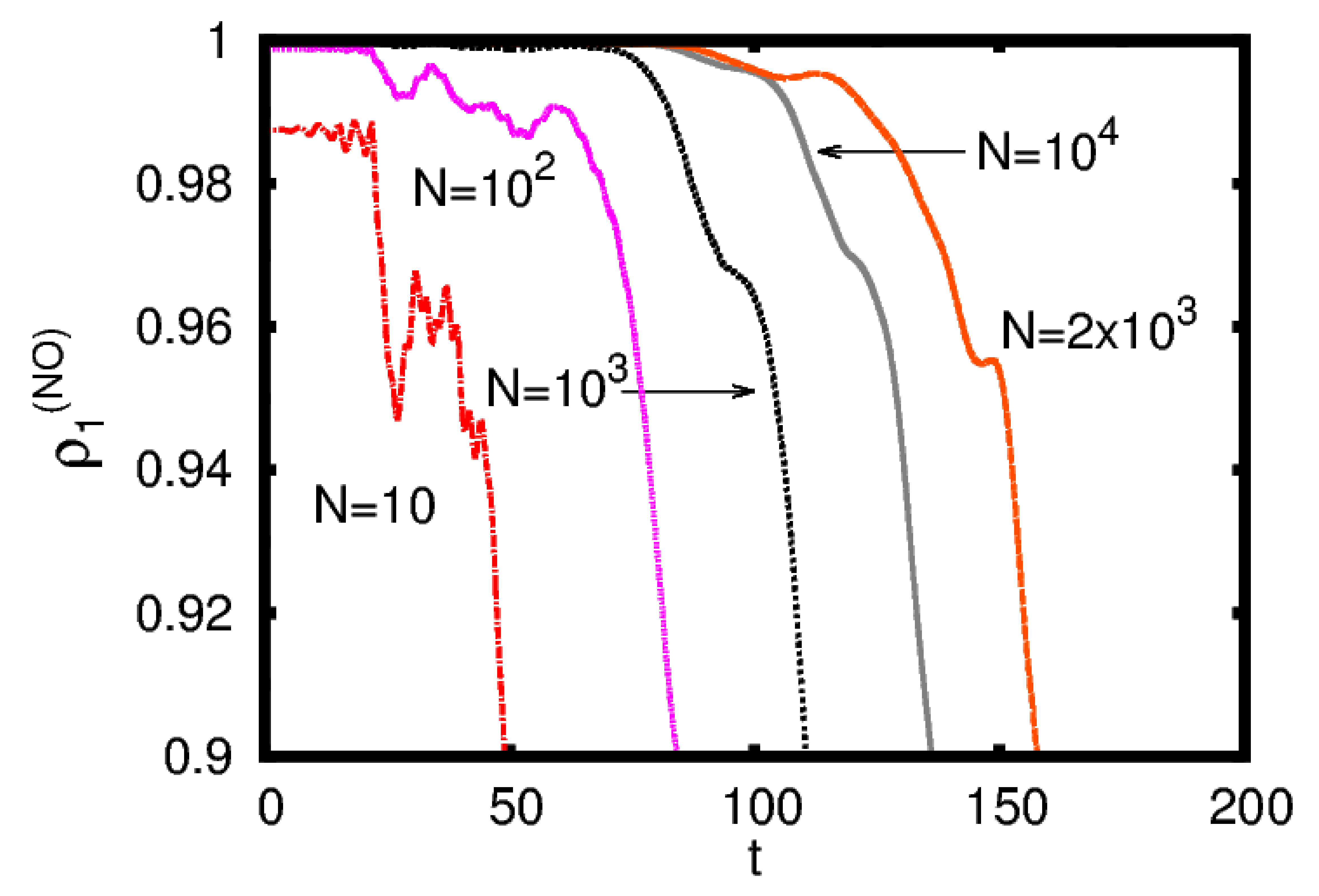}
  \caption[Dynamical fragmentation during the nucleation of phantom vortices]{Time-evolution of the fragmentation for different numbers of particles $N$. The same dynamical protocol of the trapping potential as in Fig. \ref{fig:phantom_vortices_1} is used. For all considered boson numbers, the occupation $\rho_1^{(\text{NO})}$ decreases clearly. In general, fragmentation sets in later for higher particle numbers. Results are computed using $M=2$  orbitals. The interaction parameter is kept fixed at $\Lambda=17.1$. All quantities are dimensionless. See text for details. The figure is taken from the supplementary material of Ref. \cite{Phantom_Lode}.}
  \label{fig:phantom_vortices_3}
\end{figure}

The mechanism responsible for the phantom-vortex pair in $\phi_3$ [panel (d)] is called slow orbital-orbital vorticity transfer because of the following observation. At $t\approx 220$, the pair of co-rotating phantom vortices appeared in the density of $\phi_2$. When time progressed, the two phantom vortices started to move towards the density edge and disappeared finally at $t\approx 275$. At the same time, two phantom vortices were build at the density edge of $\phi_3$, which afterwards continued to move towards the center. At $t\approx 340$, the separation between this pair is the same as before in the density of $\phi_2$. Thus, vorticity has been transferred between the two natural orbitals over a period much longer than the rotation period of $\tau=2\pi/\omega\approx 8$. This process between the second and third orbital can also be seen from the correlated orbital angular momenta $(L_z)_{22}$ and $(L_z)_{33}$ in Fig. \ref{fig:phantom_vortices_1} for $t\geq 250$. The position of the phase discontinuities shown in Figs. \ref{fig:phantom_vortices_2}(g), (i) and (j) coincide with the corresponding positions of the phantom vortex cores. Especially in panel (j) one can see from the merging three discontinuity branches in the center that the corresponding phantom vortex in panel (e) is of charge 3. Comparing the density in panel (a) and the phase $S_g(\mathbf{r}|0;450)$ in panel (f), one observes the occurrence of several so-called ghost vortices at the density edge. The latter are distinct from phantom vortices because they do not enter the bulk density of the condensate and do also not contribute much to the total angular momentum.

The dynamical fragmentation that occurs during the time-evolution, as well as its consequences for the angular momenta and the energy of the system [see again Fig. \ref{fig:phantom_vortices_1}] is now analyzed. The initial state is essentially condensed ($\rho_1^{(\text{NO})}\approx 99.7\%$), and it stays condensed for almost the entire period in which the anisotropy is ramped up to its maximal value. In this time frame, the total angular momentum and energy computed at the many-body and mean-field levels coincide as long as the BEC remains coherent. Once fragmentation sets in, the latter quantities start to deviate from each other. During the first part of the time interval in which the trap is kept maximally anisotropic ($80< t \leq 150$), $\rho_1^{(\text{NO})}$ decreases significantly to roughly $40 \%$, and the second, third and fourth natural orbitals become macroscopically occupied. At the same time, the orbital angular momenta vary the most and several of them reach their maximum values. Then, the occupations and angular momenta of $\phi_1$ and $\phi_4$ remain more or less constant. For the orbitals $\phi_2$ and $\phi_3$, both quantities change slightly in a correlated manner, i.e., when $\rho_2^{(\text{NO})}$ increases $\rho_3^{(\text{NO})}$ decreases by the same amount. This again reflects the slow orbital-orbital vorticity transfer between $\phi_2$ and $\phi_3$ discussed above. With regard to the total angular momentum and energy, both quantities vary strongly during the periods of maximal anisotropy and ramp down, and saturate only afterwards. The corresponding GP values are substantially different since they underestimate the system's gain of energy and angular momentum. In particular for the latter quantity, it stays below $0.5$, well below the threshold for the occurrence of vortices at the mean-field level at unit angular momentum per particle. Interestingly, $L_z/N$ for $M=4$ orbitals is greater than this threshold, but still no vortices are observed in the density $\rho(\mathbf{r})$. This can be rationalized because one can show analytically that a vortex in the density of a fragmented BEC, i.e., a fragmented vortex, requires more than unit angular momentum per particle [see the supplementary information of Ref. \cite{Phantom_Lode}]. The reason is that a fragmented vortex originates from coincident phantom vortices, meaning that all natural orbitals must have a phantom vortex at the same position, and all of these phantom vortices need to have different charges such that the natural orbitals are orthogonal. As seen from panels (b)-(d) in Fig. \ref{fig:phantom_vortices_2}, there are no coincident phantom vortices, and thus the fragmented system does not show any vortices in the density.

The onset of fragmentation was also observed for different numbers of bosons. Fig. \ref{fig:phantom_vortices_3} shows the evolution of $\rho_1^{(\text{NO})}$ for different $N$ between $10$ and $10^4$, utilizing $M=2$ time-adaptive orbitals. The interaction parameter $\Lambda=17.1$ is kept fixed. One generally observes that fragmentation sets in for all considered particle numbers, and that furthermore the onset of fragmentation occurs at later times for higher particle numbers. 

In conclusion, the nucleation of phantom vortices has been investigated for a rotating BEC that acquires angular momentum via a time-dependent modification of the trap. Two different mechanisms for the creation of phantom vortices were found, namely node mutation as well as slow orbital-orbital vorticity transfer. Phantom vortices, by definition many-body objects, cannot be described at the mean-field level. The initially coherent system highly fragments, also for different numbers of particles in the condensate. Further information can be found in Ref. \cite{Phantom_Lode}, in particular in the supplementary material containing videos that show, e.g., the evolution of all quantities shown in Fig. \ref{fig:phantom_vortices_2}.

\subsection{Excitation spectra}\label{Sec_Applications_LR-MCTDHB}
In this section, details on current applications of LR-MCTDHB are presented. The amount of scientific publications where it is utilized is still rather small because the theory itself has been developed only a few years ago \cite{LR-MCTDHB1,LR-MCTDHB2}. In addition to that, the efficient MPI-parallelized numerical implementation of LR-MCTDHB was a very demanding task, and all the technical challenges going along with it, extensively described in Section \ref{Sec_Num}, were solved very recently. It is stressed again that without such an implementation, a vast majority of problems cannot be treated at all, especially in $D>1$ spatial dimensions. Before developing the sophisticated and complex parallel implementation, there was a first implementation available that could treat small 1D systems. The results of Refs. \cite{LR-MCTDHB1} and \cite{1D_Theisen} were obtained with this code [see below and Section \ref{App_Theisen}]. In more recent applications, discussed in Secs. \ref{App_1D_lattice} and \ref{App_RotBEC}, the new implementation was utilized since it was impossible to obtain the same results with the former code.

\begin{figure}[h!]
 	\begin{subfigure}{0.45\textwidth}
 	  \centering	
 	  \includegraphics[angle=0,width=\textwidth]{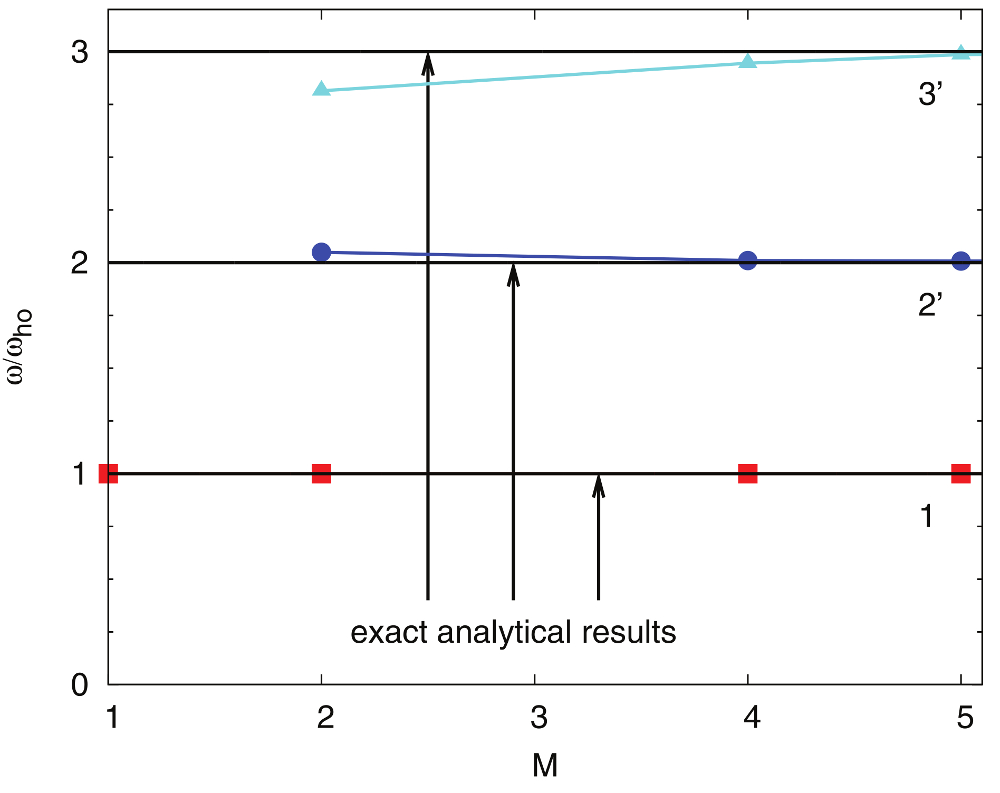} 
 	  \caption{\label{fig_Grond1}}
 	\end{subfigure}\hfill
 	\begin{subfigure}{0.55\textwidth} 
 	  \centering
 	  \includegraphics[angle=0,width=\textwidth]{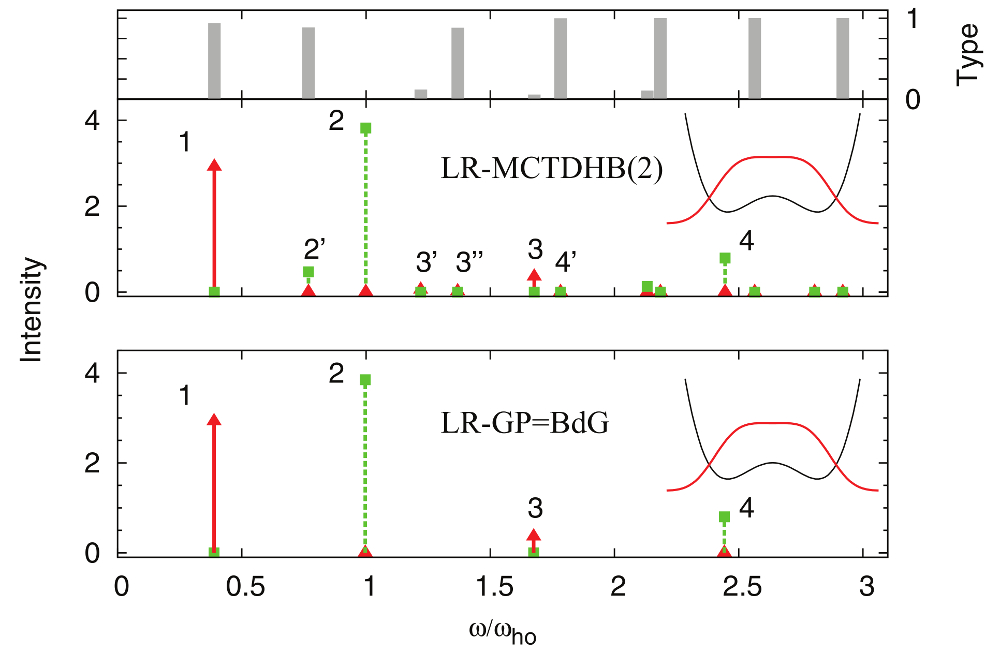}
 	  \caption{\label{fig_Grond2}}
	\end{subfigure}
\caption[Many-body excitations with LR-MCTDHB: Low-energy spectrum and convergence of the first c.m. excitations for a BEC in a harmonic trap with contact interaction potential]{(a) Numerical convergence of the first three center-of-mass excited states (labeled by 1, $2^\prime$ and $3^\prime$) with respect to the number of orbitals $M$ for $N=10$ bosons in a harmonic trap with trap frequency $\omega_{\text{ho}}=\sqrt{2}$. The strength of the contact repulsion is $\lambda_0 N=1$. For $M=5$ orbitals, the energies obtained from LR-MCTDHB coincide with the analytic predictions. The numerical implementation used is the precursor of the implementation described and benchmarked in Sections \ref{Sec_Num} and \ref{Ch_Benchmark}. (b) Comparison between the excitation spectra obtained from BdG (bottom) and LR-MCTDHB($2$) (top) for $N=10$ bosons in a shallow double well. The interaction strength is the same as in panel (a). Excitations labeled with primes refer to states solely obtained at the many-body level. The intensity of the response to a linear perturbation ($f^+=f^-=x$) is denoted by red triangles, whereas the intensity of the response to a quadratic perturbation ($f^+=f^-=x^2$) is denoted by green squares. The insets show the trap potential and the ground-state densities. One observes that the many-body low-energy spectrum contains a lot more states than the mean-field spectrum. All quantities are dimensionless. See the main text and Ref. \cite{LR-MCTDHB1}, from which both figures are taken, for further details, especially for the discussion of the excitation type shown at the top of panel (b).}
\label{fig_Grond}
\end{figure}

The main purpose of Ref. \cite{LR-MCTDHB1} is to introduce LR-MCTDHB by giving a derivation of its main equations, presented in Section \ref{Sec_LRMCTDHB} of this work. In particular, the equations for the submatrices as described in Eq. (\ref{LR_matrix}) are developed for the case of contact interaction between the bosons. As an application, excitations for BECs in harmonic and shallow double-well traps are analyzed. Fig. \ref{fig_Grond}(a) shows the numerical convergence for the first three c.m. excited states (labeled by 1, $2^\prime$ and $3^\prime$) of $N=10$ bosons in a harmonic trap with trap frequency $\omega_{\text{ho}}=\sqrt{2}$ in terms of the number of orbitals $M$. The strength of the repulsion is given by $\lambda_0 N=1$. One can observe that the c.m. excitations are accurately obtained for $M\geq 5$ orbitals. The BdG theory only yields the first c.m. excitation. In Fig. \ref{fig_Grond}(b), the low-energy spectrum as well as the response intensities for linear and quadratic perturbations ($f^+=f^-=x,\,x^2$) are presented for $N=10$ bosons in a shallow double well of the form $V(x)=b/2\cos\left( \pi x/3 \right)+\omega_{\text{ho}}^2x^2/2$ with barrier height $b=5$. The most important observation by comparing the number of obtained states for BdG and LR-MCTDHB($2$) is that the many-body spectrum contains a lot more states. This in turn means that although the degree of depletion is only about $0.2\%$, a many-body description for the lowest-in-energy excitation spectrum is unavoidable in order to obtain accurate results. In the subsequent Section \ref{App_Theisen}, the excitation spectrum for the case of BECs trapped in harmonic and double-well potentials is elaborated in more detail, especially with respect to the question of how many-body excitations can be triggered in the out-of-equilibrium dynamics of the condensates.
 
Before discussing further applications of LR-MCTDHB in the following, a remark on how mean-field and many-body excitations are defined in this work is made in order to avoid misconceptions of these important terms. Excitations calculated from Eq. (\ref{LR_eigenvalue_final}), i.e., from LR-MCTDHB ($M>1$), are termed many-body excitations since they stem from a many-body theory. On the contrary, all excitations calculated with the BdG equations in Eq. (\ref{BdG_part_conv}) of Appendix \ref{Sec_BdG} are referred to as mean-field excitations since they stem from a mean-field theory. Moreover, excitations where only a single boson is excited from the condensed mode are called single-particle excitation, whereas excitations where more than a single boson is excited are called multi-particle excitations. These definitions are utilized consistently in the applications below.

\subsubsection{One-dimensional harmonic and double-well systems}\label{App_Theisen}
As a first application of LR-MCTDHB, the many-body nature of excitations in a 1D harmonic trap as well as in shallow and deep symmetric double wells is analyzed. The low-energy spectra obtained from the BdG equation at the mean-field level and from LR-MCTDHB at the many-body level are compared. Furthermore, it is investigated which excitations are involved in the dynamics due to certain quench scenarios. The following results where published in Ref. \cite{1D_Theisen}.

Excited states are studied in a harmonic trapping potential with a Gaussian barrier given by
\begin{equation}\label{Eq_Trap_Theisen}
	V(x)=ax^2+b\,\exp\left( -cx^2 \right)
\end{equation}
where $a$ denotes the harmonic trap frequency and $B$ denotes the barrier height. For simplicity, $\hbar=m=1$ is assumed. Furthermore, the trapping parameters are set to $a=1/2$ and $c=1$. The bosons interact via the zero-ranged contact interaction potential. The interaction strength is again measured by the mean-field parameter $\Lambda=\lambda_0(N-1)$ where $\lambda_0$ is chosen to be positive to account for repulsion. Fig. \ref{fig:1D_Theisen_1} shows $V(x)$ for different barrier heights, together with the corresponding ground-state densities. For the purely harmonic trap ($b=0$), the BEC is essentially condensed with $n_1=99.9\%$. On the contrary, for the shallow ($b=5$) and deep ($b=10$) double wells, the ground state is already fragmented by approximately $5\%$ and $40\%$, respectively. One can further observe that for the deep well, the overlap between the densities in the left and right wells is marginal, whereas there is still a clear overlap observable in the shallow case.

Since the trap is reflection invariant, the excitations can be categorized in gerade ($g$) and ungerade ($u$) symmetry. Fig. \ref{fig:1D_Theisen_2} shows the low-energy spectrum of $N=10$ bosons for different barrier heights $0\leq b\leq 10$ and $\Lambda=1.0$. Excitation energies are given relative to the ground-state energy, i.e., $\Delta E=E-E_0$. One readily observes the formation of bands, which is even more pronounced for lower interaction strengths [see Fig. 2 in Ref. \cite{1D_Theisen}]. The bands are labeled by integer numbers, where the zeroth band denotes the one that is lowest in energy. Comparing the mean-field and many-body approaches, one can see that the BdG spectrum only contains 5 different excitations, whereas the spectrum computed with LR-MCTDHB shows a much richer structure where a lot more excited states appear.  In terms of numerical convergence, $M=2$ orbitals are sufficient to describe the spectrum up to $\Delta E\sim2$ for not too low barrier heights, i.e., $b\geq 4$. For states higher in energy, at least four orbitals are necessary.

\begin{figure}[h!]
  \centering
  \includegraphics[width=\textwidth]{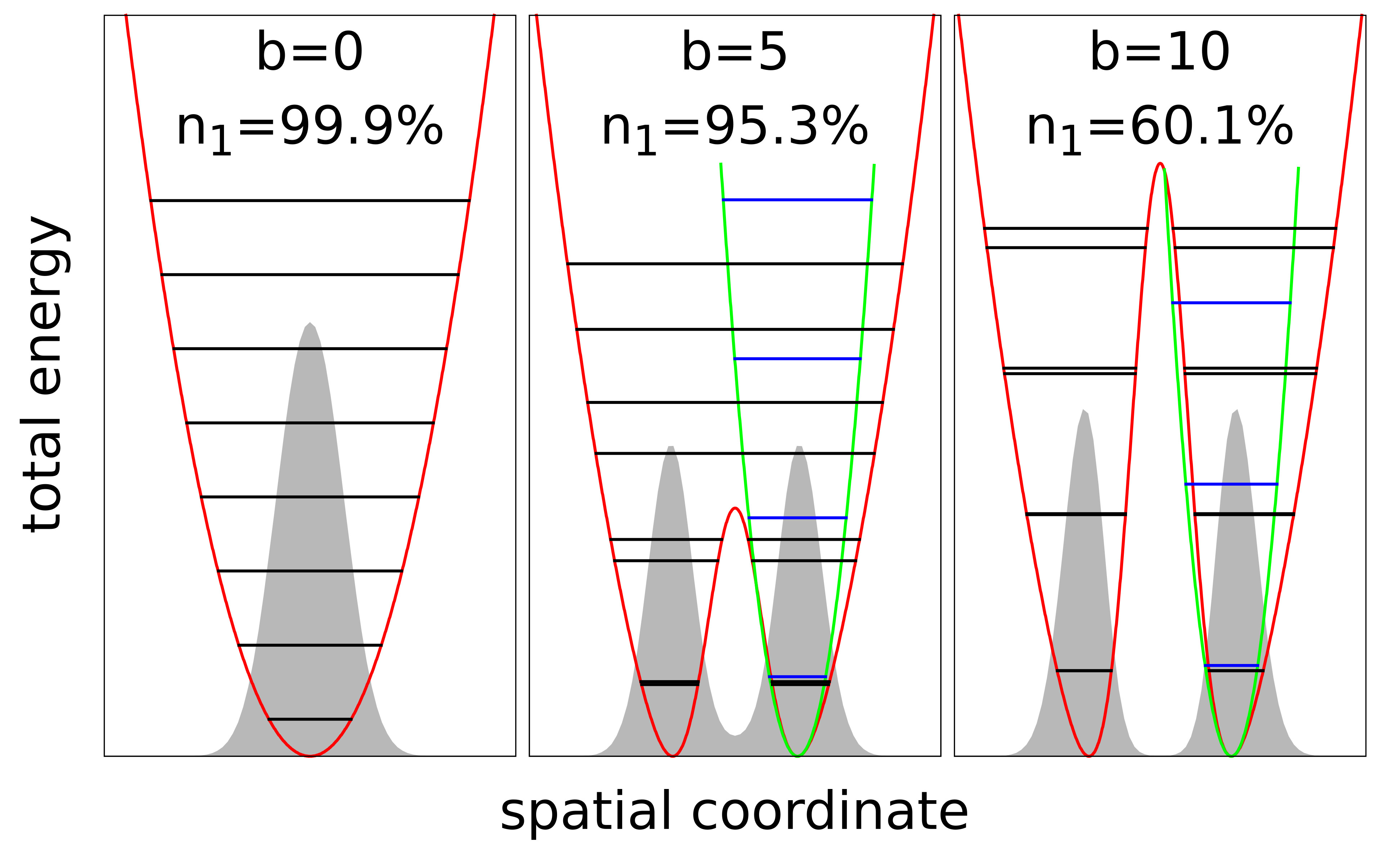}
  \caption[Different ground-state densities and trap potentials for BECs in 1D harmonic and double-well traps]{Trap potentials (solid red) and ground-state densities (gray) for different barrier heights $b$. The ground states are computed with MCTDHB($2$). The BEC consists of $N=10$ bosons and the repulsion strength is $\Lambda=1.0$. The occupation number $n_1$ of the first natural orbital shows that increasing the barrier height leads to fragmentation. Vertical lines refer to the lowest-in-energy excited states computed with LR-MCTDHB($2$). The green potentials and the blue eigenvalues refer to the harmonic approximation of the right well. All quantities are dimensionless. See text for details. The figure is taken from Ref. \cite{1D_Theisen}.}
  \label{fig:1D_Theisen_1}
\end{figure}

\begin{figure}[h!]
  \centering
  \includegraphics[width=\textwidth]{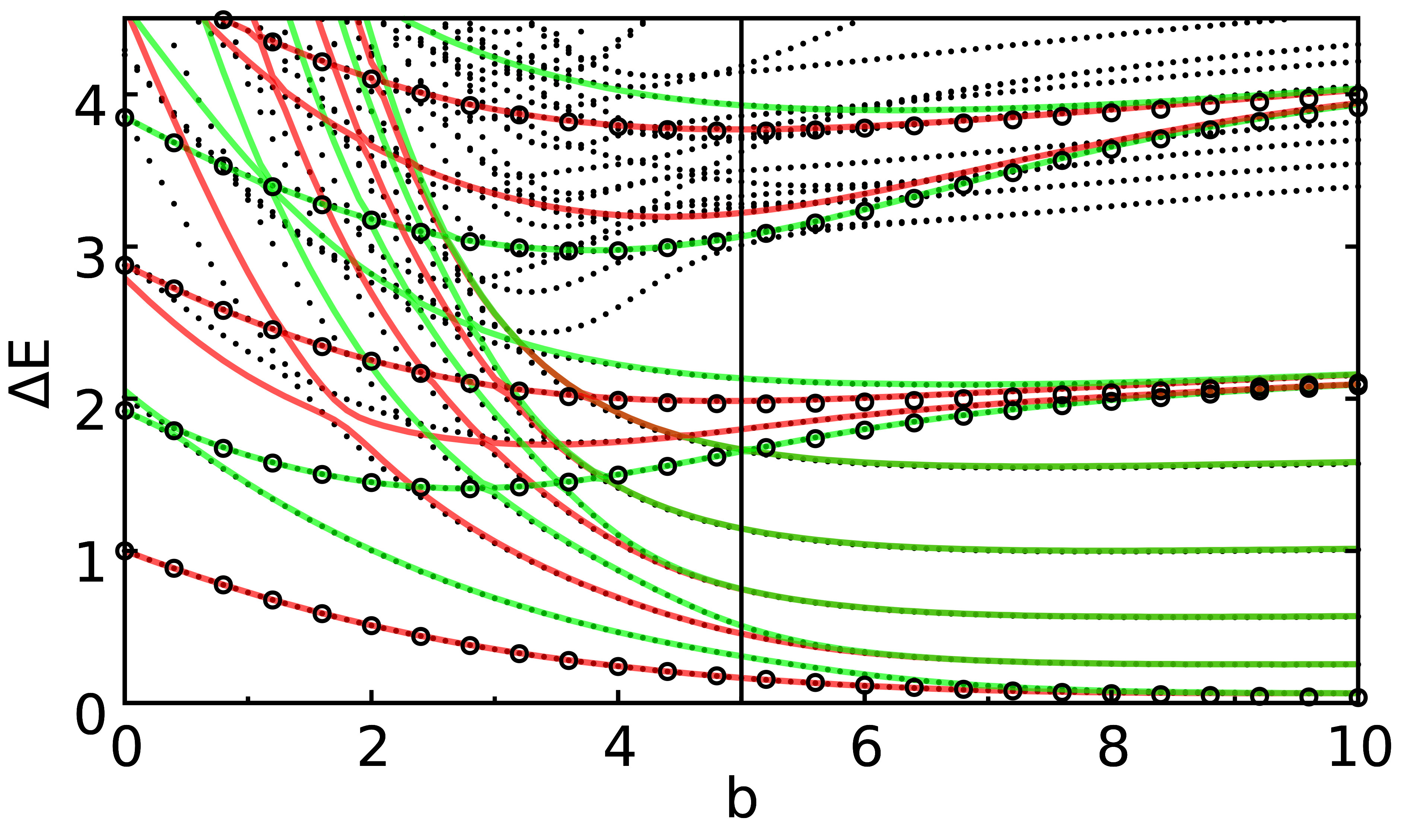}
  \caption[Many-body low-energy spectrum for a BEC in a double-well potential with different barrier heights]{Low-energy excitation spectrum of $N=10$ bosons with interaction parameter $\Lambda=1.0$ for different barrier heights $b$. Many-body results computed with LR-MCTDHB($2$) are denoted by solid lines (gerade excitations in green, ungerade excitations in red). Results for $M=4$ orbitals are given by dotted lines, and the corresponding BdG energies are given by open black circles. All states up to $\Delta E\sim 2$ are converged with two orbitals. The BdG approach misses many states in the shown energy range. All quantities are dimensionless. See text for details. The figure is taken from Ref. \cite{1D_Theisen}.}
  \label{fig:1D_Theisen_2}
\end{figure}

In the following, different shift and quench protocols are applied to trigger excitations from the low-energy spectrum. For gerade excitations, a quench of the harmonic confinement from $a=0.4\rightarrow 0.5$ is performed, whereas ungerade excitations are addressed by a shift of the trap position, meaning that $x\rightarrow x+x_{\text{shift}}$ with $x_{\text{shift}}=0.1$. The gerade excitations are referred to as breathing modes, and the ungerade excitations are referred to as dipole modes. The dynamics of the bosonic clouds for the two cases are shown in Fig. \ref{fig:1D_Theisen_3} on a short time scale. In the left panel, dipole oscillations for a BEC in the deep double well are depicted, which means that in each well, the bosons oscillate around the equilibrium position given by the minima of the wells. On the RHS, the breathing dynamics for $N=10$ bosons in the purely harmonic trap is shown. One can observe that the density spreads to the left and right in a symmetric manner, and returns to the minimum of the well. The Fourier spectra of the time-evolution of certain quantities is used to extract the spectrum of involved excitations in the dynamics. This comprises $(a)$ the expectation value $\langle x\rangle$, $(b)$ the variance per particle $\frac{1}{N}\Delta^2_{\hat{X}}$, and $(c)$ the density $\rho$ at the position $x=1$. The many-body variance is discussed in detail in Appendix \ref{Appendix_variance}. In Ref. \cite{1D_Theisen}, the Fourier spectra for $b=0\,,5$ and $10$ are investigated. Here, the discussion is restricted to the latter. To remind the reader, the system is highly fragmented by approximately $40\%$ in this case.

\begin{figure}[h!]
  \centering
  \includegraphics[width=\textwidth]{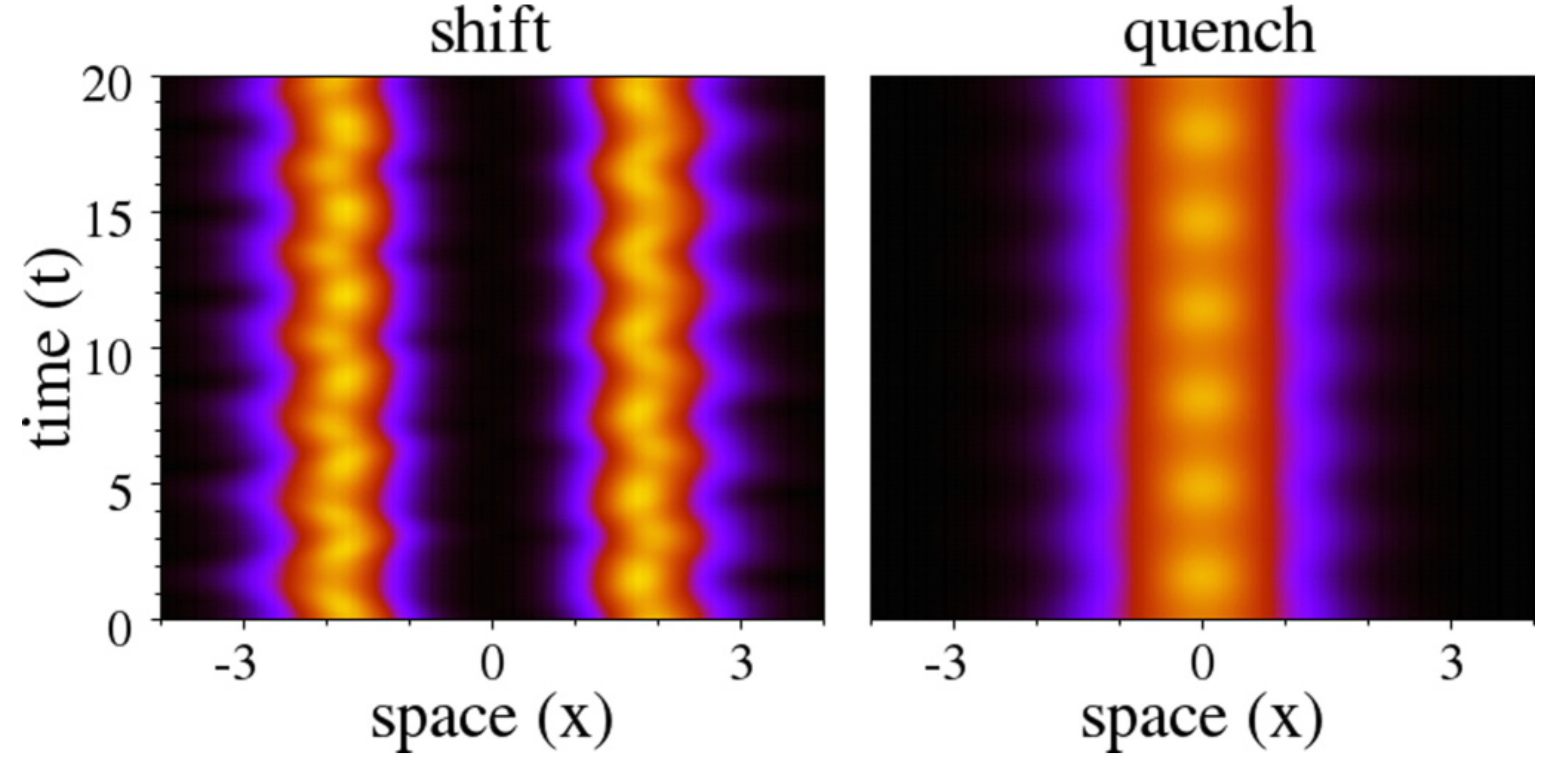}
  \caption[Shift and quench dynamics of the density of a BEC in a deep double well and a harmonic trap]{Many-body dynamics of the density of $N=10$ bosons with interaction parameter $\Lambda=1.0$ for a shift of the trap position in the deep double well (left panel) and a quench of the harmonic confinement (right panel). In the former scenario, one observes dipole oscillations of the bosonic clouds in the two wells, whereas one observes breathing oscillations for the dynamics due to the quench of the confinement. All quantities are dimensionless. See text for details. The figure is taken from Ref. \cite{1D_Theisen}.}
  \label{fig:1D_Theisen_3}
\end{figure}

Table \ref{table_Theisen} shows excitation and de-excitation energies where the obtained values from the BdG and LR-MCTDHB($2$) theories (columns 2 and 3) are compared to the obtained excitation energies of the FT spectra from the shift and quench scenarios (columns 4-7). Moreover, several de-excitations, i.e., transitions between a higher and a lower excited state, are presented (last 4 rows). Those refer to higher-order effects and are in principle not obtained directly by LR. However, since LR yields the exact excitation spectrum once it is applied to an exact eigenstate of the full many-body Hamiltonian, the de-excitations can be extracted from differences between the excitation energies of the LR spectrum. The excitation labels are shown in the first column. Many-body excitations, i.e., excitation that only appear in the many-body spectrum, are denoted by primed labels.

\begin{table}[t!]\scriptsize

\begin{tabular*}{\textwidth}{p{1.2cm} p{1.6cm} p{1.6cm} p{2.2cm} p{2.2cm} p{2.2cm} p{2.2cm}}
\hline\hline \rule{0pt}{3ex}  
			& $M=1$	& $M=2$ & Shift GP	&  Shift MB	& Quench GP & Quench MB \\ \hline \rule{0pt}{3ex}  
\hspace*{-0.1cm}$0u$	& 	0.035 (0.46)	& 0.063 (0.19) &	0.034 (0.02$^a$)	&			&				&			\\ 		
$0g^\prime$	&	 &	0.064 &		&			&	 		&	0.066 (0.36$^b$) \\ 
$1g$	&	2.085 (0.91) &	2.092 (0.81) &		&						&	2.085 (1.0$^b$,0.93$^c$)	& 2.092 (0.33$^b$,0.89$^c$)	\\ 
$1u$	&	2.102 (0.46) & 2.095 (0.57)	&	2.102 (0.94$^a$,0.66$^c$)	&		2.095 (0.93$^a$,0.7$^c$)			&		&	\\ 
$1u^\prime$	&		& 2.158 (0.08) &	 &			2.163 (0.02$^a$,0.02$^c$)		&		&	\\ 
$1g^\prime$	&	 & 2.161 (0.09)	&		&						&	& 2.161 (0.12$^b$)	\\
$2g$	&	3.912 (0.04)	& 3.934 (0.03) &	3.914 (0.04$^c$) &				&	3.914 (0.07$^c$)	&	3.934 (0.07$^b$,0.1$^c$) \\
$2u$	&	3.997 (0.07) & 3.943 (0.07) &	3.997 (0.04$^a$,0.23$^c$)	&			3.943 (0.03$^a$,0.15$^c$)	&		&	\\ 
$2u^\prime$	&	 & 4.029 (0.05)	&		&			4.032 (0.02$^a$,0.07$^c$)	&			&	\\
$2g^\prime$	&		& 4.039 &		&					&		&	\\ 
$1g^\prime\leftrightarrow 1g$	&		& 0.069 &		&					&		&		\\
$2g\leftrightarrow 1g$	&	1.827	& 1.841 &		&					&			&	1.893 (0.01$^b$)	\\ 
$2u\leftrightarrow 1u$	&	1.894	& 1.848	& 1.895 (0.01$^c$)	&	1.849 (0.02$^c$)	& 	\\
$1g\leftrightarrow 0g^\prime$	& 	& 2.028	&	&	& & 2.032 (0.02$^b$) \\

\hline\hline

\end{tabular*}
\caption[Excitations and de-excitations in the dynamics of a BEC in a deep double well]{Table of excitation and de-excitation energies for $N=10$ bosons with $\Lambda=1.0$ in the deep double well. Listed are results from BdG and LR-MCTDHB($2$), as well as from the Fourier spectra of the expectation value of the position $\langle x \rangle$ (superscript $a$), the variance per particle $\frac{1}{N}\Delta^2_{\hat{X}}$ (superscript $b$) and the real-space density $\rho(x=1)$ (superscript $c$), for both the shift and quench scenarios. GP denotes the mean-field and MB the many-body case. The intensities are given in brackets. All quantities are dimensionless. See text for more details. The table is adapted from Ref. \cite{1D_Theisen}.}
\label{table_Theisen}
\end{table}

Up to $\Delta E\sim 4$, the many-body approach yields ten excited states, whereas on the contrary the BdG spectrum only consists of five excitations. The energies of states that appear in both spectra differ slightly from each other for $1g$, $1u$, $2g$ and $2u$, but significantly for the lowest excitation $0u$. That demonstrates that strong deviations between the mean-field and many-body excitation energies can already occur for the lowest excited states. Also the intensities of several states differ clearly from each other, especially for the lowest states $0u$, $1g$ and $1u$. The intensities from the LR calculations are identical to the response weights described in Eq. (\ref{gamma_k}), with $f^+=f^-=x$ for the ungerade and $f^+=f^-=x^2$ for the gerade states.

\begin{figure}[h!]
  \centering
  \includegraphics[width=0.7\textwidth]{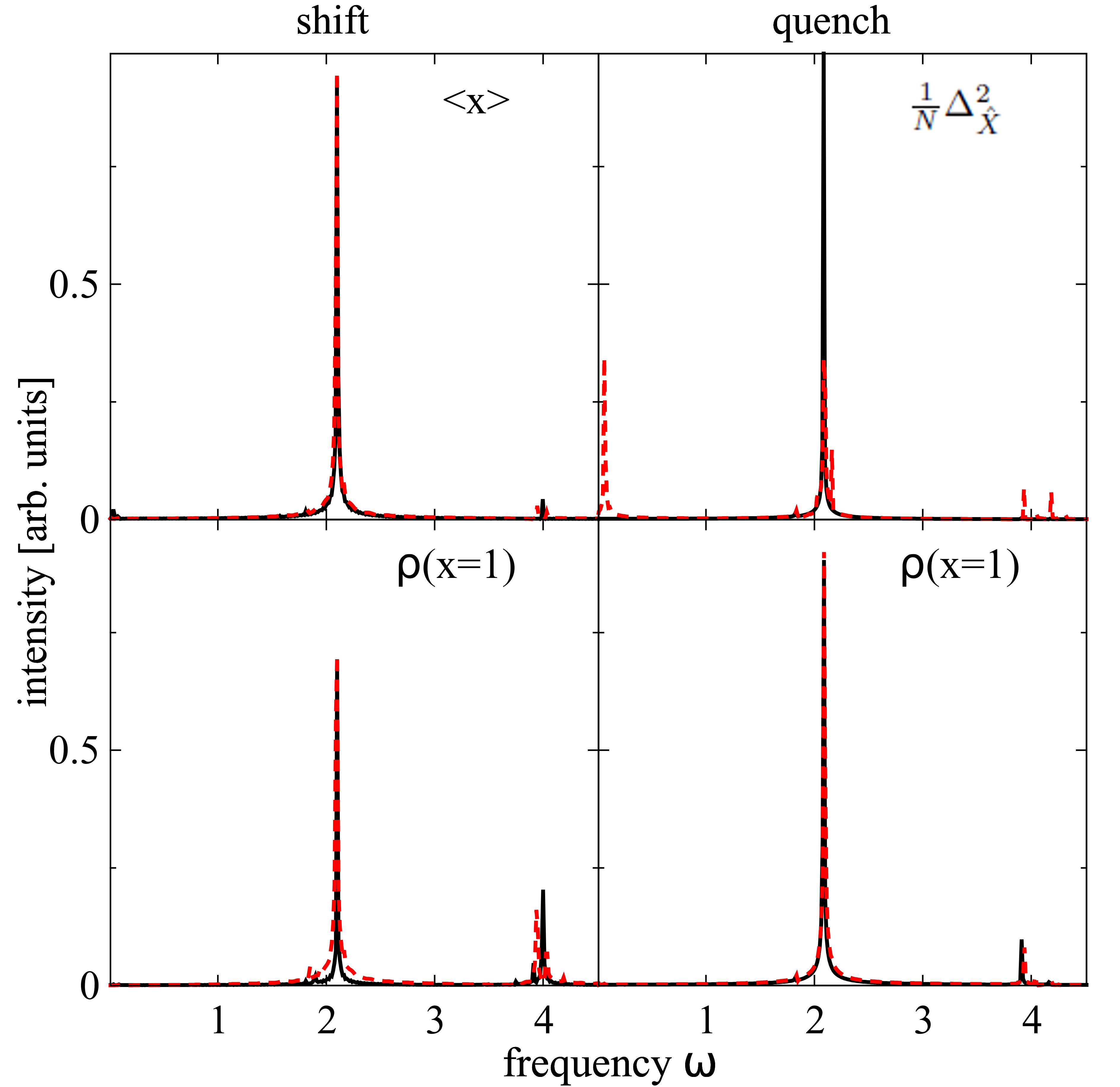}
  \caption[Fourier spectra of different quantities from the shift and quench dynamics in a deep double well]{Fourier spectra of $\langle x \rangle$, $\frac{1}{N}\Delta^2_{\hat{X}}$, and $\rho(x=1)$ from the shift and quench dynamics in the deep double well. The GP spectra are denoted by solid black and the MCTDHB($2$) spectra by dashed red lines. The intensities are normalized by the sum of the intensities of all constituent peaks, and excitations with less than $1\%$ intensity are omitted. Especially the spectrum of the many-body variance consists of several peaks that are not observed in the corresponding GP spectrum. All quantities are dimensionless. See text for details. The figure is adapted from Ref. \cite{1D_Theisen}.}
  \label{fig:1D_Theisen_4}
\end{figure}

Columns 4 and 5 show the obtained peaks in the Fourier analysis of $\langle x \rangle$ and $\rho(x=1)$ for the shift scenario, both for the GP and many-body cases. The Fourier spectra are shown in the left panels of Fig. \ref{fig:1D_Theisen_4}. Surprisingly, the shift also triggered a gerade excitation ($2g$) at the mean-field level, whereas at the many-body level only ungerade states are observed. Moreover, no peak is observed for $0u$ in the many-body dynamics, presumably because its intensity is too weak to be distinguished from the background noise. The obtained peak positions coincide with the predictions from BdG and LR-MCTDHB($2$), respectively. With regard to de-excitations, the mean-field and many-body approaches yield a very small peak for the $2u\leftrightarrow 1u$ transition in the density spectrum.

The Fourier analysis of the quench dynamics of $\frac{1}{N}\Delta^2_{\hat{X}}$ and $\rho(x=1)$ are shown in column 6 (GP) and column 7 [MCTDHB($2$)]. Only peaks corresponding to gerade excitations are observed. Both approaches yield peaks for $1g$ and $2g$. At the many-body level, two additional peaks referring to $0g^\prime$ and $1g^\prime$ are obtained. The peak positions again accurately match the predictions from the static analysis. Most importantly, the intensities of these excitations in the variance spectrum are comparatively large, and therefore $0g^\prime$ and $1g^\prime$ are the most promising many-body excited states to be detected in an experiment. Furthermore, no de-excitation peaks are found in the GP spectra, while the variance spectrum shows small peaks for the $2g\leftrightarrow 1g$ and $1g\leftrightarrow 0g^\prime$ transitions at the many-body level, coinciding with the values from LR-MCTDHB($2$). 

In general, the sensitivity of the utilized operators to the detection of excited states can be seen as follows. The spectrum of the local operator $\rho(x=1)$ is in particular useful to detect ungerade excitations in the shift scenario. The intensities of the peaks of many-body excitations are however weak. To detect gerade excitation, nonlocal quantities, especially the variance, turn out to be more sensitive than the density. In addition, the variance spectrum yields intense peaks also for many-body excitations.

To summarize, it has been demonstrated that from a methodological point of view, utilizing LR-MCTDHB in order to obtain the many-body excitation spectra of repulsive BECs in 1D harmonic and double-well trap geometries is a promising and highly accurate technique. Its predictions coincide with the Fourier spectra of the time-evolution of different quantities like the local density or the position variance. From a physical point of view, some of the obtained excited states corresponding to many-body states not included in standard mean-field approaches yield intense responses especially in the variance spectra and are therefore likely to be measured in experiments. More details on the presented results can be obtained from Ref. \cite{1D_Theisen}.

\subsubsection{Triple wells and larger lattices}\label{App_1D_lattice}
In this section, many-body effects in the excitation spectra of weakly-interacting BECs in 1D lattices are investigated. The main objective is to demonstrate with numerically converged results that a marginal depletion of the order of $1\%$ is sufficient in order to observe clear differences between the lowest-in-energy excitations computed at the mean-field and many-body levels. The results presented below were recently published in Ref. \cite{Beike_1d_lattice}.

Studying dynamics and excitations of ultracold bosons in 1D optical lattices was of high interest in the past, and the literature is extensive in this research field. Examples are the dynamics in the superfluid phase \cite{Burger}, nonlinear dynamics in periodic potentials \cite{Bloch_1D_Latt} or in fluctuation-driven binary condensates \cite{Suthar}. A review of earlier works can be found in Ref. \cite{Oberthaler_1D_Latt}. Of particular relevance for the findings discussed in this section are studies were the BdG theory does not agree with numerical and experimental observations. It has been found that the oscillation frequency of a superfluid BEC in a large optical lattice with a superimposed harmonic confinement deviates clearly from the BdG prediction \cite{Kasevich1,Kasevich2}. In addition, the strong depletion of a gaseous BEC in optical lattices cannot be fully explained by the BdG theory, not only in 1D \cite{Ketterle_1D_Latt}. 

Optical lattices, in particular triple-well systems, have also been analyzed recently utilizing MCTDHB. This includes the out-of-equilibrium dynamics of rather small BECs consisting of a few particles only, induced by quenching the interaction and the lattice depth \cite{Schmelcher_1D_Latt1,Schmelcher_1D_Latt3,Schmelcher_1D_Latt4} and periodic driving of the system \cite{Mistakidis_1D_Latt1,Schmelcher_1D_Latt2}. Of particular interest was the analysis of cradle modes, dipole oscillations corresponding to the transport of bosons over the barrier between adjacent wells. Furthermore, the many-body dynamics of BECs initially prepared in Mott-insulating states have been of interest \cite{Triple_well_dynamics}. The different phases of the BEC, i.e., the superfluid, the Mott-insulating, and the fermionized gas phase, can be detected from the many-body entropy and the Glauber correlation functions \cite{Axel_entropy_lattice}. It was even shown for a BEC in a tilted triple well with hard-wall boundary conditions how the correlations between wells can be controlled by the well depth, the interaction strength and the tilt \cite{Marios_tilted_triple_well}. Many-body effects in the ground state of a bosonic mixture in a more complex 2D lattice where studied using the multi-layer version of MCTDHB \cite{Schmelcher_honeycomb}.

Many-body excitations in optical lattices were addressed as well, either by making a Fourier analysis of certain quantities \cite{Schmelcher_1D_Latt1} or by doing an exact diagonalization. There, the system's wave function was expressed using a many-body ansatz based on a number state expansion including localized Wannier functions \cite{Mistakidis_1D_Latt1,Schmelcher_1D_Latt2,Schmelcher_1D_Latt3,Schmelcher_1D_Latt4}. The latter approach of exact diagonalization is possible for small systems with a few bosons only.

\begin{figure}[h!]
	\vspace*{-0.5cm}
 	\begin{subfigure}{0.5\textwidth}
 	  \centering	
 	  \includegraphics[angle=-90,width=\textwidth]{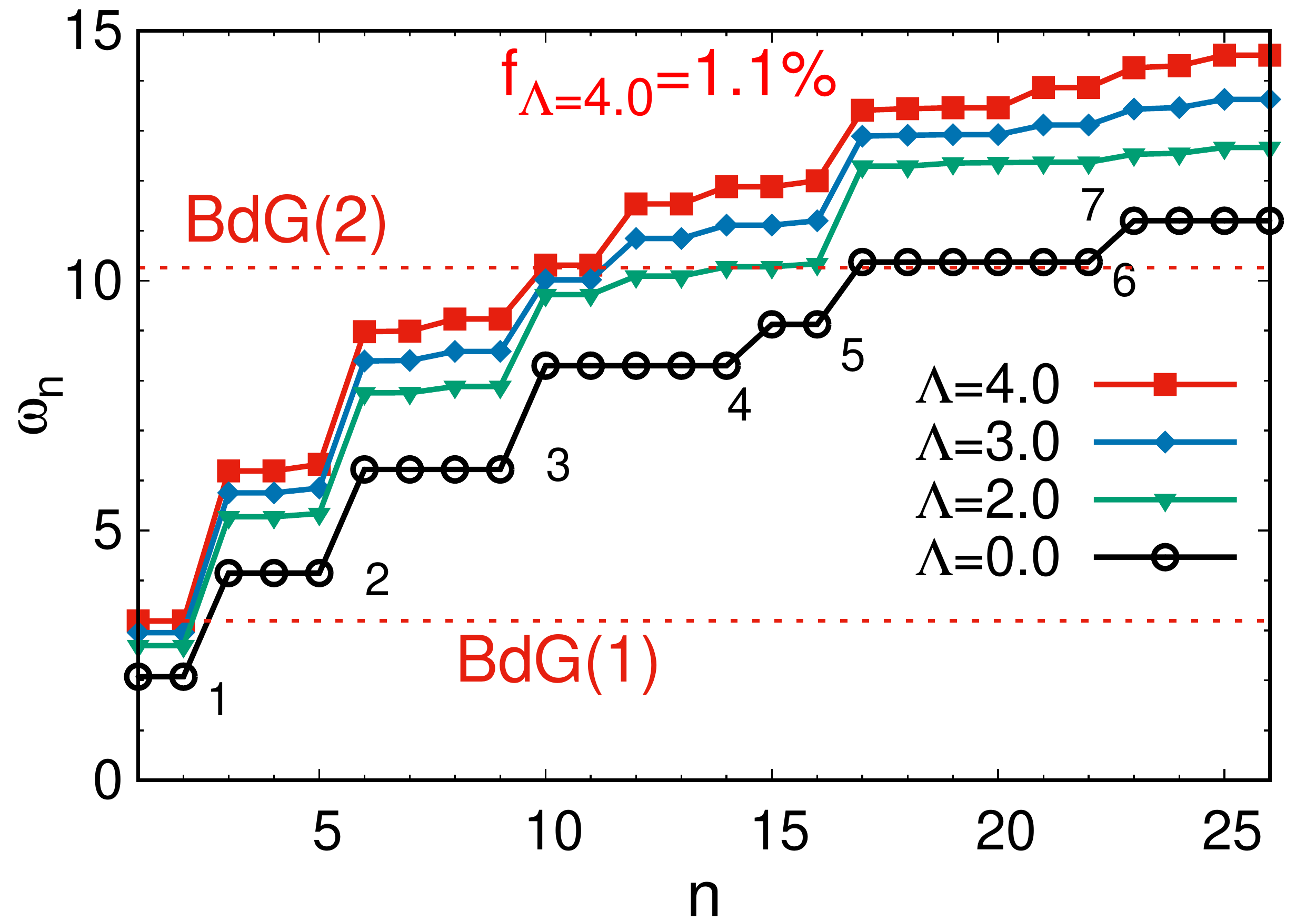} 
 	  \caption{\label{fig_1D_lattice1}}
 	\end{subfigure}
 	\begin{subfigure}{0.5\textwidth} 
 	  \centering
 	  \includegraphics[angle=-90,width=\textwidth]{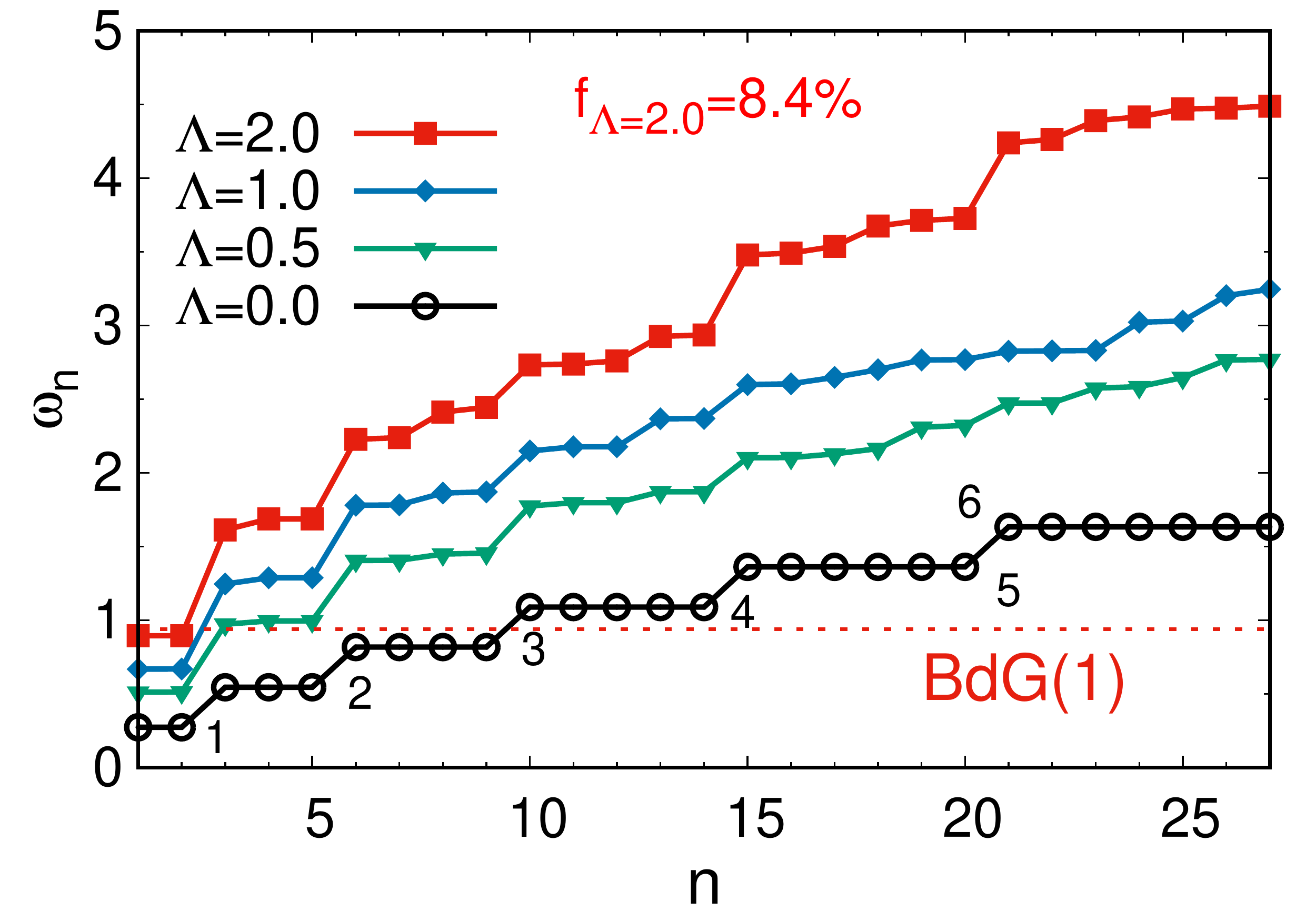}
 	  \caption{\label{fig_1D_lattice2}}
	\end{subfigure}
	\vskip\baselineskip\vspace*{-0.2cm}
	\begin{subfigure}{0.5\textwidth}
	  \centering
 	  \includegraphics[angle=-90,width=\textwidth]{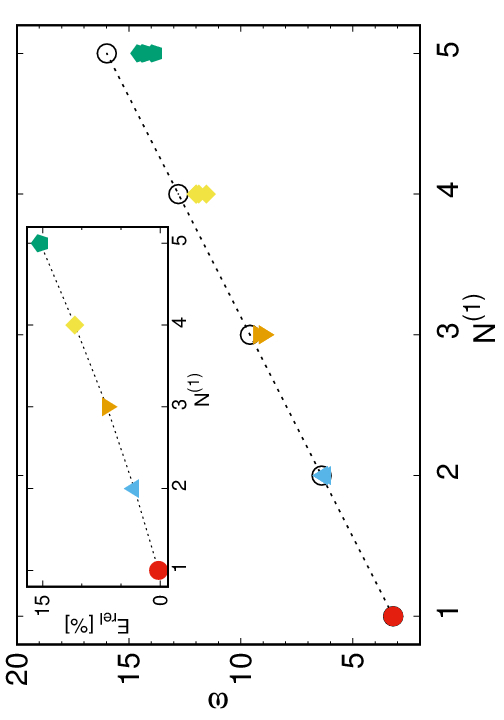} 
 	  \caption{\label{fig_1D_lattice3}}
 	\end{subfigure}
 	\begin{subfigure}{0.5\textwidth} 
 	  \centering
 	  \includegraphics[angle=-90,width=\textwidth]{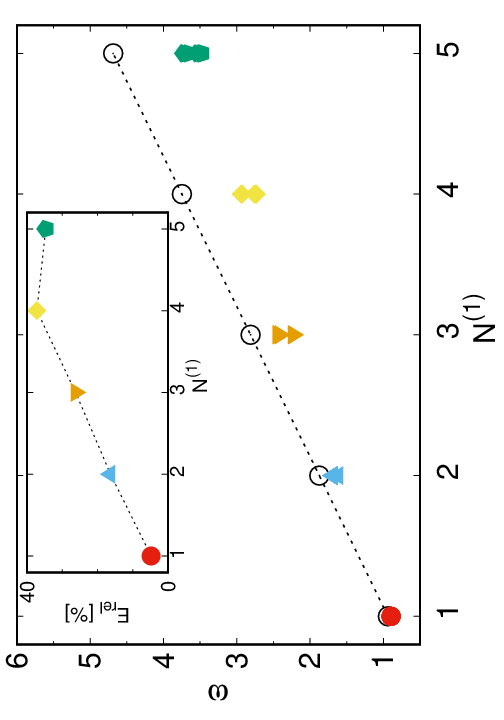}
 	  \caption{\label{fig_1D_lattice4}}
	\end{subfigure}
\caption[Low-energy spectra of $N=10$ bosons in shallow and deep triple wells]{Low-energy spectra of $N=10$ bosons in a triple well computed with LR-MCTDHB($7$). (a) Shallow case with lattice depth $V_0=1.01\,E_R$. For $\Lambda=0$, the excitations form levels of degenerate states. Degeneracies are lifted for $\Lambda>0$. The BdG results only contain the two doublets denoted by BdG(1) and BdG(2). (b) Same as in (a) but in the deep lattice with $V_0=10.13\,E_R$. The BEC for $\Lambda=2.0$ is fragmented by $8.4\%$. (c) Comparison between BdG(1) and its multiples (open circles) and the many-body results (colored closed symbols) for the system in (a) with $\Lambda=4.0$ where only the states $(+1)$ and $(-1)$ of the first single-particle band are occupied by $N^{(1)}$ bosons. The difference grows with $N^{(1)}$. Inset: Evolution of the relative error as defined in the main text. (d) Same as in (c), but for the system in (b) with $\Lambda=2.0$. All quantities are dimensionless. See text for details. The figures are adapted from Ref. \cite{Beike_1d_lattice}.}
\label{fig_1D_lattice_gen1}
\end{figure}

Below, the lowest-in-energy excitations of BECs in 1D lattice potentials with periodic boundary conditions are investigated using LR-MCTDHB. This opens the possibility to study also larger systems with more wells and particles. The utilized trap potential reads
\begin{equation}\label{eq_1D_latt_pot}
	V(x)=V_0\,\cos^2\left(\frac{\pi}{l}\,x\right)
\end{equation}
where $V_0$ denotes the lattice depth and $l$ denotes the distance between two neighboring lattice sites. The depth of the lattice is commonly expressed in terms of the recoil energy $E_R=\frac{\hbar^2 k_0^2}{2m}$ with the lattice momentum $k_0=\frac{\pi}{l}$. For simplicity, $\hbar=m=l=1$ is assumed. The bosons repel each other via the contact interaction potential and its strength is again expressed in terms of the mean-field parameter $\Lambda=\lambda_0(N-1)$.

\begin{figure}[h!]
 	\begin{subfigure}{0.5\textwidth}
 	  \centering	
 	  \includegraphics[angle=-90,width=\textwidth]{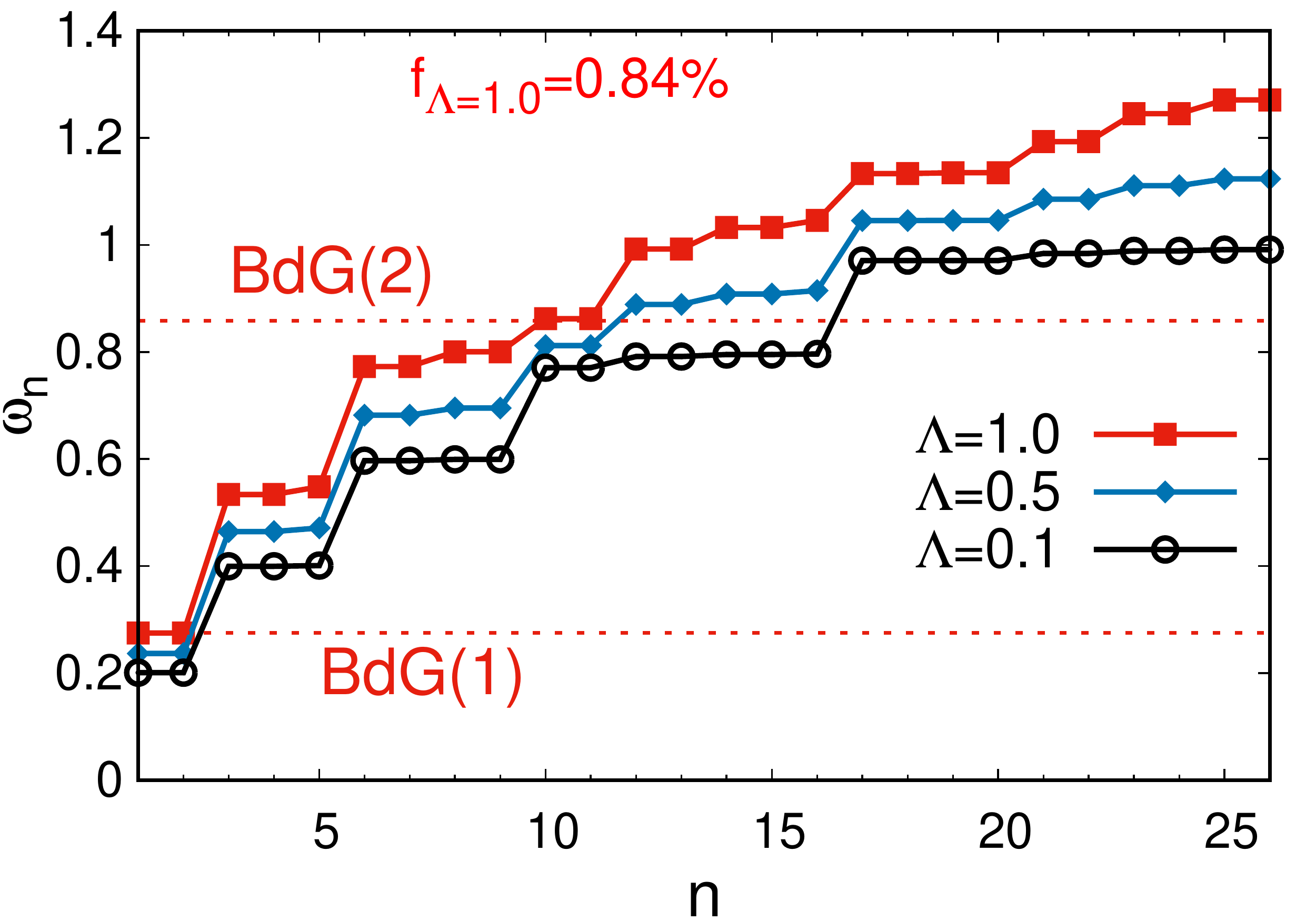} 
 	  \caption{\label{fig_1D_lattice5}}
 	\end{subfigure}
 	\begin{subfigure}{0.5\textwidth} 
 	  \centering
 	  \includegraphics[angle=-90,width=\textwidth]{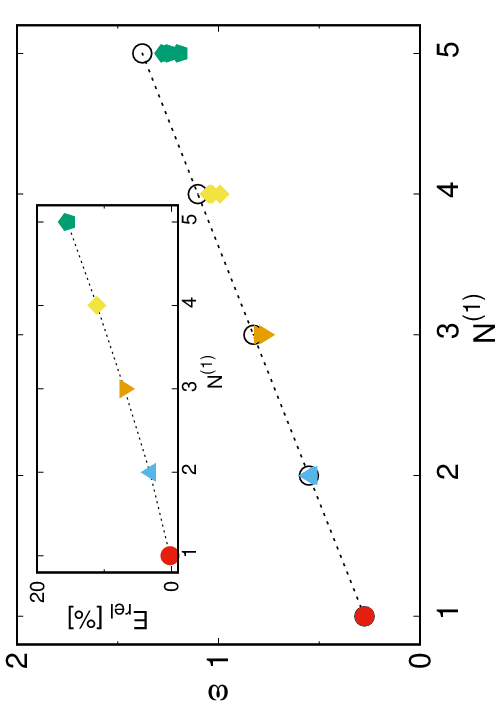}
 	  \caption{\label{fig_1D_lattice6}}
	\end{subfigure}
\caption[Low-energy spectrum of $N=10$ bosons in a shallow lattice with 10 sites]{(a) Same as in Fig. \ref{fig_1D_lattice1} but in a lattice with 10 sites. The depletion grows faster with $\Lambda$ than in the triple well, meaning that the system is even more sensitive for many-body effects. (b) Same as in Fig. \ref{fig_1D_lattice3} but for 10 sites and $\Lambda=1.0$. The relative error (inset) reaches similar values as for systems in the shallow triple well with comparable depletion. All quantities are dimensionless. See text for details. The figures are adapted from \cite{Beike_1d_lattice}.}
\label{fig_1D_lattice_gen2}
\end{figure}

As a start, $N=10$ bosons confined in a triple-well potential with periodic boundary conditions are considered. The low-energy excitation spectra for both a shallow ($V_0=5.0\approx 1.01\,E_R$) and a deep well ($V_0=50.0\approx 10.13\,E_R$) are discussed. Results are presented in Figs. \ref{fig_1D_lattice_gen1}(a) and (b). In both wells, the spectra for the non-interacting BEC are structured in sequences of levels with different degeneracies. The states and levels can be associated with appropriate quantum numbers, envisioning that for $\Lambda=0$ eigenstates of a single particle in a periodic lattice are exact quasimomentum eigenstates. The eigenstates appear in a band structure that is more pronounced for deep lattices. For the shallow lattice in panel (a), the first level consisting of two degenerate states denotes the excitations of putting one boson in either a state with quasimomentum $(+1)$ or with quasimomentum $(-1)$. Similarly, the states of level 5 denote the excitations where one boson is put to either $(+1)$ or $(-1)$ from the second single-particle band. To this end, the notation $\left(n_{+1}^{(1)},n_{-1}^{(1)};n_{+1}^{(2)},n_{-1}^{(2)}\right)$ is introduced, where the first two entries in brackets denote the occupation of $(+1)$ and $(-1)$ from the first single-particle band and the last two entries denote the occupation of $(+1)$ and $(-1)$ from the second single-particle band. As another example, level 2 contains the states $(2,0;0,0)$, $(0,2;0,0)$ and $(1,1;0,0)$. It is stressed that the BdG spectrum only contains the excitations of levels 1 and 5, and misses all multi-particle excitations in the low-energy part. However, for $\Lambda=0$, the missing levels can be anticipated by taking sums and multiples of the energies of the first and fifth levels. As soon as there are correlations between the bosons, i.e., for $\Lambda >0$, this is no longer possible, as discussed below. In panel (b), only the first single-particle band appears in the shown energy range because, as shown in Fig. \ref{fig_1D_lattice7}, the first and second band are strongly separated from each other.

\begin{figure}[ht!]
 	  \centering	
 	  \includegraphics[angle=-90,width=0.7\textwidth]{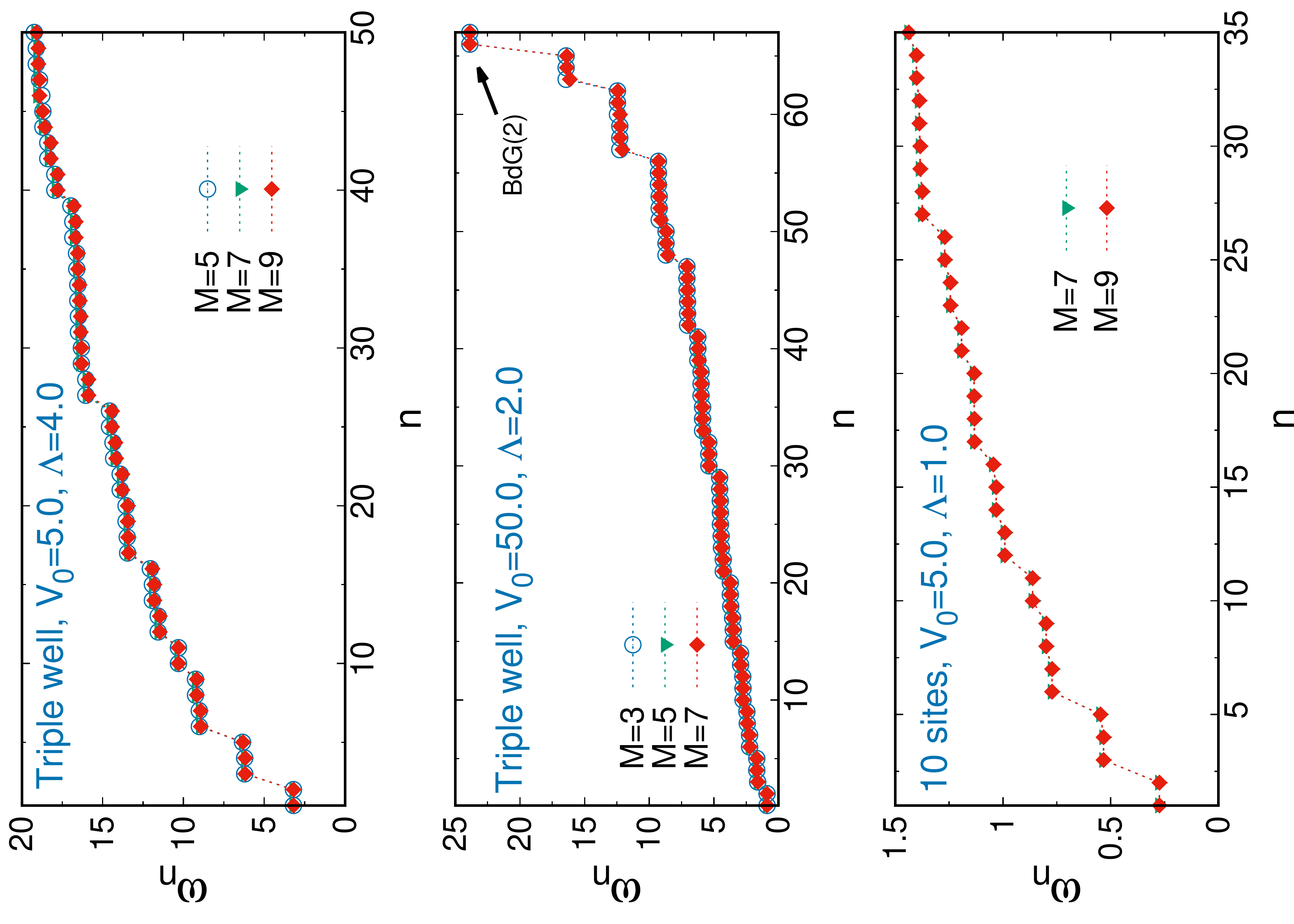} 
\caption[Numerical convergence of the computed low-energy spectra of BECs in 1D lattices]{Proof of numerical convergence for the systems discussed in Figs. \ref{fig_1D_lattice_gen1} and \ref{fig_1D_lattice_gen2}. LR-MCTDHB($7$) clearly suffices in all cases. For the BEC in the deep triple well [middle panel], all 65 single- and multi-particle excitations below the bottom of the second band [denoted by BdG(2)] are obtained. Already for $M=3$, the results are highly accurate which is due to the large separation of the first two single-particle bands. All quantities are dimensionless. See text for details. The figures are adapted from Ref. \cite{Beike_1d_lattice}.}
\label{fig_1D_lattice7}
\end{figure}

For non-zero repulsion in the shallow well in panel (a), the BECs are slightly depleted ($f=1.1\%$ for $\Lambda=4.0$), whereas for the deep well in panel (b) the BEC for $\Lambda=2.0$ is already fragmented by $8.4\%$. One can observe that all excited states grow in energy once the bosons interact, and that splittings of several states from the same level occur although they were degenerate for $\Lambda=0$. Thus, the clear level structure of the non-interacting case vanishes as $\Lambda$ increases, which makes it more difficult to identify the individual states with regard to the categorization in terms of quasimomentum states. The corresponding mean-field results contain only two doublets for the shallow and one doublet for the deep well, denoted by BdG(1) and BdG(2). All other states, representing multi-particle excitations where more than one boson at a time is excited from the ground state, are absent.

Figs. \ref{fig_1D_lattice_gen1}(c) and (d) demonstrate in how far the BdG energies can be utilized to anticipate the positions of the missing multi-particle states. As mentioned above, this yields the exact energies for the missing states in the non-interacting system. In particular, multi-particle excitations of putting in total $N^{(1)}$ bosons into the states $(+1)$ and $(-1)$ of the first single-particle band are discussed. With regard to the BEC with $\Lambda=4.0$ in the shallow triple well [panel (c)], one observes that BdG overestimates the excitation energies for the states of $1\leq N^{(1)}\leq 5$, i.e., the accurate many-body values are lower. The discrepancy between mean-field and many-body results grows with $N^{(1)}$. Especially the relative error, defined as $E_\text{rel}=\left|\frac{\omega_\text{BdG}\left(N^{(1)}\right)-\omega_\text{MB}\left(N^{(1)}\right)}{\omega_\text{MB}\left(N^{(1)}\right)}\right|$ where $\omega_\text{MB}\left(N^{(1)}\right)$ denotes the many-body energy with the largest distance to $\omega_{\text{BdG}}\left(N^{(1)}\right)=N^{(1)}\cdot \text{BdG(1)}$, grows quickly to more than $10\%$. It is emphasized again that the BEC is only slightly depleted by approximately $1\%$. For the fragmented BEC with $\Lambda=2.0$ in the deep triple well [panel (d)], the differences become even more substantial because already for the single-particle excitation ($N^{(1)}=1$) the relative error is about $5\%$. This means that in this case BdG does not only miss all multi-particle excited states in the low-energy spectrum, but it is also very inaccurate for the single-particle excitations.

Fig. \ref{fig_1D_lattice_gen2} shows a similar analysis for $N=10$ bosons in a larger shallow lattice with 10 sites. In principle, the described many-body effects observed in the triple well also appear in this extended lattice. However, with respect to the degree of fragmentation, which is $f=0.84\%$ for $\Lambda=1.0$, one can deduce that these many-body effects set in even earlier than in the triple well because a comparable depletion is achieved already for weaker repulsion. Hence, an accurate many-body description is of even larger importance if the lattice size is increased.

\begin{figure}[h!]
\includegraphics[angle=-90,width=0.5\columnwidth]{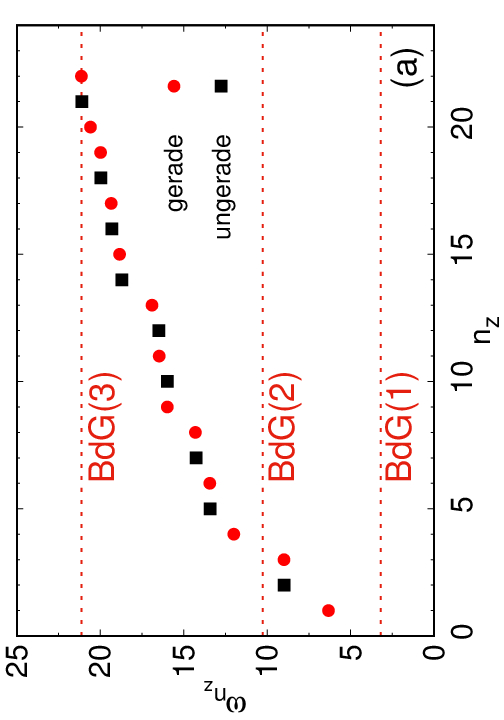}
\includegraphics[angle=-90,width=0.5\columnwidth]{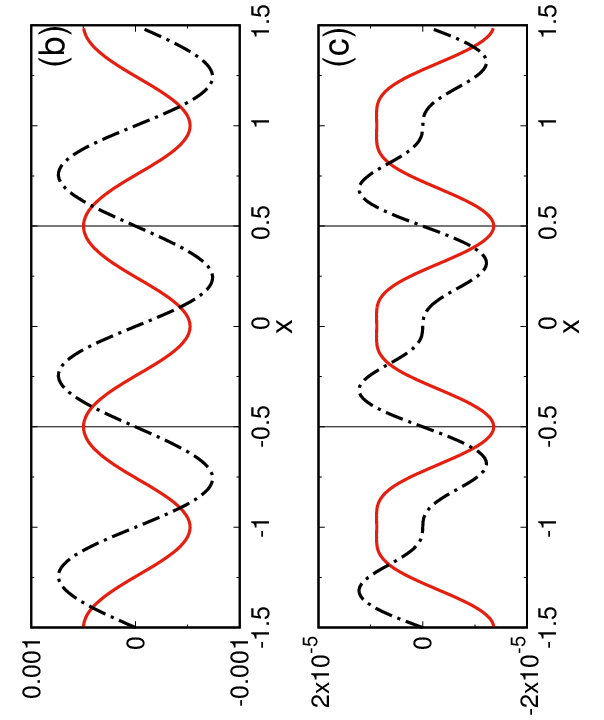}
\caption[Zero-quasimomentum modes in the low-energy spectrum of $N=10$ bosons in a shallow triple well]{(a) Energies of the zero-quasimomentum modes (ZQMs) for the system in Fig. \ref{fig_1D_lattice_gen1}(a) with $\Lambda=4.0$, up to the top of the second single-particle band [denoted by BdG(3)]. All shown states are not obtained by the BdG mean-field approach. The ZQMs are categorized by the symmetry of their density responses, which is either gerade or ungerade on each lattice site. Mind the different index $n_z$, which enumerates only the ZQMs compared to the index $n$ of the upper figures which enumerates all excitations. (b) and (c): Real part of the density responses of the states $(3,0;0,0)$ and $(0,3;0,0)$ [panel (c)] as well as $(4,1;0,0)$ and $(1,4;0,0)$ [panel (d)]. The vertical lines separate the lattice sites from each other. Notice the different scales. All quantities are dimensionless. See text for more details. The figures are taken from Ref. \cite{Beike_1d_lattice}.}
\label{fig_1D_lattice8}
\end{figure}

The results shown in Fig. \ref{fig_1D_lattice7} demonstrate numerical convergence with respect to the number of self-consistent orbitals $M$ utilized. In all cases, $M=7$ orbitals clearly suffice for the low-energy excitations. In particular, for the middle panel describing a BEC in the deep triple well, one can see that firstly, $M=3$ orbitals already give very accurate values for the excitation energies, and, secondly, all multi-particle states build up by the states $(+1)$ and $(-1)$ from the first single-particle band are obtained (65 in total). To remind the reader, it is stressed again that the BEC is fragmented in this case, and nevertheless LR-MCTDHB is capable to give converged results.

So far, it was shown that the BdG energies for a BEC in a 1D lattice can be largely inaccurate, even for weakly-depleted condensates with $f\approx1\%$. Though it has been shown analytically that for a BEC with sufficiently weak and long-ranged repulsion in the Hartree limit, i.e., where $N\rightarrow\infty$ with fixed $\Lambda$, the BdG equation yields the exact energies of all single-particle excitations, and the energies of the multi-particle states are given by multiples and sums of the BdG energies \cite{Seiringer1} (extending the findings of a previous work on homogeneous systems \cite{Seiringer3}). Whether this result is applicable also for BECs with contact interaction remains unclear. However, naturally the question arises how fast, if at all, the excitation spectrum of a BEC with arbitrary shape of the repulsion approaches the BdG energies upon increasing the number of bosons. This question is discussed further in the subsequent section, Section \ref{App_RotBEC}, where excited states in rotating BECs are analyzed.

\begin{figure}[h!]
\centering
\includegraphics[angle=-90,width=0.9\columnwidth]{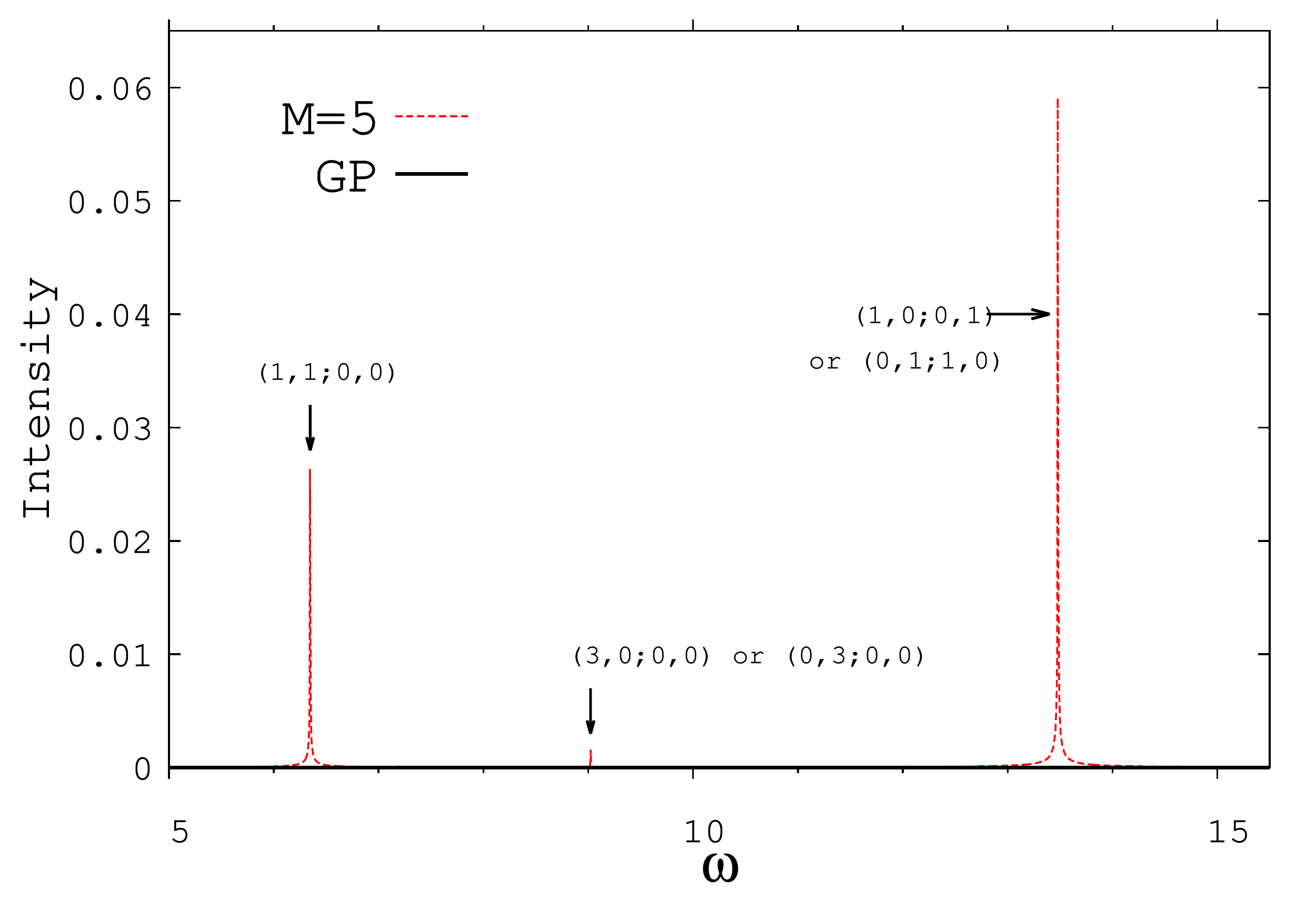}
\caption[Fourier spectrum of the time evolution of the position variance of $N=10$ bosons in a shallow triple well due to a quantum quench of the lattice depth]{Low-energy Fourier spectrum of the time evolution of the position variance per particle in $x$-direction, $\frac{1}{N}\Delta^2_{\hat{X}}$, due to a quantum quench of the lattice depth from $V_0=4.9$ to $V_0=5.0$. The system parameters are the same as in Fig. \ref{fig_1D_lattice_gen1}(a) with $\Lambda=4.0$. The many-body spectrum (dashed red), obtained from the evolution with $M=5$ orbitals, shows two sharp peaks referring to two-particle excitations and one less intense peak referring to a three-particle excitation. The first intense peak refers to the two-particle excitation $(1,1;0,0)$, whereas the second intense peak refers to $(1,0;0,1)$ [or $(0,1;1,0)$]. The less intense peak refers to $(3,0;0,0)$ [or $(0,3;0,0)$]. All obtained peaks are ZQMs with gerade symmetry. The GP spectrum (solid black) in the shown energy range is \textit{completely flat}. The depletion of the BEC is $f\sim 1.1\%$ during the entire propagation time. All quantities are dimensionless. See text for more details.}
\label{fig_1D_lattice10}
\end{figure}

Finally, the zero-quasimomentum modes (ZQMs) are briefly discussed. The latter are defined in this work as the excitations where $\text{MOD}(P,L)=0$, with $P$ denoting the total quasimomentum up to the second single-particle band given by
\begin{equation}\label{Tot_Quasi}
	P=n_{+1}^{(1)}+n_{+1}^{(2)}-n_{-1}^{(1)}-n_{-1}^{(2)},
\end{equation}
and $L$ denoting the number of lattice sites. The spectrum of the ZQMs up to BdG(3), which marks the top of the second single-particle band, is shown in Fig. \ref{fig_1D_lattice8}(a). The system parameters are the same as in Fig. \ref{fig_1D_lattice_gen1}(a) with $\Lambda=4.0$. It is worth noting that all of the shown ZQMs are multi-particle excitations, meaning that BdG does not account for a single ZQM in the low-energy spectrum although there exist more than 20. By analyzing the response densities of the ZQMs, it was found that they can be categorized by their symmetry with respect to the individual wells, which is either gerade or ungerade. Specific examples are given in Figs. \ref{fig_1D_lattice8}(b) and (c). The ZQMs are most likely the easiest states to excite in an experiment where the BEC is subject to either a quench of the potential or the interaction strength. For example, a simple quench scenario to excite the gerade ZQMs would be a sudden change of the lattice depth. To this end, the ground state of a BEC with $N=10$ bosons in the triple well with interaction parameter $\Lambda=4.0$ and lattice depth $V_0=4.9$ is calculated. Then, the depth is suddenly quenched to $V_0=5.0$, and the subsequent time-evolution is investigated. Fig. \ref{fig_1D_lattice10} shows the Fourier spectrum of the dynamical evolution of the position variance per particle, $\frac{1}{N}\Delta^2_{\hat{X}}$, obtained from both MCTDHB($5$) and GP. Shown is the energy range $5.0\leq \omega\leq 15.5$. At the many-body level, 3 distinct peaks are observed, whose positions correspond to the first, second, and fourth gerade excitations of the spectrum in Fig. \ref{fig_1D_lattice8}(a).  They can thus be labeled according to the occupation of (+1) and (-1) from the first and second single-particle band. The peaks corresponding to the two-particle excitations at $\omega\sim 6.35$ and $\omega\sim 13.48$ are more intense than the peak corresponding to the three-particle excitation at $\omega\sim 9.02$. Excitations where higher particle numbers are involved are not intense enough to be distinguished from the background noise. At the GP level, the spectrum is flat, which means that none of the ZQMs obtained at the many-body level is included in the mean-field dynamics, which was anticipated due to the multi-particle nature of the ZQMs.

To summarize, it was demonstrated that the excitation spectrum of a weakly-depleted BEC in a 1D optical lattice shows clear many-body effects which are not accounted for at the mean-field level. A depletion of $f\sim 1\%$ is sufficient to observe these effects. In larger lattices, the sensitivity to many-body features even increases. It was further shown that a special type of many-body excited states (ZQMs) are easily accessible by performing a simple quantum quench to the system. A promising quantity to detect these states is the position variance, as already seen in the previous section.

\subsubsection{Rotating Bose-Einstein condensates in an anharmonic trap}\label{App_RotBEC}

After the discussion of many-body excited states in 1D trapped BECs, an example in 2D, namely the case of rotating BECs in an anharmonic, radially-symmetric potential crater, is examined. Of particular interest is the weakly-interacting regime in which the ground-state depletion is marginal and one would naively expect mean-field approaches like the BdG theory to be valid and accurate. The results below were recently published in Ref. \cite{Beinke_ROT_BEC} and it is referred to this work for additional details.

Rotating BECs were of high interest in the past. Research dealt with the occurrence of quantized vortices \cite{Rokhsar,Matthews,Madison1,Madison2,Madison3} or with the similarity to the fractional quantum Hall effect for charged particles in the vicinity of a magnetic field \cite{Regnault,Chang,Dalibard2,Dalibard3}. Moreover, low-lying excitations in such systems were addressed, for example in vortex lattices \cite{Intro_BdG_appl_2,Cornell1,Cornell2,Dalibard1}. Furthermore, investigations were made on the decay of the counter-rotating quadrupole mode \cite{Mizushima}, on Tkachenko modes \cite{Simula2, Bigelow2}, on the twiston spectrum \cite{Chevy}, or on excitations in anharmonic traps \cite{Collin,Ancilotto}. In the latter works, the mean-field excitations were calculated by utilizing the BdG theory. Computations employing a many-body description arerather rare. An examples is the analysis of the yrast spectra in a harmonic confinement obtained by exact diagonalization using the lowest Landau level approximation \cite{Reimann1,Reimann2,Viefers2,Reimann3}.

As mentioned above, accurate many-body results for the low-energy spectrum of rotating BECs in an anharmonic, radially-symmetric trapping potential are presented. Especially, the effect of the angular momentum in the BECs' ground states on the excitation energies is discussed. As for the examples in the earlier sections, the BdG and many-body results are compared to each other. 

\begin{figure}[h!]
  \centering
  \includegraphics[width=0.8\textwidth]{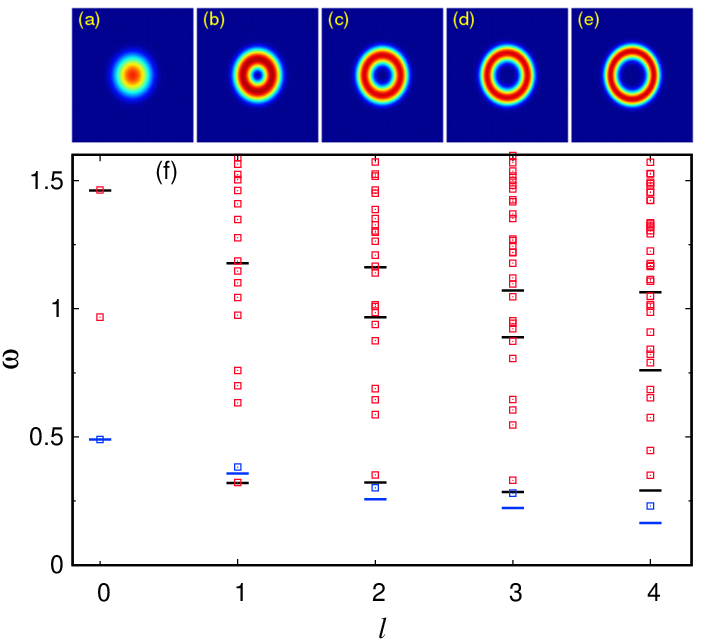}
  \caption[Ground-state densities and low-energy spectra of a rotating BEC with different vorticities]{(a)-(e) Ground-state densities of rotating BECs with $N=10$ bosons and interaction parameter $\Lambda=0.5$ for different vorticities $l$. Results are shown for MCTDHB($7$), the corresponding GP densities look alike (not shown). For non-zero vorticity, the ground state is a single vortex whose core size grows with its charge $l$. (f) Low-energy excitation spectra atop the ground states from the upper panels. Many-body results obtained by LR-MCTDHB($7$) are denoted by red and blue squares, the BdG results are denoted by black and blue lines. Once the ground state has non-zero vorticity, deviations between the mean-field and many-body excitation energies are observed, and they grow with $l$. Moreover, the density of (many-body) excited states grows strongly. All ground states are only slightly depleted, with a maximum of $f\approx 0.38\%$ for $l=4$. Symbols colored in blue refer to the (+1) excitation energies described in the main text. All quantities are dimensionless. See text for details. The figure is adapted from Ref. \cite{Beinke_ROT_BEC}.}
  \label{fig:rot_BEC_1}
\end{figure}

The crater in which the BEC is confined reads 
\begin{equation}\label{eq_trap_RotBEC}
V(r)=
\begin{cases}
	C\,e^{-0.5\,(r-R_C)^4}     \,\,  &\mbox{if } r\leq R_C \\
 	C \,\, &\mbox{if } r>R_C
\end{cases}
\end{equation}
which is essentially the same as in Eq. (\ref{eq_crater_potential}) but without the central ring-shaped barrier. Here, the radius of the crater is set to $R_C=3.0$ and the height of the wall is kept at $C=200.0$. Furthermore, the same Gaussian repulsion as in Eq. (\ref{eq_Gauss_rep}) is employed.

Excitations in this system are analyzed in the rotating frame of reference, where the Hamiltonian $\hat{H}$ in Eq. (\ref{MB_HAM1}) contains an additional contribution accounting for the rotation around the $z$-axis. The whole Hamiltonian thus reads
\begin{equation}\label{eq_Hamiltonian_rotBEC}
 \hat{H}_{\text{rot}}=\hat{H}-\Omega_\text{rot}\hat{L}_z
\end{equation} 
where $\Omega_\text{rot}$ denotes the angular velocity around the $z$-axis and $\hat{L}_z$ denotes the operator of the angular momentum in the $z$-direction.

A LR analysis is applied atop the ground states of $N=10$ bosons with weak repulsion strength $\Lambda=0.5$ for different values of $\Omega_\text{rot}$, and thus of the angular momentum per particle $l$, which is referred to as the vorticity or the charge below. The degree of condensation, $N(1-f)$, ranges from $9.999$ ($l=0$) to $9.962$ ($l=4$) out of 10 particles, meaning that the BEC is highly condensed for all considered parameter values. The upper panels in Fig. \ref{fig:rot_BEC_1} show the ground-state densities calculated with MCTDHB($7$) for vorticities $0\leq l\leq 4$. Whereas for $l=0$, the density resembles a Gaussian in the center of the crater, the shape of the densities for $l>0$ indicate that the ground state is a vortex with a density node at its core. One can further see that this core is growing with $l$. The corresponding GP densities look alike (not shown).

The lower panel in Fig. \ref{fig:rot_BEC_1} compares the low-energy excitation spectrum calculated atop these ground states, both for the mean-field BdG case (black and blue lines) and for LR-MCTDHB($7$) (red and blue squares). The excitations denoted by blue symbols are explained below. At first, the spectrum for $l=0$ is examined. Counting the number of excited states, one obtains 2 states from the BdG approach and 3 states from the many-body approach. Moreover, the energies of the two mean-field excitations coincide with the corresponding many-body results. Since the BdG theory only accounts for single-particle excitations, the third excited stated obtained from LR-MCTDHB is an excitation where more than one particle is excited from the ground state. In particular, it represents the state where two bosons are put into the orbital corresponding to the first excitation. The latter single-particle excitation is referred to as (+1) since it describes the case of putting a boson from the ground-state mode with angular momentum $l_z=0$ to an orbital with $l_z=1$. Examining the spectra for $l>0$, one observes two major changes. At first, the density of excitations in the low-energy spectrum clearly grows with $l$, which can be deduced from the large number of many-body excited states entering. The number of mean-field excitations also grows (from 2 states for $l=0$ to 4 states for $l=4$), but very slightly compared to the number of many-body states. Therefore, the amount of excited states that BdG misses is strongly growing with the vorticity of the BEC's ground state. The second observation is that the values of the single-particle energies from BdG and LR-MCTDHB do not coincide any longer, as it was the case for $l=0$. In fact, the entire BdG spectrum for vorticity $l=4$ appears to be completely unrelated to the corresponding many-body spectrum. The increasing gap between the (+1) results of the two approaches is clearly visible. 

Fig. \ref{fig:rot_BEC_2} addresses the evolution of the energy of (+1), denoted by $\omega_{(+1)}$, with respect to the angular velocity $\Omega_\text{rot}$. Shown are the results obtained from the BdG theory, as well as many-body results for different numbers of bosons ($N=10,\,100$ and $1000$). In the $l=0$ sector, all energies coincide, indicating that for this case the BdG theory is sufficient. Henceforth, this obviously changes. With growing vortex charge of the ground state, the mean-field and many-body lines for (+1) separate further from each other. Most importantly, increasing the number of bosons in the condensate only marginally changes this, meaning that the many-body and mean-field results for large $N$ approach each other very slowly (see inset). As mentioned in the previous section, it was proven analytically that for a trapped BEC with sufficiently weak and long-ranged repulsion, the excitation energies in the Hartree limit converge towards the BdG energies \cite{Seiringer1}. The same is believed to happen for a rotating BEC under certain conditions \cite{Seiringer2}.

\begin{figure}[h!]
  \centering
  \includegraphics[angle=-90,width=\textwidth]{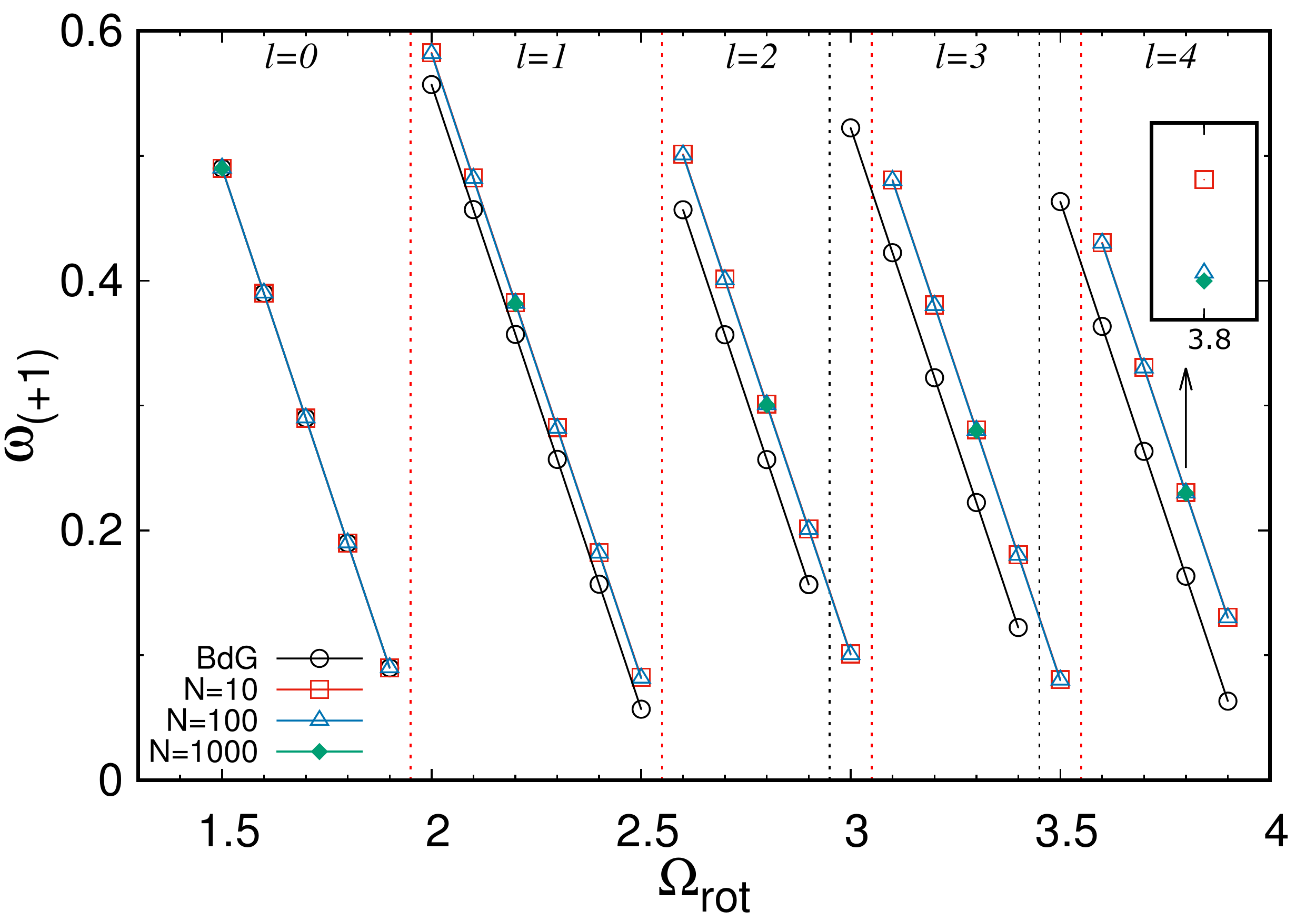}
  \caption[Excitation energies $\omega_{(+1)}$ for different angular velocities of a rotating BEC]{Excitation energies $\omega_{(+1)}$ for different angular velocities $\Omega_\text{rot}$. Compared are the BdG energies and the many-body results for different numbers of bosons $N$, obtained by LR-MCTDHB. The interaction strength is $\Lambda=0.5$. The transition from ground-state vorticity $l$ to $l+1$ between two adjacent analyzed values of $\Omega_\text{rot}$ are denoted by dotted vertical lines (mean-field in black and many-body in red). The higher the ground-state vorticity $l$, the larger is the separation between the BdG and many-body energies. Inset: Magnified view for $\Omega_\text{rot}=3.8$. All quantities are dimensionless. See text for details. The figure is adapted from Ref. \cite{Beinke_ROT_BEC}.}
  \label{fig:rot_BEC_2}
\end{figure}

\begin{figure}[h!]
  \centering
  \includegraphics[angle=-90,width=\textwidth]{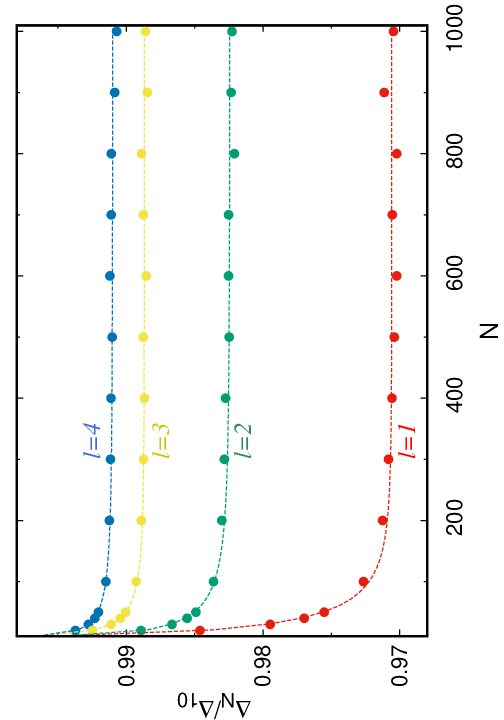}
  \caption[Evolution of the difference between the mean-field and many-body energies of $\omega_{(+1)}$ for different particle numbers of a rotating BEC]{Evolution of the energetic difference $\Delta_N$ between the mean-field and many-body results for $\omega_{(+1)}$ for different particle numbers $N$. The differences are shown relative to $\Delta_{10}$, i.e., the gap for $N=10$ bosons. Solid lines denote exponential fits. The gap becomes larger as $l$ grows. Even for $l=1$ and $N=1000$, it is approximately $97\%$ of $\Delta_{10}$, and the slopes of the fit curves do not indicate a sharp decent towards zero if $N$ is increased further. All quantities are dimensionless. See text for details. The figure is taken from Ref. \cite{Beinke_ROT_BEC}.}
  \label{fig:rot_BEC_3}
\end{figure}

In Fig. \ref{fig:rot_BEC_3}, the dependence of the difference $\Delta_N$ between the mean-field and many-body energies of (+1) on the number of bosons in the BEC is analyzed in more detail. The results are given relative to the energy for $N=10$ bosons, denoted by $\Delta_{10}$. For all shown vorticities, the gap becomes smaller with increasing $N$. However, it is obvious that even for BECs with an experimentally relevant number of bosons ($N=1000$) one is still far away from the predictions of the BdG theory. For $l=1$, one obtains that the gap size $\Delta_{1000}$ is still about 97$\%$ of $\Delta_{10}$, and correspondingly higher for $l>1$. Additionally, the shape of the exponential fit curves to the numerical data does not show any sign of a steep descend when $N$ is increased further. It can firstly be deduced that the excitation energies approach the BdG predictions for $l>0$ very slowly, and, secondly, one apparently needs a very high number of particles in order to enter the regime where BdG yields approximately the correct results. However, it is important to note that the convergence towards the BdG excitation energies in the Hartree limit is not proven for the system under consideration, since Ref. \cite{Seiringer2} only deals with the case of a rotating BEC with broken radial symmetry. The latter situation appears, e.g., for a BEC in a harmonic confinement when the rotation is fast enough to generate multiple quantized vortices in the ground state (see, e.g., Ref. \cite{Fetter}). Although the numerical results cannot give a definite answer for the Hartree limit, they at least suggest that the regime in which one is clearly in need of an accurate many-body theory for the excitation energies covers the experimentally relevant regime of rotating BECs, with $N$ being of the order of $10^3-10^6$ bosons. A further discussion on the applicability of the GP and BdG theories in the Hartree limit can be found in Appendix \ref{CH_MF_Theory}.

\begin{figure}[h!]
  \centering
  \includegraphics[angle=-90,width=\textwidth]{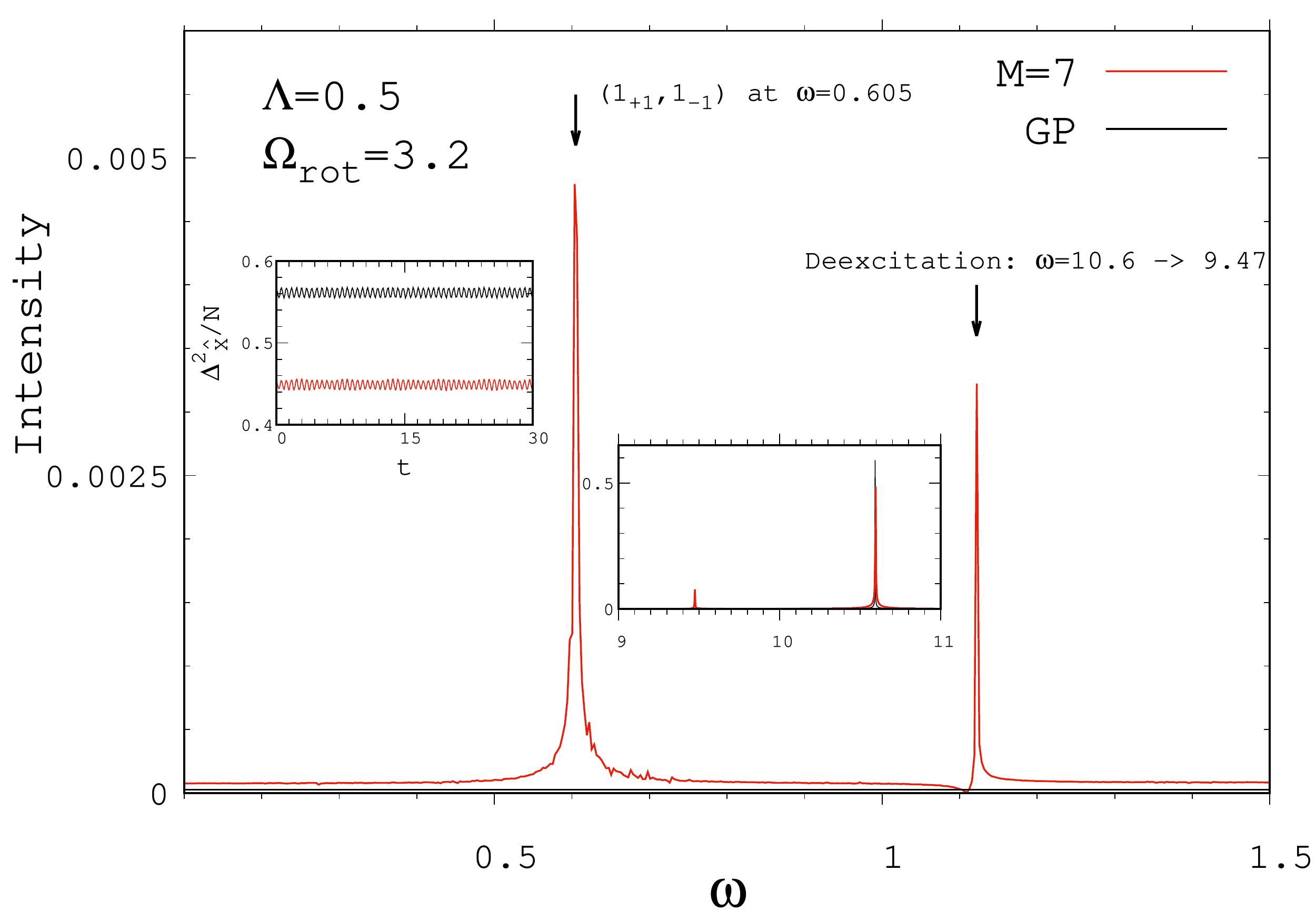}
  \caption[Fourier spectrum of the time evolution of the position variance of a rotating BEC due to a quantum quench of the radial size of the trap]{Low-energy Fourier spectrum of the time evolution of the position variance in $x$-direction, $\frac{1}{N}\Delta^2_{\hat{X}}$, due to a quantum quench of the radius $R_C$ from $3.01$ to $3.0$. The number of bosons is $N=10$. The many-body spectrum (red), obtained from the evolution with $M=7$ orbitals, shows two sharp peaks in the considered energy range, whereas the GP spectrum (black) is flat. The first peak refers to the two-particle excitation $(1_{+1},1_{-1})$, whereas the second peak indicates a de-excitation between two higher-lying states (see right inset). The left inset shows the evolution of $\frac{1}{N}\Delta^2_{\hat{X}}$ from $0 \leq t \leq 30$. The depletion of the BEC stays below $0.4\%$ during the entire propagation. All quantities are dimensionless. See text for details.}
  \label{fig:rot_BEC_4}
\end{figure}

It is finally suggested, similar to the 1D systems discussed in Section \ref{App_Theisen}, how a quench can be utilized in order to excite breathing modes. To this end, the ground state of a BEC with $N=10$ bosons and angular velocity $\Omega_\text{rot}=3.2$ in the same trap as described in Eq. (\ref{eq_trap_RotBEC}), but with a slightly larger radius of $R_C=3.01$, is calculated. The obtained ground state has the same symmetry and vorticity $l=3$ as the one for $R_C=3.0$ and $\Omega_\text{rot}=3.2$. Then, the crater is suddenly shrinked back to $R_C=3.0$ and the dynamical evolution is computed with MCTDHB($7$), which is sufficient in order to obtain converged dynamics for the position variance. As could be seen in the previous sections, the latter quantity is very sensitive for many-body effects, and therefore, a spectral analysis of its dynamics by computing the Fourier spectrum of $\frac{1}{N}\Delta^2_{\hat{X}}(t)$ is carried out. Due to the radial symmetry, one could as well analyze the position variance in the $y$-direction. The result are shown in Fig. \ref{fig:rot_BEC_4}.  The many-body spectrum shows two distinct peaks, where the first corresponds to the $(1_{+1},1_{-1})$ two-particle excitation, where one boson is excited to an orbital with $l=4$ and one is excited to an orbital with $l=2$, such that the total angular momentum is unchanged. The energy that LR-MCTDHB($7$) predicts for this state is $\omega=0.605$ which coincides with the peak position of the Fourier analysis. The second peak at $\omega\approx 1.12$ denotes a de-excitation between two states that are much higher in energy ($\omega\approx 10.6 \rightarrow 9.47$, see right inset in Fig. \ref{fig:rot_BEC_4}). The occurrence of de-excitations is an indication that the quench described above already leads to second-order effects, meaning that one has left the LR regime. With regard to the mean-field spectrum from the GP dynamics, one observes that in the shown energy range, the Fourier spectrum is completely flat. The absence of the first peak is anticipated because the GP theory does not account for two-particle excitations by construction. The absence of the second peak can be understood from the right inset where the Fourier spectrum in the interval $9\leq \omega\leq 11$ is shown. One sees that in the GP case, there is only one peak at $\omega\approx 10.6$, but no second one at $\omega \approx 9.47$ as in the many-body spectrum. Thus, very different spectra from the dynamics of the variance at the mean-field and many-body levels are obtained, and it is shown that the suggested quench leads to an excitation of a many-body breathing mode. It is stressed that the depletion stays below $0.4\%$ for the entire propagation, which means that the BEC is essentially condensed.  

To summarize, the low-energy spectrum of a rotating BEC in an anharmonic, radially-symmetric crater shows significant many-body effects, even for weakly-interacting, highly-condensed systems. These effects can be observed already for the lowest-in-energy, single-particle excited states, and remain even for mesoscopic and large BECs. Furthermore, a simple quench triggers a many-body breathing mode, which can be observed, e.g., by analyzing the Fourier spectrum of the position variance.




\section{Summary and Conclusions}\label{Ch_summary}

This review presents recently developed many-body approaches capable to describe and compute accurately the ground state and dynamics as well as the energy spectrum of trapped ultracold bosons. It is relevant to employ methods which are on the same footing. The dynamics is calculated by the well-known MCTDHB method, the ground state is obtained by imaginary-time propagation using the same method, and the energy spectrum of the system is then determined by applying the linear-response theory to the ground state computed via MCTDHB. The resulting theory is named LR-MCTDHB.

The MCTDHB theory is briefly discussed and more attention is paid to the less spread LR-MCTDHB theory. As the implementation of the MCTDHB is well documented in the literature, only references are given here, while the newly developed LR-MCTDHB implementation which makes the theory widely applicable, is presented here in some detail for the first time. The latter posed a major challenge due to both the complexity of the underlying theory and the accompanying technical difficulties like the vast memory consumption and the necessity of parallelization. The code structure is explained in detail both with respect to the construction of the linear-response matrix and its subsequent partial diagonalization which provides the desired excitation energies. A comprehensive description of the procedure used to calculate the lowest eigenvalues is also provided. The corresponding benchmarks of the LR-MCTDHB code (see also Appendix \ref{Appendix_further_benchmarks}) with an analytically solvable many-boson non-trivial model demonstrate its correct functioning and show that the theory provides numerically exact results.

The second main goal of this review is to present applications to many-body dynamics and excitations of BECs in interesting scenarios, and, of course, to discuss the underlying physics. Several systems of trapped BECs in one and two dimensions are addressed in detail and the main attention is drawn to the detection of many-body effects that are not describable by applying mean-field theory. Particularly emphasized are the tunneling dynamics in double-well systems and the development of fragmentation as a function of time in initially coherent BECs. The applications shown in this work have demonstrated that a full many-body treatment of the dynamics of trapped interacting BECs is inevitable. For instance, it is exhibited that the out-of-equilibrium tunneling dynamics of initially coherent BECs in 1D and 2D double wells are of many-body nature. After a few Rabi cycles, tunneling between the wells becomes suppressed at the many-body level, whereas unperturbed oscillations are predicted at the mean-field level. Surprisingly, the degree of fragmentation in 1D is universal for constant values of the interaction parameter. Going to 2D, the sensitivity of the tunneling dynamics to many-body effects is even enhanced as soon as the bosons carry angular momentum.
This can also be deduced from the larger amount of orbitals necessary in the limit of weak repulsion as compared to the case of bosons with zero angular momentum. Further applications deal with the nucleation of phantom vortices, which are pure many-body objects not observable in the condensate density, as well as with the tunneling of bosons to open space. In both cases, the initially coherent systems become fragmented and are, therefore, in need of an accurate many-body description.

With respect to excitations, it is observed that the ground-state depletion has a strong impact on the low-energy spectrum. Even for marginal depletion where one might expect mean-field approaches to give accurate results, substantial many-body effects like the shift of excitation energies and the appearance of complicated many-boson excitations occur. Applications in 1D explores the many-body effects in double- and triple-well systems as well as in larger lattices. In 2D, it is found that once the bosons carry angular momentum and the ground state becomes a vortex, the deviations between the mean-field and many-body excitation spectra grow strongly. Moreover, this discrepancy tends to persist when the number of particles in the condensate increases up to an experimentally relevant size. This has a potentially large impact on experiments on ultracold bosons since the dynamics are mostly determined by the lowest-in-energy excitations. Hopefully, these findings and the new insights into the physics of ultracold bosons will stimulate future experiments.

Before closing, we would like to add several remarks concerning further general and technical developments of the methods which can further broaden their range of applicability. Until now the applications of the many-body linear-response theory were for excitations on top of the system's ground state.  We emphasize that the LR-MCTDHB can also be applied to any other state obtained via MCTDHB. In other  words, it is possible to compute excitations on top of any other state. This will help to address excitations which are difficult to reach from the ground state. Another general possibility lies in the formulation of LR-MCTDHB in block-diagonal form, which is potentially
advantageous in calculating the low-energy spectrum, as discussed in Appendix \ref{Appendix_BlockDiagonal}.

Apart from these, further general extensions and developments of both the LR theory itself as well as of its numerical implementation are worth of additional investigation. With regard to theoretical developments, an extension of LR-MCTDHB to BEC mixtures, i.e., compound systems of different types of bosons, is highly  desirable as their popularity has substantially grown \cite{Ofir_BEC_mixture,Schmelcher_honeycomb,Schmelcher_BB_mixture1,Schmelcher_BB_mixture2,BB_mixture_Cikojevic,BB_mixture_Utesov,BB_mixture_Cheiney}. The underlying MCTDHB theory for bosonic mixtures, also including the possibility of internal degrees of freedom, has been formulated \cite{MCTDHB_internal_degrees,ML-MCTDHB-impl,ML-MCTDHB-extension}, but a linear-response theory is completely missing. It turns out that the so-called  multi-layer MCTDHB, introduced and described in Ref. \cite{ML-MCTDHB-impl}, is particularly advantageous numerically. To this end, an interesting goal would be to apply the linear-response theory to multi-layer MCTDHB. Moreover, mixtures of bosons and fermions might become even more relevant \cite{ML-MCTDHB-extension,BF_collective_exc,BF_mixture_Lous,Schmelcher_BF_mixture1,Schmelcher_BF_mixture2,Schmelcher_BF_mixture3,BF_mixture_efimov1}, meaning that also the equations of motion of the fermionic counterpart of MCTDHB, called MCTDHF (see, e.g., Ref. \cite{MCTDHF_Alon} and references therein, and the recently developed implementation described in Ref. \cite{MCTDHF_Lode_implementation}), should be utilized to develop the respective linear-response theory.

Further technical developments which accelerate the computations are, of course, also welcome. One interesting possible extension lies in the application of linear-response theory to the time-dependent restricted-active-space self-consistent-field (TD-RASSCF) method, introduced in Ref. \cite{Summary_Leveque}. The latter theory is similar to MCTDHB, but is an approximation to it. It is, however, capable of reducing systematically the amount of necessary coefficients significantly. This is achieved by restricting the possible excitations in the active space of the single-particle orbitals according to certain conditions or symmetries of the physical problem at hand. Thus, the dimensionalities of the linear-response matrices which would result when constructed based on the linear-response theory on top of TD-RASSCF, could also be greatly reduced. Such an approach will substantially simplify the computation of the excitation spectrum and, moreover, enable the treatment of even larger systems, i.e., with more particles and if necessary higher numbers of orbitals.

With regard to the technical development of the current implementation, the application of additional techniques to diagonalize the often huge linear-response matrix can be beneficial. A potentially very powerful method may be the FEAST algorithm \cite{FEAST_Polizzi}, which is a contour integration and density matrix-based method. It is available in a free software package \cite{FEAST_package} and also allows for high-level parallelization. In particular, it enables the specification of an energy interval in which the desired eigenvalues should be found. Hence, it might become easier to explore not just the lowest but also intermediate eigenvalues.

The above outlined avenues for future research demonstrate the large variety of perspectives, which this work will hopefully stimulate.




\begin{appendices}

\renewcommand{\thesection}{\Alph{section}}
\renewcommand{\thesubsection}{\Alph{section}.\arabic{subsection}}
\renewcommand{\theequation}{\Alph{section}.\arabic{equation}}
\renewcommand{\thetable}{\Alph{section}.\arabic{table}}
\setcounter{section}{0}
\setcounter{equation}{0}
\setcounter{table}{0}

\addtocontents{lot}{\protect\hspace*{5cm}}
\section{Mean-field theory}\label{CH_MF_Theory}
In this Appendix, several mean-field approaches to describe the ground state and dynamics as well as the excitation spectra of trapped BECs are presented. At first, the GP single-orbital mean-field theory (Section \ref{Sec_GP}) and the LR theory atop it, yielding the famous BdG equations (Section \ref{Sec_BdG}), are described. Those approaches are the most common ones in the literature, and are expected to accurately describe the physics of coherent BECs. They are also used in this work to compare mean-field and many-body results. Afterwards, a time-dependent multi-orbital mean-field approach, called TDMF (Section \ref{Sec_TDMF}), and its corresponding LR theory LR-BMF (Section. \ref{LR_BMF}), are discussed. The latter two theories are suitable for describing fragmented BECs at the mean-field level. They can be understood as bridging theories between the GP/BdG theories and the full many-body descriptions given by MCTDHB and LR-MCTDHB, respectively.

\subsection{Single-orbital approaches}

\subsubsection{The GP equation}\label{Sec_GP}
The GP equation is commonly used to obtain the ground state and the time evolution of a coherent BEC, i.e., where all $N$ bosons occupy the same single-particle state $\phi(\mathbf{r},t)$. The many-boson wave function $\Psi$ is thus given by a single Hartree product,
\begin{equation}\label{GP_wavefunction}
	\Psi(\mathbf{r}_1,...,\mathbf{r}_N;t)=\phi(\mathbf{r}_1,t)\, ... \,\phi(\mathbf{r}_N,t)\equiv  |N;t\rangle.
\end{equation}
Originally, the GP equation is derived assuming contact interaction between the bosons. However, to be more general, the two-body interaction potential is taken to be of general form $\hat{W}(\mathbf{r},\mathbf{r}^{\,\prime})$. 

The GP equation can be derived from the least action principle defined in Eq. (\ref{least_action}). The expectation value that appears in the action functional, one obtains
\begin{equation}\label{GP_expectation_value}
	\left\langle N;t \left|\hat{H}-i\frac{\partial}{\partial\,t} \right|N;t  \right\rangle=N h_{\text{\tiny{GP}}}+\frac{\lambda_0N(N-1)}{2}W_{\text{\tiny{GP}}}-iN\left( \frac{\partial}{\partial t}\right)_{\text{\tiny{GP}}}
\end{equation}
where the quantities $h_{\text{\tiny{GP}}}$ and $W_{\text{\tiny{GP}}}$ are defined in Eqs. (\ref{hij}) and (\ref{Wijkl}) utilizing $\phi(\mathbf{r},t)$. The time derivative is given by $\left(\frac{\partial}{\partial t}\right)_{\text{\tiny{GP}}}=\int\phi^\ast(\mathbf{r},t)\,\frac{\partial}{\partial t}\,\phi(\mathbf{r},t)\,d\mathbf{r}$. One arrives at
\begin{align}\label{EQ_GP_deriv}
	0&=\frac{\delta S}{\delta\phi^\ast(\mathbf{r},t)} \left[N h_{\text{\tiny{GP}}}+\frac{\lambda_0N(N-1)}{2}W_{\text{\tiny{GP}}}-iN\left( \frac{\partial}{\partial t}\right)_{\text{\tiny{GP}}} \right] \nonumber \\
	&=N  \left[ \hat{h} + \lambda_0(N-1)\int\hat{W}(\mathbf{r},\mathbf{r}^{\,\prime}) |\phi(\mathbf{r}^{\,\prime},t)|^2d\mathbf{r}^{\,\prime}-i\frac{\partial}{\partial t} \right]\phi(\mathbf{r},t).
\end{align}
Thus, the time-dependent GP equation used in this work is thus given by
\begin{equation}\label{Eq_GP_time}
	\left[\hat{h}+\Lambda\int \hat{W}(\mathbf{r},\mathbf{r}^{\,\prime})\,|\phi(\mathbf{r}^{\,\prime},t)|^2\,d\mathbf{r}^{\,\prime}\right]\phi(\mathbf{r},t)=i\frac{\partial}{\partial t}\phi(\mathbf{r},t)
\end{equation}
with the mean-field interaction parameter $\Lambda=\lambda_0(N-1)$. As mentioned above, for the contact interaction potential it takes the original form as derived in \cite{Gross,Pitaevskii} which can also be found in standard textbooks \cite{Book_Pitaevskii,Book_Pethick}. The time-independent GP equation reads
\begin{equation}\label{Eq_GP_notime}
	\left[\hat{h}+\Lambda\int\hat{W}(\mathbf{r},\mathbf{r}^{\,\prime})\,|\phi^0(\mathbf{r}^{\,\prime})|^2\, d\mathbf{r}^{\,\prime}\right]\phi^0(\mathbf{r})=\mu\phi^0(\mathbf{r})
\end{equation}
where $\mu$ is the chemical potential. Whereas Eq. (\ref{Eq_GP_time}) describes the time-evolution of a fully-condensed BEC, Eq. (\ref{Eq_GP_notime}) is used to determine the GP ground-state orbital $\phi^0(\mathbf{r})$. 

This section is closed with a brief discussion of the applicability of the GP theory in the Hartree limit of repulsive trapped BECs, meaning for $N\rightarrow\infty$ with fixed interaction parameter $\Lambda$. It has been proven analytically that in this limit, the GP theory provides the exact energy and density per particle of the system's ground state \cite{Seiringer_GP_dens_energy}. This, however, does not necessarily mean that the corresponding GP wave function, Eq. (\ref{GP_wavefunction}), coincides with the exact wave function in this limit. To this end, it has been demonstrated recently that the overlap of the exact and GP wave function in the Hartree limit can be clearly less than unity or even be vanishingly small \cite{Lenz_GP1}. Moreover, a subsequent study has shown the many-body character of the exact wave function in the Hartree limit by applying many-body perturbation theory to the mean-field Hamiltonian \cite{Lenz_GP2}, which demonstrates that the mean-field and exact wave functions can be very different from each other. Moreover, as a further interesting result, it was found that the condensate depletion in the Hartree limit becomes a constant. These findings potentially have severe consequences for the excitation spectrum of a BEC in or close to the Hartree limit, which is discussed at the end of Section \ref{Sec_BdG}.

\subsubsection{The BdG equations}\label{Sec_BdG}
The LR theory atop the time-dependent GP equation, Eq. (\ref{Eq_GP_time}), is derived below. Therefore, a weak time-dependent perturbation is added to the Hamiltonian, i.e., $\hat{H}(\mathbf{r})\rightarrow \hat{H}(\mathbf{r})+\hat{H}_\text{ext}(\mathbf{r},t)$. The perturbation $\hat{H}_\text{ext}(\mathbf{r},t)$ explicitly reads
\begin{equation}\label{pert_h}
	\hat{H}_\text{ext}(\mathbf{r},t)=f^+(\mathbf{r})e^{-i\omega t}+f^-(\mathbf{r})e^{i\omega t}
\end{equation}
with the real amplitudes $f^+(\mathbf{r})$ and $f^-(\mathbf{r})$ and the frequency $\omega$. The response of the system is calculated around the stationary GP ground-state orbital $\phi^0(\mathbf{r})$ obtained from Eq. (\ref{Eq_GP_notime}). For the perturbed orbital $\phi(\mathbf{r},t)$, the ansatz
\begin{equation}\label{Eq_BdG_Ansatz}
	\sqrt{N}\phi(\mathbf{r},t)=e^{-i\mu t}\left( \sqrt{N}\phi^0(\mathbf{r})+u(\mathbf{r})e^{-i\omega t}+v^\ast(\mathbf{r})e^{i\omega t} \right)	
\end{equation} 
is utilized. In Eq. (\ref{Eq_BdG_Ansatz}), the quantities $u(\mathbf{r})$ and $v(\mathbf{r})$ denote small and time-independent response amplitudes. Plugging in Eqs. (\ref{pert_h}) and (\ref{Eq_BdG_Ansatz}) into the time-dependent GP equation, Eq. (\ref{Eq_GP_time}), yields in zeroth order the time-independent GP equation, Eq. (\ref{Eq_GP_notime}). In first order, collecting all terms proportional to $e^{-i\omega t}$ leads to 
\begin{align}\label{Eq_BdG_u}
	(\omega+\mu)u(\mathbf{r})&= \sqrt{N}f^+(\mathbf{r})\phi^0(\mathbf{r})+\hat{H}_{\text{\tiny{GP}}}u(\mathbf{r})+\Lambda\int \hat{W}(\mathbf{r},\mathbf{r}^{\,\prime})\phi^0(\mathbf{r}^{\,\prime})v(\mathbf{r}^{\,\prime})\phi^0(\mathbf{r})d\mathbf{r}^{\,\prime} \nonumber \\ 
	&+\Lambda\int \hat{W}(\mathbf{r},\mathbf{r}^{\,\prime})\phi^{0,\ast}(\mathbf{r}^{\,\prime})u(\mathbf{r}^{\,\prime})\phi^0(\mathbf{r})d\mathbf{r}^{\,\prime}
\end{align}
where the GP Hamiltonian $\hat{H}{\text{\tiny{GP}}}$ is defined as
\begin{equation}\label{Eq_GP_Hamiltonian}
	\hat{H}{\text{\tiny{GP}}}=\hat{h}+\Lambda\int \hat{W}(\mathbf{r},\mathbf{r}^{\,\prime})\,|\phi^0(\mathbf{r}^{\,\prime},t)|^2\,d\mathbf{r}^{\,\prime}.
\end{equation}
Collecting all terms proportional to $e^{i\omega t}$ results in
\begin{align}\label{Eq_BdG_v}
	(\mu-\omega)v^\ast(\mathbf{r})&= \sqrt{N}f^-(\mathbf{r})\phi^0(\mathbf{r})+\hat{H}_{\text{\tiny{GP}}}v^\ast(\mathbf{r})+\Lambda\int \hat{W}(\mathbf{r},\mathbf{r}^{\,\prime})\phi^0(\mathbf{r}^{\,\prime})u^\ast(\mathbf{r}^{\,\prime})\phi^0(\mathbf{r})\,d\mathbf{r}^{\,\prime} \nonumber \\ 
	&+\Lambda\int \hat{W}(\mathbf{r},\mathbf{r}^{\,\prime})\phi^{0,\ast}(\mathbf{r}^{\,\prime})v^\ast(\mathbf{r}^{\,\prime})\phi^0(\mathbf{r})\,d\mathbf{r}^{\,\prime}.
\end{align}
Complex conjugation and multiplication by $(-1)$ of Eq. (\ref{Eq_BdG_v}) opens the possibility to cast Eqs. (\ref{Eq_BdG_u}) and (\ref{Eq_BdG_v}) in matrix form. The result reads
\begin{equation}\label{Inhomo_BdG_Eq}
	\left(\mathcal{L}_\text{BdG}-\omega\right) \begin{pmatrix} u(\mathbf{r})\\ v(\mathbf{r}) \end{pmatrix} = \begin{pmatrix} -\sqrt{N}f^+(\mathbf{r})\phi^0(\mathbf{r}) \\ \sqrt{N}f^{-,\ast}(\mathbf{r}) \phi^{0,\ast}(\mathbf{r}) \end{pmatrix}
\end{equation}
with the BdG matrix
\begin{equation}\label{BdG_matrix_general1}
	\mathcal{L}_\text{BdG}=\begin{pmatrix} A & B \\-B^\ast & -A^\ast     \end{pmatrix}
\end{equation}
where the quantities $A$ and $B$ are given by
\begin{equation}\label{BdG_A}
	A=\hat{H}_{\text{\tiny{GP}}}-\mu+\Lambda\int d\mathbf{r}^{\,\prime}\hat{W}(\mathbf{r},\mathbf{r}^{\,\prime})\phi^{0,\ast}(\mathbf{r}^{\,\prime})\phi^0(\mathbf{r})\hat{\mathcal{P}}_{\mathbf{r}\mathbf{r}^{\,\prime}}
\end{equation}
and
\begin{equation}\label{BdG_B}
	B=\Lambda\int d\mathbf{r}^{\,\prime}\hat{W}(\mathbf{r},\mathbf{r}^{\,\prime})\phi^0(\mathbf{r}^{\,\prime})\phi^0(\mathbf{r})\hat{\mathcal{P}}_{\mathbf{r}\mathbf{r}^{\,\prime}}\,.
\end{equation}
where the operator $\hat{\mathcal{P}}_{\mathbf{r}\mathbf{r}^{\,\prime}}$ was introduced in Eq. (\ref{K_sl}). In the literature, Eq. (\ref{Inhomo_BdG_Eq}) with contact interaction $\hat{W}(\mathbf{r},\mathbf{r}^{\,\prime})=\delta(\mathbf{r}-\mathbf{r}^{\,\prime})$ and zero RHS, i.e.,
\begin{equation}\label{BdG_Eq}
	\left(\mathcal{L}_\text{BdG}-\omega\right) \begin{pmatrix} u(\mathbf{r})\\ v(\mathbf{r}) \end{pmatrix} = 0,
\end{equation}
is referred to as the BdG equations \cite{Bogoliubov,deGennes}. In this work, however, the interaction potential is in general different from the contact interaction potential. Moreover, the particle-conserving BdG equations \cite{Ruprecht,Castin,Gardiner,Castin2}, given by
\begin{equation}\label{BdG_part_conv}
	\mathbf{\mathcal{P}}\mathcal{L}_\text{BdG}\mathbf{\mathcal{P}} \begin{pmatrix} u(\mathbf{r})\\ v(\mathbf{r}) \end{pmatrix} = \omega \begin{pmatrix} u(\mathbf{r})\\ v(\mathbf{r}) \end{pmatrix}
\end{equation}
where the projector $\mathbf{\mathcal{P}}=\mathbb{1}-|\phi^0\rangle\langle \phi^0|$ removes any contribution of the ground-state orbital $\phi^0(\mathbf{r})$ from the response amplitudes $u(\mathbf{r})$ and $v(\mathbf{r})$, are used to calculate the low-energy excitation spectra at the mean-field level.  

From the above derivation one can readily see the close connection of the GP orbital and the BdG equations. As discussed in the previous section on the GP theory, the wave function of the system in the Hartree limit can be very different from the GP wave function, although the latter gives the exact energy and density per particle. Thus, it can be questioned whether the LR theory atop the GP wave function, resulting in the BdG equations, is in general reliable in the Hartree limit. Scientific works that prove the exactness of the BdG spectrum in this limit under certain conditions were discussed in Section \ref{App_RotBEC}. Nevertheless, the example of a rotating BEC in an anharmonic trap, which does not fulfill the assumptions made in Ref. \cite{Seiringer2}, serves as an example where at least in the experimentally relevant regime of $N\approx 10^3-10^6$ bosons in the condensate the lowest-in-energy excitations substantially deviate from the BdG predictions.

\subsection{Multi-orbital approaches}
\subsubsection{Time-dependent multi-orbital mean field (TDMF)}\label{Sec_TDMF}
The TDMF theory, introduced in Ref. \cite{TDMF_main}, is briefly described. It can be seen as an intermediate step between the GP theory and MCTDHB because it allows for more than one single-particle orbital that the bosons can occupy, and is thus applicable to fragmented condensates. However, as a major difference to the full many-body description in Section \ref{CH_MB_theory}, the occupation of these single-particle orbitals is fixed. The ansatz for the $N$-boson wave function is given by
\begin{equation}\label{Eq_WF_multi_orb}
	\Psi(\mathbf{r}_1,...,\mathbf{r}_N;t)=\hat{\mathcal{S}}\phi_1(\mathbf{r}_1,t)\phi_2(\mathbf{r}_2,t)\, ... \,\phi_N(\mathbf{r}_N,t)
\end{equation}
where the operator $\hat{\mathcal{S}}$ accounts for the symmetrization of the wave function. In general, not all single-particle orbitals $\phi_i$ need to be different. The only constraint is that the number of particles is fixed such that $N=\sum_{i=1}^M n_i$ where $M$ is the number of different orbitals in Eq. (\ref{Eq_WF_multi_orb}) and the $n_i$ are their fixed occupation numbers. Applying the least action principle to 
\begin{equation}\label{Eq_TDMF_action}
	S=\int dt \left( \langle\Psi|\hat{H}-i\frac{\partial}{\partial t} |\Psi\rangle-\sum_{i,j=1}^M \mu_{ij}(t)[\langle\phi_i|\phi_j\rangle-\delta_{ij}]\right)
\end{equation}
with the time-dependent Lagrange multipliers $\mu_{ij}(t)$ yields
\begin{align}
	0=\frac{\delta S}{\delta\phi_k^\ast(\mathbf{r},t)}&=n_k\left( \hat{h}-i\frac{\partial}{\partial t}+\lambda_0(n_k-1)\hat{J}_k+\sum_{l\neq k}^M \lambda_0 n_l (\hat{J}_l+\hat{K}_l)  \right) |\phi_k\rangle \nonumber \\
	&=\sum_{j=1}^M \mu_{kj}(t)|\phi_j(t)\rangle			\quad \forall \, k=1,...,M \label{Eq_TDMF_action_variation}
\end{align}
where 
\begin{align}\label{Eq_J_def}
	\hat{J}_l(\mathbf{r},t)&=\int d\mathbf{r}^{\,\prime} \phi_l^\ast(\mathbf{r}^{\,\prime},t)\hat{W}(\mathbf{r},\mathbf{r}^{\,\prime})\phi_l(\mathbf{r}^{\,\prime},t) 
\end{align}
and
\begin{align}\label{Eq_K_def}
		\hat{K}_l(\mathbf{r},t)&=\int d\mathbf{r}^{\,\prime} \phi_l^\ast(\mathbf{r}^{\,\prime},t)\hat{W}(\mathbf{r},\mathbf{r}^{\,\prime})\hat{\mathcal{P}}_{\mathbf{r}\mathbf{r}^{\,\prime}}\phi_l(\mathbf{r}^{\,\prime},t) 
\end{align}
are the direct and exchange interaction potentials, respectively. Multiplying Eq. (\ref{Eq_TDMF_action_variation}) by $\langle\phi_l|$ gives 
\begin{equation}\label{Eq_chem_pot_TDMF}
	\mu_{kl}(t)=n_k\left(  h_{lk}-\left(\frac{\partial}{\partial t}\right)_{lk}+\lambda_0(n_k-1)\hat{J}_k W_{lkkk}+\sum_{j\neq k}^M \lambda_0 n_j (W_{ljjk}+W_{ljkj}) \right)
\end{equation}
for the Lagrange multipliers. Plugging this into Eq. (\ref{Eq_TDMF_action_variation}) results in
\begin{align}\label{Eq_TDMF_EOM_prefinal}
	i\mathbf{\hat{P}}\frac{\partial}{\partial t}|\phi_k\rangle=\mathbf{\hat{P}}\left(  \hat{h}+\lambda_0(n_k-1)\hat{J}_k+\sum_{l\neq k}^M \lambda_0 n_l (\hat{J}_l+\hat{K}_l)  \right)|\phi_k\rangle \quad \forall \, k=1,...,M
\end{align}
where the projector $\mathbf{\hat{P}}=\mathbb{1}-\sum_{j=1}^M|\phi_j\rangle\langle\phi_j|$ projects onto the tangential space of $\text{span}(\phi_1,...,\phi_M)$. To simplify the above EOMs for the orbitals, the condition 
\begin{equation}\label{Eq_TDMF_cond}
	\langle \phi_i|\dot{\phi}_j\rangle=0 \quad \forall \, i,j=1,...,M
\end{equation}
is invoked. This leads to the conservation of energy, and the projector on the LHS of Eq. (\ref{Eq_TDMF_EOM_prefinal}) can be omitted, i.e.,
\begin{equation}\label{Eq_TDMF_EOM_final}
	i\frac{\partial}{\partial t}|\phi_k\rangle=\mathbf{\hat{P}}\left(  \hat{h}+\lambda_0(n_k-1)\hat{J}_k+\sum_{l\neq k}^M \lambda_0 n_l (\hat{J}_l+\hat{K}_l)  \right)|\phi_k\rangle \quad \forall \,k=1,...,M.
\end{equation}
It is important to note that the condition in Eq. (\ref{Eq_TDMF_cond}) is a manually-invoked restriction to the orbitals within the TDMF theory, whereas the same condition in MCTDHB appears naturally and rigorously. Furthermore, one observes that for $M=1$, i.e., when only a single orbital is available, Eq. (\ref{Eq_TDMF_EOM_final}) reduces to the (projected) time-dependent GP equation, Eq. (\ref{Eq_GP_time}). The TDMF theory has been extended to describe spinor condensates and Bose-Bose or Bose-Fermi mixtures as well, see in this context Ref. \cite{TDMF_spinor_mixtures}. By imaginary-time propagation of the TDMF EOMs, one obtains the corresponding theory for the stationary equations, called the best mean field for condensates \cite{BMF1}. Applications can be found in, e.g., Refs. \cite{BMF2,BMF3,BMF4,BMF5}.

\subsubsection{Linear-response best-mean-field (LR-BMF)}\label{LR_BMF}
A possible LR approach based on a multi-orbital mean-field description is the LR-BMF theory which has been introduced in Ref. \cite{LR-BMF1}. To derive it, the EOMs of the previously described TDMF theory, Eq. (\ref{Eq_TDMF_action_variation}), are linearized for the case of adding a weak time-dependent external field to the Hamiltonian, i.e., $\hat{H}\rightarrow \hat{H}+\hat{H}_\text{ext}(t)$. The perturbation $\hat{H}_\text{ext}(t)$ is assumed to be of the same form as in Eq. (\ref{pert_h}). Plugging this into Eq. (\ref{Eq_TDMF_action_variation}) yields
\begin{align}\label{Eq_LRBMF_start}
	\left(\hat{Z}_i-i\frac{\partial}{\partial t}\right) |\phi_i\rangle -\sum_{j=1}^M \frac{\mu_{ij}(t)}{n_i}|\phi_j\rangle=-\hat{H}_\text{ext}(t)|\phi_i\rangle
\end{align}
with the replacement
\begin{equation}
	\hat{Z}_i=\hat{h}+\lambda_0(n_i-1)\hat{J}_i+\sum_{l\neq i}^M \lambda_0 n_l (\hat{J}_l+\hat{K}_l).
\end{equation}
The Lagrange multipliers, after redefining them as $\mu_{ij}(t)/n_i\rightarrow \mu_{ij}(t)$, can be expressed by
\begin{equation}
	\mu_{ij}(t)=\langle\phi_j|\hat{Z}_i+\hat{H}_\text{ext}(t)|\phi_i\rangle
\end{equation}
where the condition of Eq. (\ref{Eq_TDMF_cond}) was used. The ansatz for the perturbed orbitals is given by
\begin{equation}
	\phi_i(\mathbf{r},t)\approx \phi_i^0(\mathbf{r})+\delta\phi_i(\mathbf{r},t) \label{Eq_LRBMF_ansatz1} 
\end{equation}
Plugging Eq. (\ref{Eq_LRBMF_ansatz1})  into Eq. (\ref{Eq_LRBMF_start}) leads to
\begin{align}\label{Eq_LRBMF_inter1}
	-\hat{H}_\text{ext}(t)|\phi_i^0\rangle &= \left( \hat{Z}_i^0-i\frac{\partial}{\partial t}\right)|\delta\phi_i\rangle-\sum_{j=1}^M \mu_{ij}^0 (t) |\delta\phi_j\rangle \nonumber \\
	&+\lambda_0(n_i-1)\delta\hat{J}_i|\phi_i^0\rangle+\sum_{j\neq i}^M \lambda_0 n_j (\delta\hat{J}_j+\delta\hat{K}_j)|\phi_i^0\rangle \\
	&-\sum_{j=1}^M \delta\mu_{ij}(t)|\phi_j^0\rangle \nonumber
\end{align}
with
\begin{align}\label{Eq_LRBMF_del_J}
	\delta\hat{J}_i(\mathbf{r},t)&=\int d\mathbf{r}^{\,\prime} \delta\phi_i^\ast(\mathbf{r}^{\,\prime},t)\hat{W}(\mathbf{r},\mathbf{r}^{\,\prime})\phi_i^0(\mathbf{r}^{\,\prime})+\int d\mathbf{r}^{\,\prime} \phi_i^{0,\ast}(\mathbf{r}^{\,\prime})\hat{W}(\mathbf{r},\mathbf{r}^{\,\prime})\delta\phi_i^0(\mathbf{r}^{\,\prime},t)
\end{align}
and
\begin{align}\label{Eq_LRBMF_del_K}	
	\delta\hat{K}_i(\mathbf{r},t)&=\int d\mathbf{r}^{\,\prime} \delta\phi_i^\ast(\mathbf{r}^{\,\prime},t)\hat{W}(\mathbf{r},\mathbf{r}^{\,\prime})\hat{\mathcal{P}}_{\mathbf{r}\mathbf{r}^{\,\prime}}\phi_i^0(\mathbf{r}^{\,\prime})+\int d\mathbf{r}^{\,\prime} \phi_i^{0,\ast}(\mathbf{r}^{\,\prime})\hat{W}(\mathbf{r},\mathbf{r}^{\,\prime})\hat{\mathcal{P}}_{\mathbf{r}\mathbf{r}^{\,\prime}}\delta\phi_i^0(\mathbf{r}^{\,\prime},t).
\end{align}
The variation of the Lagrange multipliers can be expressed as
\begin{align}\label{Eq_LRBMF_Lagrange_variation}
	\delta\mu_{ij}(t)=-\sum_{l=1}^M\mu_{il}^0\langle\phi_j^0|\delta\phi_l\rangle+\langle\phi_j^0|\delta(\hat{Z}_i|\phi_i\rangle)+\langle\phi_j^0|\hat{H}_\text{ext}(t)|\phi_i^0\rangle
\end{align}
where the identity $\hat{Z}_i^0|\phi_i^0\rangle=\sum_{l=1}^M \mu_{il}^0|\phi_l^0\rangle$ and integration by parts was used in the first term. One finally obtains 
\begin{align}\label{Eq_LRBMF_inter2}
	&\mathbf{\hat{P}}\left[ \hat{Z}_i^0|\delta\phi_i\rangle -\sum_{j=1}^M\mu_{ij}^0|\delta\phi_j\rangle+\left(\lambda_0(n_i-1)\delta\hat{J}_i +\sum_{j\neq i}^M \lambda_0 n_j (\delta\hat{J}_j+\delta\hat{K}_j)\right)|\phi_i^0\rangle\right] \nonumber \\
	&=-\mathbf{\hat{P}}\hat{H}_\text{ext}(t)|\phi_i^0\rangle+i\frac{\partial}{\partial t}|\delta\phi_i\rangle
\end{align}
with the projector $\mathbf{\hat{P}}=\mathbb{1}-\sum_k|\phi_k^0\rangle\langle\phi_k^0|$. By first making the ansatz 
\begin{equation}\label{Eq_LRBMF_ansatz2}
	\delta\phi_i(\mathbf{r},t)=\frac{1}{\sqrt{n_i}}\left(u_i(\mathbf{r})e^{-i\omega t}+v_i^\ast(\mathbf{r}) e^{i\omega t}\right)
\end{equation}	
with stationary response amplitudes $u_i(\mathbf{r})$ and $v_i(\mathbf{r})$ and plugging this into Eq. (\ref{Eq_LRBMF_inter2}) afterwards, it is again possible, similar to derivation of the BdG equations in Section \ref{Sec_BdG}, to group the terms of the resulting equations with respect to their linearity to either $e^{-i\omega t}$ or $e^{i\omega t}$. As a result, one arrives at an eigenvalue equation of the form
\begin{equation}\label{Eq_LRBMF_EV1}
	(\mathbf{\mathcal{P}}\mathbf{\mathcal{L}}-\omega)\begin{pmatrix}
	|\mathbf{u}\rangle \\ |\mathbf{v}\rangle
	\end{pmatrix}
	=\mathbf{\mathcal{P}}\begin{pmatrix}
	-f^+(\mathbf{r})|\boldsymbol{\phi_n^0}\rangle \\ f^{-,\ast}(\mathbf{r})|\boldsymbol{\phi_n^{0,\ast}}\rangle
	\end{pmatrix}
\end{equation}
with the vector $|\boldsymbol{\phi_n^0}\rangle=|\sqrt{n_1}\phi_1^0,...,\sqrt{n_M}\phi_M^0\rangle$, the $2(M\times M)$ square matrix
\begin{equation}\label{Eq_LRBMF_L}
	\mathbf{\mathcal{L}}=\begin{pmatrix}
	\mathbf{Z^0}-\boldsymbol{\mu^0}+\mathbf{A} & \mathbf{B} \\
	-\mathbf{B}^\ast & -(\mathbf{Z^0}-\boldsymbol{\mu^0}+\mathbf{A})^\ast
	\end{pmatrix},
\end{equation}
and the projection matrix
\begin{equation}\label{Eq_LRBMF_P}
	\mathbf{\mathcal{P}}=\{\mathcal{P}_{ij}\}=\begin{cases} \mathbf{\hat{P}} & i=j\leq M \\ \mathbf{\hat{P}}^\ast & i=j> M \\ 0 & i\neq j
	\end{cases}
\end{equation}
of the same size. The matrix $\mathbf{Z^0}=\text{diag}(\hat{Z}_1^0,...,\hat{Z}_M^0)$ is a diagonal matrix, whereas $\boldsymbol{\mu^0}=\{\mu_{ij}^0\}$ contains the Lagrange multipliers. The matrices $\mathbf{A}$ and $\mathbf{B}$, for the case of the contact interaction potential, read
\begin{equation}\label{Eq_LRBMF_A}
	\mathbf{A}=\{A_{ij}\}=\begin{cases}
	\lambda_0(n_i-1) |\phi_i^0|^2 & i=j \\
	2\lambda_0\sqrt{n_in_j}\phi_i^0\phi_j^{0,\ast} & i\neq j
	\end{cases}
\end{equation}
and 
\begin{equation}\label{Eq_LRBMF_B}
	\mathbf{B}=\{B_{ij}\}=\begin{cases}
	\lambda_0(n_i-1) (\phi_i^0)^2 & i=j \\
	2\lambda_0\sqrt{n_in_j}\phi_i^0\phi_j^0 & i\neq j.
	\end{cases}
\end{equation}
Since the term proportional to $\omega$ is the only one without a projector in Eq. (\ref{Eq_LRBMF_EV1}), a redundant projector can be added to the RHS of $\mathbf{\mathcal{P}}\mathbf{\mathcal{L}}$, i.e., $\mathbf{\mathcal{P}}\mathbf{\mathcal{L}}\rightarrow \mathbf{\mathcal{P}}\mathbf{\mathcal{L}}\mathbf{\mathcal{P}}$. The resulting homogeneous eigenvalue equation
\begin{equation}\label{Eq_LRBMF_EV_final}
	\mathbf{\mathcal{P}}\mathbf{\mathcal{L}}\mathbf{\mathcal{P}}\begin{pmatrix}
	|\mathbf{u}\rangle \\ |\mathbf{v}\rangle
	\end{pmatrix}=
	\omega \begin{pmatrix}
	|\mathbf{u}\rangle \\ |\mathbf{v}\rangle
	\end{pmatrix}
\end{equation}
is the central equation in the LR-BMF theory. A comparison to the BdG matrix $\mathcal{L}_{\text{BdG}}$ in Eqs. (\ref{BdG_matrix_general1})-(\ref{BdG_B}) shows that for $M=1$, LR-BMF reduces to the BdG theory, i.e., LR-BMF($M=1$)$\equiv$BdG. If one however compares the LR matrix of Eq. (\ref{Eq_LRBMF_L}) to the one of LR-MCTDHB in Section \ref{Sec_LRMCTDHB}, it can be seen that the latter is clearly more complicated because it also includes the couplings between orbitals and coefficients. Again, in both mean-field approaches presented in this Appendix, there is only one possible configuration, i.e., the configuration $|N;t\rangle$ for the BdG case and the configuration $|n_1,...,n_M;t\rangle$ for the multi-orbital mean-field case. The occupations of the underlying single-particle orbitals are fixed. In fact, the LR-BMF matrix $\mathbf{\mathcal{P}}\mathbf{\mathcal{L}}\mathbf{\mathcal{P}}$ can be seen as the upper-left block $\mathcal{L}_{oo}$ of the LR matrix in Eq. (\ref{LR_matrix}). Further details on the LR-BMF theory, together with an application to a BEC in a double well, can be found in Ref. \cite{LR-BMF1}.

\setcounter{equation}{0}
\section{LR-MCTDHB in block-diagonal form}\label{Appendix_BlockDiagonal}
In this Appendix, a complex transformation that essentially halves the dimensionality of the homogeneous eigenvalue problem of Eq. (\ref{LR_eigenvalue_final}) is introduced. However, as  shown below, it involves matrix-matrix multiplications of individual blocks of the LR matrix $\mathcal{L}$. The latter products might be numerically expensive. It is therefore not in general beneficial to perform the transformation described below, which was first discussed in Ref. \cite{Ofir_L2_trick}.

One considers the transformation matrix
\begin{equation}\label{Eq_trafo_L2}
	Q=\frac{1}{\sqrt{2}}\begin{pmatrix}
		1 & 1\theta \\ 1 & -1 \theta
	\end{pmatrix}, \quad
	Q^{-1}=\frac{1}{\sqrt{2}}\begin{pmatrix}
		1 & 1 \\ 1\theta & -1 \theta
	\end{pmatrix}, \quad QQ^{-1}=Q^{-1}Q=\mathbb{1}
\end{equation}
where the operation $\theta$ is defined via its action on any operator $\hat{O}$ given by $\theta\hat{O}=\hat{O}^\ast$. A matrix of the form 
\begin{equation}\label{Eq_App_L2_trick_matrix}
	\mathcal{A}=\begin{pmatrix}
		A & B \\ -B^\ast & -A^\ast
	\end{pmatrix}
\end{equation}
behaves under the transformation in Eq. (\ref{Eq_trafo_L2}) as
\begin{align}\label{Eq_L2_trick_step1}
	Q\mathcal{A}Q^{-1}&=\frac{1}{2}\begin{pmatrix}
		1 & 1 \\ 1\theta & -1 \theta
	\end{pmatrix}
	\begin{pmatrix}
		A & B \\ -B^\ast & -A^\ast
	\end{pmatrix}
	\begin{pmatrix}
		1 & 1\theta \\ 1 & -1 \theta
	\end{pmatrix} \nonumber \\
	&=\begin{pmatrix}
		0	&	A-B\theta \\
		A+B\theta	&	0
	\end{pmatrix}.
\end{align}
Multiplying the result of Eq. (\ref{Eq_L2_trick_step1}) with itself yields the block-diagonal matrix
\begin{equation}
	\left(Q\mathcal{A}Q^{-1}\right)\left(Q\mathcal{A}Q^{-1}\right)=\begin{pmatrix}
		(A-B\theta)(A+B\theta)	&	0 \\
		0	&	(A+B\theta)(A-B\theta)
	\end{pmatrix}.
\end{equation}
This transforms the eigenvalue problem $\mathcal{A}\alpha=\lambda \alpha$ into
\begin{align}
	 \lambda^2 (Q\alpha)&=\left(Q\mathcal{A}Q^{-1}\right)\left(Q\mathcal{A}Q^{-1}\right)(Q\alpha)  \nonumber \\
	 \Leftrightarrow\,
	 \lambda^2 \frac{1}{\sqrt{2}}
	\begin{pmatrix}
		u+v^\ast \\ u-v^\ast
	\end{pmatrix}
	&=
	 \begin{pmatrix}
		(A-B\theta)(A+B\theta)	&	0 \\
		0	&	(A+B\theta)(A-B\theta)
	\end{pmatrix}	\frac{1}{\sqrt{2}}
	\begin{pmatrix}
		u+v^\ast \\ u-v^\ast
	\end{pmatrix}	
\end{align}
with $\alpha=\frac{1}{\sqrt{2}}\begin{pmatrix}u \\v\end{pmatrix}$. Thus, the original eigenvalue problem of $\mathcal{A}$ can be split up into two smaller eigenvalue problems of $\mathcal{L}^{(1)}\equiv(A-B\theta)(A+B\theta)$ and $\mathcal{L}^{(2)}\equiv(A+B\theta)(A-B\theta)$ with $\text{dim}(\mathcal{L}^{(1)})=\text{dim}(\mathcal{L}^{(2)})=\text{dim}(\mathcal{A})/2$. Moreover, both $\mathcal{L}^{(1)}$ and $\mathcal{L}^{(2)}$ have the same eigenvalues, which are the squared eigenvalues of $\mathcal{A}$.

Interchanging the second and third row as well as the second and third column of the LR matrix in Eq. (\ref{Eq_final_LR_matrix}) which explicitly reads
\small
\begin{align}
	\mathcal{L}=\begin{pmatrix}
	\boldsymbol{\rho}^{-\frac{1}{2}}\hat{\mathbf{P}} \mathcal{L}_{oo}^u \hat{\mathbf{P}}\boldsymbol{\rho}^{-\frac{1}{2}} & \boldsymbol{\rho}^{-\frac{1}{2}}\hat{\mathbf{P}}\mathcal{L}_{oo}^v \hat{\mathbf{P}}^\ast(\boldsymbol{\rho^\ast})^{-\frac{1}{2}} & \boldsymbol{\rho}^{-\frac{1}{2}}\hat{\mathbf{P}}\mathcal{L}_{oc}^u & \boldsymbol{\rho}^{-\frac{1}{2}}\hat{\mathbf{P}}\mathcal{L}_{oc}^v \\
	-(\boldsymbol{\rho^\ast})^{-\frac{1}{2}}\hat{\mathbf{P}}^\ast \mathcal{L}_{oo}^{v,\ast} \hat{\mathbf{P}}\boldsymbol{\rho}^{-\frac{1}{2}} & -(\boldsymbol{\rho^\ast})^{-\frac{1}{2}}\hat{\mathbf{P}}^\ast \mathcal{L}_{oo}^{u,\ast}\hat{\mathbf{P}}^\ast(\boldsymbol{\rho^\ast})^{-\frac{1}{2}} & -(\boldsymbol{\rho^\ast})^{-\frac{1}{2}}\hat{\mathbf{P}}^\ast \mathcal{L}_{oc}^{v,\ast} & -(\boldsymbol{\rho^\ast})^{-\frac{1}{2}}\hat{\mathbf{P}}^\ast \mathcal{L}_{oc}^{u,\ast} \\
	\mathcal{L}_{co}^u \hat{\mathbf{P}}\boldsymbol{\rho}^{-\frac{1}{2}} & \mathcal{L}_{co}^v\hat{\mathbf{P}}^\ast(\boldsymbol{\rho^\ast})^{-\frac{1}{2}} & \mathbf{\mathcal{H}}-\varepsilon^0 & 0 \\
	-\mathcal{L}_{co}^{v,\ast} \hat{\mathbf{P}}\boldsymbol{\rho}^{-\frac{1}{2}} & -\mathcal{L}_{co}^{u,\ast}\hat{\mathbf{P}}^\ast(\boldsymbol{\rho^\ast})^{-\frac{1}{2}} & 0 & -(\mathbf{\mathcal{H}}-\varepsilon^0)^\ast
	\end{pmatrix}
\end{align}\normalsize
yields a matrix of the form given in Eq. \ref{Eq_App_L2_trick_matrix}. Its submatrices are 
\begin{equation}\label{Eq_A_LR}
	A=\begin{pmatrix}
		\boldsymbol{\rho}^{-\frac{1}{2}}\hat{\mathbf{P}} \mathcal{L}_{oo}^u \hat{\mathbf{P}}\boldsymbol{\rho}^{-\frac{1}{2}} & \boldsymbol{\rho}^{-\frac{1}{2}}\hat{\mathbf{P}}\mathcal{L}_{oc}^u \\
		\mathcal{L}_{co}^u \hat{\mathbf{P}}\boldsymbol{\rho}^{-\frac{1}{2}} & \mathbf{\mathcal{H}}-\varepsilon^0
	\end{pmatrix}
\end{equation}
and 
\begin{equation}\label{Eq_B_LR}
	B=\begin{pmatrix}
		\boldsymbol{\rho}^{-\frac{1}{2}}\hat{\mathbf{P}}\mathcal{L}_{oo}^v \hat{\mathbf{P}}^\ast(\boldsymbol{\rho^\ast})^{-\frac{1}{2}} & \boldsymbol{\rho}^{-\frac{1}{2}}\hat{\mathbf{P}}\mathcal{L}_{oc}^v \\
		\mathcal{L}_{co}^v\hat{\mathbf{P}}^\ast(\boldsymbol{\rho^\ast})^{-\frac{1}{2}} & 0
	\end{pmatrix}.
\end{equation}
Mind the suppressed superscript '0' for the density matrices as compared to, e.g., Eq. (\ref{metric_oo}). It is hence possible to diagonalize either $\mathcal{L}^{(1)}$ or $\mathcal{L}^{(2)}$ with $A$ and $B$ defined in Eqs. (\ref{Eq_A_LR}) and (\ref{Eq_B_LR}) to obtain the squared eigenvalues of the original LR matrix $\mathcal{L}$. The advantage is, as mentioned above, that the dimensionality of $\mathcal{L}^{(1)}$ and $\mathcal{L}^{(2)}$ is only half of the dimensionality of $\mathcal{L}$. Moreover, for two distinct eigenvalues $\omega_1$ and $\omega_2$ of $\mathcal{L}$ with $\omega_2\geq \omega_1> 0.5$ and separation $\Delta\equiv \omega_2-\omega_1\geq 0$, the separation $\Delta^{(2)}$ in the spectrum of the squared eigenvalues is given by
\begin{align}
	\Delta^{(2)}&\equiv\omega_2^2-\omega_1^2=(\omega_1+\Delta)^2-\omega_1^2 \nonumber \\
	&=2\omega_1\Delta+\Delta^2=\Delta(2\omega_1+\Delta) \nonumber \\
	&\geq 2\omega_1\Delta \nonumber \\
	&> \Delta.
\end{align} 
Thus, the separation grows in the spectra of $\mathcal{L}^{(1)}$ and $\mathcal{L}^{(2)}$. Since in general, the IRAM converges in less iterations when the eigenvalues are more separated from each other, eigenvalues of larger magnitude than $0.5$ might be obtained with less iterations by diagonalizing either $\mathcal{L}^{(1)}$ or $\mathcal{L}^{(2)}$ instead of $\mathcal{L}$. On the contrary, for eigenvalues with $0\leq \omega_1\leq\omega_2 \leq 0.5$, one finds that
\begin{align}
	\Delta^{(2)}&=\Delta(2\omega_1+\Delta) \nonumber \\
	&=\Delta\underbrace{(\omega_2+\omega_1)}_{\leq 1} \nonumber \\
	&\leq \Delta,
\end{align} 
which means that the separation of these eigenvalues is decreased in the spectra of $\mathcal{L}^{(1)}$ and $\mathcal{L}^{(2)}$. Therefore, although each iteration of the IRAM is less expensive for the smaller matrices, more iterations might be necessary to obtain converged eigenvalues with smaller magnitude than 0.5. A possible solution to this problem is to shift $\mathcal{L}\rightarrow \mathcal{L}+0.5\cdot\mathbb{1}$, where $\mathbb{1}$ is the $2(M+N_\text{conf})$-dimensional unit matrix, before applying the transformation in Eq. (\ref{Eq_trafo_L2}).

It therefore remains a system-dependent question whether it is beneficial to use the full LR matrix $\mathcal{L}$ or one of the smaller matrices $\mathcal{L}^{(1)}$ and $\mathcal{L}^{(2)}$ to obtain the low-energy spectrum. One needs to take into account the computational cost of building the latter two matrices. Moreover, it is not in general true that, assuming the blocks $A$ and $B$ are sparse matrices, their product is also sparse. If one is in the unfortunate situation that the matrices $\mathcal{L}^{(1)}$ and $\mathcal{L}^{(2)}$ are much denser than the original matrix $\mathcal{L}$, it is potentially also much more demanding to converge to their lowest eigenvalues, although they are much smaller. It is therefore \textit{a priori} not possible to decide which approach is more efficient, and in principle, one would always need to first calculate the (shifted) matrices $\mathcal{L}^{(1)}$ and $\mathcal{L}^{(2)}$ and analyze how dense they are before making an appropriate decision. All applications of LR-MCTDHB discussed in this work, however, avoided this preliminary analysis, and used the full LR matrix $\mathcal{L}$ to obtain the low-energy spectra. For additional details, it is referred to Ref. \cite{Ofir_L2_trick}.

\setcounter{equation}{0}
\addtocontents{lot}{\protect\vspace*{4ex}}
\newpage
\section{Further benchmarks of LR-MCTDHB}\label{Appendix_further_benchmarks}
Further results on the benchmark of the newly developed implementation of LR-MCTDHB against the HIM are presented in this Appendix. In particular, comparisons of the exact and numerical results for attractive bosons in 1D as well as for repulsive bosons in the anisotropic HIM in 2D are made. In addition, the numerical convergence in the rotating frame of reference, which is important for the validity of the results presented in Section \ref{App_RotBEC}, is demonstrated.

\begin{table}[b!]
\centering
\begin{tabular*}{\textwidth}{p{2cm} p{2.4cm} p{2.4cm} p{2.4cm} p{2cm} p{2cm}}

 \hline\hline \rule{0pt}{3ex} 
 	&	$M=1$	&	$M=4	$	&	$M=6$	& $(m_x,n_x)$	&	Exact 	\\ 
 \hline  \rule{-4pt}{3ex} 
$\varepsilon^0$	&	9.\underline{137833}		&	9.03815\underline{1}	&	9.038150	&	$(0,0)$		&	9.038150		 \\ 	
$\omega_1$	&	1.000000		&	1.000000	&	1.000000		&	$(1,0)$	&	1.000000		 \\ 
$\omega_2$	&	n/a	&	2.000\underline{222}	&	2.000000		&	$(2,0)$	&	2.000000		 \\ 
$\omega_3$	&	n/a	&	3.000\underline{432}	&	3.000000	    &	$(3,0)$	&	3.000000		\\ 
$\omega_4$	&	3.\underline{655133}	&	3.7947\underline{52}	&	3.794733 	&	$(0,2)$	&	3.794733	\\ 
$\omega_5$	&	n/a	&	4.0\underline{11839}	&	4.0000\underline{12}	 &	$(4,0)$	&	4.000000 \\ 
$\omega_6$	&	n/a	&	4.794\underline{870}	& 4.794733 &	 $(1,2)$	&	4.794733	\\ 
$\omega_7$	&	n/a	&	5.0\underline{22646}	&	5.0000\underline{28}	&	$(5,0)$	&	5.000000 \\ 
$\omega_8$	&	5.\underline{482700} 	&	5.6921\underline{43}	& 5.692100 &	$(0,3)$	 &	5.692100	\\ 
$\omega_9$	&	n/a	&	5.79\underline{7912}	& 5.79473\underline{7} &	$(2,2)$	 &	5.794733	\\ 
$\omega_{10}$	&	n/a	&	6.\underline{142955}	& 6.000\underline{676} &	$(6,0)$	 &6.000000	\\ \hline\hline
\end{tabular*}
\caption[Low-energy spectrum of $N=10$ attractive bosons in the 1D HIM]{Benchmark of LR-MCTDHB to the attractive 1D HIM. Shown are the ground-state energy $\varepsilon^0$ and the energies $\omega_i$ of the first ten excitations for $N=10$ bosons and different numbers of orbitals $M$. The trapping frequency is $\Omega=1.0$ and the interaction strength is $\lambda_0=0.13$, yielding a ground-state depletion of $f=0.94\%$. This is comparable to the average depletion of the 1D lattice systems discussed in Section \ref{App_1D_lattice}. Underlined digits denote deviations from the exact values from Eqs. (\ref{HIM_GS_Energy}) and (\ref{HIM_Energy_dist}). All quantities are dimensionless. See text for more details.}
\label{table_1d_HIM_attr}
\end{table}

For the case of attractive bosons in the 1D HIM, the system parameters are chosen as $\Omega=1.0$ for the trapping frequency and $\lambda_0=0.13$ for the two-body interaction strength. The number of bosons is $N=10$. The calculations were carried out on a grid with 128 grid points in the interval $[-9,9)$. Results for the excitation energies relative to the ground-state energy, $\omega_i=E_i-\varepsilon^0$, are shown in Table \ref{table_1d_HIM_attr}. With respect to $\varepsilon^0$, one observes that $M=4$ and $M=6$ lead to highly accurate results. With respect to the excited states, it can be observed that the BdG approach, i.e., for $M=1$, misses several excitations in the low-energy spectrum. In contrast to that, the many-body calculations yield all states. The numerical accuracy clearly improves upon increasing the number of self-consistent orbitals. For $M=6$, even the higher c.m. excited states $\omega_5,\,\omega_7$, and $\omega_{10}$ are obtained to very high accuracy. Together with the benchmark from Table \ref{table_1d_HIM_rep}, it can be deduced that the implementation of LR-MCTDHB is capable of accurately describing the low-energy spectrum of both repulsive and attractive bosons.

\begin{table}[t!]
\begin{tabular*}{\textwidth}{p{2.0cm} p{2.0cm} p{2.0cm} p{2.0cm} c c}
\hline\hline \rule{0pt}{3ex}
			& $M=1$	& $M=2$ & $M=3$	&  $(m_x,m_y$ $n_x,n_y)$	& Exact \\ \hline \rule{-4pt}{3ex}
$\varepsilon^0$	& 	110.00\underline{7587}	& 110.00\underline{4507} &	110.003452	&	$(0,0,0,0)$			&			110.003452				\\ 		
$\omega_1$	&	1.000000 &	\underline{0.999999} &	1.000000	&		$(1,0,0,0)$			&	 	1.000000		\\ 
$\omega_2$	&	1.378405 &	1.378\underline{280} &	1.378405	&		$(0,1,0,0)$				&	1.378405		\\ 
$\omega_3$	&	1.7\underline{91089} & 1.78885\underline{9}	&	1.7888\underline{60}	&		$(0,0,2,0)$			&	1.788854		\\ 
$\omega_4$	&	n/a	& 2.000\underline{438} &	2.000\underline{477} &			$(2,0,0,0)$			&	2.000000	\\ 
$\omega_5$	&	2.\underline{200152} & 2.\underline{200520}	&	2.1982\underline{71}	&			$(0,0,1,1)$			&	2.198268		\\
$\omega_6$	&	n/a	& \underline{1.940430} &		2.37870\underline{4} &			$(1,1,0,0)$	&	2.378405		\\
$\omega_7$	&	2.60\underline{9215} & 2.60\underline{9214}	&	2.60768\underline{4}	&			$(0,0,0,2)$		&	2.607681		\\ 
$\omega_8$	&	2.68\underline{6634} & 2.6832\underline{92}	&	2.6832\underline{94}	&			$(0,0,3,0)$		&		2.683282		\\
$\omega_9$	&	n/a	& n/a &	2.75\underline{7078}	&			$(0,2,0,0)$			&	2.756810		\\ 
$\omega_{10}$	&	n/a	& 2.78\underline{9191} &	2.78\underline{9227}	&		$(1,0,2,0)$				&	2.788854			\\
$\omega_{11}$	&	n/a	& 3.000\underline{758} &	3.000\underline{816}	&	$(3,0,0,0)$				&	3.000000				\\ \hline\hline

\end{tabular*}
\caption[Low-energy spectrum of $N=100$ repulsive bosons in the 2D anisotropic HIM]{Benchmark of LR-MCTDHB to the anisotropic 2D HIM with $N=100$ bosons. The trapping frequencies are $\Omega_x=1.0$ and $\Omega_y=\sqrt{1.9}$ in $x$- and $y$-directions, and the two-body interaction strength id $\lambda=-0.001$. All previously obtained degeneracies in the isotropic case (see Table \ref{table_2d_HIM_iso}) are lifted. The accuracy of the energies increases with the number of orbitals. Underlined digits denote deviations from the exact values from Eqs. (\ref{HIM_GS_Energy}) and (\ref{HIM_Energy_dist}). All quantities are dimensionless. See text for more details.}
\label{table_2d_HIM_aniso}
\end{table}

Table \ref{table_2d_HIM_aniso} shows the numerical results obtained for the anisotropic HIM in 2D with trap frequencies $\Omega_x=1.0$ and $\Omega_y=\sqrt{1.9}$. The remaining system parameters are the same as in the isotropic case from Table \ref{table_2d_HIM_iso}. One first observes that the anisotropy lifts all previous degeneracies of the isotropic case. Apart from that, the numerical results are qualitatively similar. In the BdG case, only the first c.m. excitation is obtained numerically exact, all other c.m. excited states are inaccessible. The pure relative excitations are also found to a good accuracy. For $M=2$, one gets access to higher c.m. excited states in the $x$-direction, together with combined excitations of that type with excitations of relative coordinates. As in the isotropic case, all excited states obtained in the $y$-coordinate deteriorate in accuracy compared to the BdG case, see, e.g., $\omega_2$ and $\omega_5$. The reason is the same as for the isotropic case, namely that the additional orbital resembles a $p_x$-orbital and leads to preferred direction in the description. Including a third orbital that resembles a $p_y$-orbital solves this problem. It additionally improves the overall accuracy of the low-energy excitations, and moreover ensures that all low-lying excited states are obtained.

\begin{table}[t!]
\centering
\begin{tabular*}{\textwidth}{p{2.3cm} p{2.9cm} p{2.9cm} p{2.9cm} c}
\hline\hline \rule{0pt}{3ex}
			& $M=1$	&  $M=3$	&	$(n,m,l_z)$	&  Exact \\ \hline  \rule{-4pt}{3ex}
$\varepsilon^0$	& 	89.55\underline{4453}	&	89.54829\underline{3}	&	(0,0,0)	&	89.548292			\\ 
$\omega_1$	&	0.900000	 & 	0.900000		&	(0,1,1)	&	0.900000		\\ 
$\omega_2$	&	1.100000	 & 	1.100000			&	(0,1,-1)	&	1.100000			\\
$\omega_3$	&	1.5\underline{91089}	 & 	1.58885\underline{9}	&	(2,0,2)		&	1.588854			\\ 
$\omega_4$	&	1.7\underline{91089}	 & 	1.7888\underline{62}			&	(2,0,0)	&	1.788854		\\
$\omega_5$	&	n/a	 &	1.800\underline{439}				&	(0,2,2)	&	1.800000		\\ 
$\omega_6$	&	1.9\underline{91089}		&	1.98885\underline{9}	&	(2,0,-2)		&	1.988854			\\
$\omega_7$	&	n/a & 	2.000\underline{657}		&	(0,2,0)	&	2.000000			\\ 
$\omega_8$	&	n/a & 	2.200\underline{438}		&	(0,2,-2)		&	2.200000			\\ 
	
$\omega_9$	&	2.38\underline{6633}	 & 2.3832\underline{26}		&	(3,0,3)			&	2.383282			\\ 
$\omega_{10}$	&	n/a	 &	2.48\underline{7088}			&	(2,1,3)	&	2.488854			\\ 
$\omega_{11}$	&	2.58\underline{6634}	 &	2.583\underline{318}	&	(3,0,1)	&	2.583282			\\
$\omega_{12}$	&	n/a	 &	2.68\underline{9270}		&	(2,1,1)	&	2.688854			\\ 
	
$\omega_{13}$	&	n/a		&	2.6\underline{91535}		&	(2,1,1)	&	2.688854			\\ 
$\omega_{14}$	&	n/a 	&	2.700\underline{528}		&	(0,3,3)	&	2.700000		\\ 
$\omega_{15}$	&	2.78\underline{6634}	&	2.783\underline{318}		&	(3,0,-1)	&		2.783282	 	\\  \hline\hline

\end{tabular*}
\caption[Low-energy spectrum of $N=100$ repulsive bosons in the 2D isotropic HIM in the rotating frame of reference]{Benchmark of LR-MCTDHB to the isotropic 2D HIM in the rotating frame. The angular velocity around the $z$-axis is $\Omega_\text{rot}=0.1$. Results are presented for $N=100$ bosons and different numbers of orbitals $M$. The trapping frequencies are $\Omega_x=\Omega_y=1.0$, whereas the strength of the repulsive interaction is $\lambda_0=-0.001$. Shown are the energies of the ground state, $\varepsilon^0$, and of the first $15$ excited states, $\omega_i=E_i-\varepsilon^0$. Underlined digits denote deviations from the exact values from Eqs. (\ref{HIM_GS_Energy}), (\ref{HIM_Energy_dist}) and (\ref{Eq_HIM_rot}). The quantum numbers $n$ and $m$ refer to the relative and c.m. coordinates, whereas $l_z$ refers to the angular momentum in the $z$-direction. All quantities are dimensionless. See text for more details.}
\label{table_2d_ROT_BEC}
\end{table}

Finally, excitations in the rotating frame of reference are discussed. The energies of the ground and excited states, given by Eqs. (\ref{HIM_GS_Energy}) and (\ref{HIM_Energy_dist}), become shifted by the term
\begin{equation}\label{Eq_HIM_rot}
	E_{\text{rot}}=-\Omega_\text{rot}\,l_z
\end{equation}
where $\Omega_\text{rot}$ denotes the angular velocity around the $z$-axis and $l_z$ denotes the angular momentum of the exited state in the $z$-direction. In Table \ref{table_2d_HIM_iso}, all obtained excitations are labeled by corresponding quantum numbers for relative and c.m. excitations, $n$ and $m$, as well as by $l_z$. The increased numerical accuracy of LR-MCTDHB($3$) in comparison to the BdG results becomes clearly obvious. Moreover, all states are obtained for $M=3$, whereas the BdG spectrum misses several states. As a conclusion, the utilized implementation of LR-MCTDHB is also capable of accurately describing the low-energy excitation spectra in the rotating frame.

\setcounter{equation}{0}
\section{Many-particle variance: A sensitive quantity to many-body effects}\label{Appendix_variance}
In this Appendix, a brief introduction of the many-body variance, which is employed in the main text to study many-body excitations involved in the dynamics of trapped BECs, is given. The many-body variance has been of interest in recent studies on correlations between bosons in the ground state \cite{Ofir_MBVar_1} and for out-of-equilibrium BECs \cite{Ofir_MBVar_2,Ofir_MBVar_3}, in particular in a 1D bosonic Josephson junction \cite{MB_Var_Sudip} or in anharmonic and anisotropic trapping potentials \cite{Ofir_MBVar_4,Ofir_MBVar_5}. It has also been found that the variance is a sensitive measure for the numerical convergence of MCTDHB \cite{Brand_convergence}. The focus is laid on the position variance in the $x$-direction in the following.

The position operator of $N$ bosons is given by
\begin{equation}\label{Variance_pos_op}
		\hat{X}=\sum_{i=1}^N\hat{x}_i
\end{equation}
where the operator $\hat{x}_i$ denotes the position operator of the $i$-th particle. To compute its variance per particle in the state $\Psi(\mathbf{r}_1,...,\mathbf{r}_N)$, which reads 
\begin{equation}\label{Eq_MBVar}
	\frac{1}{N}\Delta^2_{\hat{X}}=\frac{1}{N}\left( \langle\Psi|\hat{X}^2|\Psi\rangle-\langle\Psi|\hat{X}|\Psi\rangle^2 \right),
\end{equation}
one also needs to evaluate the expectation value of the square of the position operator. The latter is given by
\begin{equation}\label{Eq_Squared_Pos_Var}
	\hat{X}^2=\sum_{i=1}^N\hat{x}_i^2+\sum_{i<k}2\,\hat{x}_i \hat{x}_k,
\end{equation} 
i.e., it consists of a one-body and a two-body term. For the expectation value of $\hat{X}$ one obtains
\begin{align}
	\frac{1}{N}\langle\Psi|\hat{X}|\Psi\rangle &=\frac{1}{N}\sum_{i=1}^N\int d\mathbf{r}_1...d\mathbf{r}_N\,\Psi^\ast(\mathbf{r}_1,...,\mathbf{r}_N)\, x_i \, \Psi(\mathbf{r}_1,...,\mathbf{r}_N) \nonumber \\
	&=\frac{1}{N}\sum_{i=1}^N\int d\mathbf{r}_i\,\frac{\rho^{(1)}(\mathbf{r}_i|\mathbf{r}_i)}{N}x_i \nonumber \\
	&=\int d\mathbf{r}\,\frac{\rho(\mathbf{r})}{N}x.
\end{align}
In the second step, the definition of the density $\rho(\mathbf{r})\equiv \rho^{(1)}(\mathbf{r}|\mathbf{r})$, employing the diagonal of the one-body RDM, was used. In general, the time-dependent $p$-th order RDM of a given $N$-boson system in the state $\Psi(\mathbf{r}_1,...,\mathbf{r}_N;t)$ is defined as
\begin{align}\label{Eq_pth_RDM}
	\rho^{(p)}(\mathbf{r}_1,...,\mathbf{r}_p|\mathbf{r}_1^\prime,...,\mathbf{r}_p^\prime;t)&=\frac{N!}{(N-p)!}\int d\mathbf{r}_{p+1}...d\mathbf{r}_N\,\Psi(\mathbf{r}_1,...,\mathbf{r}_p,\mathbf{r}_{p+1},...,\mathbf{r}_N;t) \nonumber \\
	&\times \Psi^\ast(\mathbf{r}_1^\prime,...,\mathbf{r}_p^\prime,\mathbf{r}_{p+1},...,\mathbf{r}_N;t).
\end{align}
From Eq. (\ref{Eq_pth_RDM}), one can define the $p$-th order correlation function $g^{(p)}$ as
\begin{align}\label{Eq_pth_corr_func}
	g^{(p)}(\mathbf{r}_1^\prime,...,\mathbf{r}_p^\prime|\mathbf{r}_1,...,\mathbf{r}_p;t)=\frac{\rho^{(p)}(\mathbf{r}_1^\prime,...,\mathbf{r}_p^\prime|\mathbf{r}_1,...,\mathbf{r}_p;t)}{\sqrt{\Pi_{i=1}^p\, \rho^{(1)}(\mathbf{r}_i|\mathbf{r}_i;t)\,\rho^{(1)}(\mathbf{r}_i^\prime|\mathbf{r}_i^\prime;t)}}.
\end{align}
Whereas the wave function $\Psi$ is assumed to be normalized to unity, i.e., $\langle\Psi(t)|\Psi(t)\rangle=1$, the $p$-th order RDM is not. 

One further obtains 
\begin{align}
	\frac{1}{N}\langle\Psi|\hat{X}^2|\Psi\rangle &=\int d\mathbf{r}\,\frac{\rho(\mathbf{r})}{N}x^2 \nonumber \\
	&+\frac{2}{N}\sum_{i<k}\int d\mathbf{r}_{1}...d\mathbf{r}_N\,\Psi^\ast(\mathbf{r}_1,...,\mathbf{r}_N)\, x_ix_k \, \Psi(\mathbf{r}_1,...,\mathbf{r}_N)
\end{align}
where the second term can be simplified due to
\begin{align}	
	&\sum_{i<k}\int d\mathbf{r}_{1}...d\mathbf{r}_N\,\Psi^\ast(\mathbf{r}_1,...,\mathbf{r}_N)\, x_ix_k \, \Psi(\mathbf{r}_1,...,\mathbf{r}_N) \nonumber \\
	&= \sum_{i<k}\int d\mathbf{r}_i d\mathbf{r}_k\,\frac{\rho^{(2)}(\mathbf{r}_i,\mathbf{r}_k,\mathbf{r}_i,\mathbf{r}_k)}{N(N-1)}x_ix_k \nonumber \\
	&= \frac{N(N-1)}{2}\int d\mathbf{r}_1 d\mathbf{r}_2\,\frac{\rho^{(2)}(\mathbf{r}_1,\mathbf{r}_2,\mathbf{r}_1,\mathbf{r}_2)}{N(N-1)}x_1x_2
\end{align}
where the definition of the two-body RDM $\rho^{(2)}$ was utilized. The choice of the coordinates of the first and second particle in the last step is arbitrary. This yields
\begin{align}
	\frac{1}{N}\langle\Psi|\hat{X}^2|\Psi\rangle &=\int d\mathbf{r}\,\frac{\rho(\mathbf{r})}{N}x^2 \nonumber \\
	&+\int d\mathbf{r}_1 d\mathbf{r}_2\,\frac{\rho^{(2)}(\mathbf{r}_1,\mathbf{r}_2,\mathbf{r}_1,\mathbf{r}_2)}{N}x_1x_2
\end{align}
such that one obtains 
\begin{align}
	\frac{1}{N}\Delta^2_{\hat{X}}&=\int d\mathbf{r}\,\frac{\rho(\mathbf{r})}{N}x^2-\frac{1}{N}\left[\int d\mathbf{r}\,\rho(\mathbf{r})\,x \right]^2 \nonumber \\
	&+ \int d\mathbf{r}_1 d\mathbf{r}_2\,\frac{\rho^{(2)}(\mathbf{r}_1,\mathbf{r}_2,\mathbf{r}_1,\mathbf{r}_2)}{N}x_1x_2
\end{align}
for the variance of the position operator $\hat{X}$. 

To arrive at an expression in the Hartree limit, we first expand $\rho^{(2)}$ by the natural orbitals $\{\alpha_i\}$, which yields
\begin{equation}
	\rho^{(2)}(\mathbf{r}_1,\mathbf{r}_2,\mathbf{r}_1,\mathbf{r}_2)=\sum_{i,j,k,l}\,\rho_{ijkl}\,\alpha^\ast_i(\mathbf{r}_1)\alpha^\ast_j(\mathbf{r}_2)\alpha_k(\mathbf{r}_1)\alpha_l(\mathbf{r}_2).
\end{equation}
Then, we can perform the limit which gives
\begin{align}
	\lim_{N\rightarrow\infty}\frac{1}{N}\Delta_{\hat{X}}^2 &= \Delta_{\text{GP}}^2+\Delta_{\text{correlations}}^2, \\
	   \Delta_{\text{GP}}^2&= \int d\mathbf{r}\,|\phi^0(\mathbf{r})|^2\,x^2-\left[\int d\mathbf{r}\,|\phi^0(\mathbf{r})|^2\,x \right]^2  \label{Var_GP} \\
	 \Delta_{\text{correlations}}^2&= \lim_{N\rightarrow\infty}\sum_{i,j,k,l\neq 1111}\int d\mathbf{r}_1 d\mathbf{r}_2\,\frac{\rho_{ijkl}}{N}\alpha^\ast_i(\mathbf{r}_1)\alpha^\ast_j(\mathbf{r}_2)\,x_1x_2\,\alpha_k(\mathbf{r}_1)\alpha_l(\mathbf{r}_2).  \label{Var_corr}
\end{align}
Eq. (\ref{Var_GP}) expresses the fact that, in the Hartree limit, the GP orbital $\phi^0$, obtained from Eq. (\ref{Eq_GP_notime}), exactly describes the ground-state density per particle, i.e., $\lim_{N\rightarrow\infty}\frac{\rho(\mathbf{r})}{N}=|\phi^0(\mathbf{r})|^2$.
To remind the reader, only one natural orbital exists in the GP case. In obtaining Eq. (\ref{Var_GP}) one makes use of the relation
$\lim_{N\rightarrow\infty}\frac{\rho_{1111}}{N(N-1)}=1$ as well.
This means that many-body contributions to the variance per particle, Eq. (\ref{Var_corr}), can only occur due to the occupation of higher natural orbitals, 
as is expressed in the excluded summation $i,j,k,l\neq 1111$ in Eq. (\ref{Var_corr}). It has been shown that even a small occupation of the higher natural orbitals, i.e., the number of depleted particles, can have a sizable effect at the infinite particle limit. For further details, also including the variance of other quantities like the momentum and angular momentum operators, see Refs. \cite{Ofir_MBVar_1,Sakmann_analytic}.

Applications to exact many-body excitations from the dynamics of the variance were discussed in section \ref{Sec_Applications_LR-MCTDHB}. Here, we show an example that demonstrates the numerical convergence of the ground states of the rotating BECs of Fig. \ref{fig:rot_BEC_1} with $M=7$ orbitals, the position variance in $x$-direction depending on different values of the vorticity $l$ is shown in Fig. \ref{fig:variance}. One readily observes that already for $M=5$ orbitals the ground state variances are essentially condensed, and the deviations are less than $\mathcal{O}(10^{-3})$ of the exact values (see inset).

\begin{figure}[h!]
  \centering
  \includegraphics[width=\textwidth]{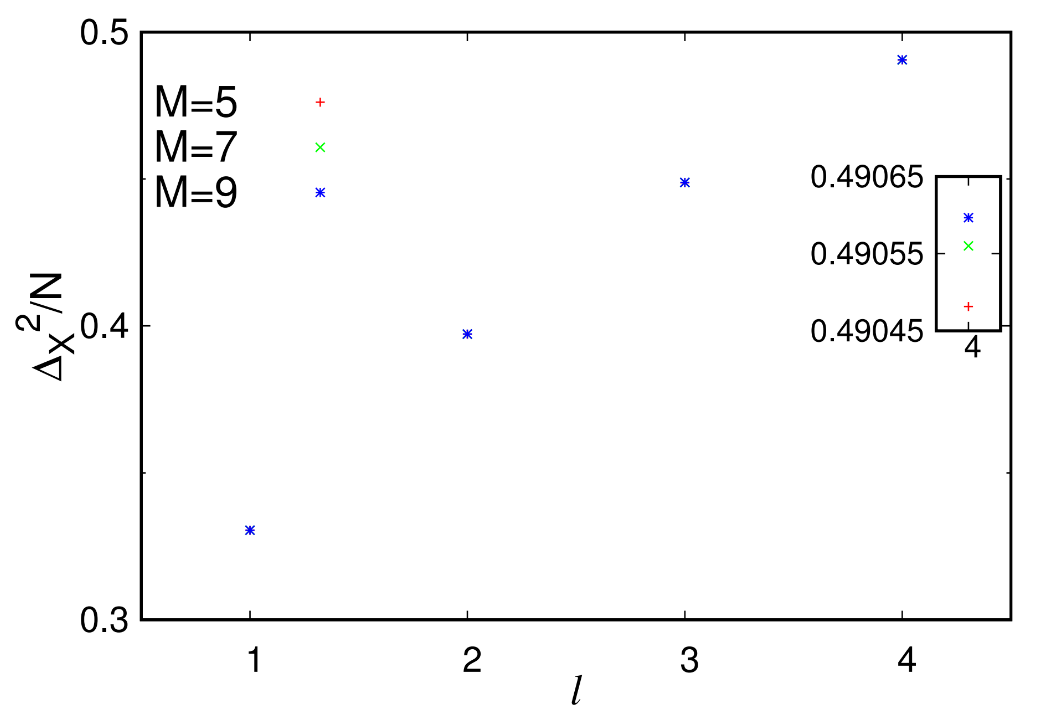}
  \caption[Position variance in the $x$-direction for different vorticities $l$.]{The position variance per particle in $x$-direction for the ground states of the rotating BECs treated in Fig. \ref{fig:rot_BEC_1}. For all vorticities $l$ shown, essentially $M=5$ self-consistent orbitals are sufficient to reach numerical convergence because the results lie atop the ones for higher values of $M$. Inset: Enlarged view for the case of $l=4$. All quantities are dimensionless. See text for details. The figure is taken from Ref. \cite{Beinke_ROT_BEC}.}
  \label{fig:variance}
\end{figure}

\end{appendices}

\newpage

\section*{Acknowledgement}
We thank Alexej I. Streltsov and Shachar Klaiman for many discussions. Computation time on the Cray XC40 cluster Hazel Hen at the High Performance Computing Center Stuttgart (HLRS) and the BwForCluster is acknowledged. RB acknowledges financial support by the IMPRS-QD (International Max Plack Research School for Quantum Dynamics), the Landesgraduiertenf{\"o}rderung Baden-W{\"u}rttemberg, and the Minerva Foundation. OEA acknowledges funding by the Israel Science Foundation (Grant No. 600/15). \\




\end{document}